\documentclass[apjl,ip,twocolumn]{aastex61}
\usepackage{comment}
\usepackage{ifthen}
\usepackage[switch]{lineno}
\modulolinenumbers[5]
\newcommand{\forloop}[5][1]%
{%
\setcounter{#2}{#3}%
\ifthenelse{#4}%
	{%
	#5%
	\addtocounter{#2}{#1}%
	\forloop[#1]{#2}{\value{#2}}{#4}{#5}%
	}%
	{%
	}%
}% 

\newcommand{\ctbd}[1]{}

\newcommand{\lcs}{light curves}
\newcommand{\Lc}{Light curve}

\newcommand{\masy}{\ensuremath{\rm mas\,yr^{-1}}}
\newcommand{\kms}{\ensuremath{\rm km\,s^{-1}}}
\newcommand{\ms}{\ensuremath{\rm m\,s^{-1}}}

\newcommand{\gcmc}{\ensuremath{\rm g\,cm^{-3}}}

\newcommand{\vsini}{\ensuremath{v \sin{i}}}
\newcommand{\feh}{\ensuremath{\rm [Fe/H]}}

\newcommand{\vmac}{\ensuremath{v_{\rm mac}}}
\newcommand{\vmic}{\ensuremath{v_{\rm mic}}}
\newcommand{\rsun}{\ensuremath{R_\sun}}
\newcommand{\msun}{\ensuremath{M_\sun}}
\newcommand{\lsun}{\ensuremath{L_\sun}}

\newcommand{\rstar}{\ensuremath{R_\star}}
\newcommand{\mstar}{\ensuremath{M_\star}}
\newcommand{\lstar}{\ensuremath{L_\star}}

\newcommand{\teffstar}{\ensuremath{T_{\rm eff\star}}}
\newcommand{\rhostar}{\ensuremath{\rho_\star}}
\newcommand{\loggstar}{\ensuremath{\log{g_{\star}}}}

\newcommand{\rpl}{\ensuremath{R_{p}}}
\newcommand{\mpl}{\ensuremath{M_{p}}}

\newcommand{\rhopl}{\ensuremath{\rho_{p}}}

\newcommand{\arstar}{\ensuremath{a/\rstar}}
\newcommand{\zrstar}{\ensuremath{\zeta/\rstar}}

\newcommand{\rjup}{\ensuremath{R_{\rm J}}}
\newcommand{\mjup}{\ensuremath{M_{\rm J}}}

\newcommand{\reffigl}[1]{Figure~\ref{fig:#1}}
\newcommand{\refsecl}[1]{\mbox{Section \ref{sec:#1}}}

\newcommand{\reftabl}[1]{Table~\ref{tab:#1}}

\newcommand{\loopand}{\ifnum\value{planetcounter}=2 and \else\fi}
\newcommand{\loopcomma}{\ifnum\value{planetcounter}<2 ,\else. \fi}
\newcommand{\loopcommanoperiod}{\ifnum\value{planetcounter}<2 ,\else \space\fi}
\newcommand{\loopcommanospace}{\ifnum\value{planetcounter}<2 ,\else \fi}

\newcommand{\hatcurhtrxxxxxA}{HATS700-017}                      % Original HTR name of target
\newcommand{\hatcurfieldxxxxxA}{\ensuremath{string}}            % HTR field
\newcommand{\hatcurCCraxxxxxA}{\ensuremath{13^{\mathrm h}22^{\mathrm m}32.3724{\mathrm s}}}                   % Right Ascension
\newcommand{\hatcurCCdecxxxxxA}{\ensuremath{-44{\arcdeg}41{\arcmin}19.6988{\arcsec}}}                 % Declination
\newcommand{\hatcurCCmagxxxxxA}{13.913}                         % apparent V-band magnitude
\newcommand{\hatcurCCtwomassxxxxxA}{2MASS~13223237-4441196}     % 2MASS identifier
\newcommand{\hatcurCCgscxxxxxA}{GSC~7799-01184}                 % GSC(1.2) identifier
\newcommand{\hatcurCCgaiaxxxxxA}{GAIA~6087996845069842176}      % GAIA DR1 identifier
\newcommand{\hatcurCCgaiadrtwoxxxxxA}{6087996849371141248} % GAIA DR2 identifier
\newcommand{\hatcurCCtassmvxxxxxA}{\ensuremath{13.913\pm0.040}} % APASS V-band magnitude
\newcommand{\hatcurCCtassmvshortxxxxxA}{\ensuremath{13.9}}      % APASS V-band magnitude
\newcommand{\hatcurCCtassmBxxxxxA}{\ensuremath{14.729\pm0.030}} % APASS B-band magnitude
\newcommand{\hatcurCCtassmBshortxxxxxA}{\ensuremath{14.7}}      % APASS B-band magnitude
\newcommand{\hatcurCCtassmIxxxxxA}{\ensuremath{nff\pmnff}}      % TASS I-band magnitude
\newcommand{\hatcurCCtassmIshortxxxxxA}{\ensuremath{0.0}}       % TASS I-band magnitude
\newcommand{\hatcurCCtassmgxxxxxA}{\ensuremath{14.301\pm0.010}} % APASS g-band magnitude
\newcommand{\hatcurCCtassmgshortxxxxxA}{\ensuremath{14.3}}      % APASS g-band magnitude
\newcommand{\hatcurCCtassmrxxxxxA}{\ensuremath{13.681\pm0.010}} % APASS r-band magnitude
\newcommand{\hatcurCCtassmrshortxxxxxA}{\ensuremath{13.7}}      % APASS r-band magnitude
\newcommand{\hatcurCCtassmixxxxxA}{\ensuremath{13.52\pm0.10}}   % APASS i-band magnitude
\newcommand{\hatcurCCtassmishortxxxxxA}{\ensuremath{13.5}}      % APASS i-band magnitude
\newcommand{\hatcurCCparallaxxxxxxA}{\ensuremath{1.308\pm0.039}} % Gaia DR2 parallax [mas]
\newcommand{\hatcurCCgaiamGxxxxxA}{\ensuremath{13.77620\pm0.00040}} % Gaia G-band magnitude
\newcommand{\hatcurCCgaiamBPxxxxxA}{\ensuremath{14.1834\pm0.0024}} % Gaia BP-band magnitude
\newcommand{\hatcurCCgaiamRPxxxxxA}{\ensuremath{13.2231\pm0.0016}} % Gaia RP-band magnitude
\newcommand{\hatcurCCtwomassJmagxxxxxA}{\ensuremath{12.611\pm0.024}} % 2MASS ORIG MAG
\newcommand{\hatcurCCtwomassHmagxxxxxA}{\ensuremath{12.273\pm0.025}} % 2MASS ORIG MAG
\newcommand{\hatcurCCtwomassKmagxxxxxA}{\ensuremath{12.170\pm0.019}} % 2MASS ORIG MAG
\newcommand{\hatcurCCcitJmagxxxxxA}{\ensuremath{12.624\pm0.024}} % 2MASS CIT MAG
\newcommand{\hatcurCCcitHmagxxxxxA}{\ensuremath{12.267\pm0.025}} % 2MASS CIT MAG
\newcommand{\hatcurCCcitKmagxxxxxA}{\ensuremath{12.194\pm0.019}} % 2MASS CIT MAG
\newcommand{\hatcurCCbbJmagxxxxxA}{\ensuremath{12.679\pm0.026}} % 2MASS BB MAG
\newcommand{\hatcurCCbbHmagxxxxxA}{\ensuremath{12.290\pm0.026}} % 2MASS BB MAG
\newcommand{\hatcurCCbbKmagxxxxxA}{\ensuremath{12.214\pm0.019}} % 2MASS BB MAG
\newcommand{\hatcurCCesoJmagxxxxxA}{\ensuremath{12.682\pm0.028}} % 2MASS ESO MAG
\newcommand{\hatcurCCesoHmagxxxxxA}{\ensuremath{12.285\pm0.030}} % 2MASS ESO MAG
\newcommand{\hatcurCCesoKmagxxxxxA}{\ensuremath{12.213\pm0.020}} % 2MASS ESO MAG
\newcommand{\hatcurCCesoJHmagxxxxxA}{\ensuremath{0.396\pm0.039}} % 2MASS ESO JH COLOR
\newcommand{\hatcurCCesoJKmagxxxxxA}{\ensuremath{0.470\pm0.034}} % 2MASS ESO JK COLOR
\newcommand{\hatcurCCesoHKmagxxxxxA}{\ensuremath{0.073\pm0.036}} % 2MASS ESO HK COLOR
\newcommand{\hatcurLCdipxxxxxA}{\ensuremath{8.9}}               % BLS detected dip (mmag)
\newcommand{\hatcurLCrprstarxxxxxA}{\ensuremath{0.0832\pm0.0025}} % Rp/R*
\newcommand{\hatcurLCbsqxxxxxA}{\ensuremath{0.548_{-0.034}^{+0.032}}} % impact parameter square
\newcommand{\hatcurLCimpxxxxxA}{\ensuremath{0.740_{-0.023}^{+0.022}}} % impact parameter
\newcommand{\hatcurLCzetaxxxxxA}{\ensuremath{22.56\pm0.50}}     % zeta/R*
\newcommand{\hatcurLCdurxxxxxA}{\ensuremath{0.1042\pm0.0019}}   % transit duration (days)
\newcommand{\hatcurLCdurshortxxxxxA}{\ensuremath{0.1042}}       % transit duration (days)
\newcommand{\hatcurLCdurhrxxxxxA}{\ensuremath{2.501\pm0.046}}   % transit duration (hours)
\newcommand{\hatcurLCdurhrshortxxxxxA}{\ensuremath{2.501}}      % transit duration (hours)
\newcommand{\hatcurLCqxxxxxA}{\ensuremath{0.04100\pm0.00075}}   % fractional transit duration (days)
\newcommand{\hatcurLCqshortxxxxxA}{\ensuremath{0.041}}          % fractional transit duration (days)
\newcommand{\hatcurLCingdurxxxxxA}{\ensuremath{0.0165\pm0.0013}} % ingress/egress duration (days)
\newcommand{\hatcurLCPxxxxxA}{\ensuremath{2.5441828\pm0.0000043}} % period (days)
\newcommand{\hatcurLCPprecxxxxxA}{\ensuremath{2.5441828}}       % period (days)
\newcommand{\hatcurLCPshortxxxxxA}{\ensuremath{2.5442}}         % period (days)
\newcommand{\hatcurLCTxxxxxA}{\ensuremath{2457780.01102\pm0.00089}} % epoch (BJD)
\newcommand{\hatcurLCTAxxxxxA}{\ensuremath{2455678.5161\pm0.0035}} % TA (BJD)
\newcommand{\hatcurLCTBxxxxxA}{\ensuremath{2457947.92710\pm0.00098}} % TB (BJD)
\newcommand{\hatcurLChatnetmxxxxxA}{\ensuremath{13.77661\pm0.00011}} % HATNet OOT level
\newcommand{\hatcurLCiblendxxxxxA}{\ensuremath{0.937\pm0.065}}  % HATNet iblend factor
\newcommand{\hatcurLCrhoxxxxxA}{\ensuremath{0.678\pm0.054}}     % stellar density no isochrone constraint (cgs)
\newcommand{\hatcurSMEiteffxxxxxA}{\ensuremath{5659\pm90}}      % Ini SME, stellar effective temperature
\newcommand{\hatcurSMEizfehxxxxxA}{\ensuremath{0.440\pm0.043}}  % Ini SME, stellar metallicity
\newcommand{\hatcurSMEizfehshortxxxxxA}{\ensuremath{0.44}}      % Ini SME, stellar metallicity
\newcommand{\hatcurSMEiloggxxxxxA}{\ensuremath{4.54\pm0.21}}    % Ini SME, stellar surface gravity
\newcommand{\hatcurSMEivsinxxxxxA}{\ensuremath{3.11\pm0.63}}    % Ini SME, stellar rotational velocity
\newcommand{\hatcurSMEivmacxxxxxA}{\ensuremath{3.81\pm0.14}}    % Ini SME, stellar macroturbulence
\newcommand{\hatcurSMEivmicxxxxxA}{\ensuremath{1.014\pm0.048}}  % Ini SME, stellar microturbulence
\newcommand{\hatcurSMEiiteffxxxxxA}{\ensuremath{5528\pm78}}     % Final SME, stellar effective temperature
\newcommand{\hatcurSMEiizfehxxxxxA}{\ensuremath{0.390\pm0.032}} % Final SME, stellar metallicity
\newcommand{\hatcurSMEiizfehshortxxxxxA}{\ensuremath{0.39}}     % Final SME, stellar metallicity
\newcommand{\hatcurSMEiiloggxxxxxA}{\ensuremath{4.095\pm0.081}} % Final SME, stellar surface gravity
\newcommand{\hatcurSMEiivsinxxxxxA}{\ensuremath{3.83\pm0.42}}   % Final SME, stellar rotational velocity
\newcommand{\hatcurSMEiivmacxxxxxA}{\ensuremath{3.61\pm0.11}}   % Final SME, stellar macroturbulence
\newcommand{\hatcurSMEiivmicxxxxxA}{\ensuremath{0.948\pm0.034}} % Final SME, stellar microturbulence
\newcommand{\hatcurLBizxxxxxA}{\ensuremath{0.2454}}             % Limb darkening parameters, Gamma1, z-band
\newcommand{\hatcurLBiizxxxxxA}{\ensuremath{0.3222}}            % Limb darkening parameters, Gamma2, z-band
\newcommand{\hatcurLBiixxxxxA}{\ensuremath{0.3222}}             % Limb darkening parameters, Gamma1, i-band
\newcommand{\hatcurLBiiixxxxxA}{\ensuremath{0.3120}}            % Limb darkening parameters, Gamma2, i-band
\newcommand{\hatcurLBiIxxxxxA}{\ensuremath{0.2968}}             % Limb darkening parameters, Gamma1, I-band
\newcommand{\hatcurLBiiIxxxxxA}{\ensuremath{0.3161}}            % Limb darkening parameters, Gamma2, I-band
\newcommand{\hatcurLBigxxxxxA}{\ensuremath{0.6588}}             % Limb darkening parameters, Gamma1, g-band
\newcommand{\hatcurLBiigxxxxxA}{\ensuremath{0.1582}}            % Limb darkening parameters, Gamma2, g-band
\newcommand{\hatcurLBirxxxxxA}{\ensuremath{0.4314}}             % Limb darkening parameters, Gamma1, r-band
\newcommand{\hatcurLBiirxxxxxA}{\ensuremath{0.2867}}            % Limb darkening parameters, Gamma2, r-band
\newcommand{\hatcurLBiRxxxxxA}{\ensuremath{0.4011}}             % Limb darkening parameters, Gamma1, R-band
\newcommand{\hatcurLBiiRxxxxxA}{\ensuremath{0.2948}}            % Limb darkening parameters, Gamma2, R-band
\newcommand{\hatcurLBikepxxxxxA}{\ensuremath{0.1000}}           % Limb darkening parameters, Gamma1, Kep-band
\newcommand{\hatcurLBiikepxxxxxA}{\ensuremath{0.1000}}          % Limb darkening parameters, Gamma2, Kep-band
\newcommand{\hatcurISOmxxxxxA}{\ensuremath{1.097\pm0.022}}      % stellar mass
\newcommand{\hatcurISOmshortxxxxxA}{\ensuremath{1.10}}          % stellar mass
\newcommand{\hatcurISOmlongxxxxxA}{\ensuremath{1.097\pm0.022}}  % stellar mass
\newcommand{\hatcurISOrxxxxxA}{\ensuremath{1.317\pm0.036}}      % stellar radius
\newcommand{\hatcurISOrshortxxxxxA}{\ensuremath{1.32}}          % stellar radius
\newcommand{\hatcurISOrlongxxxxxA}{\ensuremath{1.317\pm0.036}}  % stellar radius
\newcommand{\hatcurISOrhoxxxxxA}{\ensuremath{0.678\pm0.054}}    % stellar density (cgs)
\newcommand{\hatcurISOrholongxxxxxA}{\ensuremath{0.678\pm0.054}} % stellar density (cgs)
\newcommand{\hatcurISOloggxxxxxA}{\ensuremath{4.240\pm0.023}}   % stellar surface gravity from isochrones
\newcommand{\hatcurISOlumxxxxxA}{\ensuremath{1.631_{-0.076}^{+0.114}}} % stellar luminosity
\newcommand{\hatcurISOlumshortxxxxxA}{\ensuremath{1.63}}        % stellar luminosity
\newcommand{\hatcurISOteffxxxxxA}{\ensuremath{5702\pm26}}       % stellar effective temperature adjusted via MCMC
\newcommand{\hatcurISOzfehxxxxxA}{\ensuremath{0.396\pm0.031}}   % stellar [M/H] adjusted via MCMC
\newcommand{\hatcurISOagexxxxxA}{\ensuremath{6.60\pm0.76}}      % stellar age
\newcommand{\hatcurISOspecxxxxxA}{G}                            % stellar spectral type
\newcommand{\hatcurRVKxxxxxA}{\ensuremath{105\pm14}}            % RV semi-amplitude [m/s]
\newcommand{\hatcurRVrkxxxxxA}{\ensuremath{0\pm0}}              % sqrt(e)*cos(omega)
\newcommand{\hatcurRVrhxxxxxA}{\ensuremath{0\pm0}}              % sqrt(e)*sin(omega)
\newcommand{\hatcurRVkxxxxxA}{\ensuremath{0\pm0}}               % e*cos(omega)
\newcommand{\hatcurRVhxxxxxA}{\ensuremath{0\pm0}}               % e*sin(omega)
\newcommand{\hatcurRVtronexxxxxA}{\ensuremath{0\pm0}}           % RV linear trend tr1 factor
\newcommand{\hatcurRVtrtwoxxxxxA}{\ensuremath{0\pm0}}           % RV linear trend tr2 factor
\newcommand{\hatcurRVgammaAxxxxxA}{\ensuremath{46128\pm13}}     % RV gamma velocity, relative scale
\newcommand{\hatcurRVjitterAxxxxxA}{\ensuremath{60.5\pm8.2}}    % RV jitter (m/s)
\newcommand{\hatcurRVjittertwosiglimAxxxxxA}{\ensuremath{<76.3}} % RV jitter (m/s) 95 percent confidence upper limit
\newcommand{\hatcurRVfitrmsAxxxxxA}{\ensuremath{0.0}}           % RVfitrms
\newcommand{\hatcurRVgammaBxxxxxA}{\ensuremath{46056\pm61}}     % RV gamma velocity, relative scale
\newcommand{\hatcurRVjitterBxxxxxA}{\ensuremath{22\pm90}}       % RV jitter (m/s)
\newcommand{\hatcurRVjittertwosiglimBxxxxxA}{\ensuremath{<242.8}} % RV jitter (m/s) 95 percent confidence upper limit
\newcommand{\hatcurRVfitrmsBxxxxxA}{\ensuremath{0.0}}           % RVfitrms
\newcommand{\hatcurRVeccenxxxxxA}{\ensuremath{0\pm0}}           % eccentricity
\newcommand{\hatcurRVeccentwosiglimxxxxxA}{\ensuremath{<0.000}} % eccentricity
\newcommand{\hatcurRVomegaxxxxxA}{\ensuremath{0\pm0}}           % argument of pericenter
\newcommand{\hatcurPPixxxxxA}{\ensuremath{83.08\pm0.36}}        % orbital inclination
\newcommand{\hatcurPPgxxxxxA}{\ensuremath{16.4\pm2.9}}          % planetary surface gravity (m/s^2)
\newcommand{\hatcurPPloggxxxxxA}{\ensuremath{3.216\pm0.076}}    % planetary surface gravity (log cgs)
\newcommand{\hatcurPParxxxxxA}{\ensuremath{6.15\pm0.16}}        % relative orbital radius (a/R*)
\newcommand{\hatcurPParelxxxxxA}{\ensuremath{0.03763\pm0.00024}} % semimajor axis (AU)
\newcommand{\hatcurPPrhoxxxxxA}{\ensuremath{0.77\pm0.16}}       % planetary density (cgs)
\newcommand{\hatcurPPmxxxxxA}{\ensuremath{0.76\pm0.10}}         % planetary mass (M_jup)
\newcommand{\hatcurPPmshortxxxxxA}{\ensuremath{0.76}}           % planetary mass (M_jup)
\newcommand{\hatcurPPmlongxxxxxA}{\ensuremath{0.76\pm0.10}}     % planetary mass (M_jup)
\newcommand{\hatcurPPmexxxxxA}{\ensuremath{241\pm32}}           % planetary mass (M_earth)
\newcommand{\hatcurPPmeshortxxxxxA}{\ensuremath{240.6}}         % planetary mass (M_earth)
\newcommand{\hatcurPPmelongxxxxxA}{\ensuremath{241\pm32}}       % planetary mass (M_earth)
\newcommand{\hatcurPPrxxxxxA}{\ensuremath{1.067\pm0.052}}       % planetary radius (R_jup)
\newcommand{\hatcurPPrshortxxxxxA}{\ensuremath{1.07}}           % planetary radius (R_jup)
\newcommand{\hatcurPPrlongxxxxxA}{\ensuremath{1.067\pm0.052}}   % planetary radius (R_jup)
\newcommand{\hatcurPPrexxxxxA}{\ensuremath{11.96\pm0.59}}       % planetary radius (R_earth)
\newcommand{\hatcurPPreshortxxxxxA}{\ensuremath{12.0}}          % planetary radius (R_earth)
\newcommand{\hatcurPPrelongxxxxxA}{\ensuremath{11.96\pm0.59}}   % planetary radius (R_earth)
\newcommand{\hatcurPPmrcorrxxxxxA}{\ensuremath{-0.11}}          % mass/radius correlation
\newcommand{\hatcurPPteffxxxxxA}{\ensuremath{1625\pm22}}        % planetary temperature (K)
\newcommand{\hatcurPPthetaxxxxxA}{\ensuremath{0.0482\pm0.0071}} % Safranov number
\newcommand{\hatcurPPfluxperixxxxxA}{\ensuremath{1.573\pm0.085}} % flux @ periastron (CGS)
\newcommand{\hatcurPPfluxperidimxxxxxA}{\ensuremath{9}}         % flux @ periastron (CGS) units.
\newcommand{\hatcurPPfluxapxxxxxA}{\ensuremath{1.573\pm0.085}}  % flux @ apastron (CGS)
\newcommand{\hatcurPPfluxapdimxxxxxA}{\ensuremath{9}}           % flux @ apastron (CGS) units.
\newcommand{\hatcurPPfluxavgxxxxxA}{\ensuremath{1.573\pm0.085}} % flux on average (CGS)
\newcommand{\hatcurPPfluxavgdimxxxxxA}{\ensuremath{9}}          % flux average (CGS) units.
\newcommand{\hatcurPPfluxavglogxxxxxA}{\ensuremath{9.197\pm0.023}} % log10 flux on average (CGS)
\newcommand{\hatcurXsecphasexxxxxA}{\ensuremath{0\pm0}}         % Phase of secondary eclipse
\newcommand{\hatcurXsecondaryxxxxxA}{\ensuremath{2457781.28311\pm0.00089}} % Secondary eclipse epoch
\newcommand{\hatcurXsecdurxxxxxA}{\ensuremath{0.1042\pm0.0019}} % sec eclipse duration (days)
\newcommand{\hatcurXsecingdurxxxxxA}{\ensuremath{0.0165\pm0.0013}} % sec I/E duration (days)
\newcommand{\hatcurPPphiconjxxxxxA}{\ensuremath{0\pm0}}         % phase diff between conjunction and periastron
\newcommand{\hatcurPPperixxxxxA}{\ensuremath{2457779.37498\pm0.00089}} % time of periastron passage.
\newcommand{\hatcurPPaequivxxxxxA}{\ensuremath{0.02940\pm0.00079}} % equivalent semi-major axis
\newcommand{\hatcurPPtcircxxxxxA}{\ensuremath{101\pm30}}        % circularization timescale
\newcommand{\hatcurPPtinfallxxxxxA}{\ensuremath{314\pm56}}      % infall timescale
\newcommand{\hatcurXdistxxxxxA}{\ensuremath{769\pm21}}          % distance (pc), no reddenning correction
\newcommand{\hatcurXAvxxxxxA}{\ensuremath{0.279\pm0.018}}       % Av (mag)
\newcommand{\hatcurXdistredxxxxxA}{\ensuremath{769\pm21}}       % distance with Av correction (pc)
\newcommand{\hatcurXEBVxxxxxA}{\ensuremath{0.0900\pm0.0058}}    % E(B-V) (mag)
\newcommand{\hatcurCCpmraxxxxxA}{\ensuremath{-3.451\pm0.054}}   % proper motion, in RA
\newcommand{\hatcurCCpmdecxxxxxA}{\ensuremath{-7.915\pm0.093}}  % proper motion, in DEC
\newcommand{\hatcurCCpmxxxxxA}{\ensuremath{8.63\pm0.11}}        % proper motion

\newcommand{\hatcurhtrxxxxxB}{HATS602-072}                      % Original HTR name of target
\newcommand{\hatcurfieldxxxxxB}{\ensuremath{string}}            % HTR field
\newcommand{\hatcurCCraxxxxxB}{\ensuremath{07^{\mathrm h}37^{\mathrm m}08.0194{\mathrm s}}}                   % Right Ascension
\newcommand{\hatcurCCdecxxxxxB}{\ensuremath{-32{\arcdeg}45{\arcmin}19.5158{\arcsec}}}                 % Declination
\newcommand{\hatcurCCmagxxxxxB}{13.470}                         % apparent V-band magnitude
\newcommand{\hatcurCCtwomassxxxxxB}{2MASS~07370802-3245195}     % 2MASS identifier
\newcommand{\hatcurCCgscxxxxxB}{GSC~7109-00596}                 % GSC(1.2) identifier
\newcommand{\hatcurCCgaiaxxxxxB}{GAIA~5592019553648331776}      % GAIA DR1 identifier
\newcommand{\hatcurCCgaiadrtwoxxxxxB}{5592019557950033536} % GAIA DR2 identifier
\newcommand{\hatcurCCtassmvxxxxxB}{\ensuremath{13.470\pm0.040}} % APASS V-band magnitude
\newcommand{\hatcurCCtassmvshortxxxxxB}{\ensuremath{13.5}}      % APASS V-band magnitude
\newcommand{\hatcurCCtassmBxxxxxB}{\ensuremath{14.111\pm0.020}} % APASS B-band magnitude
\newcommand{\hatcurCCtassmBshortxxxxxB}{\ensuremath{14.1}}      % APASS B-band magnitude
\newcommand{\hatcurCCtassmIxxxxxB}{\ensuremath{nff\pmnff}}      % TASS I-band magnitude
\newcommand{\hatcurCCtassmIshortxxxxxB}{\ensuremath{0.0}}       % TASS I-band magnitude
\newcommand{\hatcurCCtassmgxxxxxB}{\ensuremath{13.746\pm0.030}} % APASS g-band magnitude
\newcommand{\hatcurCCtassmgshortxxxxxB}{\ensuremath{13.7}}      % APASS g-band magnitude
\newcommand{\hatcurCCtassmrxxxxxB}{\ensuremath{13.279\pm0.030}} % APASS r-band magnitude
\newcommand{\hatcurCCtassmrshortxxxxxB}{\ensuremath{13.3}}      % APASS r-band magnitude
\newcommand{\hatcurCCtassmixxxxxB}{\ensuremath{13.120\pm0.030}} % APASS i-band magnitude
\newcommand{\hatcurCCtassmishortxxxxxB}{\ensuremath{13.1}}      % APASS i-band magnitude
\newcommand{\hatcurCCparallaxxxxxxB}{\ensuremath{1.611\pm0.016}} % Gaia DR2 parallax [mas]
\newcommand{\hatcurCCgaiamGxxxxxB}{\ensuremath{13.34980\pm0.00030}} % Gaia G-band magnitude
\newcommand{\hatcurCCgaiamBPxxxxxB}{\ensuremath{13.6816\pm0.0010}} % Gaia BP-band magnitude
\newcommand{\hatcurCCgaiamRPxxxxxB}{\ensuremath{12.8651\pm0.0012}} % Gaia RP-band magnitude
\newcommand{\hatcurCCtwomassJmagxxxxxB}{\ensuremath{12.338\pm0.026}} % 2MASS ORIG MAG
\newcommand{\hatcurCCtwomassHmagxxxxxB}{\ensuremath{12.048\pm0.040}} % 2MASS ORIG MAG
\newcommand{\hatcurCCtwomassKmagxxxxxB}{\ensuremath{12.020\pm0.041}} % 2MASS ORIG MAG
\newcommand{\hatcurCCcitJmagxxxxxB}{\ensuremath{12.358\pm0.026}} % 2MASS CIT MAG
\newcommand{\hatcurCCcitHmagxxxxxB}{\ensuremath{12.044\pm0.039}} % 2MASS CIT MAG
\newcommand{\hatcurCCcitKmagxxxxxB}{\ensuremath{12.044\pm0.041}} % 2MASS CIT MAG
\newcommand{\hatcurCCbbJmagxxxxxB}{\ensuremath{12.403\pm0.028}} % 2MASS BB MAG
\newcommand{\hatcurCCbbHmagxxxxxB}{\ensuremath{12.064\pm0.040}} % 2MASS BB MAG
\newcommand{\hatcurCCbbKmagxxxxxB}{\ensuremath{12.064\pm0.041}} % 2MASS BB MAG
\newcommand{\hatcurCCesoJmagxxxxxB}{\ensuremath{12.404\pm0.029}} % 2MASS ESO MAG
\newcommand{\hatcurCCesoHmagxxxxxB}{\ensuremath{12.058\pm0.045}} % 2MASS ESO MAG
\newcommand{\hatcurCCesoKmagxxxxxB}{\ensuremath{12.063\pm0.041}} % 2MASS ESO MAG
\newcommand{\hatcurCCesoJHmagxxxxxB}{\ensuremath{0.346\pm0.052}} % 2MASS ESO JH COLOR
\newcommand{\hatcurCCesoJKmagxxxxxB}{\ensuremath{0.342\pm0.051}} % 2MASS ESO JK COLOR
\newcommand{\hatcurCCesoHKmagxxxxxB}{\ensuremath{-0.004\pm0.063}} % 2MASS ESO HK COLOR
\newcommand{\hatcurLCdipxxxxxB}{\ensuremath{13.7}}              % BLS detected dip (mmag)
\newcommand{\hatcurLCrprstarxxxxxB}{\ensuremath{0.1141\pm0.0020}} % Rp/R*
\newcommand{\hatcurLCbsqxxxxxB}{\ensuremath{0.439_{-0.013}^{+0.015}}} % impact parameter square
\newcommand{\hatcurLCimpxxxxxB}{\ensuremath{0.662_{-0.010}^{+0.012}}} % impact parameter
\newcommand{\hatcurLCzetaxxxxxB}{\ensuremath{20.62\pm0.20}}     % zeta/R*
\newcommand{\hatcurLCdurxxxxxB}{\ensuremath{0.11599\pm0.00097}} % transit duration (days)
\newcommand{\hatcurLCdurshortxxxxxB}{\ensuremath{0.1160}}       % transit duration (days)
\newcommand{\hatcurLCdurhrxxxxxB}{\ensuremath{2.784\pm0.023}}   % transit duration (hours)
\newcommand{\hatcurLCdurhrshortxxxxxB}{\ensuremath{2.784}}      % transit duration (hours)
\newcommand{\hatcurLCqxxxxxB}{\ensuremath{0.02760\pm0.00023}}   % fractional transit duration (days)
\newcommand{\hatcurLCqshortxxxxxB}{\ensuremath{0.028}}          % fractional transit duration (days)
\newcommand{\hatcurLCingdurxxxxxB}{\ensuremath{0.01996\pm0.00054}} % ingress/egress duration (days)
\newcommand{\hatcurLCPxxxxxB}{\ensuremath{4.2042001\pm0.0000033}} % period (days)
\newcommand{\hatcurLCPprecxxxxxB}{\ensuremath{4.2042001}}       % period (days)
\newcommand{\hatcurLCPshortxxxxxB}{\ensuremath{4.2042}}         % period (days)
\newcommand{\hatcurLCTxxxxxB}{\ensuremath{2457413.13042\pm0.00051}} % epoch (BJD)
\newcommand{\hatcurLCTAxxxxxB}{\ensuremath{2455798.7176\pm0.0013}} % TA (BJD)
\newcommand{\hatcurLCTBxxxxxB}{\ensuremath{2457854.57146\pm0.00064}} % TB (BJD)
\newcommand{\hatcurLChatnetmxxxxxB}{\ensuremath{13.330300\pm0.000089}} % HATNet OOT level
\newcommand{\hatcurLCiblendxxxxxB}{\ensuremath{0.818\pm0.052}}  % HATNet iblend factor
\newcommand{\hatcurLCrhoxxxxxB}{\ensuremath{1.179\pm0.029}}     % stellar density no isochrone constraint (cgs)
\newcommand{\hatcurSMEiteffxxxxxB}{\ensuremath{6220\pm150}}     % Ini SME, stellar effective temperature
\newcommand{\hatcurSMEizfehxxxxxB}{\ensuremath{0.300\pm0.090}}  % Ini SME, stellar metallicity
\newcommand{\hatcurSMEizfehshortxxxxxB}{\ensuremath{0.30}}      % Ini SME, stellar metallicity
\newcommand{\hatcurSMEiloggxxxxxB}{\ensuremath{4.64\pm0.15}}    % Ini SME, stellar surface gravity
\newcommand{\hatcurSMEivsinxxxxxB}{\ensuremath{5.02\pm0.18}}    % Ini SME, stellar rotational velocity
\newcommand{\hatcurSMEivmacxxxxxB}{\ensuremath{4.68\pm0.23}}    % Ini SME, stellar macroturbulence
\newcommand{\hatcurSMEivmicxxxxxB}{\ensuremath{1.43\pm0.15}}    % Ini SME, stellar microturbulence
\newcommand{\hatcurSMEiiteffxxxxxB}{\ensuremath{6095\pm92}}     % Final SME, stellar effective temperature
\newcommand{\hatcurSMEiizfehxxxxxB}{\ensuremath{0.220\pm0.049}} % Final SME, stellar metallicity
\newcommand{\hatcurSMEiizfehshortxxxxxB}{\ensuremath{0.22}}     % Final SME, stellar metallicity
\newcommand{\hatcurSMEiiloggxxxxxB}{\ensuremath{4.419\pm0.023}} % Final SME, stellar surface gravity
\newcommand{\hatcurSMEiivsinxxxxxB}{\ensuremath{5.01\pm0.15}}   % Final SME, stellar rotational velocity
\newcommand{\hatcurSMEiivmacxxxxxB}{\ensuremath{4.48\pm0.14}}   % Final SME, stellar macroturbulence
\newcommand{\hatcurSMEiivmicxxxxxB}{\ensuremath{1.308\pm0.079}} % Final SME, stellar microturbulence
\newcommand{\hatcurLBizxxxxxB}{\ensuremath{0.1716}}             % Limb darkening parameters, Gamma1, z-band
\newcommand{\hatcurLBiizxxxxxB}{\ensuremath{0.3513}}            % Limb darkening parameters, Gamma2, z-band
\newcommand{\hatcurLBiixxxxxB}{\ensuremath{0.2290}}             % Limb darkening parameters, Gamma1, i-band
\newcommand{\hatcurLBiiixxxxxB}{\ensuremath{0.3569}}            % Limb darkening parameters, Gamma2, i-band
\newcommand{\hatcurLBiIxxxxxB}{\ensuremath{0.2091}}             % Limb darkening parameters, Gamma1, I-band
\newcommand{\hatcurLBiiIxxxxxB}{\ensuremath{0.3561}}            % Limb darkening parameters, Gamma2, I-band
\newcommand{\hatcurLBigxxxxxB}{\ensuremath{0.4944}}             % Limb darkening parameters, Gamma1, g-band
\newcommand{\hatcurLBiigxxxxxB}{\ensuremath{0.2817}}            % Limb darkening parameters, Gamma2, g-band
\newcommand{\hatcurLBirxxxxxB}{\ensuremath{0.3112}}             % Limb darkening parameters, Gamma1, r-band
\newcommand{\hatcurLBiirxxxxxB}{\ensuremath{0.3574}}            % Limb darkening parameters, Gamma2, r-band
\newcommand{\hatcurLBiRxxxxxB}{\ensuremath{0.2882}}             % Limb darkening parameters, Gamma1, R-band
\newcommand{\hatcurLBiiRxxxxxB}{\ensuremath{0.3585}}            % Limb darkening parameters, Gamma2, R-band
\newcommand{\hatcurLBikepxxxxxB}{\ensuremath{0.1000}}           % Limb darkening parameters, Gamma1, Kep-band
\newcommand{\hatcurLBiikepxxxxxB}{\ensuremath{0.1000}}          % Limb darkening parameters, Gamma2, Kep-band
\newcommand{\hatcurISOmxxxxxB}{\ensuremath{1.1955_{-0.0119}^{+0.0091}}} % stellar mass
\newcommand{\hatcurISOmshortxxxxxB}{\ensuremath{1.20}}          % stellar mass
\newcommand{\hatcurISOmlongxxxxxB}{\ensuremath{1.1955_{-0.0119}^{+0.0091}}} % stellar mass
\newcommand{\hatcurISOrxxxxxB}{\ensuremath{1.126\pm0.011}}      % stellar radius
\newcommand{\hatcurISOrshortxxxxxB}{\ensuremath{1.13}}          % stellar radius
\newcommand{\hatcurISOrlongxxxxxB}{\ensuremath{1.126\pm0.011}}  % stellar radius
\newcommand{\hatcurISOrhoxxxxxB}{\ensuremath{1.179\pm0.029}}    % stellar density (cgs)
\newcommand{\hatcurISOrholongxxxxxB}{\ensuremath{1.179\pm0.029}} % stellar density (cgs)
\newcommand{\hatcurISOloggxxxxxB}{\ensuremath{4.4121\pm0.0070}} % stellar surface gravity from isochrones
\newcommand{\hatcurISOlumxxxxxB}{\ensuremath{1.694\pm0.054}}    % stellar luminosity
\newcommand{\hatcurISOlumshortxxxxxB}{\ensuremath{1.69}}        % stellar luminosity
\newcommand{\hatcurISOteffxxxxxB}{\ensuremath{6214\pm36}}       % stellar effective temperature adjusted via MCMC
\newcommand{\hatcurISOzfehxxxxxB}{\ensuremath{0.108_{-0.030}^{+0.046}}} % stellar [M/H] adjusted via MCMC
\newcommand{\hatcurISOagexxxxxB}{\ensuremath{0.40_{-0.13}^{+0.29}}} % stellar age
\newcommand{\hatcurISOspecxxxxxB}{F}                            % stellar spectral type
\newcommand{\hatcurRVKxxxxxB}{\ensuremath{102.8\pm8.4}}         % RV semi-amplitude [m/s]
\newcommand{\hatcurRVrkxxxxxB}{\ensuremath{0\pm0}}              % sqrt(e)*cos(omega)
\newcommand{\hatcurRVrhxxxxxB}{\ensuremath{0\pm0}}              % sqrt(e)*sin(omega)
\newcommand{\hatcurRVkxxxxxB}{\ensuremath{0\pm0}}               % e*cos(omega)
\newcommand{\hatcurRVhxxxxxB}{\ensuremath{0\pm0}}               % e*sin(omega)
\newcommand{\hatcurRVtronexxxxxB}{\ensuremath{0\pm0}}           % RV linear trend tr1 factor
\newcommand{\hatcurRVtrtwoxxxxxB}{\ensuremath{0\pm0}}           % RV linear trend tr2 factor
\newcommand{\hatcurRVgammaxxxxxB}{\ensuremath{-2919.9\pm6.4}}   % RV gamma velocity, relative scale
\newcommand{\hatcurRVjitterxxxxxB}{\ensuremath{0.1\pm3.7}}      % RV jitter (m/s)
\newcommand{\hatcurRVjittertwosiglimxxxxxB}{\ensuremath{<9.3}}  % RV jitter (m/s) 95 percent confidence upper limit
\newcommand{\hatcurRVfitrmsxxxxxB}{\ensuremath{.1fym}}          % 
\newcommand{\hatcurRVeccenxxxxxB}{\ensuremath{0\pm0}}           % eccentricity
\newcommand{\hatcurRVeccentwosiglimxxxxxB}{\ensuremath{<0.000}} % eccentricity
\newcommand{\hatcurRVomegaxxxxxB}{\ensuremath{0\pm0}}           % argument of pericenter
\newcommand{\hatcurPPixxxxxB}{\ensuremath{86.320\pm0.084}}      % orbital inclination
\newcommand{\hatcurPPgxxxxxB}{\ensuremath{14.6\pm1.4}}          % planetary surface gravity (m/s^2)
\newcommand{\hatcurPPloggxxxxxB}{\ensuremath{3.165\pm0.042}}    % planetary surface gravity (log cgs)
\newcommand{\hatcurPParxxxxxB}{\ensuremath{10.330\pm0.086}}     % relative orbital radius (a/R*)
\newcommand{\hatcurPParelxxxxxB}{\ensuremath{0.05412_{-0.00018}^{+0.00014}}} % semimajor axis (AU)
\newcommand{\hatcurPPrhoxxxxxB}{\ensuremath{0.587\pm0.062}}     % planetary density (cgs)
\newcommand{\hatcurPPmxxxxxB}{\ensuremath{0.921\pm0.076}}       % planetary mass (M_jup)
\newcommand{\hatcurPPmshortxxxxxB}{\ensuremath{0.92}}           % planetary mass (M_jup)
\newcommand{\hatcurPPmlongxxxxxB}{\ensuremath{0.921\pm0.076}}   % planetary mass (M_jup)
\newcommand{\hatcurPPmexxxxxB}{\ensuremath{293\pm24}}           % planetary mass (M_earth)
\newcommand{\hatcurPPmeshortxxxxxB}{\ensuremath{292.6}}         % planetary mass (M_earth)
\newcommand{\hatcurPPmelongxxxxxB}{\ensuremath{293\pm24}}       % planetary mass (M_earth)
\newcommand{\hatcurPPrxxxxxB}{\ensuremath{1.251\pm0.026}}       % planetary radius (R_jup)
\newcommand{\hatcurPPrshortxxxxxB}{\ensuremath{1.25}}           % planetary radius (R_jup)
\newcommand{\hatcurPPrlongxxxxxB}{\ensuremath{1.251\pm0.026}}   % planetary radius (R_jup)
\newcommand{\hatcurPPrexxxxxB}{\ensuremath{14.02\pm0.29}}       % planetary radius (R_earth)
\newcommand{\hatcurPPreshortxxxxxB}{\ensuremath{14.0}}          % planetary radius (R_earth)
\newcommand{\hatcurPPrelongxxxxxB}{\ensuremath{14.02\pm0.29}}   % planetary radius (R_earth)
\newcommand{\hatcurPPmrcorrxxxxxB}{\ensuremath{-0.07}}          % mass/radius correlation
\newcommand{\hatcurPPteffxxxxxB}{\ensuremath{1367\pm10}}        % planetary temperature (K)
\newcommand{\hatcurPPthetaxxxxxB}{\ensuremath{0.0666\pm0.0057}} % Safranov number
\newcommand{\hatcurPPfluxperixxxxxB}{\ensuremath{7.87\pm0.24}}  % flux @ periastron (CGS)
\newcommand{\hatcurPPfluxperidimxxxxxB}{\ensuremath{8}}         % flux @ periastron (CGS) units.
\newcommand{\hatcurPPfluxapxxxxxB}{\ensuremath{7.87\pm0.24}}    % flux @ apastron (CGS)
\newcommand{\hatcurPPfluxapdimxxxxxB}{\ensuremath{8}}           % flux @ apastron (CGS) units.
\newcommand{\hatcurPPfluxavgxxxxxB}{\ensuremath{7.87\pm0.24}}   % flux on average (CGS)
\newcommand{\hatcurPPfluxavgdimxxxxxB}{\ensuremath{8}}          % flux average (CGS) units.
\newcommand{\hatcurPPfluxavglogxxxxxB}{\ensuremath{8.896\pm0.013}} % log10 flux on average (CGS)
\newcommand{\hatcurXsecphasexxxxxB}{\ensuremath{0\pm0}}         % Phase of secondary eclipse
\newcommand{\hatcurXsecondaryxxxxxB}{\ensuremath{2457415.23252\pm0.00051}} % Secondary eclipse epoch
\newcommand{\hatcurXsecdurxxxxxB}{\ensuremath{0.11599\pm0.00097}} % sec eclipse duration (days)
\newcommand{\hatcurXsecingdurxxxxxB}{\ensuremath{0.01996\pm0.00054}} % sec I/E duration (days)
\newcommand{\hatcurPPphiconjxxxxxB}{\ensuremath{0\pm0}}         % phase diff between conjunction and periastron
\newcommand{\hatcurPPperixxxxxB}{\ensuremath{2457412.07937\pm0.00051}} % time of periastron passage.
\newcommand{\hatcurPPaequivxxxxxB}{\ensuremath{0.04160\pm0.00063}} % equivalent semi-major axis
\newcommand{\hatcurPPtcircxxxxxB}{\ensuremath{518\pm72}}        % circularization timescale
\newcommand{\hatcurPPtinfallxxxxxB}{\ensuremath{6120\pm590}}    % infall timescale
\newcommand{\hatcurXdistxxxxxB}{\ensuremath{623.6\pm6.2}}       % distance (pc), no reddenning correction
\newcommand{\hatcurXAvxxxxxB}{\ensuremath{0.350\pm0.027}}       % Av (mag)
\newcommand{\hatcurXdistredxxxxxB}{\ensuremath{623.6\pm6.2}}    % distance with Av correction (pc)
\newcommand{\hatcurXEBVxxxxxB}{\ensuremath{0.1130\pm0.0088}}    % E(B-V) (mag)
\newcommand{\hatcurCCpmraxxxxxB}{\ensuremath{-6.283\pm0.026}}   % proper motion, in RA
\newcommand{\hatcurCCpmdecxxxxxB}{\ensuremath{0.531\pm0.031}}   % proper motion, in DEC
\newcommand{\hatcurCCpmxxxxxB}{\ensuremath{6.305\pm0.040}}      % proper motion

\newcommand{\hatcurhtrxxxxxC}{HATS698-034}                      % Original HTR name of target
\newcommand{\hatcurfieldxxxxxC}{\ensuremath{string}}            % HTR field
\newcommand{\hatcurCCraxxxxxC}{\ensuremath{12^{\mathrm h}00^{\mathrm m}39.6300{\mathrm s}}}                   % Right Ascension
\newcommand{\hatcurCCdecxxxxxC}{\ensuremath{-45{\arcdeg}47{\arcmin}57.9955{\arcsec}}}                 % Declination
\newcommand{\hatcurCCmagxxxxxC}{11.578}                         % apparent V-band magnitude
\newcommand{\hatcurCCtwomassxxxxxC}{2MASS~12003962-4547579}     % 2MASS identifier
\newcommand{\hatcurCCgscxxxxxC}{GSC~8229-02228}                 % GSC(1.2) identifier
\newcommand{\hatcurCCgaiaxxxxxC}{GAIA~6144125882874203008}      % GAIA DR1 identifier
\newcommand{\hatcurCCgaiadrtwoxxxxxC}{6144125887172751232} % GAIA DR2 identifier
\newcommand{\hatcurCCtassmvxxxxxC}{\ensuremath{11.578\pm0.023}} % APASS V-band magnitude
\newcommand{\hatcurCCtassmvshortxxxxxC}{\ensuremath{11.6}}      % APASS V-band magnitude
\newcommand{\hatcurCCtassmBxxxxxC}{\ensuremath{12.097\pm0.029}} % APASS B-band magnitude
\newcommand{\hatcurCCtassmBshortxxxxxC}{\ensuremath{12.1}}      % APASS B-band magnitude
\newcommand{\hatcurCCtassmIxxxxxC}{\ensuremath{nff\pmnff}}      % TASS I-band magnitude
\newcommand{\hatcurCCtassmIshortxxxxxC}{\ensuremath{0.0}}       % TASS I-band magnitude
\newcommand{\hatcurCCtassmgxxxxxC}{\ensuremath{11.801\pm0.022}} % APASS g-band magnitude
\newcommand{\hatcurCCtassmgshortxxxxxC}{\ensuremath{11.8}}      % APASS g-band magnitude
\newcommand{\hatcurCCtassmrxxxxxC}{\ensuremath{11.473\pm0.017}} % APASS r-band magnitude
\newcommand{\hatcurCCtassmrshortxxxxxC}{\ensuremath{11.5}}      % APASS r-band magnitude
\newcommand{\hatcurCCtassmixxxxxC}{\ensuremath{11.320\pm0.046}} % APASS i-band magnitude
\newcommand{\hatcurCCtassmishortxxxxxC}{\ensuremath{11.3}}      % APASS i-band magnitude
\newcommand{\hatcurCCparallaxxxxxxC}{\ensuremath{1.744\pm0.035}} % Gaia DR2 parallax [mas]
\newcommand{\hatcurCCgaiamGxxxxxC}{\ensuremath{11.48770\pm0.00080}} % Gaia G-band magnitude
\newcommand{\hatcurCCgaiamBPxxxxxC}{\ensuremath{11.7645\pm0.0016}} % Gaia BP-band magnitude
\newcommand{\hatcurCCgaiamRPxxxxxC}{\ensuremath{11.0523\pm0.0022}} % Gaia RP-band magnitude
\newcommand{\hatcurCCtwomassJmagxxxxxC}{\ensuremath{10.514\pm0.023}} % 2MASS ORIG MAG
\newcommand{\hatcurCCtwomassHmagxxxxxC}{\ensuremath{10.325\pm0.029}} % 2MASS ORIG MAG
\newcommand{\hatcurCCtwomassKmagxxxxxC}{\ensuremath{10.251\pm0.019}} % 2MASS ORIG MAG
\newcommand{\hatcurCCcitJmagxxxxxC}{\ensuremath{10.537\pm0.023}} % 2MASS CIT MAG
\newcommand{\hatcurCCcitHmagxxxxxC}{\ensuremath{10.320\pm0.029}} % 2MASS CIT MAG
\newcommand{\hatcurCCcitKmagxxxxxC}{\ensuremath{10.275\pm0.019}} % 2MASS CIT MAG
\newcommand{\hatcurCCbbJmagxxxxxC}{\ensuremath{10.577\pm0.025}} % 2MASS BB MAG
\newcommand{\hatcurCCbbHmagxxxxxC}{\ensuremath{10.342\pm0.030}} % 2MASS BB MAG
\newcommand{\hatcurCCbbKmagxxxxxC}{\ensuremath{10.295\pm0.019}} % 2MASS BB MAG
\newcommand{\hatcurCCesoJmagxxxxxC}{\ensuremath{10.578\pm0.026}} % 2MASS ESO MAG
\newcommand{\hatcurCCesoHmagxxxxxC}{\ensuremath{10.337\pm0.033}} % 2MASS ESO MAG
\newcommand{\hatcurCCesoKmagxxxxxC}{\ensuremath{10.294\pm0.020}} % 2MASS ESO MAG
\newcommand{\hatcurCCesoJHmagxxxxxC}{\ensuremath{0.241\pm0.040}} % 2MASS ESO JH COLOR
\newcommand{\hatcurCCesoJKmagxxxxxC}{\ensuremath{0.284\pm0.032}} % 2MASS ESO JK COLOR
\newcommand{\hatcurCCesoHKmagxxxxxC}{\ensuremath{0.043\pm0.039}} % 2MASS ESO HK COLOR
\newcommand{\hatcurLCdipxxxxxC}{\ensuremath{6.7}}               % BLS detected dip (mmag)
\newcommand{\hatcurLCrprstarxxxxxC}{\ensuremath{0.0789\pm0.0018}} % Rp/R*
\newcommand{\hatcurLCbsqxxxxxC}{\ensuremath{0.475_{-0.018}^{+0.019}}} % impact parameter square
\newcommand{\hatcurLCimpxxxxxC}{\ensuremath{0.690_{-0.013}^{+0.014}}} % impact parameter
\newcommand{\hatcurLCzetaxxxxxC}{\ensuremath{11.85\pm0.15}}     % zeta/R*
\newcommand{\hatcurLCdurxxxxxC}{\ensuremath{0.1934\pm0.0022}}   % transit duration (days)
\newcommand{\hatcurLCdurshortxxxxxC}{\ensuremath{0.1934}}       % transit duration (days)
\newcommand{\hatcurLCdurhrxxxxxC}{\ensuremath{4.641\pm0.053}}   % transit duration (hours)
\newcommand{\hatcurLCdurhrshortxxxxxC}{\ensuremath{4.641}}      % transit duration (hours)
\newcommand{\hatcurLCqxxxxxC}{\ensuremath{0.04470\pm0.00051}}   % fractional transit duration (days)
\newcommand{\hatcurLCqshortxxxxxC}{\ensuremath{0.045}}          % fractional transit duration (days)
\newcommand{\hatcurLCingdurxxxxxC}{\ensuremath{0.02548\pm0.00091}} % ingress/egress duration (days)
\newcommand{\hatcurLCPxxxxxC}{\ensuremath{4.324799\pm0.000027}} % period (days)
\newcommand{\hatcurLCPprecxxxxxC}{\ensuremath{4.3247992}}       % period (days)
\newcommand{\hatcurLCPshortxxxxxC}{\ensuremath{4.3248}}         % period (days)
\newcommand{\hatcurLCTxxxxxC}{\ensuremath{2457788.0029\pm0.0012}} % epoch (BJD)
\newcommand{\hatcurLCTAxxxxxC}{\ensuremath{2457091.7102\pm0.0044}} % TA (BJD)
\newcommand{\hatcurLCTBxxxxxC}{\ensuremath{2457839.9004\pm0.0013}} % TB (BJD)
\newcommand{\hatcurLChatnetmAxxxxxC}{\ensuremath{11.402810\pm0.000081}} % HATNet OOT level
\newcommand{\hatcurLCiblendAxxxxxC}{\ensuremath{0.724\pm0.064}} % HATNet iblend factor
\newcommand{\hatcurLChatnetmBxxxxxC}{\ensuremath{11.402740\pm0.000047}} % HATNet OOT level
\newcommand{\hatcurLCiblendBxxxxxC}{\ensuremath{0.869\pm0.048}} % HATNet iblend factor
\newcommand{\hatcurLCrhoxxxxxC}{\ensuremath{0.2079\pm0.0090}}   % stellar density no isochrone constraint (cgs)
\newcommand{\hatcurSMEiteffxxxxxC}{\ensuremath{6652\pm75}}      % Ini SME, stellar effective temperature
\newcommand{\hatcurSMEizfehxxxxxC}{\ensuremath{0.240\pm0.034}}  % Ini SME, stellar metallicity
\newcommand{\hatcurSMEizfehshortxxxxxC}{\ensuremath{0.24}}      % Ini SME, stellar metallicity
\newcommand{\hatcurSMEiloggxxxxxC}{\ensuremath{4.36\pm0.14}}    % Ini SME, stellar surface gravity
\newcommand{\hatcurSMEivsinxxxxxC}{\ensuremath{6.43\pm0.20}}    % Ini SME, stellar rotational velocity
\newcommand{\hatcurSMEivmacxxxxxC}{\ensuremath{5.34\pm0.11}}    % Ini SME, stellar macroturbulence
\newcommand{\hatcurSMEivmicxxxxxC}{\ensuremath{1.99\pm0.12}}    % Ini SME, stellar microturbulence
\newcommand{\hatcurSMEiiteffxxxxxC}{\ensuremath{6552\pm61}}     % Final SME, stellar effective temperature
\newcommand{\hatcurSMEiizfehxxxxxC}{\ensuremath{0.200\pm0.025}} % Final SME, stellar metallicity
\newcommand{\hatcurSMEiizfehshortxxxxxC}{\ensuremath{0.2}}      % Final SME, stellar metallicity
\newcommand{\hatcurSMEiiloggxxxxxC}{\ensuremath{4.016\pm0.079}} % Final SME, stellar surface gravity
\newcommand{\hatcurSMEiivsinxxxxxC}{\ensuremath{6.49\pm0.19}}   % Final SME, stellar rotational velocity
\newcommand{\hatcurSMEiivmacxxxxxC}{\ensuremath{5.183\pm0.093}} % Final SME, stellar macroturbulence
\newcommand{\hatcurSMEiivmicxxxxxC}{\ensuremath{1.832\pm0.090}} % Final SME, stellar microturbulence
\newcommand{\hatcurLBizxxxxxC}{\ensuremath{0.1118}}             % Limb darkening parameters, Gamma1, z-band
\newcommand{\hatcurLBiizxxxxxC}{\ensuremath{0.3797}}            % Limb darkening parameters, Gamma2, z-band
\newcommand{\hatcurLBiixxxxxC}{\ensuremath{0.1618}}             % Limb darkening parameters, Gamma1, i-band
\newcommand{\hatcurLBiiixxxxxC}{\ensuremath{0.3903}}            % Limb darkening parameters, Gamma2, i-band
\newcommand{\hatcurLBiIxxxxxC}{\ensuremath{0.1457}}             % Limb darkening parameters, Gamma1, I-band
\newcommand{\hatcurLBiiIxxxxxC}{\ensuremath{0.3773}}            % Limb darkening parameters, Gamma2, I-band
\newcommand{\hatcurLBigxxxxxC}{\ensuremath{0.4056}}             % Limb darkening parameters, Gamma1, g-band
\newcommand{\hatcurLBiigxxxxxC}{\ensuremath{0.3422}}            % Limb darkening parameters, Gamma2, g-band
\newcommand{\hatcurLBirxxxxxC}{\ensuremath{0.2388}}             % Limb darkening parameters, Gamma1, r-band
\newcommand{\hatcurLBiirxxxxxC}{\ensuremath{0.3963}}            % Limb darkening parameters, Gamma2, r-band
\newcommand{\hatcurLBiRxxxxxC}{\ensuremath{0.2170}}             % Limb darkening parameters, Gamma1, R-band
\newcommand{\hatcurLBiiRxxxxxC}{\ensuremath{0.3871}}            % Limb darkening parameters, Gamma2, R-band
\newcommand{\hatcurLBikepxxxxxC}{\ensuremath{0.1000}}           % Limb darkening parameters, Gamma1, Kep-band
\newcommand{\hatcurLBiikepxxxxxC}{\ensuremath{0.1000}}          % Limb darkening parameters, Gamma2, Kep-band
\newcommand{\hatcurISOmxxxxxC}{\ensuremath{1.573\pm0.017}}      % stellar mass
\newcommand{\hatcurISOmshortxxxxxC}{\ensuremath{1.57}}          % stellar mass
\newcommand{\hatcurISOmlongxxxxxC}{\ensuremath{1.573\pm0.017}}  % stellar mass
\newcommand{\hatcurISOrxxxxxC}{\ensuremath{2.201\pm0.036}}      % stellar radius
\newcommand{\hatcurISOrshortxxxxxC}{\ensuremath{2.20}}          % stellar radius
\newcommand{\hatcurISOrlongxxxxxC}{\ensuremath{2.201\pm0.036}}  % stellar radius
\newcommand{\hatcurISOrhoxxxxxC}{\ensuremath{0.2079\pm0.0090}}  % stellar density (cgs)
\newcommand{\hatcurISOrholongxxxxxC}{\ensuremath{0.2079\pm0.0090}} % stellar density (cgs)
\newcommand{\hatcurISOloggxxxxxC}{\ensuremath{3.949\pm0.012}}   % stellar surface gravity from isochrones
\newcommand{\hatcurISOlumxxxxxC}{\ensuremath{7.90\pm0.31}}      % stellar luminosity
\newcommand{\hatcurISOlumshortxxxxxC}{\ensuremath{7.90}}        % stellar luminosity
\newcommand{\hatcurISOteffxxxxxC}{\ensuremath{6536\pm31}}       % stellar effective temperature adjusted via MCMC
\newcommand{\hatcurISOzfehxxxxxC}{\ensuremath{0.190\pm0.024}}   % stellar [M/H] adjusted via MCMC
\newcommand{\hatcurISOagexxxxxC}{\ensuremath{1.894\pm0.077}}    % stellar age
\newcommand{\hatcurISOspecxxxxxC}{F}                            % stellar spectral type
\newcommand{\hatcurRVKxxxxxC}{\ensuremath{55.1\pm3.2}}          % RV semi-amplitude [m/s]
\newcommand{\hatcurRVrkxxxxxC}{\ensuremath{0\pm0}}              % sqrt(e)*cos(omega)
\newcommand{\hatcurRVrhxxxxxC}{\ensuremath{0\pm0}}              % sqrt(e)*sin(omega)
\newcommand{\hatcurRVkxxxxxC}{\ensuremath{0\pm0}}               % e*cos(omega)
\newcommand{\hatcurRVhxxxxxC}{\ensuremath{0\pm0}}               % e*sin(omega)
\newcommand{\hatcurRVtronexxxxxC}{\ensuremath{6.469\pm0.044}}   % RV linear trend tr1 factor
\newcommand{\hatcurRVtrtwoxxxxxC}{\ensuremath{-0.003400\pm0.000052}} % RV linear trend tr2 factor
\newcommand{\hatcurRVgammaAxxxxxC}{\ensuremath{35148\pm15}}     % RV gamma velocity, relative scale
\newcommand{\hatcurRVjitterAxxxxxC}{\ensuremath{21.8\pm3.2}}    % RV jitter (m/s)
\newcommand{\hatcurRVjittertwosiglimAxxxxxC}{\ensuremath{<28.0}} % RV jitter (m/s) 95 percent confidence upper limit
\newcommand{\hatcurRVfitrmsAxxxxxC}{\ensuremath{0.0}}           % RVfitrms
\newcommand{\hatcurRVgammaBxxxxxC}{\ensuremath{35141\pm17}}     % RV gamma velocity, relative scale
\newcommand{\hatcurRVjitterBxxxxxC}{\ensuremath{0.2\pm3.9}}     % RV jitter (m/s)
\newcommand{\hatcurRVjittertwosiglimBxxxxxC}{\ensuremath{<10.3}} % RV jitter (m/s) 95 percent confidence upper limit
\newcommand{\hatcurRVfitrmsBxxxxxC}{\ensuremath{0.0}}           % RVfitrms
\newcommand{\hatcurRVeccenxxxxxC}{\ensuremath{0\pm0}}           % eccentricity
\newcommand{\hatcurRVeccentwosiglimxxxxxC}{\ensuremath{<0.000}} % eccentricity
\newcommand{\hatcurRVomegaxxxxxC}{\ensuremath{0\pm0}}           % argument of pericenter
\newcommand{\hatcurPPixxxxxC}{\ensuremath{83.29\pm0.21}}        % orbital inclination
\newcommand{\hatcurPPgxxxxxC}{\ensuremath{5.23\pm0.42}}         % planetary surface gravity (m/s^2)
\newcommand{\hatcurPPloggxxxxxC}{\ensuremath{2.718\pm0.034}}    % planetary surface gravity (log cgs)
\newcommand{\hatcurPParxxxxxC}{\ensuremath{5.902\pm0.085}}      % relative orbital radius (a/R*)
\newcommand{\hatcurPParelxxxxxC}{\ensuremath{0.06043\pm0.00022}} % semimajor axis (AU)
\newcommand{\hatcurPPrhoxxxxxC}{\ensuremath{0.155_{-0.013}^{+0.017}}} % planetary density (cgs)
\newcommand{\hatcurPPmxxxxxC}{\ensuremath{0.602\pm0.035}}       % planetary mass (M_jup)
\newcommand{\hatcurPPmshortxxxxxC}{\ensuremath{0.60}}           % planetary mass (M_jup)
\newcommand{\hatcurPPmlongxxxxxC}{\ensuremath{0.602\pm0.035}}   % planetary mass (M_jup)
\newcommand{\hatcurPPmexxxxxC}{\ensuremath{191\pm11}}           % planetary mass (M_earth)
\newcommand{\hatcurPPmeshortxxxxxC}{\ensuremath{191.4}}         % planetary mass (M_earth)
\newcommand{\hatcurPPmelongxxxxxC}{\ensuremath{191\pm11}}       % planetary mass (M_earth)
\newcommand{\hatcurPPrxxxxxC}{\ensuremath{1.688_{-0.055}^{+0.039}}} % planetary radius (R_jup)
\newcommand{\hatcurPPrshortxxxxxC}{\ensuremath{1.69}}           % planetary radius (R_jup)
\newcommand{\hatcurPPrlongxxxxxC}{\ensuremath{1.688_{-0.055}^{+0.039}}} % planetary radius (R_jup)
\newcommand{\hatcurPPrexxxxxC}{\ensuremath{18.92_{-0.62}^{+0.44}}} % planetary radius (R_earth)
\newcommand{\hatcurPPreshortxxxxxC}{\ensuremath{18.9}}          % planetary radius (R_earth)
\newcommand{\hatcurPPrelongxxxxxC}{\ensuremath{18.92_{-0.62}^{+0.44}}} % planetary radius (R_earth)
\newcommand{\hatcurPPmrcorrxxxxxC}{\ensuremath{0.06}}           % mass/radius correlation
\newcommand{\hatcurPPteffxxxxxC}{\ensuremath{1902\pm16}}        % planetary temperature (K)
\newcommand{\hatcurPPthetaxxxxxC}{\ensuremath{0.0271\pm0.0018}} % Safranov number
\newcommand{\hatcurPPfluxperixxxxxC}{\ensuremath{2.944\pm0.098}} % flux @ periastron (CGS)
\newcommand{\hatcurPPfluxperidimxxxxxC}{\ensuremath{9}}         % flux @ periastron (CGS) units.
\newcommand{\hatcurPPfluxapxxxxxC}{\ensuremath{2.944\pm0.098}}  % flux @ apastron (CGS)
\newcommand{\hatcurPPfluxapdimxxxxxC}{\ensuremath{9}}           % flux @ apastron (CGS) units.
\newcommand{\hatcurPPfluxavgxxxxxC}{\ensuremath{2.944\pm0.098}} % flux on average (CGS)
\newcommand{\hatcurPPfluxavgdimxxxxxC}{\ensuremath{9}}          % flux average (CGS) units.
\newcommand{\hatcurPPfluxavglogxxxxxC}{\ensuremath{9.469\pm0.014}} % log10 flux on average (CGS)
\newcommand{\hatcurXsecphasexxxxxC}{\ensuremath{0\pm0}}         % Phase of secondary eclipse
\newcommand{\hatcurXsecondaryxxxxxC}{\ensuremath{2457790.1653\pm0.0012}} % Secondary eclipse epoch
\newcommand{\hatcurXsecdurxxxxxC}{\ensuremath{0.1934\pm0.0022}} % sec eclipse duration (days)
\newcommand{\hatcurXsecingdurxxxxxC}{\ensuremath{0.02548\pm0.00091}} % sec I/E duration (days)
\newcommand{\hatcurPPphiconjxxxxxC}{\ensuremath{0\pm0}}         % phase diff between conjunction and periastron
\newcommand{\hatcurPPperixxxxxC}{\ensuremath{2457786.9217\pm0.0012}} % time of periastron passage.
\newcommand{\hatcurPPaequivxxxxxC}{\ensuremath{0.02150_{-0.00040}^{+0.00030}}} % equivalent semi-major axis
\newcommand{\hatcurPPtcircxxxxxC}{\ensuremath{103_{-12}^{+18}}} % circularization timescale
\newcommand{\hatcurPPtinfallxxxxxC}{\ensuremath{770\pm70}}      % infall timescale
\newcommand{\hatcurXdistxxxxxC}{\ensuremath{577.1\pm9.6}}       % distance (pc), no reddenning correction
\newcommand{\hatcurXAvxxxxxC}{\ensuremath{0.335\pm0.017}}       % Av (mag)
\newcommand{\hatcurXdistredxxxxxC}{\ensuremath{577.1\pm9.6}}    % distance with Av correction (pc)
\newcommand{\hatcurXEBVxxxxxC}{\ensuremath{0.1080\pm0.0054}}    % E(B-V) (mag)
\newcommand{\hatcurCCpmraxxxxxC}{\ensuremath{-8.604\pm0.046}}   % proper motion, in RA
\newcommand{\hatcurCCpmdecxxxxxC}{\ensuremath{-2.950\pm0.035}}  % proper motion, in DEC
\newcommand{\hatcurCCpmxxxxxC}{\ensuremath{9.096\pm0.058}}      % proper motion

\newcommand{\hatcurhtrxxxxxD}{HATS548-012}                      % Original HTR name of target
\newcommand{\hatcurfieldxxxxxD}{\ensuremath{string}}            % HTR field
\newcommand{\hatcurCCraxxxxxD}{\ensuremath{04^{\mathrm h}03^{\mathrm m}47.6005{\mathrm s}}}                   % Right Ascension
\newcommand{\hatcurCCdecxxxxxD}{\ensuremath{-19{\arcdeg}03{\arcmin}24.3267{\arcsec}}}                 % Declination
\newcommand{\hatcurCCmagxxxxxD}{12.344}                         % apparent V-band magnitude
\newcommand{\hatcurCCtwomassxxxxxD}{2MASS~04034760-1903242}     % 2MASS identifier
\newcommand{\hatcurCCgscxxxxxD}{GSC~5885-00663}                 % GSC(1.2) identifier
\newcommand{\hatcurCCgaiaxxxxxD}{GAIA~5094406188917950848}      % GAIA DR1 identifier
\newcommand{\hatcurCCgaiadrtwoxxxxxD}{5094406193214399616} % GAIA DR2 identifier
\newcommand{\hatcurCCtassmvxxxxxD}{\ensuremath{12.344\pm0.047}} % APASS V-band magnitude
\newcommand{\hatcurCCtassmvshortxxxxxD}{\ensuremath{12.3}}      % APASS V-band magnitude
\newcommand{\hatcurCCtassmBxxxxxD}{\ensuremath{13.094\pm0.096}} % APASS B-band magnitude
\newcommand{\hatcurCCtassmBshortxxxxxD}{\ensuremath{13.1}}      % APASS B-band magnitude
\newcommand{\hatcurCCtassmIxxxxxD}{\ensuremath{nff\pmnff}}      % TASS I-band magnitude
\newcommand{\hatcurCCtassmIshortxxxxxD}{\ensuremath{0.0}}       % TASS I-band magnitude
\newcommand{\hatcurCCtassmgxxxxxD}{\ensuremath{12.669\pm0.035}} % APASS g-band magnitude
\newcommand{\hatcurCCtassmgshortxxxxxD}{\ensuremath{12.7}}      % APASS g-band magnitude
\newcommand{\hatcurCCtassmrxxxxxD}{\ensuremath{12.129\pm0.058}} % APASS r-band magnitude
\newcommand{\hatcurCCtassmrshortxxxxxD}{\ensuremath{12.1}}      % APASS r-band magnitude
\newcommand{\hatcurCCtassmixxxxxD}{\ensuremath{11.949\pm0.063}} % APASS i-band magnitude
\newcommand{\hatcurCCtassmishortxxxxxD}{\ensuremath{11.9}}      % APASS i-band magnitude
\newcommand{\hatcurCCparallaxxxxxxD}{\ensuremath{3.550\pm0.039}} % Gaia DR2 parallax [mas]
\newcommand{\hatcurCCgaiamGxxxxxD}{\ensuremath{12.18160\pm0.00070}} % Gaia G-band magnitude
\newcommand{\hatcurCCgaiamBPxxxxxD}{\ensuremath{12.5621\pm0.0027}} % Gaia BP-band magnitude
\newcommand{\hatcurCCgaiamRPxxxxxD}{\ensuremath{11.6542\pm0.0024}} % Gaia RP-band magnitude
\newcommand{\hatcurCCtwomassJmagxxxxxD}{\ensuremath{11.071\pm0.026}} % 2MASS ORIG MAG
\newcommand{\hatcurCCtwomassHmagxxxxxD}{\ensuremath{10.738\pm0.023}} % 2MASS ORIG MAG
\newcommand{\hatcurCCtwomassKmagxxxxxD}{\ensuremath{10.707\pm0.027}} % 2MASS ORIG MAG
\newcommand{\hatcurCCcitJmagxxxxxD}{\ensuremath{11.088\pm0.026}} % 2MASS CIT MAG
\newcommand{\hatcurCCcitHmagxxxxxD}{\ensuremath{10.734\pm0.023}} % 2MASS CIT MAG
\newcommand{\hatcurCCcitKmagxxxxxD}{\ensuremath{10.731\pm0.027}} % 2MASS CIT MAG
\newcommand{\hatcurCCbbJmagxxxxxD}{\ensuremath{11.137\pm0.028}} % 2MASS BB MAG
\newcommand{\hatcurCCbbHmagxxxxxD}{\ensuremath{10.754\pm0.024}} % 2MASS BB MAG
\newcommand{\hatcurCCbbKmagxxxxxD}{\ensuremath{10.751\pm0.027}} % 2MASS BB MAG
\newcommand{\hatcurCCesoJmagxxxxxD}{\ensuremath{11.139\pm0.029}} % 2MASS ESO MAG
\newcommand{\hatcurCCesoHmagxxxxxD}{\ensuremath{10.747\pm0.027}} % 2MASS ESO MAG
\newcommand{\hatcurCCesoKmagxxxxxD}{\ensuremath{10.750\pm0.028}} % 2MASS ESO MAG
\newcommand{\hatcurCCesoJHmagxxxxxD}{\ensuremath{0.391\pm0.038}} % 2MASS ESO JH COLOR
\newcommand{\hatcurCCesoJKmagxxxxxD}{\ensuremath{0.390\pm0.040}} % 2MASS ESO JK COLOR
\newcommand{\hatcurCCesoHKmagxxxxxD}{\ensuremath{-0.002\pm0.039}} % 2MASS ESO HK COLOR
\newcommand{\hatcurLCdipxxxxxD}{\ensuremath{16.2}}              % BLS detected dip (mmag)
\newcommand{\hatcurLCrprstarxxxxxD}{\ensuremath{0.1218\pm0.0023}} % Rp/R*
\newcommand{\hatcurLCbsqxxxxxD}{\ensuremath{0.084_{-0.026}^{+0.033}}} % impact parameter square
\newcommand{\hatcurLCimpxxxxxD}{\ensuremath{0.290_{-0.049}^{+0.052}}} % impact parameter
\newcommand{\hatcurLCzetaxxxxxD}{\ensuremath{21.85\pm0.13}}     % zeta/R*
\newcommand{\hatcurLCdurxxxxxD}{\ensuremath{0.10369\pm0.00071}} % transit duration (days)
\newcommand{\hatcurLCdurshortxxxxxD}{\ensuremath{0.1037}}       % transit duration (days)
\newcommand{\hatcurLCdurhrxxxxxD}{\ensuremath{2.489\pm0.017}}   % transit duration (hours)
\newcommand{\hatcurLCdurhrshortxxxxxD}{\ensuremath{2.489}}      % transit duration (hours)
\newcommand{\hatcurLCqxxxxxD}{\ensuremath{0.04410\pm0.00030}}   % fractional transit duration (days)
\newcommand{\hatcurLCqshortxxxxxD}{\ensuremath{0.044}}          % fractional transit duration (days)
\newcommand{\hatcurLCingdurxxxxxD}{\ensuremath{0.01220\pm0.00050}} % ingress/egress duration (days)
\newcommand{\hatcurLCPxxxxxD}{\ensuremath{2.3506210\pm0.0000013}} % period (days)
\newcommand{\hatcurLCPprecxxxxxD}{\ensuremath{2.3506210}}       % period (days)
\newcommand{\hatcurLCPshortxxxxxD}{\ensuremath{2.3506}}         % period (days)
\newcommand{\hatcurLCTxxxxxD}{\ensuremath{2457778.49589\pm0.00025}} % epoch (BJD)
\newcommand{\hatcurLCTAxxxxxD}{\ensuremath{2456840.59812\pm0.00055}} % TA (BJD)
\newcommand{\hatcurLCTBxxxxxD}{\ensuremath{2458048.81731\pm0.00031}} % TB (BJD)
\newcommand{\hatcurLChatnetmxxxxxD}{\ensuremath{12.185300\pm0.000057}} % HATNet OOT level
\newcommand{\hatcurLCiblendxxxxxD}{\ensuremath{0.795\pm0.036}}  % HATNet iblend factor
\newcommand{\hatcurLCrhoxxxxxD}{\ensuremath{1.633\pm0.075}}     % stellar density no isochrone constraint (cgs)
\newcommand{\hatcurSMEiteffxxxxxD}{\ensuremath{5632\pm64}}      % Ini SME, stellar effective temperature
\newcommand{\hatcurSMEizfehxxxxxD}{\ensuremath{0.160\pm0.041}}  % Ini SME, stellar metallicity
\newcommand{\hatcurSMEizfehshortxxxxxD}{\ensuremath{0.16}}      % Ini SME, stellar metallicity
\newcommand{\hatcurSMEiloggxxxxxD}{\ensuremath{4.440\pm0.090}}  % Ini SME, stellar surface gravity
\newcommand{\hatcurSMEivsinxxxxxD}{\ensuremath{4.18\pm0.45}}    % Ini SME, stellar rotational velocity
\newcommand{\hatcurSMEivmacxxxxxD}{\ensuremath{3.768\pm0.097}}  % Ini SME, stellar macroturbulence
\newcommand{\hatcurSMEivmicxxxxxD}{\ensuremath{1.000\pm0.033}}  % Ini SME, stellar microturbulence
\newcommand{\hatcurSMEiiteffxxxxxD}{\ensuremath{5659\pm84}}     % Final SME, stellar effective temperature
\newcommand{\hatcurSMEiizfehxxxxxD}{\ensuremath{0.160\pm0.059}} % Final SME, stellar metallicity
\newcommand{\hatcurSMEiizfehshortxxxxxD}{\ensuremath{0.16}}     % Final SME, stellar metallicity
\newcommand{\hatcurSMEiiloggxxxxxD}{\ensuremath{4.504\pm0.014}} % Final SME, stellar surface gravity
\newcommand{\hatcurSMEiivsinxxxxxD}{\ensuremath{4.09\pm0.48}}   % Final SME, stellar rotational velocity
\newcommand{\hatcurSMEiivmacxxxxxD}{\ensuremath{3.81\pm0.13}}   % Final SME, stellar macroturbulence
\newcommand{\hatcurSMEiivmicxxxxxD}{\ensuremath{1.014\pm0.045}} % Final SME, stellar microturbulence
\newcommand{\hatcurLBizxxxxxD}{\ensuremath{0.2279}}             % Limb darkening parameters, Gamma1, z-band
\newcommand{\hatcurLBiizxxxxxD}{\ensuremath{0.3213}}            % Limb darkening parameters, Gamma2, z-band
\newcommand{\hatcurLBiixxxxxD}{\ensuremath{0.2951}}             % Limb darkening parameters, Gamma1, i-band
\newcommand{\hatcurLBiiixxxxxD}{\ensuremath{0.3188}}            % Limb darkening parameters, Gamma2, i-band
\newcommand{\hatcurLBiIxxxxxD}{\ensuremath{0.2727}}             % Limb darkening parameters, Gamma1, I-band
\newcommand{\hatcurLBiiIxxxxxD}{\ensuremath{0.3201}}            % Limb darkening parameters, Gamma2, I-band
\newcommand{\hatcurLBigxxxxxD}{\ensuremath{0.5973}}             % Limb darkening parameters, Gamma1, g-band
\newcommand{\hatcurLBiigxxxxxD}{\ensuremath{0.2037}}            % Limb darkening parameters, Gamma2, g-band
\newcommand{\hatcurLBirxxxxxD}{\ensuremath{0.3913}}             % Limb darkening parameters, Gamma1, r-band
\newcommand{\hatcurLBiirxxxxxD}{\ensuremath{0.3070}}            % Limb darkening parameters, Gamma2, r-band
\newcommand{\hatcurLBiRxxxxxD}{\ensuremath{0.3646}}             % Limb darkening parameters, Gamma1, R-band
\newcommand{\hatcurLBiiRxxxxxD}{\ensuremath{0.3112}}            % Limb darkening parameters, Gamma2, R-band
\newcommand{\hatcurLBikepxxxxxD}{\ensuremath{0.1000}}           % Limb darkening parameters, Gamma1, Kep-band
\newcommand{\hatcurLBiikepxxxxxD}{\ensuremath{0.1000}}          % Limb darkening parameters, Gamma2, Kep-band
\newcommand{\hatcurISOmxxxxxD}{\ensuremath{1.026_{-0.026}^{+0.019}}} % stellar mass
\newcommand{\hatcurISOmshortxxxxxD}{\ensuremath{1.03}}          % stellar mass
\newcommand{\hatcurISOmlongxxxxxD}{\ensuremath{1.026_{-0.026}^{+0.019}}} % stellar mass
\newcommand{\hatcurISOrxxxxxD}{\ensuremath{0.960\pm0.011}}      % stellar radius
\newcommand{\hatcurISOrshortxxxxxD}{\ensuremath{0.96}}          % stellar radius
\newcommand{\hatcurISOrlongxxxxxD}{\ensuremath{0.960\pm0.011}}  % stellar radius
\newcommand{\hatcurISOrhoxxxxxD}{\ensuremath{1.633\pm0.075}}    % stellar density (cgs)
\newcommand{\hatcurISOrholongxxxxxD}{\ensuremath{1.633\pm0.075}} % stellar density (cgs)
\newcommand{\hatcurISOloggxxxxxD}{\ensuremath{4.484\pm0.016}}   % stellar surface gravity from isochrones
\newcommand{\hatcurISOlumxxxxxD}{\ensuremath{0.805\pm0.018}}    % stellar luminosity
\newcommand{\hatcurISOlumshortxxxxxD}{\ensuremath{0.81}}        % stellar luminosity
\newcommand{\hatcurISOteffxxxxxD}{\ensuremath{5587\pm19}}       % stellar effective temperature adjusted via MCMC
\newcommand{\hatcurISOzfehxxxxxD}{\ensuremath{0.268\pm0.043}}   % stellar [M/H] adjusted via MCMC
\newcommand{\hatcurISOagexxxxxD}{\ensuremath{2.5_{-1.1}^{+1.5}}} % stellar age
\newcommand{\hatcurISOspecxxxxxD}{G}                            % stellar spectral type
\newcommand{\hatcurRVKxxxxxD}{\ensuremath{472.5\pm8.4}}         % RV semi-amplitude [m/s]
\newcommand{\hatcurRVrkxxxxxD}{\ensuremath{0\pm0}}              % sqrt(e)*cos(omega)
\newcommand{\hatcurRVrhxxxxxD}{\ensuremath{0\pm0}}              % sqrt(e)*sin(omega)
\newcommand{\hatcurRVkxxxxxD}{\ensuremath{0\pm0}}               % e*cos(omega)
\newcommand{\hatcurRVhxxxxxD}{\ensuremath{0\pm0}}               % e*sin(omega)
\newcommand{\hatcurRVtronexxxxxD}{\ensuremath{0\pm0}}           % RV linear trend tr1 factor
\newcommand{\hatcurRVtrtwoxxxxxD}{\ensuremath{0\pm0}}           % RV linear trend tr2 factor
\newcommand{\hatcurRVgammaxxxxxD}{\ensuremath{544.4\pm7.6}}     % RV gamma velocity, relative scale
\newcommand{\hatcurRVjitterxxxxxD}{\ensuremath{24.1\pm6.1}}     % RV jitter (m/s)
\newcommand{\hatcurRVjittertwosiglimxxxxxD}{\ensuremath{<36.0}} % RV jitter (m/s) 95 percent confidence upper limit
\newcommand{\hatcurRVfitrmsxxxxxD}{\ensuremath{.1fym}}          % 
\newcommand{\hatcurRVeccenxxxxxD}{\ensuremath{0\pm0}}           % eccentricity
\newcommand{\hatcurRVeccentwosiglimxxxxxD}{\ensuremath{<0.000}} % eccentricity
\newcommand{\hatcurRVomegaxxxxxD}{\ensuremath{0\pm0}}           % argument of pericenter
\newcommand{\hatcurPPixxxxxD}{\ensuremath{87.88\pm0.40}}        % orbital inclination
\newcommand{\hatcurPPgxxxxxD}{\ensuremath{60.2\pm3.4}}          % planetary surface gravity (m/s^2)
\newcommand{\hatcurPPloggxxxxxD}{\ensuremath{3.779\pm0.025}}    % planetary surface gravity (log cgs)
\newcommand{\hatcurPParxxxxxD}{\ensuremath{7.82\pm0.12}}        % relative orbital radius (a/R*)
\newcommand{\hatcurPParelxxxxxD}{\ensuremath{0.03493_{-0.00030}^{+0.00021}}} % semimajor axis (AU)
\newcommand{\hatcurPPrhoxxxxxD}{\ensuremath{2.65\pm0.21}}       % planetary density (cgs)
\newcommand{\hatcurPPmxxxxxD}{\ensuremath{3.147\pm0.073}}       % planetary mass (M_jup)
\newcommand{\hatcurPPmshortxxxxxD}{\ensuremath{3.15}}           % planetary mass (M_jup)
\newcommand{\hatcurPPmlongxxxxxD}{\ensuremath{3.147\pm0.073}}   % planetary mass (M_jup)
\newcommand{\hatcurPPmexxxxxD}{\ensuremath{1000\pm23}}          % planetary mass (M_earth)
\newcommand{\hatcurPPmeshortxxxxxD}{\ensuremath{1000.1}}        % planetary mass (M_earth)
\newcommand{\hatcurPPmelongxxxxxD}{\ensuremath{1000\pm23}}      % planetary mass (M_earth)
\newcommand{\hatcurPPrxxxxxD}{\ensuremath{1.139\pm0.028}}       % planetary radius (R_jup)
\newcommand{\hatcurPPrshortxxxxxD}{\ensuremath{1.14}}           % planetary radius (R_jup)
\newcommand{\hatcurPPrlongxxxxxD}{\ensuremath{1.139\pm0.028}}   % planetary radius (R_jup)
\newcommand{\hatcurPPrexxxxxD}{\ensuremath{12.76\pm0.32}}       % planetary radius (R_earth)
\newcommand{\hatcurPPreshortxxxxxD}{\ensuremath{12.8}}          % planetary radius (R_earth)
\newcommand{\hatcurPPrelongxxxxxD}{\ensuremath{12.76\pm0.32}}   % planetary radius (R_earth)
\newcommand{\hatcurPPmrcorrxxxxxD}{\ensuremath{-0.12}}          % mass/radius correlation
\newcommand{\hatcurPPteffxxxxxD}{\ensuremath{1413.4\pm9.7}}     % planetary temperature (K)
\newcommand{\hatcurPPthetaxxxxxD}{\ensuremath{0.1875\pm0.0055}} % Safranov number
\newcommand{\hatcurPPfluxperixxxxxD}{\ensuremath{8.99\pm0.25}}  % flux @ periastron (CGS)
\newcommand{\hatcurPPfluxperidimxxxxxD}{\ensuremath{8}}         % flux @ periastron (CGS) units.
\newcommand{\hatcurPPfluxapxxxxxD}{\ensuremath{8.99\pm0.25}}    % flux @ apastron (CGS)
\newcommand{\hatcurPPfluxapdimxxxxxD}{\ensuremath{8}}           % flux @ apastron (CGS) units.
\newcommand{\hatcurPPfluxavgxxxxxD}{\ensuremath{8.99\pm0.25}}   % flux on average (CGS)
\newcommand{\hatcurPPfluxavgdimxxxxxD}{\ensuremath{8}}          % flux average (CGS) units.
\newcommand{\hatcurPPfluxavglogxxxxxD}{\ensuremath{8.954\pm0.012}} % log10 flux on average (CGS)
\newcommand{\hatcurXsecphasexxxxxD}{\ensuremath{0\pm0}}         % Phase of secondary eclipse
\newcommand{\hatcurXsecondaryxxxxxD}{\ensuremath{2457779.67120\pm0.00025}} % Secondary eclipse epoch
\newcommand{\hatcurXsecdurxxxxxD}{\ensuremath{0.10369\pm0.00071}} % sec eclipse duration (days)
\newcommand{\hatcurXsecingdurxxxxxD}{\ensuremath{0.01220\pm0.00050}} % sec I/E duration (days)
\newcommand{\hatcurPPphiconjxxxxxD}{\ensuremath{0\pm0}}         % phase diff between conjunction and periastron
\newcommand{\hatcurPPperixxxxxD}{\ensuremath{2457777.90823\pm0.00025}} % time of periastron passage.
\newcommand{\hatcurPPaequivxxxxxD}{\ensuremath{0.03890\pm0.00053}} % equivalent semi-major axis
\newcommand{\hatcurPPtcircxxxxxD}{\ensuremath{208\pm28}}        % circularization timescale
\newcommand{\hatcurPPtinfallxxxxxD}{\ensuremath{213\pm18}}      % infall timescale
\newcommand{\hatcurXdistxxxxxD}{\ensuremath{280.0\pm2.9}}       % distance (pc), no reddenning correction
\newcommand{\hatcurXAvxxxxxD}{\ensuremath{0.055\pm0.011}}       % Av (mag)
\newcommand{\hatcurXdistredxxxxxD}{\ensuremath{280.0\pm2.9}}    % distance with Av correction (pc)
\newcommand{\hatcurXEBVxxxxxD}{\ensuremath{0.0180_{-0.0040}^{+0.0030}}} % E(B-V) (mag)
\newcommand{\hatcurCCpmraxxxxxD}{\ensuremath{-12.664\pm0.046}}  % proper motion, in RA
\newcommand{\hatcurCCpmdecxxxxxD}{\ensuremath{-14.115\pm0.040}} % proper motion, in DEC
\newcommand{\hatcurCCpmxxxxxD}{\ensuremath{18.963\pm0.061}}     % proper motion

\newcommand{\hatcurhtrxxxxxE}{HATS737-002}                      % Original HTR name of target
\newcommand{\hatcurfieldxxxxxE}{\ensuremath{string}}            % HTR field
\newcommand{\hatcurCCraxxxxxE}{\ensuremath{12^{\mathrm h}27^{\mathrm m}08.9729{\mathrm s}}}                   % Right Ascension
\newcommand{\hatcurCCdecxxxxxE}{\ensuremath{-48{\arcdeg}58{\arcmin}42.2278{\arcsec}}}                 % Declination
\newcommand{\hatcurCCmagxxxxxE}{11.552}                         % apparent V-band magnitude
\newcommand{\hatcurCCtwomassxxxxxE}{2MASS~12270898-4858423}     % 2MASS identifier
\newcommand{\hatcurCCgscxxxxxE}{GSC~8239-00065}                 % GSC(1.2) identifier
\newcommand{\hatcurCCgaiaxxxxxE}{GAIA~6128363662138374528}      % GAIA DR1 identifier
\newcommand{\hatcurCCgaiadrtwoxxxxxE}{6128363666439822208} % GAIA DR2 identifier
\newcommand{\hatcurCCtassmvxxxxxE}{\ensuremath{11.552\pm0.019}} % APASS V-band magnitude
\newcommand{\hatcurCCtassmvshortxxxxxE}{\ensuremath{11.6}}      % APASS V-band magnitude
\newcommand{\hatcurCCtassmBxxxxxE}{\ensuremath{12.0510\pm0.0090}} % APASS B-band magnitude
\newcommand{\hatcurCCtassmBshortxxxxxE}{\ensuremath{12.1}}      % APASS B-band magnitude
\newcommand{\hatcurCCtassmIxxxxxE}{\ensuremath{nff\pmnff}}      % TASS I-band magnitude
\newcommand{\hatcurCCtassmIshortxxxxxE}{\ensuremath{0.0}}       % TASS I-band magnitude
\newcommand{\hatcurCCtassmgxxxxxE}{\ensuremath{11.752\pm0.019}} % APASS g-band magnitude
\newcommand{\hatcurCCtassmgshortxxxxxE}{\ensuremath{11.8}}      % APASS g-band magnitude
\newcommand{\hatcurCCtassmrxxxxxE}{\ensuremath{11.466\pm0.024}} % APASS r-band magnitude
\newcommand{\hatcurCCtassmrshortxxxxxE}{\ensuremath{11.5}}      % APASS r-band magnitude
\newcommand{\hatcurCCtassmixxxxxE}{\ensuremath{11.379\pm0.070}} % APASS i-band magnitude
\newcommand{\hatcurCCtassmishortxxxxxE}{\ensuremath{11.4}}      % APASS i-band magnitude
\newcommand{\hatcurCCparallaxxxxxxE}{\ensuremath{2.35\pm0.22}}  % Gaia DR2 parallax [mas]
\newcommand{\hatcurCCgaiamGxxxxxE}{\ensuremath{11.7679\pm0.0015}} % Gaia G-band magnitude
\newcommand{\hatcurCCgaiamBPxxxxxE}{\ensuremath{0\pm10000}}     % Gaia BP-band magnitude
\newcommand{\hatcurCCgaiamRPxxxxxE}{\ensuremath{0\pm10000}}     % Gaia RP-band magnitude
\newcommand{\hatcurCCtwomassJmagxxxxxE}{\ensuremath{10.584\pm0.024}} % 2MASS ORIG MAG
\newcommand{\hatcurCCtwomassHmagxxxxxE}{\ensuremath{10.358\pm0.026}} % 2MASS ORIG MAG
\newcommand{\hatcurCCtwomassKmagxxxxxE}{\ensuremath{10.289\pm0.023}} % 2MASS ORIG MAG
\newcommand{\hatcurCCcitJmagxxxxxE}{\ensuremath{10.605\pm0.024}} % 2MASS CIT MAG
\newcommand{\hatcurCCcitHmagxxxxxE}{\ensuremath{10.353\pm0.026}} % 2MASS CIT MAG
\newcommand{\hatcurCCcitKmagxxxxxE}{\ensuremath{10.313\pm0.023}} % 2MASS CIT MAG
\newcommand{\hatcurCCbbJmagxxxxxE}{\ensuremath{10.648\pm0.026}} % 2MASS BB MAG
\newcommand{\hatcurCCbbHmagxxxxxE}{\ensuremath{10.374\pm0.027}} % 2MASS BB MAG
\newcommand{\hatcurCCbbKmagxxxxxE}{\ensuremath{10.333\pm0.023}} % 2MASS BB MAG
\newcommand{\hatcurCCesoJmagxxxxxE}{\ensuremath{10.649\pm0.027}} % 2MASS ESO MAG
\newcommand{\hatcurCCesoHmagxxxxxE}{\ensuremath{10.370\pm0.030}} % 2MASS ESO MAG
\newcommand{\hatcurCCesoKmagxxxxxE}{\ensuremath{10.332\pm0.024}} % 2MASS ESO MAG
\newcommand{\hatcurCCesoJHmagxxxxxE}{\ensuremath{0.280\pm0.038}} % 2MASS ESO JH COLOR
\newcommand{\hatcurCCesoJKmagxxxxxE}{\ensuremath{0.318\pm0.036}} % 2MASS ESO JK COLOR
\newcommand{\hatcurCCesoHKmagxxxxxE}{\ensuremath{0.038\pm0.039}} % 2MASS ESO HK COLOR
\newcommand{\hatcurLCrprstarxxxxxE}{\ensuremath{0.0786\pm0.0025}} % Rp/R*
\newcommand{\hatcurLCbsqxxxxxE}{\ensuremath{0.429_{-0.042}^{+0.041}}} % impact parameter square
\newcommand{\hatcurLCimpxxxxxE}{\ensuremath{0.655_{-0.033}^{+0.030}}} % impact parameter
\newcommand{\hatcurLCzetaxxxxxE}{\ensuremath{17.12\pm0.43}}     % zeta/R*
\newcommand{\hatcurLCdurxxxxxE}{\ensuremath{0.1325\pm0.0031}}   % transit duration (days)
\newcommand{\hatcurLCdurshortxxxxxE}{\ensuremath{0.1325}}       % transit duration (days)
\newcommand{\hatcurLCingdurxxxxxE}{\ensuremath{0.0161\pm0.0013}} % ingress/egress duration (days)
\newcommand{\hatcurLCPxxxxxE}{\ensuremath{4.2180896\pm0.0000089}} % period (days)
\newcommand{\hatcurLCPprecxxxxxE}{\ensuremath{4.2180896}}       % period (days)
\newcommand{\hatcurLCPshortxxxxxE}{\ensuremath{4.2181}}         % period (days)
\newcommand{\hatcurLCTxxxxxE}{\ensuremath{2457463.2999\pm0.0017}} % epoch (BJD)
\newcommand{\hatcurLCTAxxxxxE}{\ensuremath{2455679.0479\pm0.0039}} % TA (BJD)
\newcommand{\hatcurLCTBxxxxxE}{\ensuremath{2457939.9441\pm0.0022}} % TB (BJD)
\newcommand{\hatcurLCiblendxxxxxE}{\ensuremath{0\pm0}}          % HATNet iblend factor
\newcommand{\hatcurLCrhoxxxxxE}{\ensuremath{0.702\pm0.070}}     % stellar density no isochrone constraint (cgs)
\newcommand{\hatcurSMEiteffxxxxxE}{\ensuremath{6443\pm79}}      % Ini SME, stellar effective temperature
\newcommand{\hatcurSMEizfehxxxxxE}{\ensuremath{0.040\pm0.055}}  % Ini SME, stellar metallicity
\newcommand{\hatcurSMEizfehshortxxxxxE}{\ensuremath{0.04}}      % Ini SME, stellar metallicity
\newcommand{\hatcurSMEiloggxxxxxE}{\ensuremath{3.93\pm0.16}}    % Ini SME, stellar surface gravity
\newcommand{\hatcurSMEivsinxxxxxE}{\ensuremath{6.30\pm0.29}}    % Ini SME, stellar rotational velocity
\newcommand{\hatcurSMEivmacxxxxxE}{\ensuremath{3.768\pm0.097}}  % Ini SME, stellar macroturbulence
\newcommand{\hatcurSMEivmicxxxxxE}{\ensuremath{1.000\pm0.033}}  % Ini SME, stellar microturbulence
\newcommand{\hatcurSMEiiteffxxxxxE}{\ensuremath{6460\pm130}}    % Final SME, stellar effective temperature
\newcommand{\hatcurSMEiizfehxxxxxE}{\ensuremath{0.060\pm0.069}} % Final SME, stellar metallicity
\newcommand{\hatcurSMEiizfehshortxxxxxE}{\ensuremath{0.06}}     % Final SME, stellar metallicity
\newcommand{\hatcurSMEiiloggxxxxxE}{\ensuremath{4.15\pm0.14}}   % Final SME, stellar surface gravity
\newcommand{\hatcurSMEiivsinxxxxxE}{\ensuremath{6.22\pm0.29}}   % Final SME, stellar rotational velocity
\newcommand{\hatcurSMEiivmacxxxxxE}{\ensuremath{5.04\pm0.20}}   % Final SME, stellar macroturbulence
\newcommand{\hatcurSMEiivmicxxxxxE}{\ensuremath{1.71\pm0.17}}   % Final SME, stellar microturbulence
\newcommand{\hatcurLBizxxxxxE}{\ensuremath{0.1247}}             % Limb darkening parameters, Gamma1, z-band
\newcommand{\hatcurLBiizxxxxxE}{\ensuremath{0.3672}}            % Limb darkening parameters, Gamma2, z-band
\newcommand{\hatcurLBiixxxxxE}{\ensuremath{0.1748}}             % Limb darkening parameters, Gamma1, i-band
\newcommand{\hatcurLBiiixxxxxE}{\ensuremath{0.3768}}            % Limb darkening parameters, Gamma2, i-band
\newcommand{\hatcurLBiIxxxxxE}{\ensuremath{0.1564}}             % Limb darkening parameters, Gamma1, I-band
\newcommand{\hatcurLBiiIxxxxxE}{\ensuremath{0.3749}}            % Limb darkening parameters, Gamma2, I-band
\newcommand{\hatcurLBigxxxxxE}{\ensuremath{0.4117}}             % Limb darkening parameters, Gamma1, g-band
\newcommand{\hatcurLBiigxxxxxE}{\ensuremath{0.3345}}            % Limb darkening parameters, Gamma2, g-band
\newcommand{\hatcurLBirxxxxxE}{\ensuremath{0.2489}}             % Limb darkening parameters, Gamma1, r-band
\newcommand{\hatcurLBiirxxxxxE}{\ensuremath{0.3851}}            % Limb darkening parameters, Gamma2, r-band
\newcommand{\hatcurLBiRxxxxxE}{\ensuremath{0.2277}}             % Limb darkening parameters, Gamma1, R-band
\newcommand{\hatcurLBiiRxxxxxE}{\ensuremath{0.3845}}            % Limb darkening parameters, Gamma2, R-band
\newcommand{\hatcurLBikepxxxxxE}{\ensuremath{0.1000}}           % Limb darkening parameters, Gamma1, Kep-band
\newcommand{\hatcurLBiikepxxxxxE}{\ensuremath{0.1000}}          % Limb darkening parameters, Gamma2, Kep-band
\newcommand{\hatcurISOmxxxxxE}{\ensuremath{1.461\pm0.043}}      % stellar mass
\newcommand{\hatcurISOmshortxxxxxE}{\ensuremath{1.46}}          % stellar mass
\newcommand{\hatcurISOmlongxxxxxE}{\ensuremath{1.461\pm0.043}}  % stellar mass
\newcommand{\hatcurISOrxxxxxE}{\ensuremath{1.433\pm0.059}}      % stellar radius
\newcommand{\hatcurISOrshortxxxxxE}{\ensuremath{1.43}}          % stellar radius
\newcommand{\hatcurISOrlongxxxxxE}{\ensuremath{1.433\pm0.059}}  % stellar radius
\newcommand{\hatcurISOloggxxxxxE}{\ensuremath{4.292\pm0.028}}   % stellar surface gravity from MCMC
\newcommand{\hatcurISOlumxxxxxE}{\ensuremath{4.89\pm0.46}}      % stellar luminosity (Lsun) from MCMC
\newcommand{\hatcurISOlumshortxxxxxE}{\ensuremath{4.89}}        % stellar luminosity
\newcommand{\hatcurISOfehxxxxxE}{\ensuremath{-0.016\pm0.070}}   % stellar Fe/H from MCMC
\newcommand{\hatcurISOteffxxxxxE}{\ensuremath{7175\pm54}}       % stellar effective temperature (K) from MCMC
\newcommand{\hatcurISOagexxxxxE}{\ensuremath{0.31_{-0.20}^{+0.33}}} % stellar age
\newcommand{\hatcurISOmBxxxxxE}{\ensuremath{1.216\pm0.034}}     % stellar mass
\newcommand{\hatcurISOmshortBxxxxxE}{\ensuremath{1.22}}         % stellar mass
\newcommand{\hatcurISOmlongBxxxxxE}{\ensuremath{1.216\pm0.034}} % stellar mass
\newcommand{\hatcurISOfehBxxxxxE}{\ensuremath{-0.016\pm0.070}}  % stellar Fe/H from MCMC
\newcommand{\hatcurISOageBxxxxxE}{\ensuremath{0.31_{-0.20}^{+0.33}}} % stellar age
\newcommand{\hatcurPPixxxxxE}{\ensuremath{85.69\pm0.33}}        % orbital inclination
\newcommand{\hatcurPParxxxxxE}{\ensuremath{8.71\pm0.30}}        % relative orbital radius (a/R*)
\newcommand{\hatcurPParelxxxxxE}{\ensuremath{0.05798\pm0.00057}} % semimajor axis (AU)
\newcommand{\hatcurPPrxxxxxE}{\ensuremath{1.095\pm0.062}}       % planetary radius (R_jup)
\newcommand{\hatcurPPrshortxxxxxE}{\ensuremath{1.10}}           % planetary radius (R_jup)
\newcommand{\hatcurPPrlongxxxxxE}{\ensuremath{1.095\pm0.062}}   % planetary radius (R_jup)
\newcommand{\hatcurPPteffxxxxxE}{\ensuremath{1721\pm34}}        % planetary temperature (K)
\newcommand{\hatcurPPfluxavgxxxxxE}{\ensuremath{1980000000\pm160000000}} % flux on average (CGS)
\newcommand{\hatcurPPfluxavglogxxxxxE}{\ensuremath{9.296\pm0.034}} % log10 flux on average (CGS)
\newcommand{\hatcurXdistxxxxxE}{\ensuremath{492\pm21}}          % distance (pc), no reddenning correction
\newcommand{\hatcurXdistredxxxxxE}{\ensuremath{492\pm21}}       % distance with Av correction (pc)
\newcommand{\hatcurRVKxxxxxE}{\ensuremath{100\pm22}}            % RV semi-amplitude [m/s]
\newcommand{\hatcurRVgammaAxxxxxE}{\ensuremath{19290.0\pm8.1}}  % RV gamma velocity, relative scale
\newcommand{\hatcurRVjitterAxxxxxE}{\ensuremath{51\pm13}}       % RV jitter (m/s)
\newcommand{\hatcurRVjittertwosiglimAxxxxxE}{\ensuremath{<76.9}} % RV jitter (m/s) 95 percent confidence upper limit
\newcommand{\hatcurRVfitrmsAxxxxxE}{\ensuremath{0.0}}           % RVfitrms
\newcommand{\hatcurRVgammaBxxxxxE}{\ensuremath{19416.7\pm3.6}}  % RV gamma velocity, relative scale
\newcommand{\hatcurRVjitterBxxxxxE}{\ensuremath{5.2\pm7.1}}     % RV jitter (m/s)
\newcommand{\hatcurRVjittertwosiglimBxxxxxE}{\ensuremath{<19.2}} % RV jitter (m/s) 95 percent confidence upper limit
\newcommand{\hatcurPPloggxxxxxE}{\ensuremath{3.33\pm0.11}}      % planetary surface gravity (log cgs)
\newcommand{\hatcurPPrhoxxxxxE}{\ensuremath{0.96\pm0.27}}       % planetary density (cgs)
\newcommand{\hatcurPPmtwosiglimxxxxxE}{\ensuremath{<1.424}}     % 95 percent confidence upper limit on planetary mass (M_jup)
\newcommand{\hatcurPPmxxxxxE}{\ensuremath{1.03\pm0.23}}         % planetary mass (M_jup)
\newcommand{\hatcurPPmshortxxxxxE}{\ensuremath{1.03}}           % planetary mass (M_jup)
\newcommand{\hatcurPPmlongxxxxxE}{\ensuremath{1.03\pm0.23}}     % planetary mass (M_jup)
\newcommand{\hatcurPPthetaxxxxxE}{\ensuremath{0.074\pm0.017}}   % Safranov number
\newcommand{\hatcurCCpmraxxxxxE}{\ensuremath{-12.70\pm0.30}}    % proper motion, in RA
\newcommand{\hatcurCCpmdecxxxxxE}{\ensuremath{-3.23\pm0.16}}    % proper motion, in DEC
\newcommand{\hatcurCCpmxxxxxE}{\ensuremath{13.10\pm0.34}}       % proper motion
\newcommand{\hatcurXAvxxxxxE}{\ensuremath{0.340_{-0.017}^{+0.024}}} % Av (mag)

\newcommand{\hatcurCCbbHmag}[1]{\ifnum#1=54 %
\hatcurCCbbHmagxxxxxA
\else
\ifnum#1=55 %
\hatcurCCbbHmagxxxxxB
\else
\ifnum#1=56 %
\hatcurCCbbHmagxxxxxC
\else
\ifnum#1=57 %
\hatcurCCbbHmagxxxxxD
\else
\ifnum#1=58 %
\hatcurCCbbHmagxxxxxE
\else
??????\fi
\fi
\fi
\fi
\fi
}
\newcommand{\hatcurCCbbJmag}[1]{\ifnum#1=54 %
\hatcurCCbbJmagxxxxxA
\else
\ifnum#1=55 %
\hatcurCCbbJmagxxxxxB
\else
\ifnum#1=56 %
\hatcurCCbbJmagxxxxxC
\else
\ifnum#1=57 %
\hatcurCCbbJmagxxxxxD
\else
\ifnum#1=58 %
\hatcurCCbbJmagxxxxxE
\else
??????\fi
\fi
\fi
\fi
\fi
}
\newcommand{\hatcurCCbbKmag}[1]{\ifnum#1=54 %
\hatcurCCbbKmagxxxxxA
\else
\ifnum#1=55 %
\hatcurCCbbKmagxxxxxB
\else
\ifnum#1=56 %
\hatcurCCbbKmagxxxxxC
\else
\ifnum#1=57 %
\hatcurCCbbKmagxxxxxD
\else
\ifnum#1=58 %
\hatcurCCbbKmagxxxxxE
\else
??????\fi
\fi
\fi
\fi
\fi
}
\newcommand{\hatcurCCcitHmag}[1]{\ifnum#1=54 %
\hatcurCCcitHmagxxxxxA
\else
\ifnum#1=55 %
\hatcurCCcitHmagxxxxxB
\else
\ifnum#1=56 %
\hatcurCCcitHmagxxxxxC
\else
\ifnum#1=57 %
\hatcurCCcitHmagxxxxxD
\else
\ifnum#1=58 %
\hatcurCCcitHmagxxxxxE
\else
??????\fi
\fi
\fi
\fi
\fi
}
\newcommand{\hatcurCCcitJmag}[1]{\ifnum#1=54 %
\hatcurCCcitJmagxxxxxA
\else
\ifnum#1=55 %
\hatcurCCcitJmagxxxxxB
\else
\ifnum#1=56 %
\hatcurCCcitJmagxxxxxC
\else
\ifnum#1=57 %
\hatcurCCcitJmagxxxxxD
\else
\ifnum#1=58 %
\hatcurCCcitJmagxxxxxE
\else
??????\fi
\fi
\fi
\fi
\fi
}
\newcommand{\hatcurCCcitKmag}[1]{\ifnum#1=54 %
\hatcurCCcitKmagxxxxxA
\else
\ifnum#1=55 %
\hatcurCCcitKmagxxxxxB
\else
\ifnum#1=56 %
\hatcurCCcitKmagxxxxxC
\else
\ifnum#1=57 %
\hatcurCCcitKmagxxxxxD
\else
\ifnum#1=58 %
\hatcurCCcitKmagxxxxxE
\else
??????\fi
\fi
\fi
\fi
\fi
}
\newcommand{\hatcurCCdec}[1]{\ifnum#1=54 %
\hatcurCCdecxxxxxA
\else
\ifnum#1=55 %
\hatcurCCdecxxxxxB
\else
\ifnum#1=56 %
\hatcurCCdecxxxxxC
\else
\ifnum#1=57 %
\hatcurCCdecxxxxxD
\else
\ifnum#1=58 %
\hatcurCCdecxxxxxE
\else
??????\fi
\fi
\fi
\fi
\fi
}
\newcommand{\hatcurCCesoHKmag}[1]{\ifnum#1=54 %
\hatcurCCesoHKmagxxxxxA
\else
\ifnum#1=55 %
\hatcurCCesoHKmagxxxxxB
\else
\ifnum#1=56 %
\hatcurCCesoHKmagxxxxxC
\else
\ifnum#1=57 %
\hatcurCCesoHKmagxxxxxD
\else
\ifnum#1=58 %
\hatcurCCesoHKmagxxxxxE
\else
??????\fi
\fi
\fi
\fi
\fi
}
\newcommand{\hatcurCCesoHmag}[1]{\ifnum#1=54 %
\hatcurCCesoHmagxxxxxA
\else
\ifnum#1=55 %
\hatcurCCesoHmagxxxxxB
\else
\ifnum#1=56 %
\hatcurCCesoHmagxxxxxC
\else
\ifnum#1=57 %
\hatcurCCesoHmagxxxxxD
\else
\ifnum#1=58 %
\hatcurCCesoHmagxxxxxE
\else
??????\fi
\fi
\fi
\fi
\fi
}
\newcommand{\hatcurCCesoJHmag}[1]{\ifnum#1=54 %
\hatcurCCesoJHmagxxxxxA
\else
\ifnum#1=55 %
\hatcurCCesoJHmagxxxxxB
\else
\ifnum#1=56 %
\hatcurCCesoJHmagxxxxxC
\else
\ifnum#1=57 %
\hatcurCCesoJHmagxxxxxD
\else
\ifnum#1=58 %
\hatcurCCesoJHmagxxxxxE
\else
??????\fi
\fi
\fi
\fi
\fi
}
\newcommand{\hatcurCCesoJKmag}[1]{\ifnum#1=54 %
\hatcurCCesoJKmagxxxxxA
\else
\ifnum#1=55 %
\hatcurCCesoJKmagxxxxxB
\else
\ifnum#1=56 %
\hatcurCCesoJKmagxxxxxC
\else
\ifnum#1=57 %
\hatcurCCesoJKmagxxxxxD
\else
\ifnum#1=58 %
\hatcurCCesoJKmagxxxxxE
\else
??????\fi
\fi
\fi
\fi
\fi
}
\newcommand{\hatcurCCesoJmag}[1]{\ifnum#1=54 %
\hatcurCCesoJmagxxxxxA
\else
\ifnum#1=55 %
\hatcurCCesoJmagxxxxxB
\else
\ifnum#1=56 %
\hatcurCCesoJmagxxxxxC
\else
\ifnum#1=57 %
\hatcurCCesoJmagxxxxxD
\else
\ifnum#1=58 %
\hatcurCCesoJmagxxxxxE
\else
??????\fi
\fi
\fi
\fi
\fi
}
\newcommand{\hatcurCCesoKmag}[1]{\ifnum#1=54 %
\hatcurCCesoKmagxxxxxA
\else
\ifnum#1=55 %
\hatcurCCesoKmagxxxxxB
\else
\ifnum#1=56 %
\hatcurCCesoKmagxxxxxC
\else
\ifnum#1=57 %
\hatcurCCesoKmagxxxxxD
\else
\ifnum#1=58 %
\hatcurCCesoKmagxxxxxE
\else
??????\fi
\fi
\fi
\fi
\fi
}
\newcommand{\hatcurCCgaia}[1]{\ifnum#1=54 %
\hatcurCCgaiaxxxxxA
\else
\ifnum#1=55 %
\hatcurCCgaiaxxxxxB
\else
\ifnum#1=56 %
\hatcurCCgaiaxxxxxC
\else
\ifnum#1=57 %
\hatcurCCgaiaxxxxxD
\else
\ifnum#1=58 %
\hatcurCCgaiaxxxxxE
\else
??????\fi
\fi
\fi
\fi
\fi
}
\newcommand{\hatcurCCgaiadrtwo}[1]{\ifnum#1=54 %
\hatcurCCgaiadrtwoxxxxxA
\else
\ifnum#1=55 %
\hatcurCCgaiadrtwoxxxxxB
\else
\ifnum#1=56 %
\hatcurCCgaiadrtwoxxxxxC
\else
\ifnum#1=57 %
\hatcurCCgaiadrtwoxxxxxD
\else
\ifnum#1=58 %
\hatcurCCgaiadrtwoxxxxxE
\else
??????\fi
\fi
\fi
\fi
\fi
}
\newcommand{\hatcurCCgaiamBP}[1]{\ifnum#1=54 %
\hatcurCCgaiamBPxxxxxA
\else
\ifnum#1=55 %
\hatcurCCgaiamBPxxxxxB
\else
\ifnum#1=56 %
\hatcurCCgaiamBPxxxxxC
\else
\ifnum#1=57 %
\hatcurCCgaiamBPxxxxxD
\else
\ifnum#1=58 %
\hatcurCCgaiamBPxxxxxE
\else
??????\fi
\fi
\fi
\fi
\fi
}
\newcommand{\hatcurCCgaiamG}[1]{\ifnum#1=54 %
\hatcurCCgaiamGxxxxxA
\else
\ifnum#1=55 %
\hatcurCCgaiamGxxxxxB
\else
\ifnum#1=56 %
\hatcurCCgaiamGxxxxxC
\else
\ifnum#1=57 %
\hatcurCCgaiamGxxxxxD
\else
\ifnum#1=58 %
\hatcurCCgaiamGxxxxxE
\else
??????\fi
\fi
\fi
\fi
\fi
}
\newcommand{\hatcurCCgaiamRP}[1]{\ifnum#1=54 %
\hatcurCCgaiamRPxxxxxA
\else
\ifnum#1=55 %
\hatcurCCgaiamRPxxxxxB
\else
\ifnum#1=56 %
\hatcurCCgaiamRPxxxxxC
\else
\ifnum#1=57 %
\hatcurCCgaiamRPxxxxxD
\else
\ifnum#1=58 %
\hatcurCCgaiamRPxxxxxE
\else
??????\fi
\fi
\fi
\fi
\fi
}
\newcommand{\hatcurCCgsc}[1]{\ifnum#1=54 %
\hatcurCCgscxxxxxA
\else
\ifnum#1=55 %
\hatcurCCgscxxxxxB
\else
\ifnum#1=56 %
\hatcurCCgscxxxxxC
\else
\ifnum#1=57 %
\hatcurCCgscxxxxxD
\else
\ifnum#1=58 %
\hatcurCCgscxxxxxE
\else
??????\fi
\fi
\fi
\fi
\fi
}
\newcommand{\hatcurCCmag}[1]{\ifnum#1=54 %
\hatcurCCmagxxxxxA
\else
\ifnum#1=55 %
\hatcurCCmagxxxxxB
\else
\ifnum#1=56 %
\hatcurCCmagxxxxxC
\else
\ifnum#1=57 %
\hatcurCCmagxxxxxD
\else
\ifnum#1=58 %
\hatcurCCmagxxxxxE
\else
??????\fi
\fi
\fi
\fi
\fi
}
\newcommand{\hatcurCCparallax}[1]{\ifnum#1=54 %
\hatcurCCparallaxxxxxxA
\else
\ifnum#1=55 %
\hatcurCCparallaxxxxxxB
\else
\ifnum#1=56 %
\hatcurCCparallaxxxxxxC
\else
\ifnum#1=57 %
\hatcurCCparallaxxxxxxD
\else
\ifnum#1=58 %
\hatcurCCparallaxxxxxxE
\else
??????\fi
\fi
\fi
\fi
\fi
}
\newcommand{\hatcurCCpm}[1]{\ifnum#1=54 %
\hatcurCCpmxxxxxA
\else
\ifnum#1=55 %
\hatcurCCpmxxxxxB
\else
\ifnum#1=56 %
\hatcurCCpmxxxxxC
\else
\ifnum#1=57 %
\hatcurCCpmxxxxxD
\else
\ifnum#1=58 %
\hatcurCCpmxxxxxE
\else
??????\fi
\fi
\fi
\fi
\fi
}
\newcommand{\hatcurCCpmdec}[1]{\ifnum#1=54 %
\hatcurCCpmdecxxxxxA
\else
\ifnum#1=55 %
\hatcurCCpmdecxxxxxB
\else
\ifnum#1=56 %
\hatcurCCpmdecxxxxxC
\else
\ifnum#1=57 %
\hatcurCCpmdecxxxxxD
\else
\ifnum#1=58 %
\hatcurCCpmdecxxxxxE
\else
??????\fi
\fi
\fi
\fi
\fi
}
\newcommand{\hatcurCCpmra}[1]{\ifnum#1=54 %
\hatcurCCpmraxxxxxA
\else
\ifnum#1=55 %
\hatcurCCpmraxxxxxB
\else
\ifnum#1=56 %
\hatcurCCpmraxxxxxC
\else
\ifnum#1=57 %
\hatcurCCpmraxxxxxD
\else
\ifnum#1=58 %
\hatcurCCpmraxxxxxE
\else
??????\fi
\fi
\fi
\fi
\fi
}
\newcommand{\hatcurCCra}[1]{\ifnum#1=54 %
\hatcurCCraxxxxxA
\else
\ifnum#1=55 %
\hatcurCCraxxxxxB
\else
\ifnum#1=56 %
\hatcurCCraxxxxxC
\else
\ifnum#1=57 %
\hatcurCCraxxxxxD
\else
\ifnum#1=58 %
\hatcurCCraxxxxxE
\else
??????\fi
\fi
\fi
\fi
\fi
}
\newcommand{\hatcurCCtassmB}[1]{\ifnum#1=54 %
\hatcurCCtassmBxxxxxA
\else
\ifnum#1=55 %
\hatcurCCtassmBxxxxxB
\else
\ifnum#1=56 %
\hatcurCCtassmBxxxxxC
\else
\ifnum#1=57 %
\hatcurCCtassmBxxxxxD
\else
\ifnum#1=58 %
\hatcurCCtassmBxxxxxE
\else
??????\fi
\fi
\fi
\fi
\fi
}
\newcommand{\hatcurCCtassmBshort}[1]{\ifnum#1=54 %
\hatcurCCtassmBshortxxxxxA
\else
\ifnum#1=55 %
\hatcurCCtassmBshortxxxxxB
\else
\ifnum#1=56 %
\hatcurCCtassmBshortxxxxxC
\else
\ifnum#1=57 %
\hatcurCCtassmBshortxxxxxD
\else
\ifnum#1=58 %
\hatcurCCtassmBshortxxxxxE
\else
??????\fi
\fi
\fi
\fi
\fi
}
\newcommand{\hatcurCCtassmg}[1]{\ifnum#1=54 %
\hatcurCCtassmgxxxxxA
\else
\ifnum#1=55 %
\hatcurCCtassmgxxxxxB
\else
\ifnum#1=56 %
\hatcurCCtassmgxxxxxC
\else
\ifnum#1=57 %
\hatcurCCtassmgxxxxxD
\else
\ifnum#1=58 %
\hatcurCCtassmgxxxxxE
\else
??????\fi
\fi
\fi
\fi
\fi
}
\newcommand{\hatcurCCtassmgshort}[1]{\ifnum#1=54 %
\hatcurCCtassmgshortxxxxxA
\else
\ifnum#1=55 %
\hatcurCCtassmgshortxxxxxB
\else
\ifnum#1=56 %
\hatcurCCtassmgshortxxxxxC
\else
\ifnum#1=57 %
\hatcurCCtassmgshortxxxxxD
\else
\ifnum#1=58 %
\hatcurCCtassmgshortxxxxxE
\else
??????\fi
\fi
\fi
\fi
\fi
}
\newcommand{\hatcurCCtassmi}[1]{\ifnum#1=54 %
\hatcurCCtassmixxxxxA
\else
\ifnum#1=55 %
\hatcurCCtassmixxxxxB
\else
\ifnum#1=56 %
\hatcurCCtassmixxxxxC
\else
\ifnum#1=57 %
\hatcurCCtassmixxxxxD
\else
\ifnum#1=58 %
\hatcurCCtassmixxxxxE
\else
??????\fi
\fi
\fi
\fi
\fi
}
\newcommand{\hatcurCCtassmI}[1]{\ifnum#1=54 %
\hatcurCCtassmIxxxxxA
\else
\ifnum#1=55 %
\hatcurCCtassmIxxxxxB
\else
\ifnum#1=56 %
\hatcurCCtassmIxxxxxC
\else
\ifnum#1=57 %
\hatcurCCtassmIxxxxxD
\else
\ifnum#1=58 %
\hatcurCCtassmIxxxxxE
\else
??????\fi
\fi
\fi
\fi
\fi
}
\newcommand{\hatcurCCtassmishort}[1]{\ifnum#1=54 %
\hatcurCCtassmishortxxxxxA
\else
\ifnum#1=55 %
\hatcurCCtassmishortxxxxxB
\else
\ifnum#1=56 %
\hatcurCCtassmishortxxxxxC
\else
\ifnum#1=57 %
\hatcurCCtassmishortxxxxxD
\else
\ifnum#1=58 %
\hatcurCCtassmishortxxxxxE
\else
??????\fi
\fi
\fi
\fi
\fi
}
\newcommand{\hatcurCCtassmIshort}[1]{\ifnum#1=54 %
\hatcurCCtassmIshortxxxxxA
\else
\ifnum#1=55 %
\hatcurCCtassmIshortxxxxxB
\else
\ifnum#1=56 %
\hatcurCCtassmIshortxxxxxC
\else
\ifnum#1=57 %
\hatcurCCtassmIshortxxxxxD
\else
\ifnum#1=58 %
\hatcurCCtassmIshortxxxxxE
\else
??????\fi
\fi
\fi
\fi
\fi
}
\newcommand{\hatcurCCtassmr}[1]{\ifnum#1=54 %
\hatcurCCtassmrxxxxxA
\else
\ifnum#1=55 %
\hatcurCCtassmrxxxxxB
\else
\ifnum#1=56 %
\hatcurCCtassmrxxxxxC
\else
\ifnum#1=57 %
\hatcurCCtassmrxxxxxD
\else
\ifnum#1=58 %
\hatcurCCtassmrxxxxxE
\else
??????\fi
\fi
\fi
\fi
\fi
}
\newcommand{\hatcurCCtassmrshort}[1]{\ifnum#1=54 %
\hatcurCCtassmrshortxxxxxA
\else
\ifnum#1=55 %
\hatcurCCtassmrshortxxxxxB
\else
\ifnum#1=56 %
\hatcurCCtassmrshortxxxxxC
\else
\ifnum#1=57 %
\hatcurCCtassmrshortxxxxxD
\else
\ifnum#1=58 %
\hatcurCCtassmrshortxxxxxE
\else
??????\fi
\fi
\fi
\fi
\fi
}
\newcommand{\hatcurCCtassmv}[1]{\ifnum#1=54 %
\hatcurCCtassmvxxxxxA
\else
\ifnum#1=55 %
\hatcurCCtassmvxxxxxB
\else
\ifnum#1=56 %
\hatcurCCtassmvxxxxxC
\else
\ifnum#1=57 %
\hatcurCCtassmvxxxxxD
\else
\ifnum#1=58 %
\hatcurCCtassmvxxxxxE
\else
??????\fi
\fi
\fi
\fi
\fi
}
\newcommand{\hatcurCCtassmvshort}[1]{\ifnum#1=54 %
\hatcurCCtassmvshortxxxxxA
\else
\ifnum#1=55 %
\hatcurCCtassmvshortxxxxxB
\else
\ifnum#1=56 %
\hatcurCCtassmvshortxxxxxC
\else
\ifnum#1=57 %
\hatcurCCtassmvshortxxxxxD
\else
\ifnum#1=58 %
\hatcurCCtassmvshortxxxxxE
\else
??????\fi
\fi
\fi
\fi
\fi
}
\newcommand{\hatcurCCtwomass}[1]{\ifnum#1=54 %
\hatcurCCtwomassxxxxxA
\else
\ifnum#1=55 %
\hatcurCCtwomassxxxxxB
\else
\ifnum#1=56 %
\hatcurCCtwomassxxxxxC
\else
\ifnum#1=57 %
\hatcurCCtwomassxxxxxD
\else
\ifnum#1=58 %
\hatcurCCtwomassxxxxxE
\else
??????\fi
\fi
\fi
\fi
\fi
}
\newcommand{\hatcurCCtwomassHmag}[1]{\ifnum#1=54 %
\hatcurCCtwomassHmagxxxxxA
\else
\ifnum#1=55 %
\hatcurCCtwomassHmagxxxxxB
\else
\ifnum#1=56 %
\hatcurCCtwomassHmagxxxxxC
\else
\ifnum#1=57 %
\hatcurCCtwomassHmagxxxxxD
\else
\ifnum#1=58 %
\hatcurCCtwomassHmagxxxxxE
\else
??????\fi
\fi
\fi
\fi
\fi
}
\newcommand{\hatcurCCtwomassJmag}[1]{\ifnum#1=54 %
\hatcurCCtwomassJmagxxxxxA
\else
\ifnum#1=55 %
\hatcurCCtwomassJmagxxxxxB
\else
\ifnum#1=56 %
\hatcurCCtwomassJmagxxxxxC
\else
\ifnum#1=57 %
\hatcurCCtwomassJmagxxxxxD
\else
\ifnum#1=58 %
\hatcurCCtwomassJmagxxxxxE
\else
??????\fi
\fi
\fi
\fi
\fi
}
\newcommand{\hatcurCCtwomassKmag}[1]{\ifnum#1=54 %
\hatcurCCtwomassKmagxxxxxA
\else
\ifnum#1=55 %
\hatcurCCtwomassKmagxxxxxB
\else
\ifnum#1=56 %
\hatcurCCtwomassKmagxxxxxC
\else
\ifnum#1=57 %
\hatcurCCtwomassKmagxxxxxD
\else
\ifnum#1=58 %
\hatcurCCtwomassKmagxxxxxE
\else
??????\fi
\fi
\fi
\fi
\fi
}
\newcommand{\hatcurfield}[1]{\ifnum#1=54 %
\hatcurfieldxxxxxA
\else
\ifnum#1=55 %
\hatcurfieldxxxxxB
\else
\ifnum#1=56 %
\hatcurfieldxxxxxC
\else
\ifnum#1=57 %
\hatcurfieldxxxxxD
\else
\ifnum#1=58 %
\hatcurfieldxxxxxE
\else
??????\fi
\fi
\fi
\fi
\fi
}
\newcommand{\hatcurhtr}[1]{\ifnum#1=54 %
\hatcurhtrxxxxxA
\else
\ifnum#1=55 %
\hatcurhtrxxxxxB
\else
\ifnum#1=56 %
\hatcurhtrxxxxxC
\else
\ifnum#1=57 %
\hatcurhtrxxxxxD
\else
\ifnum#1=58 %
\hatcurhtrxxxxxE
\else
??????\fi
\fi
\fi
\fi
\fi
}
\newcommand{\hatcurISOage}[1]{\ifnum#1=54 %
\hatcurISOagexxxxxA
\else
\ifnum#1=55 %
\hatcurISOagexxxxxB
\else
\ifnum#1=56 %
\hatcurISOagexxxxxC
\else
\ifnum#1=57 %
\hatcurISOagexxxxxD
\else
\ifnum#1=58 %
\hatcurISOagexxxxxE
\else
??????\fi
\fi
\fi
\fi
\fi
}
\newcommand{\hatcurISOageB}[1]{\ifnum#1=58 %
\hatcurISOageBxxxxxE
\else
??????\fi
}
\newcommand{\hatcurISOfeh}[1]{\ifnum#1=58 %
\hatcurISOfehxxxxxE
\else
??????\fi
}
\newcommand{\hatcurISOfehB}[1]{\ifnum#1=58 %
\hatcurISOfehBxxxxxE
\else
??????\fi
}
\newcommand{\hatcurISOlogg}[1]{\ifnum#1=54 %
\hatcurISOloggxxxxxA
\else
\ifnum#1=55 %
\hatcurISOloggxxxxxB
\else
\ifnum#1=56 %
\hatcurISOloggxxxxxC
\else
\ifnum#1=57 %
\hatcurISOloggxxxxxD
\else
\ifnum#1=58 %
\hatcurISOloggxxxxxE
\else
??????\fi
\fi
\fi
\fi
\fi
}
\newcommand{\hatcurISOlum}[1]{\ifnum#1=54 %
\hatcurISOlumxxxxxA
\else
\ifnum#1=55 %
\hatcurISOlumxxxxxB
\else
\ifnum#1=56 %
\hatcurISOlumxxxxxC
\else
\ifnum#1=57 %
\hatcurISOlumxxxxxD
\else
\ifnum#1=58 %
\hatcurISOlumxxxxxE
\else
??????\fi
\fi
\fi
\fi
\fi
}
\newcommand{\hatcurISOlumshort}[1]{\ifnum#1=54 %
\hatcurISOlumshortxxxxxA
\else
\ifnum#1=55 %
\hatcurISOlumshortxxxxxB
\else
\ifnum#1=56 %
\hatcurISOlumshortxxxxxC
\else
\ifnum#1=57 %
\hatcurISOlumshortxxxxxD
\else
\ifnum#1=58 %
\hatcurISOlumshortxxxxxE
\else
??????\fi
\fi
\fi
\fi
\fi
}
\newcommand{\hatcurISOm}[1]{\ifnum#1=54 %
\hatcurISOmxxxxxA
\else
\ifnum#1=55 %
\hatcurISOmxxxxxB
\else
\ifnum#1=56 %
\hatcurISOmxxxxxC
\else
\ifnum#1=57 %
\hatcurISOmxxxxxD
\else
\ifnum#1=58 %
\hatcurISOmxxxxxE
\else
??????\fi
\fi
\fi
\fi
\fi
}
\newcommand{\hatcurISOmB}[1]{\ifnum#1=58 %
\hatcurISOmBxxxxxE
\else
??????\fi
}
\newcommand{\hatcurISOmlong}[1]{\ifnum#1=54 %
\hatcurISOmlongxxxxxA
\else
\ifnum#1=55 %
\hatcurISOmlongxxxxxB
\else
\ifnum#1=56 %
\hatcurISOmlongxxxxxC
\else
\ifnum#1=57 %
\hatcurISOmlongxxxxxD
\else
\ifnum#1=58 %
\hatcurISOmlongxxxxxE
\else
??????\fi
\fi
\fi
\fi
\fi
}
\newcommand{\hatcurISOmlongB}[1]{\ifnum#1=58 %
\hatcurISOmlongBxxxxxE
\else
??????\fi
}
\newcommand{\hatcurISOmshort}[1]{\ifnum#1=54 %
\hatcurISOmshortxxxxxA
\else
\ifnum#1=55 %
\hatcurISOmshortxxxxxB
\else
\ifnum#1=56 %
\hatcurISOmshortxxxxxC
\else
\ifnum#1=57 %
\hatcurISOmshortxxxxxD
\else
\ifnum#1=58 %
\hatcurISOmshortxxxxxE
\else
??????\fi
\fi
\fi
\fi
\fi
}
\newcommand{\hatcurISOmshortB}[1]{\ifnum#1=58 %
\hatcurISOmshortBxxxxxE
\else
??????\fi
}
\newcommand{\hatcurISOr}[1]{\ifnum#1=54 %
\hatcurISOrxxxxxA
\else
\ifnum#1=55 %
\hatcurISOrxxxxxB
\else
\ifnum#1=56 %
\hatcurISOrxxxxxC
\else
\ifnum#1=57 %
\hatcurISOrxxxxxD
\else
\ifnum#1=58 %
\hatcurISOrxxxxxE
\else
??????\fi
\fi
\fi
\fi
\fi
}
\newcommand{\hatcurISOrho}[1]{\ifnum#1=54 %
\hatcurISOrhoxxxxxA
\else
\ifnum#1=55 %
\hatcurISOrhoxxxxxB
\else
\ifnum#1=56 %
\hatcurISOrhoxxxxxC
\else
\ifnum#1=57 %
\hatcurISOrhoxxxxxD
\else
??????\fi
\fi
\fi
\fi
}
\newcommand{\hatcurISOrholong}[1]{\ifnum#1=54 %
\hatcurISOrholongxxxxxA
\else
\ifnum#1=55 %
\hatcurISOrholongxxxxxB
\else
\ifnum#1=56 %
\hatcurISOrholongxxxxxC
\else
\ifnum#1=57 %
\hatcurISOrholongxxxxxD
\else
??????\fi
\fi
\fi
\fi
}
\newcommand{\hatcurISOrlong}[1]{\ifnum#1=54 %
\hatcurISOrlongxxxxxA
\else
\ifnum#1=55 %
\hatcurISOrlongxxxxxB
\else
\ifnum#1=56 %
\hatcurISOrlongxxxxxC
\else
\ifnum#1=57 %
\hatcurISOrlongxxxxxD
\else
\ifnum#1=58 %
\hatcurISOrlongxxxxxE
\else
??????\fi
\fi
\fi
\fi
\fi
}
\newcommand{\hatcurISOrshort}[1]{\ifnum#1=54 %
\hatcurISOrshortxxxxxA
\else
\ifnum#1=55 %
\hatcurISOrshortxxxxxB
\else
\ifnum#1=56 %
\hatcurISOrshortxxxxxC
\else
\ifnum#1=57 %
\hatcurISOrshortxxxxxD
\else
\ifnum#1=58 %
\hatcurISOrshortxxxxxE
\else
??????\fi
\fi
\fi
\fi
\fi
}
\newcommand{\hatcurISOspec}[1]{\ifnum#1=54 %
\hatcurISOspecxxxxxA
\else
\ifnum#1=55 %
\hatcurISOspecxxxxxB
\else
\ifnum#1=56 %
\hatcurISOspecxxxxxC
\else
\ifnum#1=57 %
\hatcurISOspecxxxxxD
\else
??????\fi
\fi
\fi
\fi
}
\newcommand{\hatcurISOteff}[1]{\ifnum#1=54 %
\hatcurISOteffxxxxxA
\else
\ifnum#1=55 %
\hatcurISOteffxxxxxB
\else
\ifnum#1=56 %
\hatcurISOteffxxxxxC
\else
\ifnum#1=57 %
\hatcurISOteffxxxxxD
\else
\ifnum#1=58 %
\hatcurISOteffxxxxxE
\else
??????\fi
\fi
\fi
\fi
\fi
}
\newcommand{\hatcurISOzfeh}[1]{\ifnum#1=54 %
\hatcurISOzfehxxxxxA
\else
\ifnum#1=55 %
\hatcurISOzfehxxxxxB
\else
\ifnum#1=56 %
\hatcurISOzfehxxxxxC
\else
\ifnum#1=57 %
\hatcurISOzfehxxxxxD
\else
??????\fi
\fi
\fi
\fi
}
\newcommand{\hatcurLBig}[1]{\ifnum#1=54 %
\hatcurLBigxxxxxA
\else
\ifnum#1=55 %
\hatcurLBigxxxxxB
\else
\ifnum#1=56 %
\hatcurLBigxxxxxC
\else
\ifnum#1=57 %
\hatcurLBigxxxxxD
\else
\ifnum#1=58 %
\hatcurLBigxxxxxE
\else
??????\fi
\fi
\fi
\fi
\fi
}
\newcommand{\hatcurLBii}[1]{\ifnum#1=54 %
\hatcurLBiixxxxxA
\else
\ifnum#1=55 %
\hatcurLBiixxxxxB
\else
\ifnum#1=56 %
\hatcurLBiixxxxxC
\else
\ifnum#1=57 %
\hatcurLBiixxxxxD
\else
\ifnum#1=58 %
\hatcurLBiixxxxxE
\else
??????\fi
\fi
\fi
\fi
\fi
}
\newcommand{\hatcurLBiI}[1]{\ifnum#1=54 %
\hatcurLBiIxxxxxA
\else
\ifnum#1=55 %
\hatcurLBiIxxxxxB
\else
\ifnum#1=56 %
\hatcurLBiIxxxxxC
\else
\ifnum#1=57 %
\hatcurLBiIxxxxxD
\else
\ifnum#1=58 %
\hatcurLBiIxxxxxE
\else
??????\fi
\fi
\fi
\fi
\fi
}
\newcommand{\hatcurLBiig}[1]{\ifnum#1=54 %
\hatcurLBiigxxxxxA
\else
\ifnum#1=55 %
\hatcurLBiigxxxxxB
\else
\ifnum#1=56 %
\hatcurLBiigxxxxxC
\else
\ifnum#1=57 %
\hatcurLBiigxxxxxD
\else
\ifnum#1=58 %
\hatcurLBiigxxxxxE
\else
??????\fi
\fi
\fi
\fi
\fi
}
\newcommand{\hatcurLBiii}[1]{\ifnum#1=54 %
\hatcurLBiiixxxxxA
\else
\ifnum#1=55 %
\hatcurLBiiixxxxxB
\else
\ifnum#1=56 %
\hatcurLBiiixxxxxC
\else
\ifnum#1=57 %
\hatcurLBiiixxxxxD
\else
\ifnum#1=58 %
\hatcurLBiiixxxxxE
\else
??????\fi
\fi
\fi
\fi
\fi
}
\newcommand{\hatcurLBiiI}[1]{\ifnum#1=54 %
\hatcurLBiiIxxxxxA
\else
\ifnum#1=55 %
\hatcurLBiiIxxxxxB
\else
\ifnum#1=56 %
\hatcurLBiiIxxxxxC
\else
\ifnum#1=57 %
\hatcurLBiiIxxxxxD
\else
\ifnum#1=58 %
\hatcurLBiiIxxxxxE
\else
??????\fi
\fi
\fi
\fi
\fi
}
\newcommand{\hatcurLBiikep}[1]{\ifnum#1=54 %
\hatcurLBiikepxxxxxA
\else
\ifnum#1=55 %
\hatcurLBiikepxxxxxB
\else
\ifnum#1=56 %
\hatcurLBiikepxxxxxC
\else
\ifnum#1=57 %
\hatcurLBiikepxxxxxD
\else
\ifnum#1=58 %
\hatcurLBiikepxxxxxE
\else
??????\fi
\fi
\fi
\fi
\fi
}
\newcommand{\hatcurLBiir}[1]{\ifnum#1=54 %
\hatcurLBiirxxxxxA
\else
\ifnum#1=55 %
\hatcurLBiirxxxxxB
\else
\ifnum#1=56 %
\hatcurLBiirxxxxxC
\else
\ifnum#1=57 %
\hatcurLBiirxxxxxD
\else
\ifnum#1=58 %
\hatcurLBiirxxxxxE
\else
??????\fi
\fi
\fi
\fi
\fi
}
\newcommand{\hatcurLBiiR}[1]{\ifnum#1=54 %
\hatcurLBiiRxxxxxA
\else
\ifnum#1=55 %
\hatcurLBiiRxxxxxB
\else
\ifnum#1=56 %
\hatcurLBiiRxxxxxC
\else
\ifnum#1=57 %
\hatcurLBiiRxxxxxD
\else
\ifnum#1=58 %
\hatcurLBiiRxxxxxE
\else
??????\fi
\fi
\fi
\fi
\fi
}
\newcommand{\hatcurLBiiz}[1]{\ifnum#1=54 %
\hatcurLBiizxxxxxA
\else
\ifnum#1=55 %
\hatcurLBiizxxxxxB
\else
\ifnum#1=56 %
\hatcurLBiizxxxxxC
\else
\ifnum#1=57 %
\hatcurLBiizxxxxxD
\else
\ifnum#1=58 %
\hatcurLBiizxxxxxE
\else
??????\fi
\fi
\fi
\fi
\fi
}
\newcommand{\hatcurLBikep}[1]{\ifnum#1=54 %
\hatcurLBikepxxxxxA
\else
\ifnum#1=55 %
\hatcurLBikepxxxxxB
\else
\ifnum#1=56 %
\hatcurLBikepxxxxxC
\else
\ifnum#1=57 %
\hatcurLBikepxxxxxD
\else
\ifnum#1=58 %
\hatcurLBikepxxxxxE
\else
??????\fi
\fi
\fi
\fi
\fi
}
\newcommand{\hatcurLBir}[1]{\ifnum#1=54 %
\hatcurLBirxxxxxA
\else
\ifnum#1=55 %
\hatcurLBirxxxxxB
\else
\ifnum#1=56 %
\hatcurLBirxxxxxC
\else
\ifnum#1=57 %
\hatcurLBirxxxxxD
\else
\ifnum#1=58 %
\hatcurLBirxxxxxE
\else
??????\fi
\fi
\fi
\fi
\fi
}
\newcommand{\hatcurLBiR}[1]{\ifnum#1=54 %
\hatcurLBiRxxxxxA
\else
\ifnum#1=55 %
\hatcurLBiRxxxxxB
\else
\ifnum#1=56 %
\hatcurLBiRxxxxxC
\else
\ifnum#1=57 %
\hatcurLBiRxxxxxD
\else
\ifnum#1=58 %
\hatcurLBiRxxxxxE
\else
??????\fi
\fi
\fi
\fi
\fi
}
\newcommand{\hatcurLBiz}[1]{\ifnum#1=54 %
\hatcurLBizxxxxxA
\else
\ifnum#1=55 %
\hatcurLBizxxxxxB
\else
\ifnum#1=56 %
\hatcurLBizxxxxxC
\else
\ifnum#1=57 %
\hatcurLBizxxxxxD
\else
\ifnum#1=58 %
\hatcurLBizxxxxxE
\else
??????\fi
\fi
\fi
\fi
\fi
}
\newcommand{\hatcurLCbsq}[1]{\ifnum#1=54 %
\hatcurLCbsqxxxxxA
\else
\ifnum#1=55 %
\hatcurLCbsqxxxxxB
\else
\ifnum#1=56 %
\hatcurLCbsqxxxxxC
\else
\ifnum#1=57 %
\hatcurLCbsqxxxxxD
\else
\ifnum#1=58 %
\hatcurLCbsqxxxxxE
\else
??????\fi
\fi
\fi
\fi
\fi
}
\newcommand{\hatcurLCdip}[1]{\ifnum#1=54 %
\hatcurLCdipxxxxxA
\else
\ifnum#1=55 %
\hatcurLCdipxxxxxB
\else
\ifnum#1=56 %
\hatcurLCdipxxxxxC
\else
\ifnum#1=57 %
\hatcurLCdipxxxxxD
\else
??????\fi
\fi
\fi
\fi
}
\newcommand{\hatcurLCdur}[1]{\ifnum#1=54 %
\hatcurLCdurxxxxxA
\else
\ifnum#1=55 %
\hatcurLCdurxxxxxB
\else
\ifnum#1=56 %
\hatcurLCdurxxxxxC
\else
\ifnum#1=57 %
\hatcurLCdurxxxxxD
\else
\ifnum#1=58 %
\hatcurLCdurxxxxxE
\else
??????\fi
\fi
\fi
\fi
\fi
}
\newcommand{\hatcurLCdurhr}[1]{\ifnum#1=54 %
\hatcurLCdurhrxxxxxA
\else
\ifnum#1=55 %
\hatcurLCdurhrxxxxxB
\else
\ifnum#1=56 %
\hatcurLCdurhrxxxxxC
\else
\ifnum#1=57 %
\hatcurLCdurhrxxxxxD
\else
??????\fi
\fi
\fi
\fi
}
\newcommand{\hatcurLCdurhrshort}[1]{\ifnum#1=54 %
\hatcurLCdurhrshortxxxxxA
\else
\ifnum#1=55 %
\hatcurLCdurhrshortxxxxxB
\else
\ifnum#1=56 %
\hatcurLCdurhrshortxxxxxC
\else
\ifnum#1=57 %
\hatcurLCdurhrshortxxxxxD
\else
??????\fi
\fi
\fi
\fi
}
\newcommand{\hatcurLCdurshort}[1]{\ifnum#1=54 %
\hatcurLCdurshortxxxxxA
\else
\ifnum#1=55 %
\hatcurLCdurshortxxxxxB
\else
\ifnum#1=56 %
\hatcurLCdurshortxxxxxC
\else
\ifnum#1=57 %
\hatcurLCdurshortxxxxxD
\else
\ifnum#1=58 %
\hatcurLCdurshortxxxxxE
\else
??????\fi
\fi
\fi
\fi
\fi
}
\newcommand{\hatcurLChatnetm}[1]{\ifnum#1=54 %
\hatcurLChatnetmxxxxxA
\else
\ifnum#1=55 %
\hatcurLChatnetmxxxxxB
\else
\ifnum#1=57 %
\hatcurLChatnetmxxxxxD
\else
??????\fi
\fi
\fi
}
\newcommand{\hatcurLChatnetmA}[1]{\ifnum#1=56 %
\hatcurLChatnetmAxxxxxC
\else
??????\fi
}
\newcommand{\hatcurLChatnetmB}[1]{\ifnum#1=56 %
\hatcurLChatnetmBxxxxxC
\else
??????\fi
}
\newcommand{\hatcurLCiblend}[1]{\ifnum#1=54 %
\hatcurLCiblendxxxxxA
\else
\ifnum#1=55 %
\hatcurLCiblendxxxxxB
\else
\ifnum#1=57 %
\hatcurLCiblendxxxxxD
\else
\ifnum#1=58 %
\hatcurLCiblendxxxxxE
\else
??????\fi
\fi
\fi
\fi
}
\newcommand{\hatcurLCiblendA}[1]{\ifnum#1=56 %
\hatcurLCiblendAxxxxxC
\else
??????\fi
}
\newcommand{\hatcurLCiblendB}[1]{\ifnum#1=56 %
\hatcurLCiblendBxxxxxC
\else
??????\fi
}
\newcommand{\hatcurLCimp}[1]{\ifnum#1=54 %
\hatcurLCimpxxxxxA
\else
\ifnum#1=55 %
\hatcurLCimpxxxxxB
\else
\ifnum#1=56 %
\hatcurLCimpxxxxxC
\else
\ifnum#1=57 %
\hatcurLCimpxxxxxD
\else
\ifnum#1=58 %
\hatcurLCimpxxxxxE
\else
??????\fi
\fi
\fi
\fi
\fi
}
\newcommand{\hatcurLCingdur}[1]{\ifnum#1=54 %
\hatcurLCingdurxxxxxA
\else
\ifnum#1=55 %
\hatcurLCingdurxxxxxB
\else
\ifnum#1=56 %
\hatcurLCingdurxxxxxC
\else
\ifnum#1=57 %
\hatcurLCingdurxxxxxD
\else
\ifnum#1=58 %
\hatcurLCingdurxxxxxE
\else
??????\fi
\fi
\fi
\fi
\fi
}
\newcommand{\hatcurLCP}[1]{\ifnum#1=54 %
\hatcurLCPxxxxxA
\else
\ifnum#1=55 %
\hatcurLCPxxxxxB
\else
\ifnum#1=56 %
\hatcurLCPxxxxxC
\else
\ifnum#1=57 %
\hatcurLCPxxxxxD
\else
\ifnum#1=58 %
\hatcurLCPxxxxxE
\else
??????\fi
\fi
\fi
\fi
\fi
}
\newcommand{\hatcurLCPprec}[1]{\ifnum#1=54 %
\hatcurLCPprecxxxxxA
\else
\ifnum#1=55 %
\hatcurLCPprecxxxxxB
\else
\ifnum#1=56 %
\hatcurLCPprecxxxxxC
\else
\ifnum#1=57 %
\hatcurLCPprecxxxxxD
\else
\ifnum#1=58 %
\hatcurLCPprecxxxxxE
\else
??????\fi
\fi
\fi
\fi
\fi
}
\newcommand{\hatcurLCPshort}[1]{\ifnum#1=54 %
\hatcurLCPshortxxxxxA
\else
\ifnum#1=55 %
\hatcurLCPshortxxxxxB
\else
\ifnum#1=56 %
\hatcurLCPshortxxxxxC
\else
\ifnum#1=57 %
\hatcurLCPshortxxxxxD
\else
\ifnum#1=58 %
\hatcurLCPshortxxxxxE
\else
??????\fi
\fi
\fi
\fi
\fi
}
\newcommand{\hatcurLCq}[1]{\ifnum#1=54 %
\hatcurLCqxxxxxA
\else
\ifnum#1=55 %
\hatcurLCqxxxxxB
\else
\ifnum#1=56 %
\hatcurLCqxxxxxC
\else
\ifnum#1=57 %
\hatcurLCqxxxxxD
\else
??????\fi
\fi
\fi
\fi
}
\newcommand{\hatcurLCqshort}[1]{\ifnum#1=54 %
\hatcurLCqshortxxxxxA
\else
\ifnum#1=55 %
\hatcurLCqshortxxxxxB
\else
\ifnum#1=56 %
\hatcurLCqshortxxxxxC
\else
\ifnum#1=57 %
\hatcurLCqshortxxxxxD
\else
??????\fi
\fi
\fi
\fi
}
\newcommand{\hatcurLCrho}[1]{\ifnum#1=54 %
\hatcurLCrhoxxxxxA
\else
\ifnum#1=55 %
\hatcurLCrhoxxxxxB
\else
\ifnum#1=56 %
\hatcurLCrhoxxxxxC
\else
\ifnum#1=57 %
\hatcurLCrhoxxxxxD
\else
\ifnum#1=58 %
\hatcurLCrhoxxxxxE
\else
??????\fi
\fi
\fi
\fi
\fi
}
\newcommand{\hatcurLCrprstar}[1]{\ifnum#1=54 %
\hatcurLCrprstarxxxxxA
\else
\ifnum#1=55 %
\hatcurLCrprstarxxxxxB
\else
\ifnum#1=56 %
\hatcurLCrprstarxxxxxC
\else
\ifnum#1=57 %
\hatcurLCrprstarxxxxxD
\else
\ifnum#1=58 %
\hatcurLCrprstarxxxxxE
\else
??????\fi
\fi
\fi
\fi
\fi
}
\newcommand{\hatcurLCT}[1]{\ifnum#1=54 %
\hatcurLCTxxxxxA
\else
\ifnum#1=55 %
\hatcurLCTxxxxxB
\else
\ifnum#1=56 %
\hatcurLCTxxxxxC
\else
\ifnum#1=57 %
\hatcurLCTxxxxxD
\else
\ifnum#1=58 %
\hatcurLCTxxxxxE
\else
??????\fi
\fi
\fi
\fi
\fi
}
\newcommand{\hatcurLCTA}[1]{\ifnum#1=54 %
\hatcurLCTAxxxxxA
\else
\ifnum#1=55 %
\hatcurLCTAxxxxxB
\else
\ifnum#1=56 %
\hatcurLCTAxxxxxC
\else
\ifnum#1=57 %
\hatcurLCTAxxxxxD
\else
\ifnum#1=58 %
\hatcurLCTAxxxxxE
\else
??????\fi
\fi
\fi
\fi
\fi
}
\newcommand{\hatcurLCTB}[1]{\ifnum#1=54 %
\hatcurLCTBxxxxxA
\else
\ifnum#1=55 %
\hatcurLCTBxxxxxB
\else
\ifnum#1=56 %
\hatcurLCTBxxxxxC
\else
\ifnum#1=57 %
\hatcurLCTBxxxxxD
\else
\ifnum#1=58 %
\hatcurLCTBxxxxxE
\else
??????\fi
\fi
\fi
\fi
\fi
}
\newcommand{\hatcurLCzeta}[1]{\ifnum#1=54 %
\hatcurLCzetaxxxxxA
\else
\ifnum#1=55 %
\hatcurLCzetaxxxxxB
\else
\ifnum#1=56 %
\hatcurLCzetaxxxxxC
\else
\ifnum#1=57 %
\hatcurLCzetaxxxxxD
\else
\ifnum#1=58 %
\hatcurLCzetaxxxxxE
\else
??????\fi
\fi
\fi
\fi
\fi
}
\newcommand{\hatcurPPaequiv}[1]{\ifnum#1=54 %
\hatcurPPaequivxxxxxA
\else
\ifnum#1=55 %
\hatcurPPaequivxxxxxB
\else
\ifnum#1=56 %
\hatcurPPaequivxxxxxC
\else
\ifnum#1=57 %
\hatcurPPaequivxxxxxD
\else
??????\fi
\fi
\fi
\fi
}
\newcommand{\hatcurPPar}[1]{\ifnum#1=54 %
\hatcurPParxxxxxA
\else
\ifnum#1=55 %
\hatcurPParxxxxxB
\else
\ifnum#1=56 %
\hatcurPParxxxxxC
\else
\ifnum#1=57 %
\hatcurPParxxxxxD
\else
\ifnum#1=58 %
\hatcurPParxxxxxE
\else
??????\fi
\fi
\fi
\fi
\fi
}
\newcommand{\hatcurPParel}[1]{\ifnum#1=54 %
\hatcurPParelxxxxxA
\else
\ifnum#1=55 %
\hatcurPParelxxxxxB
\else
\ifnum#1=56 %
\hatcurPParelxxxxxC
\else
\ifnum#1=57 %
\hatcurPParelxxxxxD
\else
\ifnum#1=58 %
\hatcurPParelxxxxxE
\else
??????\fi
\fi
\fi
\fi
\fi
}
\newcommand{\hatcurPPfluxap}[1]{\ifnum#1=54 %
\hatcurPPfluxapxxxxxA
\else
\ifnum#1=55 %
\hatcurPPfluxapxxxxxB
\else
\ifnum#1=56 %
\hatcurPPfluxapxxxxxC
\else
\ifnum#1=57 %
\hatcurPPfluxapxxxxxD
\else
??????\fi
\fi
\fi
\fi
}
\newcommand{\hatcurPPfluxapdim}[1]{\ifnum#1=54 %
\hatcurPPfluxapdimxxxxxA
\else
\ifnum#1=55 %
\hatcurPPfluxapdimxxxxxB
\else
\ifnum#1=56 %
\hatcurPPfluxapdimxxxxxC
\else
\ifnum#1=57 %
\hatcurPPfluxapdimxxxxxD
\else
??????\fi
\fi
\fi
\fi
}
\newcommand{\hatcurPPfluxavg}[1]{\ifnum#1=54 %
\hatcurPPfluxavgxxxxxA
\else
\ifnum#1=55 %
\hatcurPPfluxavgxxxxxB
\else
\ifnum#1=56 %
\hatcurPPfluxavgxxxxxC
\else
\ifnum#1=57 %
\hatcurPPfluxavgxxxxxD
\else
\ifnum#1=58 %
\hatcurPPfluxavgxxxxxE
\else
??????\fi
\fi
\fi
\fi
\fi
}
\newcommand{\hatcurPPfluxavgdim}[1]{\ifnum#1=54 %
\hatcurPPfluxavgdimxxxxxA
\else
\ifnum#1=55 %
\hatcurPPfluxavgdimxxxxxB
\else
\ifnum#1=56 %
\hatcurPPfluxavgdimxxxxxC
\else
\ifnum#1=57 %
\hatcurPPfluxavgdimxxxxxD
\else
??????\fi
\fi
\fi
\fi
}
\newcommand{\hatcurPPfluxavglog}[1]{\ifnum#1=54 %
\hatcurPPfluxavglogxxxxxA
\else
\ifnum#1=55 %
\hatcurPPfluxavglogxxxxxB
\else
\ifnum#1=56 %
\hatcurPPfluxavglogxxxxxC
\else
\ifnum#1=57 %
\hatcurPPfluxavglogxxxxxD
\else
\ifnum#1=58 %
\hatcurPPfluxavglogxxxxxE
\else
??????\fi
\fi
\fi
\fi
\fi
}
\newcommand{\hatcurPPfluxperi}[1]{\ifnum#1=54 %
\hatcurPPfluxperixxxxxA
\else
\ifnum#1=55 %
\hatcurPPfluxperixxxxxB
\else
\ifnum#1=56 %
\hatcurPPfluxperixxxxxC
\else
\ifnum#1=57 %
\hatcurPPfluxperixxxxxD
\else
??????\fi
\fi
\fi
\fi
}
\newcommand{\hatcurPPfluxperidim}[1]{\ifnum#1=54 %
\hatcurPPfluxperidimxxxxxA
\else
\ifnum#1=55 %
\hatcurPPfluxperidimxxxxxB
\else
\ifnum#1=56 %
\hatcurPPfluxperidimxxxxxC
\else
\ifnum#1=57 %
\hatcurPPfluxperidimxxxxxD
\else
??????\fi
\fi
\fi
\fi
}
\newcommand{\hatcurPPg}[1]{\ifnum#1=54 %
\hatcurPPgxxxxxA
\else
\ifnum#1=55 %
\hatcurPPgxxxxxB
\else
\ifnum#1=56 %
\hatcurPPgxxxxxC
\else
\ifnum#1=57 %
\hatcurPPgxxxxxD
\else
??????\fi
\fi
\fi
\fi
}
\newcommand{\hatcurPPi}[1]{\ifnum#1=54 %
\hatcurPPixxxxxA
\else
\ifnum#1=55 %
\hatcurPPixxxxxB
\else
\ifnum#1=56 %
\hatcurPPixxxxxC
\else
\ifnum#1=57 %
\hatcurPPixxxxxD
\else
\ifnum#1=58 %
\hatcurPPixxxxxE
\else
??????\fi
\fi
\fi
\fi
\fi
}
\newcommand{\hatcurPPlogg}[1]{\ifnum#1=54 %
\hatcurPPloggxxxxxA
\else
\ifnum#1=55 %
\hatcurPPloggxxxxxB
\else
\ifnum#1=56 %
\hatcurPPloggxxxxxC
\else
\ifnum#1=57 %
\hatcurPPloggxxxxxD
\else
\ifnum#1=58 %
\hatcurPPloggxxxxxE
\else
??????\fi
\fi
\fi
\fi
\fi
}
\newcommand{\hatcurPPm}[1]{\ifnum#1=54 %
\hatcurPPmxxxxxA
\else
\ifnum#1=55 %
\hatcurPPmxxxxxB
\else
\ifnum#1=56 %
\hatcurPPmxxxxxC
\else
\ifnum#1=57 %
\hatcurPPmxxxxxD
\else
\ifnum#1=58 %
\hatcurPPmxxxxxE
\else
??????\fi
\fi
\fi
\fi
\fi
}
\newcommand{\hatcurPPme}[1]{\ifnum#1=54 %
\hatcurPPmexxxxxA
\else
\ifnum#1=55 %
\hatcurPPmexxxxxB
\else
\ifnum#1=56 %
\hatcurPPmexxxxxC
\else
\ifnum#1=57 %
\hatcurPPmexxxxxD
\else
??????\fi
\fi
\fi
\fi
}
\newcommand{\hatcurPPmelong}[1]{\ifnum#1=54 %
\hatcurPPmelongxxxxxA
\else
\ifnum#1=55 %
\hatcurPPmelongxxxxxB
\else
\ifnum#1=56 %
\hatcurPPmelongxxxxxC
\else
\ifnum#1=57 %
\hatcurPPmelongxxxxxD
\else
??????\fi
\fi
\fi
\fi
}
\newcommand{\hatcurPPmeshort}[1]{\ifnum#1=54 %
\hatcurPPmeshortxxxxxA
\else
\ifnum#1=55 %
\hatcurPPmeshortxxxxxB
\else
\ifnum#1=56 %
\hatcurPPmeshortxxxxxC
\else
\ifnum#1=57 %
\hatcurPPmeshortxxxxxD
\else
??????\fi
\fi
\fi
\fi
}
\newcommand{\hatcurPPmlong}[1]{\ifnum#1=54 %
\hatcurPPmlongxxxxxA
\else
\ifnum#1=55 %
\hatcurPPmlongxxxxxB
\else
\ifnum#1=56 %
\hatcurPPmlongxxxxxC
\else
\ifnum#1=57 %
\hatcurPPmlongxxxxxD
\else
\ifnum#1=58 %
\hatcurPPmlongxxxxxE
\else
??????\fi
\fi
\fi
\fi
\fi
}
\newcommand{\hatcurPPmrcorr}[1]{\ifnum#1=54 %
\hatcurPPmrcorrxxxxxA
\else
\ifnum#1=55 %
\hatcurPPmrcorrxxxxxB
\else
\ifnum#1=56 %
\hatcurPPmrcorrxxxxxC
\else
\ifnum#1=57 %
\hatcurPPmrcorrxxxxxD
\else
??????\fi
\fi
\fi
\fi
}
\newcommand{\hatcurPPmshort}[1]{\ifnum#1=54 %
\hatcurPPmshortxxxxxA
\else
\ifnum#1=55 %
\hatcurPPmshortxxxxxB
\else
\ifnum#1=56 %
\hatcurPPmshortxxxxxC
\else
\ifnum#1=57 %
\hatcurPPmshortxxxxxD
\else
\ifnum#1=58 %
\hatcurPPmshortxxxxxE
\else
??????\fi
\fi
\fi
\fi
\fi
}
\newcommand{\hatcurPPmtwosiglim}[1]{\ifnum#1=58 %
\hatcurPPmtwosiglimxxxxxE
\else
??????\fi
}
\newcommand{\hatcurPPperi}[1]{\ifnum#1=54 %
\hatcurPPperixxxxxA
\else
\ifnum#1=55 %
\hatcurPPperixxxxxB
\else
\ifnum#1=56 %
\hatcurPPperixxxxxC
\else
\ifnum#1=57 %
\hatcurPPperixxxxxD
\else
??????\fi
\fi
\fi
\fi
}
\newcommand{\hatcurPPphiconj}[1]{\ifnum#1=54 %
\hatcurPPphiconjxxxxxA
\else
\ifnum#1=55 %
\hatcurPPphiconjxxxxxB
\else
\ifnum#1=56 %
\hatcurPPphiconjxxxxxC
\else
\ifnum#1=57 %
\hatcurPPphiconjxxxxxD
\else
??????\fi
\fi
\fi
\fi
}
\newcommand{\hatcurPPr}[1]{\ifnum#1=54 %
\hatcurPPrxxxxxA
\else
\ifnum#1=55 %
\hatcurPPrxxxxxB
\else
\ifnum#1=56 %
\hatcurPPrxxxxxC
\else
\ifnum#1=57 %
\hatcurPPrxxxxxD
\else
\ifnum#1=58 %
\hatcurPPrxxxxxE
\else
??????\fi
\fi
\fi
\fi
\fi
}
\newcommand{\hatcurPPre}[1]{\ifnum#1=54 %
\hatcurPPrexxxxxA
\else
\ifnum#1=55 %
\hatcurPPrexxxxxB
\else
\ifnum#1=56 %
\hatcurPPrexxxxxC
\else
\ifnum#1=57 %
\hatcurPPrexxxxxD
\else
??????\fi
\fi
\fi
\fi
}
\newcommand{\hatcurPPrelong}[1]{\ifnum#1=54 %
\hatcurPPrelongxxxxxA
\else
\ifnum#1=55 %
\hatcurPPrelongxxxxxB
\else
\ifnum#1=56 %
\hatcurPPrelongxxxxxC
\else
\ifnum#1=57 %
\hatcurPPrelongxxxxxD
\else
??????\fi
\fi
\fi
\fi
}
\newcommand{\hatcurPPreshort}[1]{\ifnum#1=54 %
\hatcurPPreshortxxxxxA
\else
\ifnum#1=55 %
\hatcurPPreshortxxxxxB
\else
\ifnum#1=56 %
\hatcurPPreshortxxxxxC
\else
\ifnum#1=57 %
\hatcurPPreshortxxxxxD
\else
??????\fi
\fi
\fi
\fi
}
\newcommand{\hatcurPPrho}[1]{\ifnum#1=54 %
\hatcurPPrhoxxxxxA
\else
\ifnum#1=55 %
\hatcurPPrhoxxxxxB
\else
\ifnum#1=56 %
\hatcurPPrhoxxxxxC
\else
\ifnum#1=57 %
\hatcurPPrhoxxxxxD
\else
\ifnum#1=58 %
\hatcurPPrhoxxxxxE
\else
??????\fi
\fi
\fi
\fi
\fi
}
\newcommand{\hatcurPPrlong}[1]{\ifnum#1=54 %
\hatcurPPrlongxxxxxA
\else
\ifnum#1=55 %
\hatcurPPrlongxxxxxB
\else
\ifnum#1=56 %
\hatcurPPrlongxxxxxC
\else
\ifnum#1=57 %
\hatcurPPrlongxxxxxD
\else
\ifnum#1=58 %
\hatcurPPrlongxxxxxE
\else
??????\fi
\fi
\fi
\fi
\fi
}
\newcommand{\hatcurPPrshort}[1]{\ifnum#1=54 %
\hatcurPPrshortxxxxxA
\else
\ifnum#1=55 %
\hatcurPPrshortxxxxxB
\else
\ifnum#1=56 %
\hatcurPPrshortxxxxxC
\else
\ifnum#1=57 %
\hatcurPPrshortxxxxxD
\else
\ifnum#1=58 %
\hatcurPPrshortxxxxxE
\else
??????\fi
\fi
\fi
\fi
\fi
}
\newcommand{\hatcurPPtcirc}[1]{\ifnum#1=54 %
\hatcurPPtcircxxxxxA
\else
\ifnum#1=55 %
\hatcurPPtcircxxxxxB
\else
\ifnum#1=56 %
\hatcurPPtcircxxxxxC
\else
\ifnum#1=57 %
\hatcurPPtcircxxxxxD
\else
??????\fi
\fi
\fi
\fi
}
\newcommand{\hatcurPPteff}[1]{\ifnum#1=54 %
\hatcurPPteffxxxxxA
\else
\ifnum#1=55 %
\hatcurPPteffxxxxxB
\else
\ifnum#1=56 %
\hatcurPPteffxxxxxC
\else
\ifnum#1=57 %
\hatcurPPteffxxxxxD
\else
\ifnum#1=58 %
\hatcurPPteffxxxxxE
\else
??????\fi
\fi
\fi
\fi
\fi
}
\newcommand{\hatcurPPtheta}[1]{\ifnum#1=54 %
\hatcurPPthetaxxxxxA
\else
\ifnum#1=55 %
\hatcurPPthetaxxxxxB
\else
\ifnum#1=56 %
\hatcurPPthetaxxxxxC
\else
\ifnum#1=57 %
\hatcurPPthetaxxxxxD
\else
\ifnum#1=58 %
\hatcurPPthetaxxxxxE
\else
??????\fi
\fi
\fi
\fi
\fi
}
\newcommand{\hatcurPPtinfall}[1]{\ifnum#1=54 %
\hatcurPPtinfallxxxxxA
\else
\ifnum#1=55 %
\hatcurPPtinfallxxxxxB
\else
\ifnum#1=56 %
\hatcurPPtinfallxxxxxC
\else
\ifnum#1=57 %
\hatcurPPtinfallxxxxxD
\else
??????\fi
\fi
\fi
\fi
}
\newcommand{\hatcurRVeccen}[1]{\ifnum#1=54 %
\hatcurRVeccenxxxxxA
\else
\ifnum#1=55 %
\hatcurRVeccenxxxxxB
\else
\ifnum#1=56 %
\hatcurRVeccenxxxxxC
\else
\ifnum#1=57 %
\hatcurRVeccenxxxxxD
\else
??????\fi
\fi
\fi
\fi
}
\newcommand{\hatcurRVeccentwosiglim}[1]{\ifnum#1=54 %
\hatcurRVeccentwosiglimxxxxxA
\else
\ifnum#1=55 %
\hatcurRVeccentwosiglimxxxxxB
\else
\ifnum#1=56 %
\hatcurRVeccentwosiglimxxxxxC
\else
\ifnum#1=57 %
\hatcurRVeccentwosiglimxxxxxD
\else
??????\fi
\fi
\fi
\fi
}
\newcommand{\hatcurRVfitrms}[1]{\ifnum#1=55 %
\hatcurRVfitrmsxxxxxB
\else
\ifnum#1=57 %
\hatcurRVfitrmsxxxxxD
\else
??????\fi
\fi
}
\newcommand{\hatcurRVfitrmsA}[1]{\ifnum#1=54 %
\hatcurRVfitrmsAxxxxxA
\else
\ifnum#1=56 %
\hatcurRVfitrmsAxxxxxC
\else
\ifnum#1=58 %
\hatcurRVfitrmsAxxxxxE
\else
??????\fi
\fi
\fi
}
\newcommand{\hatcurRVfitrmsB}[1]{\ifnum#1=54 %
\hatcurRVfitrmsBxxxxxA
\else
\ifnum#1=56 %
\hatcurRVfitrmsBxxxxxC
\else
??????\fi
\fi
}
\newcommand{\hatcurRVgamma}[1]{\ifnum#1=55 %
\hatcurRVgammaxxxxxB
\else
\ifnum#1=57 %
\hatcurRVgammaxxxxxD
\else
??????\fi
\fi
}
\newcommand{\hatcurRVgammaA}[1]{\ifnum#1=54 %
\hatcurRVgammaAxxxxxA
\else
\ifnum#1=56 %
\hatcurRVgammaAxxxxxC
\else
\ifnum#1=58 %
\hatcurRVgammaAxxxxxE
\else
??????\fi
\fi
\fi
}
\newcommand{\hatcurRVgammaB}[1]{\ifnum#1=54 %
\hatcurRVgammaBxxxxxA
\else
\ifnum#1=56 %
\hatcurRVgammaBxxxxxC
\else
\ifnum#1=58 %
\hatcurRVgammaBxxxxxE
\else
??????\fi
\fi
\fi
}
\newcommand{\hatcurRVh}[1]{\ifnum#1=54 %
\hatcurRVhxxxxxA
\else
\ifnum#1=55 %
\hatcurRVhxxxxxB
\else
\ifnum#1=56 %
\hatcurRVhxxxxxC
\else
\ifnum#1=57 %
\hatcurRVhxxxxxD
\else
??????\fi
\fi
\fi
\fi
}
\newcommand{\hatcurRVjitter}[1]{\ifnum#1=55 %
\hatcurRVjitterxxxxxB
\else
\ifnum#1=57 %
\hatcurRVjitterxxxxxD
\else
??????\fi
\fi
}
\newcommand{\hatcurRVjitterA}[1]{\ifnum#1=54 %
\hatcurRVjitterAxxxxxA
\else
\ifnum#1=56 %
\hatcurRVjitterAxxxxxC
\else
\ifnum#1=58 %
\hatcurRVjitterAxxxxxE
\else
??????\fi
\fi
\fi
}
\newcommand{\hatcurRVjitterB}[1]{\ifnum#1=54 %
\hatcurRVjitterBxxxxxA
\else
\ifnum#1=56 %
\hatcurRVjitterBxxxxxC
\else
\ifnum#1=58 %
\hatcurRVjitterBxxxxxE
\else
??????\fi
\fi
\fi
}
\newcommand{\hatcurRVjittertwosiglim}[1]{\ifnum#1=55 %
\hatcurRVjittertwosiglimxxxxxB
\else
\ifnum#1=57 %
\hatcurRVjittertwosiglimxxxxxD
\else
??????\fi
\fi
}
\newcommand{\hatcurRVjittertwosiglimA}[1]{\ifnum#1=54 %
\hatcurRVjittertwosiglimAxxxxxA
\else
\ifnum#1=56 %
\hatcurRVjittertwosiglimAxxxxxC
\else
\ifnum#1=58 %
\hatcurRVjittertwosiglimAxxxxxE
\else
??????\fi
\fi
\fi
}
\newcommand{\hatcurRVjittertwosiglimB}[1]{\ifnum#1=54 %
\hatcurRVjittertwosiglimBxxxxxA
\else
\ifnum#1=56 %
\hatcurRVjittertwosiglimBxxxxxC
\else
\ifnum#1=58 %
\hatcurRVjittertwosiglimBxxxxxE
\else
??????\fi
\fi
\fi
}
\newcommand{\hatcurRVk}[1]{\ifnum#1=54 %
\hatcurRVkxxxxxA
\else
\ifnum#1=55 %
\hatcurRVkxxxxxB
\else
\ifnum#1=56 %
\hatcurRVkxxxxxC
\else
\ifnum#1=57 %
\hatcurRVkxxxxxD
\else
??????\fi
\fi
\fi
\fi
}
\newcommand{\hatcurRVK}[1]{\ifnum#1=54 %
\hatcurRVKxxxxxA
\else
\ifnum#1=55 %
\hatcurRVKxxxxxB
\else
\ifnum#1=56 %
\hatcurRVKxxxxxC
\else
\ifnum#1=57 %
\hatcurRVKxxxxxD
\else
\ifnum#1=58 %
\hatcurRVKxxxxxE
\else
??????\fi
\fi
\fi
\fi
\fi
}
\newcommand{\hatcurRVomega}[1]{\ifnum#1=54 %
\hatcurRVomegaxxxxxA
\else
\ifnum#1=55 %
\hatcurRVomegaxxxxxB
\else
\ifnum#1=56 %
\hatcurRVomegaxxxxxC
\else
\ifnum#1=57 %
\hatcurRVomegaxxxxxD
\else
??????\fi
\fi
\fi
\fi
}
\newcommand{\hatcurRVrh}[1]{\ifnum#1=54 %
\hatcurRVrhxxxxxA
\else
\ifnum#1=55 %
\hatcurRVrhxxxxxB
\else
\ifnum#1=56 %
\hatcurRVrhxxxxxC
\else
\ifnum#1=57 %
\hatcurRVrhxxxxxD
\else
??????\fi
\fi
\fi
\fi
}
\newcommand{\hatcurRVrk}[1]{\ifnum#1=54 %
\hatcurRVrkxxxxxA
\else
\ifnum#1=55 %
\hatcurRVrkxxxxxB
\else
\ifnum#1=56 %
\hatcurRVrkxxxxxC
\else
\ifnum#1=57 %
\hatcurRVrkxxxxxD
\else
??????\fi
\fi
\fi
\fi
}
\newcommand{\hatcurRVtrone}[1]{\ifnum#1=54 %
\hatcurRVtronexxxxxA
\else
\ifnum#1=55 %
\hatcurRVtronexxxxxB
\else
\ifnum#1=56 %
\hatcurRVtronexxxxxC
\else
\ifnum#1=57 %
\hatcurRVtronexxxxxD
\else
??????\fi
\fi
\fi
\fi
}
\newcommand{\hatcurRVtrtwo}[1]{\ifnum#1=54 %
\hatcurRVtrtwoxxxxxA
\else
\ifnum#1=55 %
\hatcurRVtrtwoxxxxxB
\else
\ifnum#1=56 %
\hatcurRVtrtwoxxxxxC
\else
\ifnum#1=57 %
\hatcurRVtrtwoxxxxxD
\else
??????\fi
\fi
\fi
\fi
}
\newcommand{\hatcurSMEiilogg}[1]{\ifnum#1=54 %
\hatcurSMEiiloggxxxxxA
\else
\ifnum#1=55 %
\hatcurSMEiiloggxxxxxB
\else
\ifnum#1=56 %
\hatcurSMEiiloggxxxxxC
\else
\ifnum#1=57 %
\hatcurSMEiiloggxxxxxD
\else
\ifnum#1=58 %
\hatcurSMEiiloggxxxxxE
\else
??????\fi
\fi
\fi
\fi
\fi
}
\newcommand{\hatcurSMEiiteff}[1]{\ifnum#1=54 %
\hatcurSMEiiteffxxxxxA
\else
\ifnum#1=55 %
\hatcurSMEiiteffxxxxxB
\else
\ifnum#1=56 %
\hatcurSMEiiteffxxxxxC
\else
\ifnum#1=57 %
\hatcurSMEiiteffxxxxxD
\else
\ifnum#1=58 %
\hatcurSMEiiteffxxxxxE
\else
??????\fi
\fi
\fi
\fi
\fi
}
\newcommand{\hatcurSMEiivmac}[1]{\ifnum#1=54 %
\hatcurSMEiivmacxxxxxA
\else
\ifnum#1=55 %
\hatcurSMEiivmacxxxxxB
\else
\ifnum#1=56 %
\hatcurSMEiivmacxxxxxC
\else
\ifnum#1=57 %
\hatcurSMEiivmacxxxxxD
\else
\ifnum#1=58 %
\hatcurSMEiivmacxxxxxE
\else
??????\fi
\fi
\fi
\fi
\fi
}
\newcommand{\hatcurSMEiivmic}[1]{\ifnum#1=54 %
\hatcurSMEiivmicxxxxxA
\else
\ifnum#1=55 %
\hatcurSMEiivmicxxxxxB
\else
\ifnum#1=56 %
\hatcurSMEiivmicxxxxxC
\else
\ifnum#1=57 %
\hatcurSMEiivmicxxxxxD
\else
\ifnum#1=58 %
\hatcurSMEiivmicxxxxxE
\else
??????\fi
\fi
\fi
\fi
\fi
}
\newcommand{\hatcurSMEiivsin}[1]{\ifnum#1=54 %
\hatcurSMEiivsinxxxxxA
\else
\ifnum#1=55 %
\hatcurSMEiivsinxxxxxB
\else
\ifnum#1=56 %
\hatcurSMEiivsinxxxxxC
\else
\ifnum#1=57 %
\hatcurSMEiivsinxxxxxD
\else
\ifnum#1=58 %
\hatcurSMEiivsinxxxxxE
\else
??????\fi
\fi
\fi
\fi
\fi
}
\newcommand{\hatcurSMEiizfeh}[1]{\ifnum#1=54 %
\hatcurSMEiizfehxxxxxA
\else
\ifnum#1=55 %
\hatcurSMEiizfehxxxxxB
\else
\ifnum#1=56 %
\hatcurSMEiizfehxxxxxC
\else
\ifnum#1=57 %
\hatcurSMEiizfehxxxxxD
\else
\ifnum#1=58 %
\hatcurSMEiizfehxxxxxE
\else
??????\fi
\fi
\fi
\fi
\fi
}
\newcommand{\hatcurSMEiizfehshort}[1]{\ifnum#1=54 %
\hatcurSMEiizfehshortxxxxxA
\else
\ifnum#1=55 %
\hatcurSMEiizfehshortxxxxxB
\else
\ifnum#1=56 %
\hatcurSMEiizfehshortxxxxxC
\else
\ifnum#1=57 %
\hatcurSMEiizfehshortxxxxxD
\else
\ifnum#1=58 %
\hatcurSMEiizfehshortxxxxxE
\else
??????\fi
\fi
\fi
\fi
\fi
}
\newcommand{\hatcurSMEilogg}[1]{\ifnum#1=54 %
\hatcurSMEiloggxxxxxA
\else
\ifnum#1=55 %
\hatcurSMEiloggxxxxxB
\else
\ifnum#1=56 %
\hatcurSMEiloggxxxxxC
\else
\ifnum#1=57 %
\hatcurSMEiloggxxxxxD
\else
\ifnum#1=58 %
\hatcurSMEiloggxxxxxE
\else
??????\fi
\fi
\fi
\fi
\fi
}
\newcommand{\hatcurSMEiteff}[1]{\ifnum#1=54 %
\hatcurSMEiteffxxxxxA
\else
\ifnum#1=55 %
\hatcurSMEiteffxxxxxB
\else
\ifnum#1=56 %
\hatcurSMEiteffxxxxxC
\else
\ifnum#1=57 %
\hatcurSMEiteffxxxxxD
\else
\ifnum#1=58 %
\hatcurSMEiteffxxxxxE
\else
??????\fi
\fi
\fi
\fi
\fi
}
\newcommand{\hatcurSMEivmac}[1]{\ifnum#1=54 %
\hatcurSMEivmacxxxxxA
\else
\ifnum#1=55 %
\hatcurSMEivmacxxxxxB
\else
\ifnum#1=56 %
\hatcurSMEivmacxxxxxC
\else
\ifnum#1=57 %
\hatcurSMEivmacxxxxxD
\else
\ifnum#1=58 %
\hatcurSMEivmacxxxxxE
\else
??????\fi
\fi
\fi
\fi
\fi
}
\newcommand{\hatcurSMEivmic}[1]{\ifnum#1=54 %
\hatcurSMEivmicxxxxxA
\else
\ifnum#1=55 %
\hatcurSMEivmicxxxxxB
\else
\ifnum#1=56 %
\hatcurSMEivmicxxxxxC
\else
\ifnum#1=57 %
\hatcurSMEivmicxxxxxD
\else
\ifnum#1=58 %
\hatcurSMEivmicxxxxxE
\else
??????\fi
\fi
\fi
\fi
\fi
}
\newcommand{\hatcurSMEivsin}[1]{\ifnum#1=54 %
\hatcurSMEivsinxxxxxA
\else
\ifnum#1=55 %
\hatcurSMEivsinxxxxxB
\else
\ifnum#1=56 %
\hatcurSMEivsinxxxxxC
\else
\ifnum#1=57 %
\hatcurSMEivsinxxxxxD
\else
\ifnum#1=58 %
\hatcurSMEivsinxxxxxE
\else
??????\fi
\fi
\fi
\fi
\fi
}
\newcommand{\hatcurSMEizfeh}[1]{\ifnum#1=54 %
\hatcurSMEizfehxxxxxA
\else
\ifnum#1=55 %
\hatcurSMEizfehxxxxxB
\else
\ifnum#1=56 %
\hatcurSMEizfehxxxxxC
\else
\ifnum#1=57 %
\hatcurSMEizfehxxxxxD
\else
\ifnum#1=58 %
\hatcurSMEizfehxxxxxE
\else
??????\fi
\fi
\fi
\fi
\fi
}
\newcommand{\hatcurSMEizfehshort}[1]{\ifnum#1=54 %
\hatcurSMEizfehshortxxxxxA
\else
\ifnum#1=55 %
\hatcurSMEizfehshortxxxxxB
\else
\ifnum#1=56 %
\hatcurSMEizfehshortxxxxxC
\else
\ifnum#1=57 %
\hatcurSMEizfehshortxxxxxD
\else
\ifnum#1=58 %
\hatcurSMEizfehshortxxxxxE
\else
??????\fi
\fi
\fi
\fi
\fi
}
\newcommand{\hatcurXAv}[1]{\ifnum#1=54 %
\hatcurXAvxxxxxA
\else
\ifnum#1=55 %
\hatcurXAvxxxxxB
\else
\ifnum#1=56 %
\hatcurXAvxxxxxC
\else
\ifnum#1=57 %
\hatcurXAvxxxxxD
\else
\ifnum#1=58 %
\hatcurXAvxxxxxE
\else
??????\fi
\fi
\fi
\fi
\fi
}
\newcommand{\hatcurXdist}[1]{\ifnum#1=54 %
\hatcurXdistxxxxxA
\else
\ifnum#1=55 %
\hatcurXdistxxxxxB
\else
\ifnum#1=56 %
\hatcurXdistxxxxxC
\else
\ifnum#1=57 %
\hatcurXdistxxxxxD
\else
\ifnum#1=58 %
\hatcurXdistxxxxxE
\else
??????\fi
\fi
\fi
\fi
\fi
}
\newcommand{\hatcurXdistred}[1]{\ifnum#1=54 %
\hatcurXdistredxxxxxA
\else
\ifnum#1=55 %
\hatcurXdistredxxxxxB
\else
\ifnum#1=56 %
\hatcurXdistredxxxxxC
\else
\ifnum#1=57 %
\hatcurXdistredxxxxxD
\else
\ifnum#1=58 %
\hatcurXdistredxxxxxE
\else
??????\fi
\fi
\fi
\fi
\fi
}
\newcommand{\hatcurXEBV}[1]{\ifnum#1=54 %
\hatcurXEBVxxxxxA
\else
\ifnum#1=55 %
\hatcurXEBVxxxxxB
\else
\ifnum#1=56 %
\hatcurXEBVxxxxxC
\else
\ifnum#1=57 %
\hatcurXEBVxxxxxD
\else
??????\fi
\fi
\fi
\fi
}
\newcommand{\hatcurXsecdur}[1]{\ifnum#1=54 %
\hatcurXsecdurxxxxxA
\else
\ifnum#1=55 %
\hatcurXsecdurxxxxxB
\else
\ifnum#1=56 %
\hatcurXsecdurxxxxxC
\else
\ifnum#1=57 %
\hatcurXsecdurxxxxxD
\else
??????\fi
\fi
\fi
\fi
}
\newcommand{\hatcurXsecingdur}[1]{\ifnum#1=54 %
\hatcurXsecingdurxxxxxA
\else
\ifnum#1=55 %
\hatcurXsecingdurxxxxxB
\else
\ifnum#1=56 %
\hatcurXsecingdurxxxxxC
\else
\ifnum#1=57 %
\hatcurXsecingdurxxxxxD
\else
??????\fi
\fi
\fi
\fi
}
\newcommand{\hatcurXsecondary}[1]{\ifnum#1=54 %
\hatcurXsecondaryxxxxxA
\else
\ifnum#1=55 %
\hatcurXsecondaryxxxxxB
\else
\ifnum#1=56 %
\hatcurXsecondaryxxxxxC
\else
\ifnum#1=57 %
\hatcurXsecondaryxxxxxD
\else
??????\fi
\fi
\fi
\fi
}
\newcommand{\hatcurXsecphase}[1]{\ifnum#1=54 %
\hatcurXsecphasexxxxxA
\else
\ifnum#1=55 %
\hatcurXsecphasexxxxxB
\else
\ifnum#1=56 %
\hatcurXsecphasexxxxxC
\else
\ifnum#1=57 %
\hatcurXsecphasexxxxxD
\else
??????\fi
\fi
\fi
\fi
}

\newcommand{\hatcurhtreccenxxxxxA}{HATS700-017}                      % Original HTR name of target
\newcommand{\hatcurfieldeccenxxxxxA}{\ensuremath{string}}            % HTR field
\newcommand{\hatcurCCraeccenxxxxxA}{\ensuremath{13^{\mathrm h}22^{\mathrm m}32.3724{\mathrm s}}}                   % Right Ascension
\newcommand{\hatcurCCdececcenxxxxxA}{\ensuremath{-44{\arcdeg}41{\arcmin}19.6988{\arcsec}}}                 % Declination
\newcommand{\hatcurCCmageccenxxxxxA}{13.913}                         % apparent V-band magnitude
\newcommand{\hatcurCCtwomasseccenxxxxxA}{2MASS~13223237-4441196}     % 2MASS identifier
\newcommand{\hatcurCCgsceccenxxxxxA}{GSC~7799-01184}                 % GSC(1.2) identifier
\newcommand{\hatcurCCgaiaeccenxxxxxA}{GAIA~6087996845069842176}      % GAIA DR1 identifier
\newcommand{\hatcurCCgaiadrtwoeccenxxxxxA}{GAIA~DR2~6087996849371141248} % GAIA DR2 identifier
\newcommand{\hatcurCCtassmveccenxxxxxA}{\ensuremath{13.913\pm0.040}} % APASS V-band magnitude
\newcommand{\hatcurCCtassmvshorteccenxxxxxA}{\ensuremath{13.9}}      % APASS V-band magnitude
\newcommand{\hatcurCCtassmBeccenxxxxxA}{\ensuremath{14.729\pm0.030}} % APASS B-band magnitude
\newcommand{\hatcurCCtassmBshorteccenxxxxxA}{\ensuremath{14.7}}      % APASS B-band magnitude
\newcommand{\hatcurCCtassmIeccenxxxxxA}{\ensuremath{nff\pmnff}}      % TASS I-band magnitude
\newcommand{\hatcurCCtassmIshorteccenxxxxxA}{\ensuremath{0.0}}       % TASS I-band magnitude
\newcommand{\hatcurCCtassmgeccenxxxxxA}{\ensuremath{14.301\pm0.010}} % APASS g-band magnitude
\newcommand{\hatcurCCtassmgshorteccenxxxxxA}{\ensuremath{14.3}}      % APASS g-band magnitude
\newcommand{\hatcurCCtassmreccenxxxxxA}{\ensuremath{13.681\pm0.010}} % APASS r-band magnitude
\newcommand{\hatcurCCtassmrshorteccenxxxxxA}{\ensuremath{13.7}}      % APASS r-band magnitude
\newcommand{\hatcurCCtassmieccenxxxxxA}{\ensuremath{13.52\pm0.10}}   % APASS i-band magnitude
\newcommand{\hatcurCCtassmishorteccenxxxxxA}{\ensuremath{13.5}}      % APASS i-band magnitude
\newcommand{\hatcurCCparallaxeccenxxxxxA}{\ensuremath{1.308\pm0.039}} % Gaia DR2 parallax [mas]
\newcommand{\hatcurCCgaiamGeccenxxxxxA}{\ensuremath{13.77620\pm0.00040}} % Gaia G-band magnitude
\newcommand{\hatcurCCgaiamBPeccenxxxxxA}{\ensuremath{14.1834\pm0.0024}} % Gaia BP-band magnitude
\newcommand{\hatcurCCgaiamRPeccenxxxxxA}{\ensuremath{13.2231\pm0.0016}} % Gaia RP-band magnitude
\newcommand{\hatcurCCtwomassJmageccenxxxxxA}{\ensuremath{12.611\pm0.024}} % 2MASS ORIG MAG
\newcommand{\hatcurCCtwomassHmageccenxxxxxA}{\ensuremath{12.273\pm0.025}} % 2MASS ORIG MAG
\newcommand{\hatcurCCtwomassKmageccenxxxxxA}{\ensuremath{12.170\pm0.019}} % 2MASS ORIG MAG
\newcommand{\hatcurCCcitJmageccenxxxxxA}{\ensuremath{12.624\pm0.024}} % 2MASS CIT MAG
\newcommand{\hatcurCCcitHmageccenxxxxxA}{\ensuremath{12.267\pm0.025}} % 2MASS CIT MAG
\newcommand{\hatcurCCcitKmageccenxxxxxA}{\ensuremath{12.194\pm0.019}} % 2MASS CIT MAG
\newcommand{\hatcurCCbbJmageccenxxxxxA}{\ensuremath{12.679\pm0.026}} % 2MASS BB MAG
\newcommand{\hatcurCCbbHmageccenxxxxxA}{\ensuremath{12.290\pm0.026}} % 2MASS BB MAG
\newcommand{\hatcurCCbbKmageccenxxxxxA}{\ensuremath{12.214\pm0.019}} % 2MASS BB MAG
\newcommand{\hatcurCCesoJmageccenxxxxxA}{\ensuremath{12.682\pm0.028}} % 2MASS ESO MAG
\newcommand{\hatcurCCesoHmageccenxxxxxA}{\ensuremath{12.285\pm0.030}} % 2MASS ESO MAG
\newcommand{\hatcurCCesoKmageccenxxxxxA}{\ensuremath{12.213\pm0.020}} % 2MASS ESO MAG
\newcommand{\hatcurCCesoJHmageccenxxxxxA}{\ensuremath{0.396\pm0.039}} % 2MASS ESO JH COLOR
\newcommand{\hatcurCCesoJKmageccenxxxxxA}{\ensuremath{0.470\pm0.034}} % 2MASS ESO JK COLOR
\newcommand{\hatcurCCesoHKmageccenxxxxxA}{\ensuremath{0.073\pm0.036}} % 2MASS ESO HK COLOR
\newcommand{\hatcurLCdipeccenxxxxxA}{\ensuremath{8.9}}               % BLS detected dip (mmag)
\newcommand{\hatcurLCrprstareccenxxxxxA}{\ensuremath{0.0842\pm0.0029}} % Rp/R*
\newcommand{\hatcurLCbsqeccenxxxxxA}{\ensuremath{0.597_{-0.047}^{+0.035}}} % impact parameter square
\newcommand{\hatcurLCimpeccenxxxxxA}{\ensuremath{0.773_{-0.031}^{+0.022}}} % impact parameter
\newcommand{\hatcurLCzetaeccenxxxxxA}{\ensuremath{22.74_{-0.42}^{+0.77}}} % zeta/R*
\newcommand{\hatcurLCdureccenxxxxxA}{\ensuremath{0.1049\pm0.0023}}   % transit duration (days)
\newcommand{\hatcurLCdurshorteccenxxxxxA}{\ensuremath{0.1049}}       % transit duration (days)
\newcommand{\hatcurLCdurhreccenxxxxxA}{\ensuremath{2.518\pm0.055}}   % transit duration (hours)
\newcommand{\hatcurLCdurhrshorteccenxxxxxA}{\ensuremath{2.518}}      % transit duration (hours)
\newcommand{\hatcurLCqeccenxxxxxA}{\ensuremath{0.04120\pm0.00090}}   % fractional transit duration (days)
\newcommand{\hatcurLCqshorteccenxxxxxA}{\ensuremath{0.041}}          % fractional transit duration (days)
\newcommand{\hatcurLCingdureccenxxxxxA}{\ensuremath{0.0185\pm0.0029}} % ingress/egress duration (days)
\newcommand{\hatcurLCPeccenxxxxxA}{\ensuremath{2.5441837\pm0.0000040}} % period (days)
\newcommand{\hatcurLCPprececcenxxxxxA}{\ensuremath{2.5441837}}       % period (days)
\newcommand{\hatcurLCPshorteccenxxxxxA}{\ensuremath{2.5442}}         % period (days)
\newcommand{\hatcurLCTeccenxxxxxA}{\ensuremath{2457769.83472\pm0.00078}} % epoch (BJD)
\newcommand{\hatcurLCTAeccenxxxxxA}{\ensuremath{2455678.5160\pm0.0033}} % TA (BJD)
\newcommand{\hatcurLCTBeccenxxxxxA}{\ensuremath{2457947.92758\pm0.00083}} % TB (BJD)
\newcommand{\hatcurLChatnetmeccenxxxxxA}{\ensuremath{13.77662\pm0.00010}} % HATNet OOT level
\newcommand{\hatcurLCiblendeccenxxxxxA}{\ensuremath{0.953\pm0.064}}  % HATNet iblend factor
\newcommand{\hatcurLCrhoeccenxxxxxA}{\ensuremath{0.392\pm0.031}}     % stellar density no isochrone constraint (cgs)
\newcommand{\hatcurSMEiteffeccenxxxxxA}{\ensuremath{5659\pm90}}      % Ini SME, stellar effective temperature
\newcommand{\hatcurSMEizfeheccenxxxxxA}{\ensuremath{0.440\pm0.043}}  % Ini SME, stellar metallicity
\newcommand{\hatcurSMEizfehshorteccenxxxxxA}{\ensuremath{0.44}}      % Ini SME, stellar metallicity
\newcommand{\hatcurSMEiloggeccenxxxxxA}{\ensuremath{4.54\pm0.21}}    % Ini SME, stellar surface gravity
\newcommand{\hatcurSMEivsineccenxxxxxA}{\ensuremath{3.11\pm0.63}}    % Ini SME, stellar rotational velocity
\newcommand{\hatcurSMEivmaceccenxxxxxA}{\ensuremath{3.81\pm0.14}}    % Ini SME, stellar macroturbulence
\newcommand{\hatcurSMEivmiceccenxxxxxA}{\ensuremath{1.014\pm0.048}}  % Ini SME, stellar microturbulence
\newcommand{\hatcurSMEiiteffeccenxxxxxA}{\ensuremath{5528\pm78}}     % Final SME, stellar effective temperature
\newcommand{\hatcurSMEiizfeheccenxxxxxA}{\ensuremath{0.390\pm0.032}} % Final SME, stellar metallicity
\newcommand{\hatcurSMEiizfehshorteccenxxxxxA}{\ensuremath{0.39}}     % Final SME, stellar metallicity
\newcommand{\hatcurSMEiiloggeccenxxxxxA}{\ensuremath{4.095\pm0.081}} % Final SME, stellar surface gravity
\newcommand{\hatcurSMEiivsineccenxxxxxA}{\ensuremath{3.83\pm0.42}}   % Final SME, stellar rotational velocity
\newcommand{\hatcurSMEiivmaceccenxxxxxA}{\ensuremath{3.61\pm0.11}}   % Final SME, stellar macroturbulence
\newcommand{\hatcurSMEiivmiceccenxxxxxA}{\ensuremath{0.948\pm0.034}} % Final SME, stellar microturbulence
\newcommand{\hatcurLBizeccenxxxxxA}{\ensuremath{0.2454}}             % Limb darkening parameters, Gamma1, z-band
\newcommand{\hatcurLBiizeccenxxxxxA}{\ensuremath{0.3222}}            % Limb darkening parameters, Gamma2, z-band
\newcommand{\hatcurLBiieccenxxxxxA}{\ensuremath{0.3222}}             % Limb darkening parameters, Gamma1, i-band
\newcommand{\hatcurLBiiieccenxxxxxA}{\ensuremath{0.3120}}            % Limb darkening parameters, Gamma2, i-band
\newcommand{\hatcurLBiIeccenxxxxxA}{\ensuremath{0.2968}}             % Limb darkening parameters, Gamma1, I-band
\newcommand{\hatcurLBiiIeccenxxxxxA}{\ensuremath{0.3161}}            % Limb darkening parameters, Gamma2, I-band
\newcommand{\hatcurLBigeccenxxxxxA}{\ensuremath{0.6588}}             % Limb darkening parameters, Gamma1, g-band
\newcommand{\hatcurLBiigeccenxxxxxA}{\ensuremath{0.1582}}            % Limb darkening parameters, Gamma2, g-band
\newcommand{\hatcurLBireccenxxxxxA}{\ensuremath{0.4314}}             % Limb darkening parameters, Gamma1, r-band
\newcommand{\hatcurLBiireccenxxxxxA}{\ensuremath{0.2867}}            % Limb darkening parameters, Gamma2, r-band
\newcommand{\hatcurLBiReccenxxxxxA}{\ensuremath{0.4011}}             % Limb darkening parameters, Gamma1, R-band
\newcommand{\hatcurLBiiReccenxxxxxA}{\ensuremath{0.2948}}            % Limb darkening parameters, Gamma2, R-band
\newcommand{\hatcurLBikepeccenxxxxxA}{\ensuremath{0.1000}}           % Limb darkening parameters, Gamma1, Kep-band
\newcommand{\hatcurLBiikepeccenxxxxxA}{\ensuremath{0.1000}}          % Limb darkening parameters, Gamma2, Kep-band
\newcommand{\hatcurISOmeccenxxxxxA}{\ensuremath{1.345_{-0.031}^{+0.040}}} % stellar mass
\newcommand{\hatcurISOmshorteccenxxxxxA}{\ensuremath{1.34}}          % stellar mass
\newcommand{\hatcurISOmlongeccenxxxxxA}{\ensuremath{1.345_{-0.031}^{+0.040}}} % stellar mass
\newcommand{\hatcurISOreccenxxxxxA}{\ensuremath{4.224_{-0.028}^{+0.021}}} % stellar radius
\newcommand{\hatcurISOrshorteccenxxxxxA}{\ensuremath{4.22}}          % stellar radius
\newcommand{\hatcurISOrlongeccenxxxxxA}{\ensuremath{4.224_{-0.028}^{+0.021}}} % stellar radius
\newcommand{\hatcurISOrhoeccenxxxxxA}{\ensuremath{0.392\pm0.031}}    % stellar density (cgs)
\newcommand{\hatcurISOrholongeccenxxxxxA}{\ensuremath{0.392\pm0.031}} % stellar density (cgs)
\newcommand{\hatcurISOloggeccenxxxxxA}{\ensuremath{6.70\pm0.76}}     % stellar surface gravity from isochrones
\newcommand{\hatcurISOlumeccenxxxxxA}{\ensuremath{4.058\pm0.061}}    % stellar luminosity
\newcommand{\hatcurISOlumshorteccenxxxxxA}{\ensuremath{4.06}}        % stellar luminosity
\newcommand{\hatcurISOteffeccenxxxxxA}{\ensuremath{1.102_{-0.015}^{+0.020}}} % stellar effective temperature adjusted via MCMC
\newcommand{\hatcurISOzfeheccenxxxxxA}{\ensuremath{5708\pm27}}       % stellar [M/H] adjusted via MCMC
\newcommand{\hatcurISOageeccenxxxxxA}{\ensuremath{1.724\pm0.097}}    % stellar age
\newcommand{\hatcurISOspececcenxxxxxA}{G}                            % stellar spectral type
\newcommand{\hatcurRVKeccenxxxxxA}{\ensuremath{112\pm10}}            % RV semi-amplitude [m/s]
\newcommand{\hatcurRVrkeccenxxxxxA}{\ensuremath{-0.03\pm0.16}}       % sqrt(e)*cos(omega)
\newcommand{\hatcurRVrheccenxxxxxA}{\ensuremath{-0.12_{-0.10}^{+0.17}}} % sqrt(e)*sin(omega)
\newcommand{\hatcurRVkeccenxxxxxA}{\ensuremath{-0.003_{-0.044}^{+0.032}}} % e*cos(omega)
\newcommand{\hatcurRVheccenxxxxxA}{\ensuremath{-0.026\pm0.035}}      % e*sin(omega)
\newcommand{\hatcurRVtroneeccenxxxxxA}{\ensuremath{0\pm0}}           % RV linear trend tr1 factor
\newcommand{\hatcurRVtrtwoeccenxxxxxA}{\ensuremath{0\pm0}}           % RV linear trend tr2 factor
\newcommand{\hatcurRVgammaAeccenxxxxxA}{\ensuremath{46127\pm11}}     % RV gamma velocity, relative scale
\newcommand{\hatcurRVjitterAeccenxxxxxA}{\ensuremath{62.8\pm9.4}}    % RV jitter (m/s)
\newcommand{\hatcurRVjittertwosiglimAeccenxxxxxA}{\ensuremath{<80.8}} % RV jitter (m/s) 95 percent confidence upper limit
\newcommand{\hatcurRVfitrmsAeccenxxxxxA}{\ensuremath{0.0}}           % RVfitrms
\newcommand{\hatcurRVgammaBeccenxxxxxA}{\ensuremath{46034\pm22}}     % RV gamma velocity, relative scale
\newcommand{\hatcurRVjitterBeccenxxxxxA}{\ensuremath{1\pm60}}        % RV jitter (m/s)
\newcommand{\hatcurRVjittertwosiglimBeccenxxxxxA}{\ensuremath{<133.2}} % RV jitter (m/s) 95 percent confidence upper limit
\newcommand{\hatcurRVfitrmsBeccenxxxxxA}{\ensuremath{0.0}}           % RVfitrms
\newcommand{\hatcurRVecceneccenxxxxxA}{\ensuremath{0.046\pm0.037}}   % eccentricity
\newcommand{\hatcurRVeccentwosiglimeccenxxxxxA}{\ensuremath{<0.126}} % eccentricity
\newcommand{\hatcurRVomegaeccenxxxxxA}{\ensuremath{235\pm90}}        % argument of pericenter
\newcommand{\hatcurPPieccenxxxxxA}{\ensuremath{68.47\pm0.47}}        % orbital inclination
\newcommand{\hatcurPPgeccenxxxxxA}{\ensuremath{2.03\pm0.21}}         % planetary surface gravity (m/s^2)
\newcommand{\hatcurPPloggeccenxxxxxA}{\ensuremath{2.308\pm0.046}}    % planetary surface gravity (log cgs)
\newcommand{\hatcurPPareccenxxxxxA}{\ensuremath{2.050_{-0.024}^{+0.035}}} % relative orbital radius (a/R*)
\newcommand{\hatcurPPareleccenxxxxxA}{\ensuremath{0.04027\pm0.00038}} % semimajor axis (AU)
\newcommand{\hatcurPPrhoeccenxxxxxA}{\ensuremath{0.0300_{-0.0040}^{+0.0030}}} % planetary density (cgs)
\newcommand{\hatcurPPmeccenxxxxxA}{\ensuremath{0.985\pm0.089}}       % planetary mass (M_jup)
\newcommand{\hatcurPPmshorteccenxxxxxA}{\ensuremath{0.99}}           % planetary mass (M_jup)
\newcommand{\hatcurPPmlongeccenxxxxxA}{\ensuremath{0.985\pm0.089}}   % planetary mass (M_jup)
\newcommand{\hatcurPPmeeccenxxxxxA}{\ensuremath{313\pm28}}           % planetary mass (M_earth)
\newcommand{\hatcurPPmeshorteccenxxxxxA}{\ensuremath{313.2}}         % planetary mass (M_earth)
\newcommand{\hatcurPPmelongeccenxxxxxA}{\ensuremath{313\pm28}}       % planetary mass (M_earth)
\newcommand{\hatcurPPreccenxxxxxA}{\ensuremath{3.46\pm0.12}}         % planetary radius (R_jup)
\newcommand{\hatcurPPrshorteccenxxxxxA}{\ensuremath{3.46}}           % planetary radius (R_jup)
\newcommand{\hatcurPPrlongeccenxxxxxA}{\ensuremath{3.46\pm0.12}}     % planetary radius (R_jup)
\newcommand{\hatcurPPreeccenxxxxxA}{\ensuremath{38.7\pm1.3}}         % planetary radius (R_earth)
\newcommand{\hatcurPPreshorteccenxxxxxA}{\ensuremath{38.7}}          % planetary radius (R_earth)
\newcommand{\hatcurPPrelongeccenxxxxxA}{\ensuremath{38.7\pm1.3}}     % planetary radius (R_earth)
\newcommand{\hatcurPPmrcorreccenxxxxxA}{\ensuremath{0.12}}           % mass/radius correlation
\newcommand{\hatcurPPteffeccenxxxxxA}{\ensuremath{0.500\pm0.053}}    % planetary temperature (K)
\newcommand{\hatcurPPthetaeccenxxxxxA}{\ensuremath{0.0158\pm0.0015}} % Safranov number
\newcommand{\hatcurPPfluxperieccenxxxxxA}{\ensuremath{3.74_{-0.23}^{+0.35}}} % flux @ periastron (CGS)
\newcommand{\hatcurPPfluxperidimeccenxxxxxA}{\ensuremath{9}}         % flux @ periastron (CGS) units.
\newcommand{\hatcurPPfluxapeccenxxxxxA}{\ensuremath{3.09\pm0.24}}    % flux @ apastron (CGS)
\newcommand{\hatcurPPfluxapdimeccenxxxxxA}{\ensuremath{9}}           % flux @ apastron (CGS) units.
\newcommand{\hatcurPPfluxavgeccenxxxxxA}{\ensuremath{3.42\pm0.11}}   % flux on average (CGS)
\newcommand{\hatcurPPfluxavgdimeccenxxxxxA}{\ensuremath{9}}          % flux average (CGS) units.
\newcommand{\hatcurPPfluxavglogeccenxxxxxA}{\ensuremath{9.534_{-0.016}^{+0.012}}} % log10 flux on average (CGS)
\newcommand{\hatcurXsecphaseeccenxxxxxA}{\ensuremath{0.498\pm0.030}} % Phase of secondary eclipse
\newcommand{\hatcurXsecondaryeccenxxxxxA}{\ensuremath{2457771.101\pm0.077}} % Secondary eclipse epoch
\newcommand{\hatcurXsecdureccenxxxxxA}{\ensuremath{0.1044\pm0.0020}} % sec eclipse duration (days)
\newcommand{\hatcurXsecingdureccenxxxxxA}{\ensuremath{0.0165\pm0.0019}} % sec I/E duration (days)
\newcommand{\hatcurPPphiconjeccenxxxxxA}{\ensuremath{-0.17_{-0.24}^{+0.58}}} % phase diff between conjunction and periastron
\newcommand{\hatcurPPperieccenxxxxxA}{\ensuremath{2457770.28\pm0.94}} % time of periastron passage.
\newcommand{\hatcurPPaequiveccenxxxxxA}{\ensuremath{0.02000\pm0.00033}} % equivalent semi-major axis
\newcommand{\hatcurPPtcirceccenxxxxxA}{\ensuremath{0.400\pm0.083}}   % circularization timescale
\newcommand{\hatcurPPtinfalleccenxxxxxA}{\ensuremath{1.200_{-0.100}^{+0.200}}} % infall timescale
\newcommand{\hatcurXdisteccenxxxxxA}{\ensuremath{11.400\pm0.092}}    % distance (pc), no reddenning correction
\newcommand{\hatcurXAveccenxxxxxA}{\ensuremath{5708\pm27}}           % Av (mag)
\newcommand{\hatcurXdistredeccenxxxxxA}{\ensuremath{11.397\pm0.087}} % distance with Av correction (pc)
\newcommand{\hatcurXEBVeccenxxxxxA}{\ensuremath{1841.4\pm8.6}}       % E(B-V) (mag)
\newcommand{\hatcurCCpmraeccenxxxxxA}{\ensuremath{-3.451\pm0.054}}   % proper motion, in RA
\newcommand{\hatcurCCpmdececcenxxxxxA}{\ensuremath{-7.915\pm0.093}}  % proper motion, in DEC
\newcommand{\hatcurCCpmeccenxxxxxA}{\ensuremath{8.63\pm0.11}}        % proper motion

\newcommand{\hatcurhtreccenxxxxxB}{HATS602-072}                      % Original HTR name of target
\newcommand{\hatcurfieldeccenxxxxxB}{\ensuremath{string}}            % HTR field
\newcommand{\hatcurCCraeccenxxxxxB}{\ensuremath{07^{\mathrm h}37^{\mathrm m}08.0194{\mathrm s}}}                   % Right Ascension
\newcommand{\hatcurCCdececcenxxxxxB}{\ensuremath{-32{\arcdeg}45{\arcmin}19.5158{\arcsec}}}                 % Declination
\newcommand{\hatcurCCmageccenxxxxxB}{13.470}                         % apparent V-band magnitude
\newcommand{\hatcurCCtwomasseccenxxxxxB}{2MASS~07370802-3245195}     % 2MASS identifier
\newcommand{\hatcurCCgsceccenxxxxxB}{GSC~7109-00596}                 % GSC(1.2) identifier
\newcommand{\hatcurCCgaiaeccenxxxxxB}{GAIA~5592019553648331776}      % GAIA DR1 identifier
\newcommand{\hatcurCCgaiadrtwoeccenxxxxxB}{GAIA~DR2~5592019557950033536} % GAIA DR2 identifier
\newcommand{\hatcurCCtassmveccenxxxxxB}{\ensuremath{13.470\pm0.040}} % APASS V-band magnitude
\newcommand{\hatcurCCtassmvshorteccenxxxxxB}{\ensuremath{13.5}}      % APASS V-band magnitude
\newcommand{\hatcurCCtassmBeccenxxxxxB}{\ensuremath{14.111\pm0.020}} % APASS B-band magnitude
\newcommand{\hatcurCCtassmBshorteccenxxxxxB}{\ensuremath{14.1}}      % APASS B-band magnitude
\newcommand{\hatcurCCtassmIeccenxxxxxB}{\ensuremath{nff\pmnff}}      % TASS I-band magnitude
\newcommand{\hatcurCCtassmIshorteccenxxxxxB}{\ensuremath{0.0}}       % TASS I-band magnitude
\newcommand{\hatcurCCtassmgeccenxxxxxB}{\ensuremath{13.746\pm0.030}} % APASS g-band magnitude
\newcommand{\hatcurCCtassmgshorteccenxxxxxB}{\ensuremath{13.7}}      % APASS g-band magnitude
\newcommand{\hatcurCCtassmreccenxxxxxB}{\ensuremath{13.279\pm0.030}} % APASS r-band magnitude
\newcommand{\hatcurCCtassmrshorteccenxxxxxB}{\ensuremath{13.3}}      % APASS r-band magnitude
\newcommand{\hatcurCCtassmieccenxxxxxB}{\ensuremath{13.120\pm0.030}} % APASS i-band magnitude
\newcommand{\hatcurCCtassmishorteccenxxxxxB}{\ensuremath{13.1}}      % APASS i-band magnitude
\newcommand{\hatcurCCparallaxeccenxxxxxB}{\ensuremath{1.611\pm0.016}} % Gaia DR2 parallax [mas]
\newcommand{\hatcurCCgaiamGeccenxxxxxB}{\ensuremath{13.34980\pm0.00030}} % Gaia G-band magnitude
\newcommand{\hatcurCCgaiamBPeccenxxxxxB}{\ensuremath{13.6816\pm0.0010}} % Gaia BP-band magnitude
\newcommand{\hatcurCCgaiamRPeccenxxxxxB}{\ensuremath{12.8651\pm0.0012}} % Gaia RP-band magnitude
\newcommand{\hatcurCCtwomassJmageccenxxxxxB}{\ensuremath{12.338\pm0.026}} % 2MASS ORIG MAG
\newcommand{\hatcurCCtwomassHmageccenxxxxxB}{\ensuremath{12.048\pm0.040}} % 2MASS ORIG MAG
\newcommand{\hatcurCCtwomassKmageccenxxxxxB}{\ensuremath{12.020\pm0.041}} % 2MASS ORIG MAG
\newcommand{\hatcurCCcitJmageccenxxxxxB}{\ensuremath{12.358\pm0.026}} % 2MASS CIT MAG
\newcommand{\hatcurCCcitHmageccenxxxxxB}{\ensuremath{12.044\pm0.039}} % 2MASS CIT MAG
\newcommand{\hatcurCCcitKmageccenxxxxxB}{\ensuremath{12.044\pm0.041}} % 2MASS CIT MAG
\newcommand{\hatcurCCbbJmageccenxxxxxB}{\ensuremath{12.403\pm0.028}} % 2MASS BB MAG
\newcommand{\hatcurCCbbHmageccenxxxxxB}{\ensuremath{12.064\pm0.040}} % 2MASS BB MAG
\newcommand{\hatcurCCbbKmageccenxxxxxB}{\ensuremath{12.064\pm0.041}} % 2MASS BB MAG
\newcommand{\hatcurCCesoJmageccenxxxxxB}{\ensuremath{12.404\pm0.029}} % 2MASS ESO MAG
\newcommand{\hatcurCCesoHmageccenxxxxxB}{\ensuremath{12.058\pm0.045}} % 2MASS ESO MAG
\newcommand{\hatcurCCesoKmageccenxxxxxB}{\ensuremath{12.063\pm0.041}} % 2MASS ESO MAG
\newcommand{\hatcurCCesoJHmageccenxxxxxB}{\ensuremath{0.346\pm0.052}} % 2MASS ESO JH COLOR
\newcommand{\hatcurCCesoJKmageccenxxxxxB}{\ensuremath{0.342\pm0.051}} % 2MASS ESO JK COLOR
\newcommand{\hatcurCCesoHKmageccenxxxxxB}{\ensuremath{-0.004\pm0.063}} % 2MASS ESO HK COLOR
\newcommand{\hatcurLCdipeccenxxxxxB}{\ensuremath{13.7}}              % BLS detected dip (mmag)
\newcommand{\hatcurLCrprstareccenxxxxxB}{\ensuremath{0.1135\pm0.0024}} % Rp/R*
\newcommand{\hatcurLCbsqeccenxxxxxB}{\ensuremath{0.436_{-0.029}^{+0.026}}} % impact parameter square
\newcommand{\hatcurLCimpeccenxxxxxB}{\ensuremath{0.660_{-0.022}^{+0.019}}} % impact parameter
\newcommand{\hatcurLCzetaeccenxxxxxB}{\ensuremath{20.75\pm0.24}}     % zeta/R*
\newcommand{\hatcurLCdureccenxxxxxB}{\ensuremath{0.1151\pm0.0015}}   % transit duration (days)
\newcommand{\hatcurLCdurshorteccenxxxxxB}{\ensuremath{0.1151}}       % transit duration (days)
\newcommand{\hatcurLCdurhreccenxxxxxB}{\ensuremath{2.763\pm0.035}}   % transit duration (hours)
\newcommand{\hatcurLCdurhrshorteccenxxxxxB}{\ensuremath{2.763}}      % transit duration (hours)
\newcommand{\hatcurLCqeccenxxxxxB}{\ensuremath{0.02740\pm0.00035}}   % fractional transit duration (days)
\newcommand{\hatcurLCqshorteccenxxxxxB}{\ensuremath{0.027}}          % fractional transit duration (days)
\newcommand{\hatcurLCingdureccenxxxxxB}{\ensuremath{0.01960\pm0.00099}} % ingress/egress duration (days)
\newcommand{\hatcurLCPeccenxxxxxB}{\ensuremath{4.2042001\pm0.0000028}} % period (days)
\newcommand{\hatcurLCPprececcenxxxxxB}{\ensuremath{4.2042001}}       % period (days)
\newcommand{\hatcurLCPshorteccenxxxxxB}{\ensuremath{4.2042}}         % period (days)
\newcommand{\hatcurLCTeccenxxxxxB}{\ensuremath{2457471.98908\pm0.00049}} % epoch (BJD)
\newcommand{\hatcurLCTAeccenxxxxxB}{\ensuremath{2455798.7175\pm0.0012}} % TA (BJD)
\newcommand{\hatcurLCTBeccenxxxxxB}{\ensuremath{2457854.57128\pm0.00056}} % TB (BJD)
\newcommand{\hatcurLChatnetmeccenxxxxxB}{\ensuremath{13.330290\pm0.000094}} % HATNet OOT level
\newcommand{\hatcurLCiblendeccenxxxxxB}{\ensuremath{0.850\pm0.054}}  % HATNet iblend factor
\newcommand{\hatcurLCrhoeccenxxxxxB}{\ensuremath{0.098\pm0.035}}     % stellar density no isochrone constraint (cgs)
\newcommand{\hatcurSMEiteffeccenxxxxxB}{\ensuremath{6220\pm150}}     % Ini SME, stellar effective temperature
\newcommand{\hatcurSMEizfeheccenxxxxxB}{\ensuremath{0.300\pm0.090}}  % Ini SME, stellar metallicity
\newcommand{\hatcurSMEizfehshorteccenxxxxxB}{\ensuremath{0.30}}      % Ini SME, stellar metallicity
\newcommand{\hatcurSMEiloggeccenxxxxxB}{\ensuremath{4.64\pm0.15}}    % Ini SME, stellar surface gravity
\newcommand{\hatcurSMEivsineccenxxxxxB}{\ensuremath{5.02\pm0.18}}    % Ini SME, stellar rotational velocity
\newcommand{\hatcurSMEivmaceccenxxxxxB}{\ensuremath{4.68\pm0.23}}    % Ini SME, stellar macroturbulence
\newcommand{\hatcurSMEivmiceccenxxxxxB}{\ensuremath{1.43\pm0.15}}    % Ini SME, stellar microturbulence
\newcommand{\hatcurSMEiiteffeccenxxxxxB}{\ensuremath{6095\pm92}}     % Final SME, stellar effective temperature
\newcommand{\hatcurSMEiizfeheccenxxxxxB}{\ensuremath{0.220\pm0.049}} % Final SME, stellar metallicity
\newcommand{\hatcurSMEiizfehshorteccenxxxxxB}{\ensuremath{0.22}}     % Final SME, stellar metallicity
\newcommand{\hatcurSMEiiloggeccenxxxxxB}{\ensuremath{4.419\pm0.023}} % Final SME, stellar surface gravity
\newcommand{\hatcurSMEiivsineccenxxxxxB}{\ensuremath{5.01\pm0.15}}   % Final SME, stellar rotational velocity
\newcommand{\hatcurSMEiivmaceccenxxxxxB}{\ensuremath{4.48\pm0.14}}   % Final SME, stellar macroturbulence
\newcommand{\hatcurSMEiivmiceccenxxxxxB}{\ensuremath{1.308\pm0.079}} % Final SME, stellar microturbulence
\newcommand{\hatcurLBizeccenxxxxxB}{\ensuremath{0.1716}}             % Limb darkening parameters, Gamma1, z-band
\newcommand{\hatcurLBiizeccenxxxxxB}{\ensuremath{0.3513}}            % Limb darkening parameters, Gamma2, z-band
\newcommand{\hatcurLBiieccenxxxxxB}{\ensuremath{0.2290}}             % Limb darkening parameters, Gamma1, i-band
\newcommand{\hatcurLBiiieccenxxxxxB}{\ensuremath{0.3569}}            % Limb darkening parameters, Gamma2, i-band
\newcommand{\hatcurLBiIeccenxxxxxB}{\ensuremath{0.2091}}             % Limb darkening parameters, Gamma1, I-band
\newcommand{\hatcurLBiiIeccenxxxxxB}{\ensuremath{0.3561}}            % Limb darkening parameters, Gamma2, I-band
\newcommand{\hatcurLBigeccenxxxxxB}{\ensuremath{0.4944}}             % Limb darkening parameters, Gamma1, g-band
\newcommand{\hatcurLBiigeccenxxxxxB}{\ensuremath{0.2817}}            % Limb darkening parameters, Gamma2, g-band
\newcommand{\hatcurLBireccenxxxxxB}{\ensuremath{0.3112}}             % Limb darkening parameters, Gamma1, r-band
\newcommand{\hatcurLBiireccenxxxxxB}{\ensuremath{0.3574}}            % Limb darkening parameters, Gamma2, r-band
\newcommand{\hatcurLBiReccenxxxxxB}{\ensuremath{0.2882}}             % Limb darkening parameters, Gamma1, R-band
\newcommand{\hatcurLBiiReccenxxxxxB}{\ensuremath{0.3585}}            % Limb darkening parameters, Gamma2, R-band
\newcommand{\hatcurLBikepeccenxxxxxB}{\ensuremath{0.1000}}           % Limb darkening parameters, Gamma1, Kep-band
\newcommand{\hatcurLBiikepeccenxxxxxB}{\ensuremath{0.1000}}          % Limb darkening parameters, Gamma2, Kep-band
\newcommand{\hatcurISOmeccenxxxxxB}{\ensuremath{1.127\pm0.010}}      % stellar mass
\newcommand{\hatcurISOmshorteccenxxxxxB}{\ensuremath{1.13}}          % stellar mass
\newcommand{\hatcurISOmlongeccenxxxxxB}{\ensuremath{1.127\pm0.010}}  % stellar mass
\newcommand{\hatcurISOreccenxxxxxB}{\ensuremath{4.4104\pm0.0075}}    % stellar radius
\newcommand{\hatcurISOrshorteccenxxxxxB}{\ensuremath{4.41}}          % stellar radius
\newcommand{\hatcurISOrlongeccenxxxxxB}{\ensuremath{4.4104\pm0.0075}} % stellar radius
\newcommand{\hatcurISOrhoeccenxxxxxB}{\ensuremath{0.098\pm0.035}}    % stellar density (cgs)
\newcommand{\hatcurISOrholongeccenxxxxxB}{\ensuremath{0.098\pm0.035}} % stellar density (cgs)
\newcommand{\hatcurISOloggeccenxxxxxB}{\ensuremath{0.46\pm0.26}}     % stellar surface gravity from isochrones
\newcommand{\hatcurISOlumeccenxxxxxB}{\ensuremath{4.062\pm0.036}}    % stellar luminosity
\newcommand{\hatcurISOlumshorteccenxxxxxB}{\ensuremath{4.06}}        % stellar luminosity
\newcommand{\hatcurISOteffeccenxxxxxB}{\ensuremath{1.194\pm0.012}}   % stellar effective temperature adjusted via MCMC
\newcommand{\hatcurISOzfeheccenxxxxxB}{\ensuremath{6229\pm38}}       % stellar [M/H] adjusted via MCMC
\newcommand{\hatcurISOageeccenxxxxxB}{\ensuremath{1.712\pm0.058}}    % stellar age
\newcommand{\hatcurISOspececcenxxxxxB}{F}                            % stellar spectral type
\newcommand{\hatcurRVKeccenxxxxxB}{\ensuremath{105.8\pm9.1}}         % RV semi-amplitude [m/s]
\newcommand{\hatcurRVrkeccenxxxxxB}{\ensuremath{-0.13_{-0.11}^{+0.16}}} % sqrt(e)*cos(omega)
\newcommand{\hatcurRVrheccenxxxxxB}{\ensuremath{0.05\pm0.10}}        % sqrt(e)*sin(omega)
\newcommand{\hatcurRVkeccenxxxxxB}{\ensuremath{-0.024_{-0.036}^{+0.027}}} % e*cos(omega)
\newcommand{\hatcurRVheccenxxxxxB}{\ensuremath{0.009_{-0.020}^{+0.027}}} % e*sin(omega)
\newcommand{\hatcurRVtroneeccenxxxxxB}{\ensuremath{0\pm0}}           % RV linear trend tr1 factor
\newcommand{\hatcurRVtrtwoeccenxxxxxB}{\ensuremath{0\pm0}}           % RV linear trend tr2 factor
\newcommand{\hatcurRVgammaeccenxxxxxB}{\ensuremath{-2919.2\pm7.6}}   % RV gamma velocity, relative scale
\newcommand{\hatcurRVjittereccenxxxxxB}{\ensuremath{0.4\pm6.4}}      % RV jitter (m/s)
\newcommand{\hatcurRVjittertwosiglimeccenxxxxxB}{\ensuremath{<17.2}} % RV jitter (m/s) 95 percent confidence upper limit
\newcommand{\hatcurRVfitrmseccenxxxxxB}{\ensuremath{.1fym}}          % 
\newcommand{\hatcurRVecceneccenxxxxxB}{\ensuremath{0.039\pm0.027}}   % eccentricity
\newcommand{\hatcurRVeccentwosiglimeccenxxxxxB}{\ensuremath{<0.092}} % eccentricity
\newcommand{\hatcurRVomegaeccenxxxxxB}{\ensuremath{165\pm68}}        % argument of pericenter
\newcommand{\hatcurPPieccenxxxxxB}{\ensuremath{75.05\pm0.26}}        % orbital inclination
\newcommand{\hatcurPPgeccenxxxxxB}{\ensuremath{0.977\pm0.095}}       % planetary surface gravity (m/s^2)
\newcommand{\hatcurPPloggeccenxxxxxB}{\ensuremath{1.990\pm0.042}}    % planetary surface gravity (log cgs)
\newcommand{\hatcurPPareccenxxxxxB}{\ensuremath{2.588\pm0.012}}      % relative orbital radius (a/R*)
\newcommand{\hatcurPPareleccenxxxxxB}{\ensuremath{0.05307\pm0.00016}} % semimajor axis (AU)
\newcommand{\hatcurPPrhoeccenxxxxxB}{\ensuremath{0.0100\pm0.0012}}   % planetary density (cgs)
\newcommand{\hatcurPPmeccenxxxxxB}{\ensuremath{0.941\pm0.080}}       % planetary mass (M_jup)
\newcommand{\hatcurPPmshorteccenxxxxxB}{\ensuremath{0.94}}           % planetary mass (M_jup)
\newcommand{\hatcurPPmlongeccenxxxxxB}{\ensuremath{0.941\pm0.080}}   % planetary mass (M_jup)
\newcommand{\hatcurPPmeeccenxxxxxB}{\ensuremath{299\pm25}}           % planetary mass (M_earth)
\newcommand{\hatcurPPmeshorteccenxxxxxB}{\ensuremath{299.0}}         % planetary mass (M_earth)
\newcommand{\hatcurPPmelongeccenxxxxxB}{\ensuremath{299\pm25}}       % planetary mass (M_earth)
\newcommand{\hatcurPPreccenxxxxxB}{\ensuremath{4.87\pm0.10}}         % planetary radius (R_jup)
\newcommand{\hatcurPPrshorteccenxxxxxB}{\ensuremath{4.87}}           % planetary radius (R_jup)
\newcommand{\hatcurPPrlongeccenxxxxxB}{\ensuremath{4.87\pm0.10}}     % planetary radius (R_jup)
\newcommand{\hatcurPPreeccenxxxxxB}{\ensuremath{54.6\pm1.2}}         % planetary radius (R_earth)
\newcommand{\hatcurPPreshorteccenxxxxxB}{\ensuremath{54.6}}          % planetary radius (R_earth)
\newcommand{\hatcurPPrelongeccenxxxxxB}{\ensuremath{54.6\pm1.2}}     % planetary radius (R_earth)
\newcommand{\hatcurPPmrcorreccenxxxxxB}{\ensuremath{-0.02}}          % mass/radius correlation
\newcommand{\hatcurPPteffeccenxxxxxB}{\ensuremath{0\pm0}}            % planetary temperature (K)
\newcommand{\hatcurPPthetaeccenxxxxxB}{\ensuremath{0.0176\pm0.0016}} % Safranov number
\newcommand{\hatcurPPfluxperieccenxxxxxB}{\ensuremath{2.12_{-0.10}^{+0.15}}} % flux @ periastron (CGS)
\newcommand{\hatcurPPfluxperidimeccenxxxxxB}{\ensuremath{9}}         % flux @ periastron (CGS) units.
\newcommand{\hatcurPPfluxapeccenxxxxxB}{\ensuremath{1.822\pm0.095}}  % flux @ apastron (CGS)
\newcommand{\hatcurPPfluxapdimeccenxxxxxB}{\ensuremath{9}}           % flux @ apastron (CGS) units.
\newcommand{\hatcurPPfluxavgeccenxxxxxB}{\ensuremath{1.965\pm0.028}} % flux on average (CGS)
\newcommand{\hatcurPPfluxavgdimeccenxxxxxB}{\ensuremath{9}}          % flux average (CGS) units.
\newcommand{\hatcurPPfluxavglogeccenxxxxxB}{\ensuremath{9.2934\pm0.0062}} % log10 flux on average (CGS)
\newcommand{\hatcurXsecphaseeccenxxxxxB}{\ensuremath{0.485\pm0.021}} % Phase of secondary eclipse
\newcommand{\hatcurXsecondaryeccenxxxxxB}{\ensuremath{2457474.026\pm0.089}} % Secondary eclipse epoch
\newcommand{\hatcurXsecdureccenxxxxxB}{\ensuremath{0.1160\pm0.0020}} % sec eclipse duration (days)
\newcommand{\hatcurXsecingdureccenxxxxxB}{\ensuremath{0.0203\pm0.0010}} % sec I/E duration (days)
\newcommand{\hatcurPPphiconjeccenxxxxxB}{\ensuremath{-0.16_{-0.12}^{+0.23}}} % phase diff between conjunction and periastron
\newcommand{\hatcurPPperieccenxxxxxB}{\ensuremath{2457472.66\pm0.90}} % time of periastron passage.
\newcommand{\hatcurPPaequiveccenxxxxxB}{\ensuremath{0.026300_{-0.000100}^{+0.000200}}} % equivalent semi-major axis
\newcommand{\hatcurPPtcirceccenxxxxxB}{\ensuremath{0.60_{-0.10}^{+1.00}}} % circularization timescale
\newcommand{\hatcurPPtinfalleccenxxxxxB}{\ensuremath{5.50_{-0.40}^{+0.60}}} % infall timescale
\newcommand{\hatcurXdisteccenxxxxxB}{\ensuremath{11.80_{-0.10}^{+0.20}}} % distance (pc), no reddenning correction
\newcommand{\hatcurXAveccenxxxxxB}{\ensuremath{6229\pm38}}           % Av (mag)
\newcommand{\hatcurXdistredeccenxxxxxB}{\ensuremath{11.82\pm0.15}}   % distance with Av correction (pc)
\newcommand{\hatcurXEBVeccenxxxxxB}{\ensuremath{2009\pm12}}          % E(B-V) (mag)
\newcommand{\hatcurCCpmraeccenxxxxxB}{\ensuremath{-6.283\pm0.026}}   % proper motion, in RA
\newcommand{\hatcurCCpmdececcenxxxxxB}{\ensuremath{0.531\pm0.031}}   % proper motion, in DEC
\newcommand{\hatcurCCpmeccenxxxxxB}{\ensuremath{6.305\pm0.040}}      % proper motion

\newcommand{\hatcurhtreccenxxxxxC}{HATS698-034}                      % Original HTR name of target
\newcommand{\hatcurfieldeccenxxxxxC}{\ensuremath{string}}            % HTR field
\newcommand{\hatcurCCraeccenxxxxxC}{\ensuremath{12^{\mathrm h}00^{\mathrm m}39.6300{\mathrm s}}}                   % Right Ascension
\newcommand{\hatcurCCdececcenxxxxxC}{\ensuremath{-45{\arcdeg}47{\arcmin}57.9955{\arcsec}}}                 % Declination
\newcommand{\hatcurCCmageccenxxxxxC}{11.578}                         % apparent V-band magnitude
\newcommand{\hatcurCCtwomasseccenxxxxxC}{2MASS~12003962-4547579}     % 2MASS identifier
\newcommand{\hatcurCCgsceccenxxxxxC}{GSC~8229-02228}                 % GSC(1.2) identifier
\newcommand{\hatcurCCgaiaeccenxxxxxC}{GAIA~6144125882874203008}      % GAIA DR1 identifier
\newcommand{\hatcurCCgaiadrtwoeccenxxxxxC}{GAIA~DR2~6144125887172751232} % GAIA DR2 identifier
\newcommand{\hatcurCCtassmveccenxxxxxC}{\ensuremath{11.578\pm0.023}} % APASS V-band magnitude
\newcommand{\hatcurCCtassmvshorteccenxxxxxC}{\ensuremath{11.6}}      % APASS V-band magnitude
\newcommand{\hatcurCCtassmBeccenxxxxxC}{\ensuremath{12.097\pm0.029}} % APASS B-band magnitude
\newcommand{\hatcurCCtassmBshorteccenxxxxxC}{\ensuremath{12.1}}      % APASS B-band magnitude
\newcommand{\hatcurCCtassmIeccenxxxxxC}{\ensuremath{nff\pmnff}}      % TASS I-band magnitude
\newcommand{\hatcurCCtassmIshorteccenxxxxxC}{\ensuremath{0.0}}       % TASS I-band magnitude
\newcommand{\hatcurCCtassmgeccenxxxxxC}{\ensuremath{11.801\pm0.022}} % APASS g-band magnitude
\newcommand{\hatcurCCtassmgshorteccenxxxxxC}{\ensuremath{11.8}}      % APASS g-band magnitude
\newcommand{\hatcurCCtassmreccenxxxxxC}{\ensuremath{11.473\pm0.017}} % APASS r-band magnitude
\newcommand{\hatcurCCtassmrshorteccenxxxxxC}{\ensuremath{11.5}}      % APASS r-band magnitude
\newcommand{\hatcurCCtassmieccenxxxxxC}{\ensuremath{11.320\pm0.046}} % APASS i-band magnitude
\newcommand{\hatcurCCtassmishorteccenxxxxxC}{\ensuremath{11.3}}      % APASS i-band magnitude
\newcommand{\hatcurCCparallaxeccenxxxxxC}{\ensuremath{1.744\pm0.035}} % Gaia DR2 parallax [mas]
\newcommand{\hatcurCCgaiamGeccenxxxxxC}{\ensuremath{11.48770\pm0.00080}} % Gaia G-band magnitude
\newcommand{\hatcurCCgaiamBPeccenxxxxxC}{\ensuremath{11.7645\pm0.0016}} % Gaia BP-band magnitude
\newcommand{\hatcurCCgaiamRPeccenxxxxxC}{\ensuremath{11.0523\pm0.0022}} % Gaia RP-band magnitude
\newcommand{\hatcurCCtwomassJmageccenxxxxxC}{\ensuremath{10.514\pm0.023}} % 2MASS ORIG MAG
\newcommand{\hatcurCCtwomassHmageccenxxxxxC}{\ensuremath{10.325\pm0.029}} % 2MASS ORIG MAG
\newcommand{\hatcurCCtwomassKmageccenxxxxxC}{\ensuremath{10.251\pm0.019}} % 2MASS ORIG MAG
\newcommand{\hatcurCCcitJmageccenxxxxxC}{\ensuremath{10.537\pm0.023}} % 2MASS CIT MAG
\newcommand{\hatcurCCcitHmageccenxxxxxC}{\ensuremath{10.320\pm0.029}} % 2MASS CIT MAG
\newcommand{\hatcurCCcitKmageccenxxxxxC}{\ensuremath{10.275\pm0.019}} % 2MASS CIT MAG
\newcommand{\hatcurCCbbJmageccenxxxxxC}{\ensuremath{10.577\pm0.025}} % 2MASS BB MAG
\newcommand{\hatcurCCbbHmageccenxxxxxC}{\ensuremath{10.342\pm0.030}} % 2MASS BB MAG
\newcommand{\hatcurCCbbKmageccenxxxxxC}{\ensuremath{10.295\pm0.019}} % 2MASS BB MAG
\newcommand{\hatcurCCesoJmageccenxxxxxC}{\ensuremath{10.578\pm0.026}} % 2MASS ESO MAG
\newcommand{\hatcurCCesoHmageccenxxxxxC}{\ensuremath{10.337\pm0.033}} % 2MASS ESO MAG
\newcommand{\hatcurCCesoKmageccenxxxxxC}{\ensuremath{10.294\pm0.020}} % 2MASS ESO MAG
\newcommand{\hatcurCCesoJHmageccenxxxxxC}{\ensuremath{0.241\pm0.040}} % 2MASS ESO JH COLOR
\newcommand{\hatcurCCesoJKmageccenxxxxxC}{\ensuremath{0.284\pm0.032}} % 2MASS ESO JK COLOR
\newcommand{\hatcurCCesoHKmageccenxxxxxC}{\ensuremath{0.043\pm0.039}} % 2MASS ESO HK COLOR
\newcommand{\hatcurLCdipeccenxxxxxC}{\ensuremath{6.7}}               % BLS detected dip (mmag)
\newcommand{\hatcurLCrprstareccenxxxxxC}{\ensuremath{0.0796\pm0.0011}} % Rp/R*
\newcommand{\hatcurLCbsqeccenxxxxxC}{\ensuremath{0.456_{-0.018}^{+0.017}}} % impact parameter square
\newcommand{\hatcurLCimpeccenxxxxxC}{\ensuremath{0.675_{-0.014}^{+0.012}}} % impact parameter
\newcommand{\hatcurLCzetaeccenxxxxxC}{\ensuremath{11.695\pm0.090}}   % zeta/R*
\newcommand{\hatcurLCdureccenxxxxxC}{\ensuremath{0.1954\pm0.0016}}   % transit duration (days)
\newcommand{\hatcurLCdurshorteccenxxxxxC}{\ensuremath{0.1954}}       % transit duration (days)
\newcommand{\hatcurLCdurhreccenxxxxxC}{\ensuremath{4.690\pm0.039}}   % transit duration (hours)
\newcommand{\hatcurLCdurhrshorteccenxxxxxC}{\ensuremath{4.690}}      % transit duration (hours)
\newcommand{\hatcurLCqeccenxxxxxC}{\ensuremath{0.04520\pm0.00038}}   % fractional transit duration (days)
\newcommand{\hatcurLCqshorteccenxxxxxC}{\ensuremath{0.045}}          % fractional transit duration (days)
\newcommand{\hatcurLCingdureccenxxxxxC}{\ensuremath{0.02515\pm0.00077}} % ingress/egress duration (days)
\newcommand{\hatcurLCPeccenxxxxxC}{\ensuremath{4.324813\pm0.000020}} % period (days)
\newcommand{\hatcurLCPprececcenxxxxxC}{\ensuremath{4.3248126}}       % period (days)
\newcommand{\hatcurLCPshorteccenxxxxxC}{\ensuremath{4.3248}}         % period (days)
\newcommand{\hatcurLCTeccenxxxxxC}{\ensuremath{2457788.00210\pm0.00079}} % epoch (BJD)
\newcommand{\hatcurLCTAeccenxxxxxC}{\ensuremath{2457091.7073\pm0.0031}} % TA (BJD)
\newcommand{\hatcurLCTBeccenxxxxxC}{\ensuremath{2457839.89983\pm0.00091}} % TB (BJD)
\newcommand{\hatcurLChatnetmAeccenxxxxxC}{\ensuremath{11.402810\pm0.000095}} % HATNet OOT level
\newcommand{\hatcurLCiblendAeccenxxxxxC}{\ensuremath{0.679\pm0.032}} % HATNet iblend factor
\newcommand{\hatcurLChatnetmBeccenxxxxxC}{\ensuremath{11.402720\pm0.000045}} % HATNet OOT level
\newcommand{\hatcurLCiblendBeccenxxxxxC}{\ensuremath{0.846\pm0.034}} % HATNet iblend factor
\newcommand{\hatcurLCrhoeccenxxxxxC}{\ensuremath{0.191\pm0.023}}     % stellar density no isochrone constraint (cgs)
\newcommand{\hatcurSMEiteffeccenxxxxxC}{\ensuremath{6652\pm75}}      % Ini SME, stellar effective temperature
\newcommand{\hatcurSMEizfeheccenxxxxxC}{\ensuremath{0.240\pm0.034}}  % Ini SME, stellar metallicity
\newcommand{\hatcurSMEizfehshorteccenxxxxxC}{\ensuremath{0.24}}      % Ini SME, stellar metallicity
\newcommand{\hatcurSMEiloggeccenxxxxxC}{\ensuremath{4.36\pm0.14}}    % Ini SME, stellar surface gravity
\newcommand{\hatcurSMEivsineccenxxxxxC}{\ensuremath{6.43\pm0.20}}    % Ini SME, stellar rotational velocity
\newcommand{\hatcurSMEivmaceccenxxxxxC}{\ensuremath{5.34\pm0.11}}    % Ini SME, stellar macroturbulence
\newcommand{\hatcurSMEivmiceccenxxxxxC}{\ensuremath{1.99\pm0.12}}    % Ini SME, stellar microturbulence
\newcommand{\hatcurSMEiiteffeccenxxxxxC}{\ensuremath{6552\pm61}}     % Final SME, stellar effective temperature
\newcommand{\hatcurSMEiizfeheccenxxxxxC}{\ensuremath{0.200\pm0.025}} % Final SME, stellar metallicity
\newcommand{\hatcurSMEiizfehshorteccenxxxxxC}{\ensuremath{0.2}}      % Final SME, stellar metallicity
\newcommand{\hatcurSMEiiloggeccenxxxxxC}{\ensuremath{4.016\pm0.079}} % Final SME, stellar surface gravity
\newcommand{\hatcurSMEiivsineccenxxxxxC}{\ensuremath{6.49\pm0.19}}   % Final SME, stellar rotational velocity
\newcommand{\hatcurSMEiivmaceccenxxxxxC}{\ensuremath{5.183\pm0.093}} % Final SME, stellar macroturbulence
\newcommand{\hatcurSMEiivmiceccenxxxxxC}{\ensuremath{1.832\pm0.090}} % Final SME, stellar microturbulence
\newcommand{\hatcurLBizeccenxxxxxC}{\ensuremath{0.1118}}             % Limb darkening parameters, Gamma1, z-band
\newcommand{\hatcurLBiizeccenxxxxxC}{\ensuremath{0.3797}}            % Limb darkening parameters, Gamma2, z-band
\newcommand{\hatcurLBiieccenxxxxxC}{\ensuremath{0.1618}}             % Limb darkening parameters, Gamma1, i-band
\newcommand{\hatcurLBiiieccenxxxxxC}{\ensuremath{0.3903}}            % Limb darkening parameters, Gamma2, i-band
\newcommand{\hatcurLBiIeccenxxxxxC}{\ensuremath{0.1457}}             % Limb darkening parameters, Gamma1, I-band
\newcommand{\hatcurLBiiIeccenxxxxxC}{\ensuremath{0.3773}}            % Limb darkening parameters, Gamma2, I-band
\newcommand{\hatcurLBigeccenxxxxxC}{\ensuremath{0.4056}}             % Limb darkening parameters, Gamma1, g-band
\newcommand{\hatcurLBiigeccenxxxxxC}{\ensuremath{0.3422}}            % Limb darkening parameters, Gamma2, g-band
\newcommand{\hatcurLBireccenxxxxxC}{\ensuremath{0.2388}}             % Limb darkening parameters, Gamma1, r-band
\newcommand{\hatcurLBiireccenxxxxxC}{\ensuremath{0.3963}}            % Limb darkening parameters, Gamma2, r-band
\newcommand{\hatcurLBiReccenxxxxxC}{\ensuremath{0.2170}}             % Limb darkening parameters, Gamma1, R-band
\newcommand{\hatcurLBiiReccenxxxxxC}{\ensuremath{0.3871}}            % Limb darkening parameters, Gamma2, R-band
\newcommand{\hatcurLBikepeccenxxxxxC}{\ensuremath{0.1000}}           % Limb darkening parameters, Gamma1, Kep-band
\newcommand{\hatcurLBiikepeccenxxxxxC}{\ensuremath{0.1000}}          % Limb darkening parameters, Gamma2, Kep-band
\newcommand{\hatcurISOmeccenxxxxxC}{\ensuremath{2.186\pm0.042}}      % stellar mass
\newcommand{\hatcurISOmshorteccenxxxxxC}{\ensuremath{2.19}}          % stellar mass
\newcommand{\hatcurISOmlongeccenxxxxxC}{\ensuremath{2.186\pm0.042}}  % stellar mass
\newcommand{\hatcurISOreccenxxxxxC}{\ensuremath{3.954\pm0.015}}      % stellar radius
\newcommand{\hatcurISOrshorteccenxxxxxC}{\ensuremath{3.95}}          % stellar radius
\newcommand{\hatcurISOrlongeccenxxxxxC}{\ensuremath{3.954\pm0.015}}  % stellar radius
\newcommand{\hatcurISOrhoeccenxxxxxC}{\ensuremath{0.191\pm0.023}}    % stellar density (cgs)
\newcommand{\hatcurISOrholongeccenxxxxxC}{\ensuremath{0.191\pm0.023}} % stellar density (cgs)
\newcommand{\hatcurISOloggeccenxxxxxC}{\ensuremath{1.909\pm0.064}}   % stellar surface gravity from isochrones
\newcommand{\hatcurISOlumeccenxxxxxC}{\ensuremath{2.400\pm0.040}}    % stellar luminosity
\newcommand{\hatcurISOlumshorteccenxxxxxC}{\ensuremath{2.40}}        % stellar luminosity
\newcommand{\hatcurISOteffeccenxxxxxC}{\ensuremath{1.569\pm0.015}}   % stellar effective temperature adjusted via MCMC
\newcommand{\hatcurISOzfeheccenxxxxxC}{\ensuremath{6535\pm25}}       % stellar [M/H] adjusted via MCMC
\newcommand{\hatcurISOageeccenxxxxxC}{\ensuremath{7.80\pm0.29}}      % stellar age
\newcommand{\hatcurISOspececcenxxxxxC}{F}                            % stellar spectral type
\newcommand{\hatcurRVKeccenxxxxxC}{\ensuremath{55.5\pm1.5}}          % RV semi-amplitude [m/s]
\newcommand{\hatcurRVrkeccenxxxxxC}{\ensuremath{0.025_{-0.072}^{+0.053}}} % sqrt(e)*cos(omega)
\newcommand{\hatcurRVrheccenxxxxxC}{\ensuremath{-0.018\pm0.047}}     % sqrt(e)*sin(omega)
\newcommand{\hatcurRVkeccenxxxxxC}{\ensuremath{0.0012_{-0.0043}^{+0.0063}}} % e*cos(omega)
\newcommand{\hatcurRVheccenxxxxxC}{\ensuremath{-0.0010_{-0.0048}^{+0.0034}}} % e*sin(omega)
\newcommand{\hatcurRVtroneeccenxxxxxC}{\ensuremath{6.462\pm0.026}}   % RV linear trend tr1 factor
\newcommand{\hatcurRVtrtwoeccenxxxxxC}{\ensuremath{-0.003400\pm0.000038}} % RV linear trend tr2 factor
\newcommand{\hatcurRVgammaAeccenxxxxxC}{\ensuremath{35148.2\pm5.3}}  % RV gamma velocity, relative scale
\newcommand{\hatcurRVjitterAeccenxxxxxC}{\ensuremath{21.3\pm3.2}}    % RV jitter (m/s)
\newcommand{\hatcurRVjittertwosiglimAeccenxxxxxC}{\ensuremath{<27.7}} % RV jitter (m/s) 95 percent confidence upper limit
\newcommand{\hatcurRVfitrmsAeccenxxxxxC}{\ensuremath{0.0}}           % RVfitrms
\newcommand{\hatcurRVgammaBeccenxxxxxC}{\ensuremath{35142.2\pm6.2}}  % RV gamma velocity, relative scale
\newcommand{\hatcurRVjitterBeccenxxxxxC}{\ensuremath{0.1\pm2.4}}     % RV jitter (m/s)
\newcommand{\hatcurRVjittertwosiglimBeccenxxxxxC}{\ensuremath{<6.2}} % RV jitter (m/s) 95 percent confidence upper limit
\newcommand{\hatcurRVfitrmsBeccenxxxxxC}{\ensuremath{0.0}}           % RVfitrms
\newcommand{\hatcurRVecceneccenxxxxxC}{\ensuremath{0.0050\pm0.0070}} % eccentricity
\newcommand{\hatcurRVeccentwosiglimeccenxxxxxC}{\ensuremath{<0.019}} % eccentricity
\newcommand{\hatcurRVomegaeccenxxxxxC}{\ensuremath{240\pm120}}       % argument of pericenter
\newcommand{\hatcurPPieccenxxxxxC}{\ensuremath{79.40\pm0.11}}        % orbital inclination
\newcommand{\hatcurPPgeccenxxxxxC}{\ensuremath{2.011\pm0.094}}       % planetary surface gravity (m/s^2)
\newcommand{\hatcurPPloggeccenxxxxxC}{\ensuremath{2.303\pm0.020}}    % planetary surface gravity (log cgs)
\newcommand{\hatcurPPareccenxxxxxC}{\ensuremath{3.667\pm0.037}}      % relative orbital radius (a/R*)
\newcommand{\hatcurPPareleccenxxxxxC}{\ensuremath{0.06743\pm0.00043}} % semimajor axis (AU)
\newcommand{\hatcurPPrhoeccenxxxxxC}{\ensuremath{0.0330\pm0.0020}}   % planetary density (cgs)
\newcommand{\hatcurPPmeccenxxxxxC}{\ensuremath{0.760\pm0.021}}       % planetary mass (M_jup)
\newcommand{\hatcurPPmshorteccenxxxxxC}{\ensuremath{0.76}}           % planetary mass (M_jup)
\newcommand{\hatcurPPmlongeccenxxxxxC}{\ensuremath{0.760\pm0.021}}   % planetary mass (M_jup)
\newcommand{\hatcurPPmeeccenxxxxxC}{\ensuremath{241.7\pm6.6}}        % planetary mass (M_earth)
\newcommand{\hatcurPPmeshorteccenxxxxxC}{\ensuremath{241.7}}         % planetary mass (M_earth)
\newcommand{\hatcurPPmelongeccenxxxxxC}{\ensuremath{241.7\pm6.6}}    % planetary mass (M_earth)
\newcommand{\hatcurPPreccenxxxxxC}{\ensuremath{3.064\pm0.046}}       % planetary radius (R_jup)
\newcommand{\hatcurPPrshorteccenxxxxxC}{\ensuremath{3.06}}           % planetary radius (R_jup)
\newcommand{\hatcurPPrlongeccenxxxxxC}{\ensuremath{3.064\pm0.046}}   % planetary radius (R_jup)
\newcommand{\hatcurPPreeccenxxxxxC}{\ensuremath{34.34\pm0.51}}       % planetary radius (R_earth)
\newcommand{\hatcurPPreshorteccenxxxxxC}{\ensuremath{34.3}}          % planetary radius (R_earth)
\newcommand{\hatcurPPrelongeccenxxxxxC}{\ensuremath{34.34\pm0.51}}   % planetary radius (R_earth)
\newcommand{\hatcurPPmrcorreccenxxxxxC}{\ensuremath{-0.32}}          % mass/radius correlation
\newcommand{\hatcurPPteffeccenxxxxxC}{\ensuremath{1\pm0}}            % planetary temperature (K)
\newcommand{\hatcurPPthetaeccenxxxxxC}{\ensuremath{0.01500\pm0.00046}} % Safranov number
\newcommand{\hatcurPPfluxperieccenxxxxxC}{\ensuremath{7.28\pm0.22}}  % flux @ periastron (CGS)
\newcommand{\hatcurPPfluxperidimeccenxxxxxC}{\ensuremath{8}}         % flux @ periastron (CGS) units.
\newcommand{\hatcurPPfluxapeccenxxxxxC}{\ensuremath{7.10\pm0.24}}    % flux @ apastron (CGS)
\newcommand{\hatcurPPfluxapdimeccenxxxxxC}{\ensuremath{8}}           % flux @ apastron (CGS) units.
\newcommand{\hatcurPPfluxavgeccenxxxxxC}{\ensuremath{7.18\pm0.21}}   % flux on average (CGS)
\newcommand{\hatcurPPfluxavgdimeccenxxxxxC}{\ensuremath{8}}          % flux average (CGS) units.
\newcommand{\hatcurPPfluxavglogeccenxxxxxC}{\ensuremath{8.856\pm0.013}} % log10 flux on average (CGS)
\newcommand{\hatcurXsecphaseeccenxxxxxC}{\ensuremath{0.5008\pm0.0054}} % Phase of secondary eclipse
\newcommand{\hatcurXsecondaryeccenxxxxxC}{\ensuremath{2457790.168\pm0.023}} % Secondary eclipse epoch
\newcommand{\hatcurXsecdureccenxxxxxC}{\ensuremath{0.1953\pm0.0018}} % sec eclipse duration (days)
\newcommand{\hatcurXsecingdureccenxxxxxC}{\ensuremath{0.02502\pm0.00078}} % sec I/E duration (days)
\newcommand{\hatcurPPphiconjeccenxxxxxC}{\ensuremath{0.18_{-0.47}^{+0.18}}} % phase diff between conjunction and periastron
\newcommand{\hatcurPPperieccenxxxxxC}{\ensuremath{2457787.2\pm1.3}}  % time of periastron passage.
\newcommand{\hatcurPPaequiveccenxxxxxC}{\ensuremath{0.04350\pm0.00063}} % equivalent semi-major axis
\newcommand{\hatcurPPtcirceccenxxxxxC}{\ensuremath{8.20\pm0.81}}     % circularization timescale
\newcommand{\hatcurPPtinfalleccenxxxxxC}{\ensuremath{78.3\pm5.3}}    % infall timescale
\newcommand{\hatcurXdisteccenxxxxxC}{\ensuremath{11.700_{-0.100}^{+1.000}}} % distance (pc), no reddenning correction
\newcommand{\hatcurXAveccenxxxxxC}{\ensuremath{6535\pm25}}           % Av (mag)
\newcommand{\hatcurXdistredeccenxxxxxC}{\ensuremath{11.674\pm0.061}} % distance with Av correction (pc)
\newcommand{\hatcurXEBVeccenxxxxxC}{\ensuremath{2108.2\pm8.2}}       % E(B-V) (mag)
\newcommand{\hatcurCCpmraeccenxxxxxC}{\ensuremath{-8.604\pm0.046}}   % proper motion, in RA
\newcommand{\hatcurCCpmdececcenxxxxxC}{\ensuremath{-2.950\pm0.035}}  % proper motion, in DEC
\newcommand{\hatcurCCpmeccenxxxxxC}{\ensuremath{9.096\pm0.058}}      % proper motion

\newcommand{\hatcurhtreccenxxxxxD}{HATS548-012}                      % Original HTR name of target
\newcommand{\hatcurfieldeccenxxxxxD}{\ensuremath{string}}            % HTR field
\newcommand{\hatcurCCraeccenxxxxxD}{\ensuremath{04^{\mathrm h}03^{\mathrm m}47.6005{\mathrm s}}}                   % Right Ascension
\newcommand{\hatcurCCdececcenxxxxxD}{\ensuremath{-19{\arcdeg}03{\arcmin}24.3267{\arcsec}}}                 % Declination
\newcommand{\hatcurCCmageccenxxxxxD}{12.344}                         % apparent V-band magnitude
\newcommand{\hatcurCCtwomasseccenxxxxxD}{2MASS~04034760-1903242}     % 2MASS identifier
\newcommand{\hatcurCCgsceccenxxxxxD}{GSC~5885-00663}                 % GSC(1.2) identifier
\newcommand{\hatcurCCgaiaeccenxxxxxD}{GAIA~5094406188917950848}      % GAIA DR1 identifier
\newcommand{\hatcurCCgaiadrtwoeccenxxxxxD}{GAIA~DR2~5094406193214399616} % GAIA DR2 identifier
\newcommand{\hatcurCCtassmveccenxxxxxD}{\ensuremath{12.344\pm0.047}} % APASS V-band magnitude
\newcommand{\hatcurCCtassmvshorteccenxxxxxD}{\ensuremath{12.3}}      % APASS V-band magnitude
\newcommand{\hatcurCCtassmBeccenxxxxxD}{\ensuremath{13.094\pm0.096}} % APASS B-band magnitude
\newcommand{\hatcurCCtassmBshorteccenxxxxxD}{\ensuremath{13.1}}      % APASS B-band magnitude
\newcommand{\hatcurCCtassmIeccenxxxxxD}{\ensuremath{nff\pmnff}}      % TASS I-band magnitude
\newcommand{\hatcurCCtassmIshorteccenxxxxxD}{\ensuremath{0.0}}       % TASS I-band magnitude
\newcommand{\hatcurCCtassmgeccenxxxxxD}{\ensuremath{12.669\pm0.035}} % APASS g-band magnitude
\newcommand{\hatcurCCtassmgshorteccenxxxxxD}{\ensuremath{12.7}}      % APASS g-band magnitude
\newcommand{\hatcurCCtassmreccenxxxxxD}{\ensuremath{12.129\pm0.058}} % APASS r-band magnitude
\newcommand{\hatcurCCtassmrshorteccenxxxxxD}{\ensuremath{12.1}}      % APASS r-band magnitude
\newcommand{\hatcurCCtassmieccenxxxxxD}{\ensuremath{11.949\pm0.063}} % APASS i-band magnitude
\newcommand{\hatcurCCtassmishorteccenxxxxxD}{\ensuremath{11.9}}      % APASS i-band magnitude
\newcommand{\hatcurCCparallaxeccenxxxxxD}{\ensuremath{3.550\pm0.039}} % Gaia DR2 parallax [mas]
\newcommand{\hatcurCCgaiamGeccenxxxxxD}{\ensuremath{12.18160\pm0.00070}} % Gaia G-band magnitude
\newcommand{\hatcurCCgaiamBPeccenxxxxxD}{\ensuremath{12.5621\pm0.0027}} % Gaia BP-band magnitude
\newcommand{\hatcurCCgaiamRPeccenxxxxxD}{\ensuremath{11.6542\pm0.0024}} % Gaia RP-band magnitude
\newcommand{\hatcurCCtwomassJmageccenxxxxxD}{\ensuremath{11.071\pm0.026}} % 2MASS ORIG MAG
\newcommand{\hatcurCCtwomassHmageccenxxxxxD}{\ensuremath{10.738\pm0.023}} % 2MASS ORIG MAG
\newcommand{\hatcurCCtwomassKmageccenxxxxxD}{\ensuremath{10.707\pm0.027}} % 2MASS ORIG MAG
\newcommand{\hatcurCCcitJmageccenxxxxxD}{\ensuremath{11.088\pm0.026}} % 2MASS CIT MAG
\newcommand{\hatcurCCcitHmageccenxxxxxD}{\ensuremath{10.734\pm0.023}} % 2MASS CIT MAG
\newcommand{\hatcurCCcitKmageccenxxxxxD}{\ensuremath{10.731\pm0.027}} % 2MASS CIT MAG
\newcommand{\hatcurCCbbJmageccenxxxxxD}{\ensuremath{11.137\pm0.028}} % 2MASS BB MAG
\newcommand{\hatcurCCbbHmageccenxxxxxD}{\ensuremath{10.754\pm0.024}} % 2MASS BB MAG
\newcommand{\hatcurCCbbKmageccenxxxxxD}{\ensuremath{10.751\pm0.027}} % 2MASS BB MAG
\newcommand{\hatcurCCesoJmageccenxxxxxD}{\ensuremath{11.139\pm0.029}} % 2MASS ESO MAG
\newcommand{\hatcurCCesoHmageccenxxxxxD}{\ensuremath{10.747\pm0.027}} % 2MASS ESO MAG
\newcommand{\hatcurCCesoKmageccenxxxxxD}{\ensuremath{10.750\pm0.028}} % 2MASS ESO MAG
\newcommand{\hatcurCCesoJHmageccenxxxxxD}{\ensuremath{0.391\pm0.038}} % 2MASS ESO JH COLOR
\newcommand{\hatcurCCesoJKmageccenxxxxxD}{\ensuremath{0.390\pm0.040}} % 2MASS ESO JK COLOR
\newcommand{\hatcurCCesoHKmageccenxxxxxD}{\ensuremath{-0.002\pm0.039}} % 2MASS ESO HK COLOR
\newcommand{\hatcurLCdipeccenxxxxxD}{\ensuremath{16.2}}              % BLS detected dip (mmag)
\newcommand{\hatcurLCrprstareccenxxxxxD}{\ensuremath{0.1238\pm0.0019}} % Rp/R*
\newcommand{\hatcurLCbsqeccenxxxxxD}{\ensuremath{0.072_{-0.037}^{+0.041}}} % impact parameter square
\newcommand{\hatcurLCimpeccenxxxxxD}{\ensuremath{0.269_{-0.080}^{+0.067}}} % impact parameter
\newcommand{\hatcurLCzetaeccenxxxxxD}{\ensuremath{21.81\pm0.11}}     % zeta/R*
\newcommand{\hatcurLCdureccenxxxxxD}{\ensuremath{0.10393\pm0.00072}} % transit duration (days)
\newcommand{\hatcurLCdurshorteccenxxxxxD}{\ensuremath{0.1039}}       % transit duration (days)
\newcommand{\hatcurLCdurhreccenxxxxxD}{\ensuremath{2.494\pm0.017}}   % transit duration (hours)
\newcommand{\hatcurLCdurhrshorteccenxxxxxD}{\ensuremath{2.494}}      % transit duration (hours)
\newcommand{\hatcurLCqeccenxxxxxD}{\ensuremath{0.04420\pm0.00031}}   % fractional transit duration (days)
\newcommand{\hatcurLCqshorteccenxxxxxD}{\ensuremath{0.044}}          % fractional transit duration (days)
\newcommand{\hatcurLCingdureccenxxxxxD}{\ensuremath{0.01226\pm0.00056}} % ingress/egress duration (days)
\newcommand{\hatcurLCPeccenxxxxxD}{\ensuremath{2.3506214\pm0.0000012}} % period (days)
\newcommand{\hatcurLCPprececcenxxxxxD}{\ensuremath{2.3506214}}       % period (days)
\newcommand{\hatcurLCPshorteccenxxxxxD}{\ensuremath{2.3506}}         % period (days)
\newcommand{\hatcurLCTeccenxxxxxD}{\ensuremath{2457818.45646\pm0.00026}} % epoch (BJD)
\newcommand{\hatcurLCTAeccenxxxxxD}{\ensuremath{2456840.59793\pm0.00052}} % TA (BJD)
\newcommand{\hatcurLCTBeccenxxxxxD}{\ensuremath{2458048.81736\pm0.00031}} % TB (BJD)
\newcommand{\hatcurLChatnetmeccenxxxxxD}{\ensuremath{12.185290\pm0.000055}} % HATNet OOT level
\newcommand{\hatcurLCiblendeccenxxxxxD}{\ensuremath{0.765\pm0.028}}  % HATNet iblend factor
\newcommand{\hatcurLCrhoeccenxxxxxD}{\ensuremath{0.252\pm0.040}}     % stellar density no isochrone constraint (cgs)
\newcommand{\hatcurSMEiteffeccenxxxxxD}{\ensuremath{5632\pm64}}      % Ini SME, stellar effective temperature
\newcommand{\hatcurSMEizfeheccenxxxxxD}{\ensuremath{0.160\pm0.041}}  % Ini SME, stellar metallicity
\newcommand{\hatcurSMEizfehshorteccenxxxxxD}{\ensuremath{0.16}}      % Ini SME, stellar metallicity
\newcommand{\hatcurSMEiloggeccenxxxxxD}{\ensuremath{4.440\pm0.090}}  % Ini SME, stellar surface gravity
\newcommand{\hatcurSMEivsineccenxxxxxD}{\ensuremath{4.18\pm0.45}}    % Ini SME, stellar rotational velocity
\newcommand{\hatcurSMEivmaceccenxxxxxD}{\ensuremath{3.768\pm0.097}}  % Ini SME, stellar macroturbulence
\newcommand{\hatcurSMEivmiceccenxxxxxD}{\ensuremath{1.000\pm0.033}}  % Ini SME, stellar microturbulence
\newcommand{\hatcurSMEiiteffeccenxxxxxD}{\ensuremath{5659\pm84}}     % Final SME, stellar effective temperature
\newcommand{\hatcurSMEiizfeheccenxxxxxD}{\ensuremath{0.160\pm0.059}} % Final SME, stellar metallicity
\newcommand{\hatcurSMEiizfehshorteccenxxxxxD}{\ensuremath{0.16}}     % Final SME, stellar metallicity
\newcommand{\hatcurSMEiiloggeccenxxxxxD}{\ensuremath{4.504\pm0.014}} % Final SME, stellar surface gravity
\newcommand{\hatcurSMEiivsineccenxxxxxD}{\ensuremath{4.09\pm0.48}}   % Final SME, stellar rotational velocity
\newcommand{\hatcurSMEiivmaceccenxxxxxD}{\ensuremath{3.81\pm0.13}}   % Final SME, stellar macroturbulence
\newcommand{\hatcurSMEiivmiceccenxxxxxD}{\ensuremath{1.014\pm0.045}} % Final SME, stellar microturbulence
\newcommand{\hatcurLBizeccenxxxxxD}{\ensuremath{0.2279}}             % Limb darkening parameters, Gamma1, z-band
\newcommand{\hatcurLBiizeccenxxxxxD}{\ensuremath{0.3213}}            % Limb darkening parameters, Gamma2, z-band
\newcommand{\hatcurLBiieccenxxxxxD}{\ensuremath{0.2951}}             % Limb darkening parameters, Gamma1, i-band
\newcommand{\hatcurLBiiieccenxxxxxD}{\ensuremath{0.3188}}            % Limb darkening parameters, Gamma2, i-band
\newcommand{\hatcurLBiIeccenxxxxxD}{\ensuremath{0.2727}}             % Limb darkening parameters, Gamma1, I-band
\newcommand{\hatcurLBiiIeccenxxxxxD}{\ensuremath{0.3201}}            % Limb darkening parameters, Gamma2, I-band
\newcommand{\hatcurLBigeccenxxxxxD}{\ensuremath{0.5973}}             % Limb darkening parameters, Gamma1, g-band
\newcommand{\hatcurLBiigeccenxxxxxD}{\ensuremath{0.2037}}            % Limb darkening parameters, Gamma2, g-band
\newcommand{\hatcurLBireccenxxxxxD}{\ensuremath{0.3913}}             % Limb darkening parameters, Gamma1, r-band
\newcommand{\hatcurLBiireccenxxxxxD}{\ensuremath{0.3070}}            % Limb darkening parameters, Gamma2, r-band
\newcommand{\hatcurLBiReccenxxxxxD}{\ensuremath{0.3646}}             % Limb darkening parameters, Gamma1, R-band
\newcommand{\hatcurLBiiReccenxxxxxD}{\ensuremath{0.3112}}            % Limb darkening parameters, Gamma2, R-band
\newcommand{\hatcurLBikepeccenxxxxxD}{\ensuremath{0.1000}}           % Limb darkening parameters, Gamma1, Kep-band
\newcommand{\hatcurLBiikepeccenxxxxxD}{\ensuremath{0.1000}}          % Limb darkening parameters, Gamma2, Kep-band
\newcommand{\hatcurISOmeccenxxxxxD}{\ensuremath{0.960\pm0.011}}      % stellar mass
\newcommand{\hatcurISOmshorteccenxxxxxD}{\ensuremath{0.96}}          % stellar mass
\newcommand{\hatcurISOmlongeccenxxxxxD}{\ensuremath{0.960\pm0.011}}  % stellar mass
\newcommand{\hatcurISOreccenxxxxxD}{\ensuremath{4.481\pm0.018}}      % stellar radius
\newcommand{\hatcurISOrshorteccenxxxxxD}{\ensuremath{4.48}}          % stellar radius
\newcommand{\hatcurISOrlongeccenxxxxxD}{\ensuremath{4.481\pm0.018}}  % stellar radius
\newcommand{\hatcurISOrhoeccenxxxxxD}{\ensuremath{0.252\pm0.040}}    % stellar density (cgs)
\newcommand{\hatcurISOrholongeccenxxxxxD}{\ensuremath{0.252\pm0.040}} % stellar density (cgs)
\newcommand{\hatcurISOloggeccenxxxxxD}{\ensuremath{2.8\pm1.6}}       % stellar surface gravity from isochrones
\newcommand{\hatcurISOlumeccenxxxxxD}{\ensuremath{4.908\pm0.023}}    % stellar luminosity
\newcommand{\hatcurISOlumshorteccenxxxxxD}{\ensuremath{4.91}}        % stellar luminosity
\newcommand{\hatcurISOteffeccenxxxxxD}{\ensuremath{1.018_{-0.030}^{+0.020}}} % stellar effective temperature adjusted via MCMC
\newcommand{\hatcurISOzfeheccenxxxxxD}{\ensuremath{5584\pm19}}       % stellar [M/H] adjusted via MCMC
\newcommand{\hatcurISOageeccenxxxxxD}{\ensuremath{0.801\pm0.017}}    % stellar age
\newcommand{\hatcurISOspececcenxxxxxD}{G}                            % stellar spectral type
\newcommand{\hatcurRVKeccenxxxxxD}{\ensuremath{481.6\pm7.1}}         % RV semi-amplitude [m/s]
\newcommand{\hatcurRVrkeccenxxxxxD}{\ensuremath{-0.010\pm0.067}}     % sqrt(e)*cos(omega)
\newcommand{\hatcurRVrheccenxxxxxD}{\ensuremath{0.060_{-0.095}^{+0.068}}} % sqrt(e)*sin(omega)
\newcommand{\hatcurRVkeccenxxxxxD}{\ensuremath{-0.0006_{-0.0092}^{+0.0066}}} % e*cos(omega)
\newcommand{\hatcurRVheccenxxxxxD}{\ensuremath{0.0056_{-0.0080}^{+0.0127}}} % e*sin(omega)
\newcommand{\hatcurRVtroneeccenxxxxxD}{\ensuremath{0\pm0}}           % RV linear trend tr1 factor
\newcommand{\hatcurRVtrtwoeccenxxxxxD}{\ensuremath{0\pm0}}           % RV linear trend tr2 factor
\newcommand{\hatcurRVgammaeccenxxxxxD}{\ensuremath{544.9\pm6.9}}     % RV gamma velocity, relative scale
\newcommand{\hatcurRVjittereccenxxxxxD}{\ensuremath{25.6\pm6.2}}     % RV jitter (m/s)
\newcommand{\hatcurRVjittertwosiglimeccenxxxxxD}{\ensuremath{<36.7}} % RV jitter (m/s) 95 percent confidence upper limit
\newcommand{\hatcurRVfitrmseccenxxxxxD}{\ensuremath{.1fym}}          % 
\newcommand{\hatcurRVecceneccenxxxxxD}{\ensuremath{0.0120\pm0.0086}} % eccentricity
\newcommand{\hatcurRVeccentwosiglimeccenxxxxxD}{\ensuremath{<0.028}} % eccentricity
\newcommand{\hatcurRVomegaeccenxxxxxD}{\ensuremath{120\pm92}}        % argument of pericenter
\newcommand{\hatcurPPieccenxxxxxD}{\ensuremath{80.5\pm2.5}}          % orbital inclination
\newcommand{\hatcurPPgeccenxxxxxD}{\ensuremath{2.65\pm0.12}}         % planetary surface gravity (m/s^2)
\newcommand{\hatcurPPloggeccenxxxxxD}{\ensuremath{2.424\pm0.019}}    % planetary surface gravity (log cgs)
\newcommand{\hatcurPPareccenxxxxxD}{\ensuremath{1.640\pm0.013}}      % relative orbital radius (a/R*)
\newcommand{\hatcurPPareleccenxxxxxD}{\ensuremath{0.03416\pm0.00014}} % semimajor axis (AU)
\newcommand{\hatcurPPrhoeccenxxxxxD}{\ensuremath{0.02500_{-0.00200}^{+0.00100}}} % planetary density (cgs)
\newcommand{\hatcurPPmeccenxxxxxD}{\ensuremath{3.112\pm0.059}}       % planetary mass (M_jup)
\newcommand{\hatcurPPmshorteccenxxxxxD}{\ensuremath{3.11}}           % planetary mass (M_jup)
\newcommand{\hatcurPPmlongeccenxxxxxD}{\ensuremath{3.112\pm0.059}}   % planetary mass (M_jup)
\newcommand{\hatcurPPmeeccenxxxxxD}{\ensuremath{989\pm19}}           % planetary mass (M_earth)
\newcommand{\hatcurPPmeshorteccenxxxxxD}{\ensuremath{989.1}}         % planetary mass (M_earth)
\newcommand{\hatcurPPmelongeccenxxxxxD}{\ensuremath{989\pm19}}       % planetary mass (M_earth)
\newcommand{\hatcurPPreccenxxxxxD}{\ensuremath{5.394\pm0.085}}       % planetary radius (R_jup)
\newcommand{\hatcurPPrshorteccenxxxxxD}{\ensuremath{5.39}}           % planetary radius (R_jup)
\newcommand{\hatcurPPrlongeccenxxxxxD}{\ensuremath{5.394\pm0.085}}   % planetary radius (R_jup)
\newcommand{\hatcurPPreeccenxxxxxD}{\ensuremath{60.46\pm0.96}}       % planetary radius (R_earth)
\newcommand{\hatcurPPreshorteccenxxxxxD}{\ensuremath{60.5}}          % planetary radius (R_earth)
\newcommand{\hatcurPPrelongeccenxxxxxD}{\ensuremath{60.46\pm0.96}}   % planetary radius (R_earth)
\newcommand{\hatcurPPmrcorreccenxxxxxD}{\ensuremath{-0.45}}          % mass/radius correlation
\newcommand{\hatcurPPteffeccenxxxxxD}{\ensuremath{0.60_{-0.10}^{+1.00}}} % planetary temperature (K)
\newcommand{\hatcurPPthetaeccenxxxxxD}{\ensuremath{0.0404\pm0.0011}} % Safranov number
\newcommand{\hatcurPPfluxperieccenxxxxxD}{\ensuremath{5.86\pm0.13}}  % flux @ periastron (CGS)
\newcommand{\hatcurPPfluxperidimeccenxxxxxD}{\ensuremath{9}}         % flux @ periastron (CGS) units.
\newcommand{\hatcurPPfluxapeccenxxxxxD}{\ensuremath{5.58\pm0.11}}    % flux @ apastron (CGS)
\newcommand{\hatcurPPfluxapdimeccenxxxxxD}{\ensuremath{9}}           % flux @ apastron (CGS) units.
\newcommand{\hatcurPPfluxavgeccenxxxxxD}{\ensuremath{5.725\pm0.068}} % flux on average (CGS)
\newcommand{\hatcurPPfluxavgdimeccenxxxxxD}{\ensuremath{9}}          % flux average (CGS) units.
\newcommand{\hatcurPPfluxavglogeccenxxxxxD}{\ensuremath{9.7578\pm0.0051}} % log10 flux on average (CGS)
\newcommand{\hatcurXsecphaseeccenxxxxxD}{\ensuremath{0.4996\pm0.0055}} % Phase of secondary eclipse
\newcommand{\hatcurXsecondaryeccenxxxxxD}{\ensuremath{2457819.631\pm0.013}} % Secondary eclipse epoch
\newcommand{\hatcurXsecdureccenxxxxxD}{\ensuremath{0.1052\pm0.0019}} % sec eclipse duration (days)
\newcommand{\hatcurXsecingdureccenxxxxxD}{\ensuremath{0.01241\pm0.00051}} % sec I/E duration (days)
\newcommand{\hatcurPPphiconjeccenxxxxxD}{\ensuremath{-0.02\pm0.22}}  % phase diff between conjunction and periastron
\newcommand{\hatcurPPperieccenxxxxxD}{\ensuremath{2457818.50\pm0.51}} % time of periastron passage.
\newcommand{\hatcurPPaequiveccenxxxxxD}{\ensuremath{0.015400\pm0.000098}} % equivalent semi-major axis
\newcommand{\hatcurPPtcirceccenxxxxxD}{\ensuremath{0\pm0}}           % circularization timescale
\newcommand{\hatcurPPtinfalleccenxxxxxD}{\ensuremath{0\pm0}}         % infall timescale
\newcommand{\hatcurXdisteccenxxxxxD}{\ensuremath{10.200\pm0.075}}    % distance (pc), no reddenning correction
\newcommand{\hatcurXAveccenxxxxxD}{\ensuremath{5584\pm19}}           % Av (mag)
\newcommand{\hatcurXdistredeccenxxxxxD}{\ensuremath{10.250\pm0.049}} % distance with Av correction (pc)
\newcommand{\hatcurXEBVeccenxxxxxD}{\ensuremath{1801.2\pm6.0}}       % E(B-V) (mag)
\newcommand{\hatcurCCpmraeccenxxxxxD}{\ensuremath{-12.664\pm0.046}}  % proper motion, in RA
\newcommand{\hatcurCCpmdececcenxxxxxD}{\ensuremath{-14.115\pm0.040}} % proper motion, in DEC
\newcommand{\hatcurCCpmeccenxxxxxD}{\ensuremath{18.963\pm0.061}}     % proper motion

\newcommand{\hatcurhtreccenxxxxxE}{HATS737-002}                      % Original HTR name of target
\newcommand{\hatcurfieldeccenxxxxxE}{\ensuremath{string}}            % HTR field
\newcommand{\hatcurCCraeccenxxxxxE}{\ensuremath{12^{\mathrm h}27^{\mathrm m}09.0000{\mathrm s}}}                   % Right Ascension
\newcommand{\hatcurCCdececcenxxxxxE}{\ensuremath{-48{\arcdeg}58{\arcmin}42.3000{\arcsec}}}                 % Declination
\newcommand{\hatcurCCmageccenxxxxxE}{11.552}                         % apparent V-band magnitude
\newcommand{\hatcurCCtwomasseccenxxxxxE}{2MASS~12270898-4858423}     % 2MASS identifier
\newcommand{\hatcurCCgsceccenxxxxxE}{GSC~8239-00065}                 % GSC(1.2) identifier
\newcommand{\hatcurCCgaiaeccenxxxxxE}{GAIA~6128363662138374528}      % GAIA DR1 identifier
\newcommand{\hatcurCCtassmveccenxxxxxE}{\ensuremath{11.552\pm0.019}} % APASS V-band magnitude
\newcommand{\hatcurCCtassmvshorteccenxxxxxE}{\ensuremath{11.6}}      % APASS V-band magnitude
\newcommand{\hatcurCCtassmBeccenxxxxxE}{\ensuremath{12.0510\pm0.0090}} % APASS B-band magnitude
\newcommand{\hatcurCCtassmBshorteccenxxxxxE}{\ensuremath{12.1}}      % APASS B-band magnitude
\newcommand{\hatcurCCtassmIeccenxxxxxE}{\ensuremath{nff\pmnff}}      % TASS I-band magnitude
\newcommand{\hatcurCCtassmIshorteccenxxxxxE}{\ensuremath{0.0}}       % TASS I-band magnitude
\newcommand{\hatcurCCtassmgeccenxxxxxE}{\ensuremath{11.752\pm0.019}} % APASS g-band magnitude
\newcommand{\hatcurCCtassmgshorteccenxxxxxE}{\ensuremath{11.8}}      % APASS g-band magnitude
\newcommand{\hatcurCCtassmreccenxxxxxE}{\ensuremath{11.466\pm0.024}} % APASS r-band magnitude
\newcommand{\hatcurCCtassmrshorteccenxxxxxE}{\ensuremath{11.5}}      % APASS r-band magnitude
\newcommand{\hatcurCCtassmieccenxxxxxE}{\ensuremath{11.379\pm0.070}} % APASS i-band magnitude
\newcommand{\hatcurCCtassmishorteccenxxxxxE}{\ensuremath{11.4}}      % APASS i-band magnitude
\newcommand{\hatcurCCgaiamGeccenxxxxxE}{\ensuremath{11.657\pm0.011}} % Gaia G-band magnitude
\newcommand{\hatcurCCtwomassJmageccenxxxxxE}{\ensuremath{10.584\pm0.024}} % 2MASS ORIG MAG
\newcommand{\hatcurCCtwomassHmageccenxxxxxE}{\ensuremath{10.358\pm0.026}} % 2MASS ORIG MAG
\newcommand{\hatcurCCtwomassKmageccenxxxxxE}{\ensuremath{10.289\pm0.023}} % 2MASS ORIG MAG
\newcommand{\hatcurCCcitJmageccenxxxxxE}{\ensuremath{10.605\pm0.024}} % 2MASS CIT MAG
\newcommand{\hatcurCCcitHmageccenxxxxxE}{\ensuremath{10.353\pm0.026}} % 2MASS CIT MAG
\newcommand{\hatcurCCcitKmageccenxxxxxE}{\ensuremath{10.313\pm0.023}} % 2MASS CIT MAG
\newcommand{\hatcurCCbbJmageccenxxxxxE}{\ensuremath{10.648\pm0.026}} % 2MASS BB MAG
\newcommand{\hatcurCCbbHmageccenxxxxxE}{\ensuremath{10.374\pm0.027}} % 2MASS BB MAG
\newcommand{\hatcurCCbbKmageccenxxxxxE}{\ensuremath{10.333\pm0.023}} % 2MASS BB MAG
\newcommand{\hatcurCCesoJmageccenxxxxxE}{\ensuremath{10.649\pm0.027}} % 2MASS ESO MAG
\newcommand{\hatcurCCesoHmageccenxxxxxE}{\ensuremath{10.370\pm0.030}} % 2MASS ESO MAG
\newcommand{\hatcurCCesoKmageccenxxxxxE}{\ensuremath{10.332\pm0.024}} % 2MASS ESO MAG
\newcommand{\hatcurCCesoJHmageccenxxxxxE}{\ensuremath{0.280\pm0.038}} % 2MASS ESO JH COLOR
\newcommand{\hatcurCCesoJKmageccenxxxxxE}{\ensuremath{0.318\pm0.036}} % 2MASS ESO JK COLOR
\newcommand{\hatcurCCesoHKmageccenxxxxxE}{\ensuremath{0.038\pm0.039}} % 2MASS ESO HK COLOR
\newcommand{\hatcurLCdipeccenxxxxxE}{\ensuremath{5.5}}               % BLS detected dip (mmag)
\newcommand{\hatcurLCrprstareccenxxxxxE}{\ensuremath{0.0714\pm0.0051}} % Rp/R*
\newcommand{\hatcurLCbsqeccenxxxxxE}{\ensuremath{0.668_{-0.145}^{+0.096}}} % impact parameter square
\newcommand{\hatcurLCimpeccenxxxxxE}{\ensuremath{0.817_{-0.094}^{+0.057}}} % impact parameter
\newcommand{\hatcurLCzetaeccenxxxxxE}{\ensuremath{18.28_{-0.58}^{+0.78}}} % zeta/R*
\newcommand{\hatcurLCdureccenxxxxxE}{\ensuremath{0.1313\pm0.0092}}   % transit duration (days)
\newcommand{\hatcurLCdurshorteccenxxxxxE}{\ensuremath{0.1313}}       % transit duration (days)
\newcommand{\hatcurLCdurhreccenxxxxxE}{\ensuremath{3.15\pm0.22}}     % transit duration (hours)
\newcommand{\hatcurLCdurhrshorteccenxxxxxE}{\ensuremath{3.152}}      % transit duration (hours)
\newcommand{\hatcurLCqeccenxxxxxE}{\ensuremath{0.0311\pm0.0022}}     % fractional transit duration (days)
\newcommand{\hatcurLCqshorteccenxxxxxE}{\ensuremath{0.031}}          % fractional transit duration (days)
\newcommand{\hatcurLCingdureccenxxxxxE}{\ensuremath{0.024\pm0.031}}  % ingress/egress duration (days)
\newcommand{\hatcurLCPeccenxxxxxE}{\ensuremath{4.2180881\pm0.0000098}} % period (days)
\newcommand{\hatcurLCPprececcenxxxxxE}{\ensuremath{4.2180881}}       % period (days)
\newcommand{\hatcurLCPshorteccenxxxxxE}{\ensuremath{4.2181}}         % period (days)
\newcommand{\hatcurLCTeccenxxxxxE}{\ensuremath{2457480.1715\pm0.0019}} % epoch (BJD)
\newcommand{\hatcurLCTAeccenxxxxxE}{\ensuremath{2455679.0478\pm0.0045}} % TA (BJD)
\newcommand{\hatcurLCTBeccenxxxxxE}{\ensuremath{2457939.9433\pm0.0022}} % TB (BJD)
\newcommand{\hatcurLChatnetmeccenxxxxxE}{\ensuremath{11.539620\pm0.000061}} % HATNet OOT level
\newcommand{\hatcurLCiblendeccenxxxxxE}{\ensuremath{0.900\pm0.079}}  % HATNet iblend factor
\newcommand{\hatcurLCrhoeccenxxxxxE}{\ensuremath{0.49_{-0.21}^{+0.48}}} % stellar density no isochrone constraint (cgs)
\newcommand{\hatcurSMEiteffeccenxxxxxE}{\ensuremath{6443\pm79}}      % Ini SME, stellar effective temperature
\newcommand{\hatcurSMEizfeheccenxxxxxE}{\ensuremath{0.040\pm0.055}}  % Ini SME, stellar metallicity
\newcommand{\hatcurSMEizfehshorteccenxxxxxE}{\ensuremath{0.04}}      % Ini SME, stellar metallicity
\newcommand{\hatcurSMEiloggeccenxxxxxE}{\ensuremath{3.93\pm0.16}}    % Ini SME, stellar surface gravity
\newcommand{\hatcurSMEivsineccenxxxxxE}{\ensuremath{6.30\pm0.29}}    % Ini SME, stellar rotational velocity
\newcommand{\hatcurSMEivmaceccenxxxxxE}{\ensuremath{3.768\pm0.097}}  % Ini SME, stellar macroturbulence
\newcommand{\hatcurSMEivmiceccenxxxxxE}{\ensuremath{1.000\pm0.033}}  % Ini SME, stellar microturbulence
\newcommand{\hatcurSMEiiteffeccenxxxxxE}{\ensuremath{6460\pm130}}    % Final SME, stellar effective temperature
\newcommand{\hatcurSMEiizfeheccenxxxxxE}{\ensuremath{0.060\pm0.069}} % Final SME, stellar metallicity
\newcommand{\hatcurSMEiizfehshorteccenxxxxxE}{\ensuremath{0.06}}     % Final SME, stellar metallicity
\newcommand{\hatcurSMEiiloggeccenxxxxxE}{\ensuremath{4.15\pm0.14}}   % Final SME, stellar surface gravity
\newcommand{\hatcurSMEiivsineccenxxxxxE}{\ensuremath{6.22\pm0.29}}   % Final SME, stellar rotational velocity
\newcommand{\hatcurSMEiivmaceccenxxxxxE}{\ensuremath{5.04\pm0.20}}   % Final SME, stellar macroturbulence
\newcommand{\hatcurSMEiivmiceccenxxxxxE}{\ensuremath{1.71\pm0.17}}   % Final SME, stellar microturbulence
\newcommand{\hatcurLBizeccenxxxxxE}{\ensuremath{0.1247}}             % Limb darkening parameters, Gamma1, z-band
\newcommand{\hatcurLBiizeccenxxxxxE}{\ensuremath{0.3672}}            % Limb darkening parameters, Gamma2, z-band
\newcommand{\hatcurLBiieccenxxxxxE}{\ensuremath{0.1748}}             % Limb darkening parameters, Gamma1, i-band
\newcommand{\hatcurLBiiieccenxxxxxE}{\ensuremath{0.3768}}            % Limb darkening parameters, Gamma2, i-band
\newcommand{\hatcurLBiIeccenxxxxxE}{\ensuremath{0.1564}}             % Limb darkening parameters, Gamma1, I-band
\newcommand{\hatcurLBiiIeccenxxxxxE}{\ensuremath{0.3749}}            % Limb darkening parameters, Gamma2, I-band
\newcommand{\hatcurLBigeccenxxxxxE}{\ensuremath{0.4117}}             % Limb darkening parameters, Gamma1, g-band
\newcommand{\hatcurLBiigeccenxxxxxE}{\ensuremath{0.3345}}            % Limb darkening parameters, Gamma2, g-band
\newcommand{\hatcurLBireccenxxxxxE}{\ensuremath{0.2489}}             % Limb darkening parameters, Gamma1, r-band
\newcommand{\hatcurLBiireccenxxxxxE}{\ensuremath{0.3851}}            % Limb darkening parameters, Gamma2, r-band
\newcommand{\hatcurLBiReccenxxxxxE}{\ensuremath{0.2277}}             % Limb darkening parameters, Gamma1, R-band
\newcommand{\hatcurLBiiReccenxxxxxE}{\ensuremath{0.3845}}            % Limb darkening parameters, Gamma2, R-band
\newcommand{\hatcurLBikepeccenxxxxxE}{\ensuremath{0.1000}}           % Limb darkening parameters, Gamma1, Kep-band
\newcommand{\hatcurLBiikepeccenxxxxxE}{\ensuremath{0.1000}}          % Limb darkening parameters, Gamma2, Kep-band
\newcommand{\hatcurISOmeccenxxxxxE}{\ensuremath{1.39\pm0.11}}        % stellar mass
\newcommand{\hatcurISOmshorteccenxxxxxE}{\ensuremath{1.39}}          % stellar mass
\newcommand{\hatcurISOmlongeccenxxxxxE}{\ensuremath{1.39\pm0.11}}    % stellar mass
\newcommand{\hatcurISOreccenxxxxxE}{\ensuremath{1.67\pm0.33}}        % stellar radius
\newcommand{\hatcurISOrshorteccenxxxxxE}{\ensuremath{1.67}}          % stellar radius
\newcommand{\hatcurISOrlongeccenxxxxxE}{\ensuremath{1.67\pm0.33}}    % stellar radius
\newcommand{\hatcurISOrhoeccenxxxxxE}{\ensuremath{0.42_{-0.16}^{+0.28}}} % stellar density (cgs)
\newcommand{\hatcurISOrholongeccenxxxxxE}{\ensuremath{0.42_{-0.16}^{+0.28}}} % stellar density (cgs)
\newcommand{\hatcurISOloggeccenxxxxxE}{\ensuremath{4.14\pm0.13}}     % stellar surface gravity from isochrones
\newcommand{\hatcurISOlumeccenxxxxxE}{\ensuremath{4.4_{-1.4}^{+2.2}}} % stellar luminosity
\newcommand{\hatcurISOlumshorteccenxxxxxE}{\ensuremath{4.36}}        % stellar luminosity
\newcommand{\hatcurISOmveccenxxxxxE}{\ensuremath{3.14\pm0.43}}       % stellar absolute magnitude
\newcommand{\hatcurISOvieccenxxxxxE}{\ensuremath{0.508\pm0.032}}     % stellar V-I index
\newcommand{\hatcurISOageeccenxxxxxE}{\ensuremath{2.10\pm0.61}}      % stellar age
\newcommand{\hatcurISOsigmaeccenxxxxxE}{\ensuremath{0.00030\pm0.00020}} % system mass-correction sigma parameter
\newcommand{\hatcurISOMJeccenxxxxxE}{\ensuremath{2.33\pm0.41}}       % stellar absolute J magnitude
\newcommand{\hatcurISOMHeccenxxxxxE}{\ensuremath{2.11\pm0.40}}       % stellar absolute H magnitude
\newcommand{\hatcurISOMKeccenxxxxxE}{\ensuremath{2.07\pm0.40}}       % stellar absolute K magnitude
\newcommand{\hatcurISOJKeccenxxxxxE}{\ensuremath{0.260\pm0.050}}     % J-K color index from isochrones.
\newcommand{\hatcurISOspececcenxxxxxE}{F}                            % stellar spectral type
\newcommand{\hatcurRVKeccenxxxxxE}{\ensuremath{63.8\pm5.4}}          % RV semi-amplitude [m/s]
\newcommand{\hatcurRVrkeccenxxxxxE}{\ensuremath{-0.153_{-0.086}^{+0.123}}} % sqrt(e)*cos(omega)
\newcommand{\hatcurRVrheccenxxxxxE}{\ensuremath{-0.14\pm0.18}}       % sqrt(e)*sin(omega)
\newcommand{\hatcurRVkeccenxxxxxE}{\ensuremath{-0.035\pm0.033}}      % e*cos(omega)
\newcommand{\hatcurRVheccenxxxxxE}{\ensuremath{-0.030_{-0.070}^{+0.040}}} % e*sin(omega)
\newcommand{\hatcurRVtroneeccenxxxxxE}{\ensuremath{0\pm0}}           % RV linear trend tr1 factor
\newcommand{\hatcurRVtrtwoeccenxxxxxE}{\ensuremath{0\pm0}}           % RV linear trend tr2 factor
\newcommand{\hatcurRVgammaAeccenxxxxxE}{\ensuremath{19296\pm17}}     % RV gamma velocity, relative scale
\newcommand{\hatcurRVjitterAeccenxxxxxE}{\ensuremath{57\pm13}}       % RV jitter (m/s)
\newcommand{\hatcurRVjittertwosiglimAeccenxxxxxE}{\ensuremath{<82.5}} % RV jitter (m/s) 95 percent confidence upper limit
\newcommand{\hatcurRVfitrmsAeccenxxxxxE}{\ensuremath{0.0}}           % RVfitrms
\newcommand{\hatcurRVgammaBeccenxxxxxE}{\ensuremath{19415.0\pm3.8}}  % RV gamma velocity, relative scale
\newcommand{\hatcurRVjitterBeccenxxxxxE}{\ensuremath{0.0\pm5.1}}     % RV jitter (m/s)
\newcommand{\hatcurRVjittertwosiglimBeccenxxxxxE}{\ensuremath{<13.3}} % RV jitter (m/s) 95 percent confidence upper limit
\newcommand{\hatcurRVfitrmsBeccenxxxxxE}{\ensuremath{0.0}}           % RVfitrms
\newcommand{\hatcurRVecceneccenxxxxxE}{\ensuremath{0.069\pm0.049}}   % eccentricity
\newcommand{\hatcurRVeccentwosiglimeccenxxxxxE}{\ensuremath{<0.168}} % eccentricity
\newcommand{\hatcurRVomegaeccenxxxxxE}{\ensuremath{218\pm52}}        % argument of pericenter
\newcommand{\hatcurPPieccenxxxxxE}{\ensuremath{83.8\pm1.6}}          % orbital inclination
\newcommand{\hatcurPPgeccenxxxxxE}{\ensuremath{11.8_{-4.1}^{+6.1}}}  % planetary surface gravity (m/s^2)
\newcommand{\hatcurPPloggeccenxxxxxE}{\ensuremath{3.07\pm0.19}}      % planetary surface gravity (log cgs)
\newcommand{\hatcurPPareccenxxxxxE}{\ensuremath{7.4\pm1.2}}          % relative orbital radius (a/R*)
\newcommand{\hatcurPPareleccenxxxxxE}{\ensuremath{0.0570\pm0.0015}}  % semimajor axis (AU)
\newcommand{\hatcurPPrhoeccenxxxxxE}{\ensuremath{0.51_{-0.24}^{+0.47}}} % planetary density (cgs)
\newcommand{\hatcurPPmeccenxxxxxE}{\ensuremath{0.633\pm0.063}}       % planetary mass (M_jup)
\newcommand{\hatcurPPmshorteccenxxxxxE}{\ensuremath{0.63}}           % planetary mass (M_jup)
\newcommand{\hatcurPPmlongeccenxxxxxE}{\ensuremath{0.633\pm0.063}}   % planetary mass (M_jup)
\newcommand{\hatcurPPmeeccenxxxxxE}{\ensuremath{201\pm20}}           % planetary mass (M_earth)
\newcommand{\hatcurPPmeshorteccenxxxxxE}{\ensuremath{201.1}}         % planetary mass (M_earth)
\newcommand{\hatcurPPmelongeccenxxxxxE}{\ensuremath{201\pm20}}       % planetary mass (M_earth)
\newcommand{\hatcurPPreccenxxxxxE}{\ensuremath{1.16\pm0.31}}         % planetary radius (R_jup)
\newcommand{\hatcurPPrshorteccenxxxxxE}{\ensuremath{1.16}}           % planetary radius (R_jup)
\newcommand{\hatcurPPrlongeccenxxxxxE}{\ensuremath{1.16\pm0.31}}     % planetary radius (R_jup)
\newcommand{\hatcurPPreeccenxxxxxE}{\ensuremath{13.0\pm3.4}}         % planetary radius (R_earth)
\newcommand{\hatcurPPreshorteccenxxxxxE}{\ensuremath{13.0}}          % planetary radius (R_earth)
\newcommand{\hatcurPPrelongeccenxxxxxE}{\ensuremath{13.0\pm3.4}}     % planetary radius (R_earth)
\newcommand{\hatcurPPmrcorreccenxxxxxE}{\ensuremath{0.41}}           % mass/radius correlation
\newcommand{\hatcurPPteffeccenxxxxxE}{\ensuremath{1690\pm140}}       % planetary temperature (K)
\newcommand{\hatcurPPthetaeccenxxxxxE}{\ensuremath{0.045\pm0.012}}   % Safranov number
\newcommand{\hatcurPPfluxperieccenxxxxxE}{\ensuremath{2.15_{-0.65}^{+0.87}}} % flux @ periastron (CGS)
\newcommand{\hatcurPPfluxperidimeccenxxxxxE}{\ensuremath{9}}         % flux @ periastron (CGS) units.
\newcommand{\hatcurPPfluxapeccenxxxxxE}{\ensuremath{1.59_{-0.50}^{+0.68}}} % flux @ apastron (CGS)
\newcommand{\hatcurPPfluxapdimeccenxxxxxE}{\ensuremath{9}}           % flux @ apastron (CGS) units.
\newcommand{\hatcurPPfluxavgeccenxxxxxE}{\ensuremath{1.84_{-0.55}^{+0.76}}} % flux on average (CGS)
\newcommand{\hatcurPPfluxavgdimeccenxxxxxE}{\ensuremath{9}}          % flux average (CGS) units.
\newcommand{\hatcurPPfluxavglogeccenxxxxxE}{\ensuremath{9.26\pm0.15}} % log10 flux on average (CGS)
\newcommand{\hatcurXsecphaseeccenxxxxxE}{\ensuremath{0.478\pm0.021}} % Phase of secondary eclipse
\newcommand{\hatcurXsecondaryeccenxxxxxE}{\ensuremath{2457482.188\pm0.089}} % Secondary eclipse epoch
\newcommand{\hatcurXsecdureccenxxxxxE}{\ensuremath{0.131\pm0.014}}   % sec eclipse duration (days)
\newcommand{\hatcurXsecingdureccenxxxxxE}{\ensuremath{0.019\pm0.013}} % sec I/E duration (days)
\newcommand{\hatcurPPphiconjeccenxxxxxE}{\ensuremath{-0.30_{-0.13}^{+0.21}}} % phase diff between conjunction and periastron
\newcommand{\hatcurPPperieccenxxxxxE}{\ensuremath{2457481.4\pm1.1}}  % time of periastron passage.
\newcommand{\hatcurPPaequiveccenxxxxxE}{\ensuremath{0.0273\pm0.0047}} % equivalent semi-major axis
\newcommand{\hatcurPPtcirceccenxxxxxE}{\ensuremath{560_{-370}^{+1120}}} % circularization timescale
\newcommand{\hatcurPPtinfalleccenxxxxxE}{\ensuremath{1830_{-970}^{+2570}}} % infall timescale
\newcommand{\hatcurXdisteccenxxxxxE}{\ensuremath{450\pm88}}          % distance (pc), no reddenning correction
\newcommand{\hatcurXAveccenxxxxxE}{\ensuremath{0.158\pm0.099}}       % Av (mag)
\newcommand{\hatcurXdistredeccenxxxxxE}{\ensuremath{446\pm87}}       % distance with Av correction (pc)
\newcommand{\hatcurXEBVeccenxxxxxE}{\ensuremath{0.051\pm0.032}}      % E(B-V) (mag)
\newcommand{\hatcurXmvisoredeccenxxxxxE}{\ensuremath{11.555\pm0.020}} % Expected m_v with reddening (mag)
\newcommand{\hatcurXmiisoredeccenxxxxxE}{\ensuremath{10.963\pm0.023}} % Expected m_i with reddening (mag)
\newcommand{\hatcurXmjisoredeccenxxxxxE}{\ensuremath{10.626\pm0.015}} % Expected m_j with reddening (mag)
\newcommand{\hatcurXmhisoredeccenxxxxxE}{\ensuremath{10.388\pm0.018}} % Expected m_h with reddening (mag)
\newcommand{\hatcurXmkisoredeccenxxxxxE}{\ensuremath{10.334\pm0.018}} % Expected m_k with reddening (mag)
\newcommand{\hatcurXviisoredeccenxxxxxE}{\ensuremath{0.591\pm0.024}} % Expected V-I with reddening (mag)
\newcommand{\hatcurXvkisoredeccenxxxxxE}{\ensuremath{1.220\pm0.029}} % Expected V-K with reddening (mag)
\newcommand{\hatcurXjhisoredeccenxxxxxE}{\ensuremath{0.238\pm0.016}} % Expected J-H with reddening (mag)
\newcommand{\hatcurXjkisoredeccenxxxxxE}{\ensuremath{0.291\pm0.012}} % Expected J-K with reddening (mag)
\newcommand{\hatcurCCpmraeccenxxxxxE}{\ensuremath{-13.7\pm1.0}}      % proper motion, in RA
\newcommand{\hatcurCCpmdececcenxxxxxE}{\ensuremath{-1.3\pm1.2}}      % proper motion, in DEC
\newcommand{\hatcurCCpmeccenxxxxxE}{\ensuremath{13.8\pm1.6}}         % proper motion

\newcommand{\hatcurCCbbHmageccen}[1]{\ifnum#1=54 %
\hatcurCCbbHmageccenxxxxxA
\else
\ifnum#1=55 %
\hatcurCCbbHmageccenxxxxxB
\else
\ifnum#1=56 %
\hatcurCCbbHmageccenxxxxxC
\else
\ifnum#1=57 %
\hatcurCCbbHmageccenxxxxxD
\else
\ifnum#1=58 %
\hatcurCCbbHmageccenxxxxxE
\else
??????\fi
\fi
\fi
\fi
\fi
}
\newcommand{\hatcurCCbbJmageccen}[1]{\ifnum#1=54 %
\hatcurCCbbJmageccenxxxxxA
\else
\ifnum#1=55 %
\hatcurCCbbJmageccenxxxxxB
\else
\ifnum#1=56 %
\hatcurCCbbJmageccenxxxxxC
\else
\ifnum#1=57 %
\hatcurCCbbJmageccenxxxxxD
\else
\ifnum#1=58 %
\hatcurCCbbJmageccenxxxxxE
\else
??????\fi
\fi
\fi
\fi
\fi
}
\newcommand{\hatcurCCbbKmageccen}[1]{\ifnum#1=54 %
\hatcurCCbbKmageccenxxxxxA
\else
\ifnum#1=55 %
\hatcurCCbbKmageccenxxxxxB
\else
\ifnum#1=56 %
\hatcurCCbbKmageccenxxxxxC
\else
\ifnum#1=57 %
\hatcurCCbbKmageccenxxxxxD
\else
\ifnum#1=58 %
\hatcurCCbbKmageccenxxxxxE
\else
??????\fi
\fi
\fi
\fi
\fi
}
\newcommand{\hatcurCCcitHmageccen}[1]{\ifnum#1=54 %
\hatcurCCcitHmageccenxxxxxA
\else
\ifnum#1=55 %
\hatcurCCcitHmageccenxxxxxB
\else
\ifnum#1=56 %
\hatcurCCcitHmageccenxxxxxC
\else
\ifnum#1=57 %
\hatcurCCcitHmageccenxxxxxD
\else
\ifnum#1=58 %
\hatcurCCcitHmageccenxxxxxE
\else
??????\fi
\fi
\fi
\fi
\fi
}
\newcommand{\hatcurCCcitJmageccen}[1]{\ifnum#1=54 %
\hatcurCCcitJmageccenxxxxxA
\else
\ifnum#1=55 %
\hatcurCCcitJmageccenxxxxxB
\else
\ifnum#1=56 %
\hatcurCCcitJmageccenxxxxxC
\else
\ifnum#1=57 %
\hatcurCCcitJmageccenxxxxxD
\else
\ifnum#1=58 %
\hatcurCCcitJmageccenxxxxxE
\else
??????\fi
\fi
\fi
\fi
\fi
}
\newcommand{\hatcurCCcitKmageccen}[1]{\ifnum#1=54 %
\hatcurCCcitKmageccenxxxxxA
\else
\ifnum#1=55 %
\hatcurCCcitKmageccenxxxxxB
\else
\ifnum#1=56 %
\hatcurCCcitKmageccenxxxxxC
\else
\ifnum#1=57 %
\hatcurCCcitKmageccenxxxxxD
\else
\ifnum#1=58 %
\hatcurCCcitKmageccenxxxxxE
\else
??????\fi
\fi
\fi
\fi
\fi
}
\newcommand{\hatcurCCdececcen}[1]{\ifnum#1=54 %
\hatcurCCdececcenxxxxxA
\else
\ifnum#1=55 %
\hatcurCCdececcenxxxxxB
\else
\ifnum#1=56 %
\hatcurCCdececcenxxxxxC
\else
\ifnum#1=57 %
\hatcurCCdececcenxxxxxD
\else
\ifnum#1=58 %
\hatcurCCdececcenxxxxxE
\else
??????\fi
\fi
\fi
\fi
\fi
}
\newcommand{\hatcurCCesoHKmageccen}[1]{\ifnum#1=54 %
\hatcurCCesoHKmageccenxxxxxA
\else
\ifnum#1=55 %
\hatcurCCesoHKmageccenxxxxxB
\else
\ifnum#1=56 %
\hatcurCCesoHKmageccenxxxxxC
\else
\ifnum#1=57 %
\hatcurCCesoHKmageccenxxxxxD
\else
\ifnum#1=58 %
\hatcurCCesoHKmageccenxxxxxE
\else
??????\fi
\fi
\fi
\fi
\fi
}
\newcommand{\hatcurCCesoHmageccen}[1]{\ifnum#1=54 %
\hatcurCCesoHmageccenxxxxxA
\else
\ifnum#1=55 %
\hatcurCCesoHmageccenxxxxxB
\else
\ifnum#1=56 %
\hatcurCCesoHmageccenxxxxxC
\else
\ifnum#1=57 %
\hatcurCCesoHmageccenxxxxxD
\else
\ifnum#1=58 %
\hatcurCCesoHmageccenxxxxxE
\else
??????\fi
\fi
\fi
\fi
\fi
}
\newcommand{\hatcurCCesoJHmageccen}[1]{\ifnum#1=54 %
\hatcurCCesoJHmageccenxxxxxA
\else
\ifnum#1=55 %
\hatcurCCesoJHmageccenxxxxxB
\else
\ifnum#1=56 %
\hatcurCCesoJHmageccenxxxxxC
\else
\ifnum#1=57 %
\hatcurCCesoJHmageccenxxxxxD
\else
\ifnum#1=58 %
\hatcurCCesoJHmageccenxxxxxE
\else
??????\fi
\fi
\fi
\fi
\fi
}
\newcommand{\hatcurCCesoJKmageccen}[1]{\ifnum#1=54 %
\hatcurCCesoJKmageccenxxxxxA
\else
\ifnum#1=55 %
\hatcurCCesoJKmageccenxxxxxB
\else
\ifnum#1=56 %
\hatcurCCesoJKmageccenxxxxxC
\else
\ifnum#1=57 %
\hatcurCCesoJKmageccenxxxxxD
\else
\ifnum#1=58 %
\hatcurCCesoJKmageccenxxxxxE
\else
??????\fi
\fi
\fi
\fi
\fi
}
\newcommand{\hatcurCCesoJmageccen}[1]{\ifnum#1=54 %
\hatcurCCesoJmageccenxxxxxA
\else
\ifnum#1=55 %
\hatcurCCesoJmageccenxxxxxB
\else
\ifnum#1=56 %
\hatcurCCesoJmageccenxxxxxC
\else
\ifnum#1=57 %
\hatcurCCesoJmageccenxxxxxD
\else
\ifnum#1=58 %
\hatcurCCesoJmageccenxxxxxE
\else
??????\fi
\fi
\fi
\fi
\fi
}
\newcommand{\hatcurCCesoKmageccen}[1]{\ifnum#1=54 %
\hatcurCCesoKmageccenxxxxxA
\else
\ifnum#1=55 %
\hatcurCCesoKmageccenxxxxxB
\else
\ifnum#1=56 %
\hatcurCCesoKmageccenxxxxxC
\else
\ifnum#1=57 %
\hatcurCCesoKmageccenxxxxxD
\else
\ifnum#1=58 %
\hatcurCCesoKmageccenxxxxxE
\else
??????\fi
\fi
\fi
\fi
\fi
}
\newcommand{\hatcurCCgaiadrtwoeccen}[1]{\ifnum#1=54 %
\hatcurCCgaiadrtwoeccenxxxxxA
\else
\ifnum#1=55 %
\hatcurCCgaiadrtwoeccenxxxxxB
\else
\ifnum#1=56 %
\hatcurCCgaiadrtwoeccenxxxxxC
\else
\ifnum#1=57 %
\hatcurCCgaiadrtwoeccenxxxxxD
\else
??????\fi
\fi
\fi
\fi
}
\newcommand{\hatcurCCgaiaeccen}[1]{\ifnum#1=54 %
\hatcurCCgaiaeccenxxxxxA
\else
\ifnum#1=55 %
\hatcurCCgaiaeccenxxxxxB
\else
\ifnum#1=56 %
\hatcurCCgaiaeccenxxxxxC
\else
\ifnum#1=57 %
\hatcurCCgaiaeccenxxxxxD
\else
\ifnum#1=58 %
\hatcurCCgaiaeccenxxxxxE
\else
??????\fi
\fi
\fi
\fi
\fi
}
\newcommand{\hatcurCCgaiamBPeccen}[1]{\ifnum#1=54 %
\hatcurCCgaiamBPeccenxxxxxA
\else
\ifnum#1=55 %
\hatcurCCgaiamBPeccenxxxxxB
\else
\ifnum#1=56 %
\hatcurCCgaiamBPeccenxxxxxC
\else
\ifnum#1=57 %
\hatcurCCgaiamBPeccenxxxxxD
\else
??????\fi
\fi
\fi
\fi
}
\newcommand{\hatcurCCgaiamGeccen}[1]{\ifnum#1=54 %
\hatcurCCgaiamGeccenxxxxxA
\else
\ifnum#1=55 %
\hatcurCCgaiamGeccenxxxxxB
\else
\ifnum#1=56 %
\hatcurCCgaiamGeccenxxxxxC
\else
\ifnum#1=57 %
\hatcurCCgaiamGeccenxxxxxD
\else
\ifnum#1=58 %
\hatcurCCgaiamGeccenxxxxxE
\else
??????\fi
\fi
\fi
\fi
\fi
}
\newcommand{\hatcurCCgaiamRPeccen}[1]{\ifnum#1=54 %
\hatcurCCgaiamRPeccenxxxxxA
\else
\ifnum#1=55 %
\hatcurCCgaiamRPeccenxxxxxB
\else
\ifnum#1=56 %
\hatcurCCgaiamRPeccenxxxxxC
\else
\ifnum#1=57 %
\hatcurCCgaiamRPeccenxxxxxD
\else
??????\fi
\fi
\fi
\fi
}
\newcommand{\hatcurCCgsceccen}[1]{\ifnum#1=54 %
\hatcurCCgsceccenxxxxxA
\else
\ifnum#1=55 %
\hatcurCCgsceccenxxxxxB
\else
\ifnum#1=56 %
\hatcurCCgsceccenxxxxxC
\else
\ifnum#1=57 %
\hatcurCCgsceccenxxxxxD
\else
\ifnum#1=58 %
\hatcurCCgsceccenxxxxxE
\else
??????\fi
\fi
\fi
\fi
\fi
}
\newcommand{\hatcurCCmageccen}[1]{\ifnum#1=54 %
\hatcurCCmageccenxxxxxA
\else
\ifnum#1=55 %
\hatcurCCmageccenxxxxxB
\else
\ifnum#1=56 %
\hatcurCCmageccenxxxxxC
\else
\ifnum#1=57 %
\hatcurCCmageccenxxxxxD
\else
\ifnum#1=58 %
\hatcurCCmageccenxxxxxE
\else
??????\fi
\fi
\fi
\fi
\fi
}
\newcommand{\hatcurCCparallaxeccen}[1]{\ifnum#1=54 %
\hatcurCCparallaxeccenxxxxxA
\else
\ifnum#1=55 %
\hatcurCCparallaxeccenxxxxxB
\else
\ifnum#1=56 %
\hatcurCCparallaxeccenxxxxxC
\else
\ifnum#1=57 %
\hatcurCCparallaxeccenxxxxxD
\else
??????\fi
\fi
\fi
\fi
}
\newcommand{\hatcurCCpmdececcen}[1]{\ifnum#1=54 %
\hatcurCCpmdececcenxxxxxA
\else
\ifnum#1=55 %
\hatcurCCpmdececcenxxxxxB
\else
\ifnum#1=56 %
\hatcurCCpmdececcenxxxxxC
\else
\ifnum#1=57 %
\hatcurCCpmdececcenxxxxxD
\else
\ifnum#1=58 %
\hatcurCCpmdececcenxxxxxE
\else
??????\fi
\fi
\fi
\fi
\fi
}
\newcommand{\hatcurCCpmeccen}[1]{\ifnum#1=54 %
\hatcurCCpmeccenxxxxxA
\else
\ifnum#1=55 %
\hatcurCCpmeccenxxxxxB
\else
\ifnum#1=56 %
\hatcurCCpmeccenxxxxxC
\else
\ifnum#1=57 %
\hatcurCCpmeccenxxxxxD
\else
\ifnum#1=58 %
\hatcurCCpmeccenxxxxxE
\else
??????\fi
\fi
\fi
\fi
\fi
}
\newcommand{\hatcurCCpmraeccen}[1]{\ifnum#1=54 %
\hatcurCCpmraeccenxxxxxA
\else
\ifnum#1=55 %
\hatcurCCpmraeccenxxxxxB
\else
\ifnum#1=56 %
\hatcurCCpmraeccenxxxxxC
\else
\ifnum#1=57 %
\hatcurCCpmraeccenxxxxxD
\else
\ifnum#1=58 %
\hatcurCCpmraeccenxxxxxE
\else
??????\fi
\fi
\fi
\fi
\fi
}
\newcommand{\hatcurCCraeccen}[1]{\ifnum#1=54 %
\hatcurCCraeccenxxxxxA
\else
\ifnum#1=55 %
\hatcurCCraeccenxxxxxB
\else
\ifnum#1=56 %
\hatcurCCraeccenxxxxxC
\else
\ifnum#1=57 %
\hatcurCCraeccenxxxxxD
\else
\ifnum#1=58 %
\hatcurCCraeccenxxxxxE
\else
??????\fi
\fi
\fi
\fi
\fi
}
\newcommand{\hatcurCCtassmBeccen}[1]{\ifnum#1=54 %
\hatcurCCtassmBeccenxxxxxA
\else
\ifnum#1=55 %
\hatcurCCtassmBeccenxxxxxB
\else
\ifnum#1=56 %
\hatcurCCtassmBeccenxxxxxC
\else
\ifnum#1=57 %
\hatcurCCtassmBeccenxxxxxD
\else
\ifnum#1=58 %
\hatcurCCtassmBeccenxxxxxE
\else
??????\fi
\fi
\fi
\fi
\fi
}
\newcommand{\hatcurCCtassmBshorteccen}[1]{\ifnum#1=54 %
\hatcurCCtassmBshorteccenxxxxxA
\else
\ifnum#1=55 %
\hatcurCCtassmBshorteccenxxxxxB
\else
\ifnum#1=56 %
\hatcurCCtassmBshorteccenxxxxxC
\else
\ifnum#1=57 %
\hatcurCCtassmBshorteccenxxxxxD
\else
\ifnum#1=58 %
\hatcurCCtassmBshorteccenxxxxxE
\else
??????\fi
\fi
\fi
\fi
\fi
}
\newcommand{\hatcurCCtassmgeccen}[1]{\ifnum#1=54 %
\hatcurCCtassmgeccenxxxxxA
\else
\ifnum#1=55 %
\hatcurCCtassmgeccenxxxxxB
\else
\ifnum#1=56 %
\hatcurCCtassmgeccenxxxxxC
\else
\ifnum#1=57 %
\hatcurCCtassmgeccenxxxxxD
\else
\ifnum#1=58 %
\hatcurCCtassmgeccenxxxxxE
\else
??????\fi
\fi
\fi
\fi
\fi
}
\newcommand{\hatcurCCtassmgshorteccen}[1]{\ifnum#1=54 %
\hatcurCCtassmgshorteccenxxxxxA
\else
\ifnum#1=55 %
\hatcurCCtassmgshorteccenxxxxxB
\else
\ifnum#1=56 %
\hatcurCCtassmgshorteccenxxxxxC
\else
\ifnum#1=57 %
\hatcurCCtassmgshorteccenxxxxxD
\else
\ifnum#1=58 %
\hatcurCCtassmgshorteccenxxxxxE
\else
??????\fi
\fi
\fi
\fi
\fi
}
\newcommand{\hatcurCCtassmieccen}[1]{\ifnum#1=54 %
\hatcurCCtassmieccenxxxxxA
\else
\ifnum#1=55 %
\hatcurCCtassmieccenxxxxxB
\else
\ifnum#1=56 %
\hatcurCCtassmieccenxxxxxC
\else
\ifnum#1=57 %
\hatcurCCtassmieccenxxxxxD
\else
\ifnum#1=58 %
\hatcurCCtassmieccenxxxxxE
\else
??????\fi
\fi
\fi
\fi
\fi
}
\newcommand{\hatcurCCtassmIeccen}[1]{\ifnum#1=54 %
\hatcurCCtassmIeccenxxxxxA
\else
\ifnum#1=55 %
\hatcurCCtassmIeccenxxxxxB
\else
\ifnum#1=56 %
\hatcurCCtassmIeccenxxxxxC
\else
\ifnum#1=57 %
\hatcurCCtassmIeccenxxxxxD
\else
\ifnum#1=58 %
\hatcurCCtassmIeccenxxxxxE
\else
??????\fi
\fi
\fi
\fi
\fi
}
\newcommand{\hatcurCCtassmishorteccen}[1]{\ifnum#1=54 %
\hatcurCCtassmishorteccenxxxxxA
\else
\ifnum#1=55 %
\hatcurCCtassmishorteccenxxxxxB
\else
\ifnum#1=56 %
\hatcurCCtassmishorteccenxxxxxC
\else
\ifnum#1=57 %
\hatcurCCtassmishorteccenxxxxxD
\else
\ifnum#1=58 %
\hatcurCCtassmishorteccenxxxxxE
\else
??????\fi
\fi
\fi
\fi
\fi
}
\newcommand{\hatcurCCtassmIshorteccen}[1]{\ifnum#1=54 %
\hatcurCCtassmIshorteccenxxxxxA
\else
\ifnum#1=55 %
\hatcurCCtassmIshorteccenxxxxxB
\else
\ifnum#1=56 %
\hatcurCCtassmIshorteccenxxxxxC
\else
\ifnum#1=57 %
\hatcurCCtassmIshorteccenxxxxxD
\else
\ifnum#1=58 %
\hatcurCCtassmIshorteccenxxxxxE
\else
??????\fi
\fi
\fi
\fi
\fi
}
\newcommand{\hatcurCCtassmreccen}[1]{\ifnum#1=54 %
\hatcurCCtassmreccenxxxxxA
\else
\ifnum#1=55 %
\hatcurCCtassmreccenxxxxxB
\else
\ifnum#1=56 %
\hatcurCCtassmreccenxxxxxC
\else
\ifnum#1=57 %
\hatcurCCtassmreccenxxxxxD
\else
\ifnum#1=58 %
\hatcurCCtassmreccenxxxxxE
\else
??????\fi
\fi
\fi
\fi
\fi
}
\newcommand{\hatcurCCtassmrshorteccen}[1]{\ifnum#1=54 %
\hatcurCCtassmrshorteccenxxxxxA
\else
\ifnum#1=55 %
\hatcurCCtassmrshorteccenxxxxxB
\else
\ifnum#1=56 %
\hatcurCCtassmrshorteccenxxxxxC
\else
\ifnum#1=57 %
\hatcurCCtassmrshorteccenxxxxxD
\else
\ifnum#1=58 %
\hatcurCCtassmrshorteccenxxxxxE
\else
??????\fi
\fi
\fi
\fi
\fi
}
\newcommand{\hatcurCCtassmveccen}[1]{\ifnum#1=54 %
\hatcurCCtassmveccenxxxxxA
\else
\ifnum#1=55 %
\hatcurCCtassmveccenxxxxxB
\else
\ifnum#1=56 %
\hatcurCCtassmveccenxxxxxC
\else
\ifnum#1=57 %
\hatcurCCtassmveccenxxxxxD
\else
\ifnum#1=58 %
\hatcurCCtassmveccenxxxxxE
\else
??????\fi
\fi
\fi
\fi
\fi
}
\newcommand{\hatcurCCtassmvshorteccen}[1]{\ifnum#1=54 %
\hatcurCCtassmvshorteccenxxxxxA
\else
\ifnum#1=55 %
\hatcurCCtassmvshorteccenxxxxxB
\else
\ifnum#1=56 %
\hatcurCCtassmvshorteccenxxxxxC
\else
\ifnum#1=57 %
\hatcurCCtassmvshorteccenxxxxxD
\else
\ifnum#1=58 %
\hatcurCCtassmvshorteccenxxxxxE
\else
??????\fi
\fi
\fi
\fi
\fi
}
\newcommand{\hatcurCCtwomasseccen}[1]{\ifnum#1=54 %
\hatcurCCtwomasseccenxxxxxA
\else
\ifnum#1=55 %
\hatcurCCtwomasseccenxxxxxB
\else
\ifnum#1=56 %
\hatcurCCtwomasseccenxxxxxC
\else
\ifnum#1=57 %
\hatcurCCtwomasseccenxxxxxD
\else
\ifnum#1=58 %
\hatcurCCtwomasseccenxxxxxE
\else
??????\fi
\fi
\fi
\fi
\fi
}
\newcommand{\hatcurCCtwomassHmageccen}[1]{\ifnum#1=54 %
\hatcurCCtwomassHmageccenxxxxxA
\else
\ifnum#1=55 %
\hatcurCCtwomassHmageccenxxxxxB
\else
\ifnum#1=56 %
\hatcurCCtwomassHmageccenxxxxxC
\else
\ifnum#1=57 %
\hatcurCCtwomassHmageccenxxxxxD
\else
\ifnum#1=58 %
\hatcurCCtwomassHmageccenxxxxxE
\else
??????\fi
\fi
\fi
\fi
\fi
}
\newcommand{\hatcurCCtwomassJmageccen}[1]{\ifnum#1=54 %
\hatcurCCtwomassJmageccenxxxxxA
\else
\ifnum#1=55 %
\hatcurCCtwomassJmageccenxxxxxB
\else
\ifnum#1=56 %
\hatcurCCtwomassJmageccenxxxxxC
\else
\ifnum#1=57 %
\hatcurCCtwomassJmageccenxxxxxD
\else
\ifnum#1=58 %
\hatcurCCtwomassJmageccenxxxxxE
\else
??????\fi
\fi
\fi
\fi
\fi
}
\newcommand{\hatcurCCtwomassKmageccen}[1]{\ifnum#1=54 %
\hatcurCCtwomassKmageccenxxxxxA
\else
\ifnum#1=55 %
\hatcurCCtwomassKmageccenxxxxxB
\else
\ifnum#1=56 %
\hatcurCCtwomassKmageccenxxxxxC
\else
\ifnum#1=57 %
\hatcurCCtwomassKmageccenxxxxxD
\else
\ifnum#1=58 %
\hatcurCCtwomassKmageccenxxxxxE
\else
??????\fi
\fi
\fi
\fi
\fi
}
\newcommand{\hatcurfieldeccen}[1]{\ifnum#1=54 %
\hatcurfieldeccenxxxxxA
\else
\ifnum#1=55 %
\hatcurfieldeccenxxxxxB
\else
\ifnum#1=56 %
\hatcurfieldeccenxxxxxC
\else
\ifnum#1=57 %
\hatcurfieldeccenxxxxxD
\else
\ifnum#1=58 %
\hatcurfieldeccenxxxxxE
\else
??????\fi
\fi
\fi
\fi
\fi
}
\newcommand{\hatcurhtreccen}[1]{\ifnum#1=54 %
\hatcurhtreccenxxxxxA
\else
\ifnum#1=55 %
\hatcurhtreccenxxxxxB
\else
\ifnum#1=56 %
\hatcurhtreccenxxxxxC
\else
\ifnum#1=57 %
\hatcurhtreccenxxxxxD
\else
\ifnum#1=58 %
\hatcurhtreccenxxxxxE
\else
??????\fi
\fi
\fi
\fi
\fi
}
\newcommand{\hatcurISOageeccen}[1]{\ifnum#1=54 %
\hatcurISOageeccenxxxxxA
\else
\ifnum#1=55 %
\hatcurISOageeccenxxxxxB
\else
\ifnum#1=56 %
\hatcurISOageeccenxxxxxC
\else
\ifnum#1=57 %
\hatcurISOageeccenxxxxxD
\else
\ifnum#1=58 %
\hatcurISOageeccenxxxxxE
\else
??????\fi
\fi
\fi
\fi
\fi
}
\newcommand{\hatcurISOJKeccen}[1]{\ifnum#1=58 %
\hatcurISOJKeccenxxxxxE
\else
??????\fi
}
\newcommand{\hatcurISOloggeccen}[1]{\ifnum#1=54 %
\hatcurISOloggeccenxxxxxA
\else
\ifnum#1=55 %
\hatcurISOloggeccenxxxxxB
\else
\ifnum#1=56 %
\hatcurISOloggeccenxxxxxC
\else
\ifnum#1=57 %
\hatcurISOloggeccenxxxxxD
\else
\ifnum#1=58 %
\hatcurISOloggeccenxxxxxE
\else
??????\fi
\fi
\fi
\fi
\fi
}
\newcommand{\hatcurISOlumeccen}[1]{\ifnum#1=54 %
\hatcurISOlumeccenxxxxxA
\else
\ifnum#1=55 %
\hatcurISOlumeccenxxxxxB
\else
\ifnum#1=56 %
\hatcurISOlumeccenxxxxxC
\else
\ifnum#1=57 %
\hatcurISOlumeccenxxxxxD
\else
\ifnum#1=58 %
\hatcurISOlumeccenxxxxxE
\else
??????\fi
\fi
\fi
\fi
\fi
}
\newcommand{\hatcurISOlumshorteccen}[1]{\ifnum#1=54 %
\hatcurISOlumshorteccenxxxxxA
\else
\ifnum#1=55 %
\hatcurISOlumshorteccenxxxxxB
\else
\ifnum#1=56 %
\hatcurISOlumshorteccenxxxxxC
\else
\ifnum#1=57 %
\hatcurISOlumshorteccenxxxxxD
\else
\ifnum#1=58 %
\hatcurISOlumshorteccenxxxxxE
\else
??????\fi
\fi
\fi
\fi
\fi
}
\newcommand{\hatcurISOmeccen}[1]{\ifnum#1=54 %
\hatcurISOmeccenxxxxxA
\else
\ifnum#1=55 %
\hatcurISOmeccenxxxxxB
\else
\ifnum#1=56 %
\hatcurISOmeccenxxxxxC
\else
\ifnum#1=57 %
\hatcurISOmeccenxxxxxD
\else
\ifnum#1=58 %
\hatcurISOmeccenxxxxxE
\else
??????\fi
\fi
\fi
\fi
\fi
}
\newcommand{\hatcurISOMHeccen}[1]{\ifnum#1=58 %
\hatcurISOMHeccenxxxxxE
\else
??????\fi
}
\newcommand{\hatcurISOMJeccen}[1]{\ifnum#1=58 %
\hatcurISOMJeccenxxxxxE
\else
??????\fi
}
\newcommand{\hatcurISOMKeccen}[1]{\ifnum#1=58 %
\hatcurISOMKeccenxxxxxE
\else
??????\fi
}
\newcommand{\hatcurISOmlongeccen}[1]{\ifnum#1=54 %
\hatcurISOmlongeccenxxxxxA
\else
\ifnum#1=55 %
\hatcurISOmlongeccenxxxxxB
\else
\ifnum#1=56 %
\hatcurISOmlongeccenxxxxxC
\else
\ifnum#1=57 %
\hatcurISOmlongeccenxxxxxD
\else
\ifnum#1=58 %
\hatcurISOmlongeccenxxxxxE
\else
??????\fi
\fi
\fi
\fi
\fi
}
\newcommand{\hatcurISOmshorteccen}[1]{\ifnum#1=54 %
\hatcurISOmshorteccenxxxxxA
\else
\ifnum#1=55 %
\hatcurISOmshorteccenxxxxxB
\else
\ifnum#1=56 %
\hatcurISOmshorteccenxxxxxC
\else
\ifnum#1=57 %
\hatcurISOmshorteccenxxxxxD
\else
\ifnum#1=58 %
\hatcurISOmshorteccenxxxxxE
\else
??????\fi
\fi
\fi
\fi
\fi
}
\newcommand{\hatcurISOmveccen}[1]{\ifnum#1=58 %
\hatcurISOmveccenxxxxxE
\else
??????\fi
}
\newcommand{\hatcurISOreccen}[1]{\ifnum#1=54 %
\hatcurISOreccenxxxxxA
\else
\ifnum#1=55 %
\hatcurISOreccenxxxxxB
\else
\ifnum#1=56 %
\hatcurISOreccenxxxxxC
\else
\ifnum#1=57 %
\hatcurISOreccenxxxxxD
\else
\ifnum#1=58 %
\hatcurISOreccenxxxxxE
\else
??????\fi
\fi
\fi
\fi
\fi
}
\newcommand{\hatcurISOrhoeccen}[1]{\ifnum#1=54 %
\hatcurISOrhoeccenxxxxxA
\else
\ifnum#1=55 %
\hatcurISOrhoeccenxxxxxB
\else
\ifnum#1=56 %
\hatcurISOrhoeccenxxxxxC
\else
\ifnum#1=57 %
\hatcurISOrhoeccenxxxxxD
\else
\ifnum#1=58 %
\hatcurISOrhoeccenxxxxxE
\else
??????\fi
\fi
\fi
\fi
\fi
}
\newcommand{\hatcurISOrholongeccen}[1]{\ifnum#1=54 %
\hatcurISOrholongeccenxxxxxA
\else
\ifnum#1=55 %
\hatcurISOrholongeccenxxxxxB
\else
\ifnum#1=56 %
\hatcurISOrholongeccenxxxxxC
\else
\ifnum#1=57 %
\hatcurISOrholongeccenxxxxxD
\else
\ifnum#1=58 %
\hatcurISOrholongeccenxxxxxE
\else
??????\fi
\fi
\fi
\fi
\fi
}
\newcommand{\hatcurISOrlongeccen}[1]{\ifnum#1=54 %
\hatcurISOrlongeccenxxxxxA
\else
\ifnum#1=55 %
\hatcurISOrlongeccenxxxxxB
\else
\ifnum#1=56 %
\hatcurISOrlongeccenxxxxxC
\else
\ifnum#1=57 %
\hatcurISOrlongeccenxxxxxD
\else
\ifnum#1=58 %
\hatcurISOrlongeccenxxxxxE
\else
??????\fi
\fi
\fi
\fi
\fi
}
\newcommand{\hatcurISOrshorteccen}[1]{\ifnum#1=54 %
\hatcurISOrshorteccenxxxxxA
\else
\ifnum#1=55 %
\hatcurISOrshorteccenxxxxxB
\else
\ifnum#1=56 %
\hatcurISOrshorteccenxxxxxC
\else
\ifnum#1=57 %
\hatcurISOrshorteccenxxxxxD
\else
\ifnum#1=58 %
\hatcurISOrshorteccenxxxxxE
\else
??????\fi
\fi
\fi
\fi
\fi
}
\newcommand{\hatcurISOsigmaeccen}[1]{\ifnum#1=58 %
\hatcurISOsigmaeccenxxxxxE
\else
??????\fi
}
\newcommand{\hatcurISOspececcen}[1]{\ifnum#1=54 %
\hatcurISOspececcenxxxxxA
\else
\ifnum#1=55 %
\hatcurISOspececcenxxxxxB
\else
\ifnum#1=56 %
\hatcurISOspececcenxxxxxC
\else
\ifnum#1=57 %
\hatcurISOspececcenxxxxxD
\else
\ifnum#1=58 %
\hatcurISOspececcenxxxxxE
\else
??????\fi
\fi
\fi
\fi
\fi
}
\newcommand{\hatcurISOteffeccen}[1]{\ifnum#1=54 %
\hatcurISOteffeccenxxxxxA
\else
\ifnum#1=55 %
\hatcurISOteffeccenxxxxxB
\else
\ifnum#1=56 %
\hatcurISOteffeccenxxxxxC
\else
\ifnum#1=57 %
\hatcurISOteffeccenxxxxxD
\else
??????\fi
\fi
\fi
\fi
}
\newcommand{\hatcurISOvieccen}[1]{\ifnum#1=58 %
\hatcurISOvieccenxxxxxE
\else
??????\fi
}
\newcommand{\hatcurISOzfeheccen}[1]{\ifnum#1=54 %
\hatcurISOzfeheccenxxxxxA
\else
\ifnum#1=55 %
\hatcurISOzfeheccenxxxxxB
\else
\ifnum#1=56 %
\hatcurISOzfeheccenxxxxxC
\else
\ifnum#1=57 %
\hatcurISOzfeheccenxxxxxD
\else
??????\fi
\fi
\fi
\fi
}
\newcommand{\hatcurLBigeccen}[1]{\ifnum#1=54 %
\hatcurLBigeccenxxxxxA
\else
\ifnum#1=55 %
\hatcurLBigeccenxxxxxB
\else
\ifnum#1=56 %
\hatcurLBigeccenxxxxxC
\else
\ifnum#1=57 %
\hatcurLBigeccenxxxxxD
\else
\ifnum#1=58 %
\hatcurLBigeccenxxxxxE
\else
??????\fi
\fi
\fi
\fi
\fi
}
\newcommand{\hatcurLBiieccen}[1]{\ifnum#1=54 %
\hatcurLBiieccenxxxxxA
\else
\ifnum#1=55 %
\hatcurLBiieccenxxxxxB
\else
\ifnum#1=56 %
\hatcurLBiieccenxxxxxC
\else
\ifnum#1=57 %
\hatcurLBiieccenxxxxxD
\else
\ifnum#1=58 %
\hatcurLBiieccenxxxxxE
\else
??????\fi
\fi
\fi
\fi
\fi
}
\newcommand{\hatcurLBiIeccen}[1]{\ifnum#1=54 %
\hatcurLBiIeccenxxxxxA
\else
\ifnum#1=55 %
\hatcurLBiIeccenxxxxxB
\else
\ifnum#1=56 %
\hatcurLBiIeccenxxxxxC
\else
\ifnum#1=57 %
\hatcurLBiIeccenxxxxxD
\else
\ifnum#1=58 %
\hatcurLBiIeccenxxxxxE
\else
??????\fi
\fi
\fi
\fi
\fi
}
\newcommand{\hatcurLBiigeccen}[1]{\ifnum#1=54 %
\hatcurLBiigeccenxxxxxA
\else
\ifnum#1=55 %
\hatcurLBiigeccenxxxxxB
\else
\ifnum#1=56 %
\hatcurLBiigeccenxxxxxC
\else
\ifnum#1=57 %
\hatcurLBiigeccenxxxxxD
\else
\ifnum#1=58 %
\hatcurLBiigeccenxxxxxE
\else
??????\fi
\fi
\fi
\fi
\fi
}
\newcommand{\hatcurLBiiieccen}[1]{\ifnum#1=54 %
\hatcurLBiiieccenxxxxxA
\else
\ifnum#1=55 %
\hatcurLBiiieccenxxxxxB
\else
\ifnum#1=56 %
\hatcurLBiiieccenxxxxxC
\else
\ifnum#1=57 %
\hatcurLBiiieccenxxxxxD
\else
\ifnum#1=58 %
\hatcurLBiiieccenxxxxxE
\else
??????\fi
\fi
\fi
\fi
\fi
}
\newcommand{\hatcurLBiiIeccen}[1]{\ifnum#1=54 %
\hatcurLBiiIeccenxxxxxA
\else
\ifnum#1=55 %
\hatcurLBiiIeccenxxxxxB
\else
\ifnum#1=56 %
\hatcurLBiiIeccenxxxxxC
\else
\ifnum#1=57 %
\hatcurLBiiIeccenxxxxxD
\else
\ifnum#1=58 %
\hatcurLBiiIeccenxxxxxE
\else
??????\fi
\fi
\fi
\fi
\fi
}
\newcommand{\hatcurLBiikepeccen}[1]{\ifnum#1=54 %
\hatcurLBiikepeccenxxxxxA
\else
\ifnum#1=55 %
\hatcurLBiikepeccenxxxxxB
\else
\ifnum#1=56 %
\hatcurLBiikepeccenxxxxxC
\else
\ifnum#1=57 %
\hatcurLBiikepeccenxxxxxD
\else
\ifnum#1=58 %
\hatcurLBiikepeccenxxxxxE
\else
??????\fi
\fi
\fi
\fi
\fi
}
\newcommand{\hatcurLBiireccen}[1]{\ifnum#1=54 %
\hatcurLBiireccenxxxxxA
\else
\ifnum#1=55 %
\hatcurLBiireccenxxxxxB
\else
\ifnum#1=56 %
\hatcurLBiireccenxxxxxC
\else
\ifnum#1=57 %
\hatcurLBiireccenxxxxxD
\else
\ifnum#1=58 %
\hatcurLBiireccenxxxxxE
\else
??????\fi
\fi
\fi
\fi
\fi
}
\newcommand{\hatcurLBiiReccen}[1]{\ifnum#1=54 %
\hatcurLBiiReccenxxxxxA
\else
\ifnum#1=55 %
\hatcurLBiiReccenxxxxxB
\else
\ifnum#1=56 %
\hatcurLBiiReccenxxxxxC
\else
\ifnum#1=57 %
\hatcurLBiiReccenxxxxxD
\else
\ifnum#1=58 %
\hatcurLBiiReccenxxxxxE
\else
??????\fi
\fi
\fi
\fi
\fi
}
\newcommand{\hatcurLBiizeccen}[1]{\ifnum#1=54 %
\hatcurLBiizeccenxxxxxA
\else
\ifnum#1=55 %
\hatcurLBiizeccenxxxxxB
\else
\ifnum#1=56 %
\hatcurLBiizeccenxxxxxC
\else
\ifnum#1=57 %
\hatcurLBiizeccenxxxxxD
\else
\ifnum#1=58 %
\hatcurLBiizeccenxxxxxE
\else
??????\fi
\fi
\fi
\fi
\fi
}
\newcommand{\hatcurLBikepeccen}[1]{\ifnum#1=54 %
\hatcurLBikepeccenxxxxxA
\else
\ifnum#1=55 %
\hatcurLBikepeccenxxxxxB
\else
\ifnum#1=56 %
\hatcurLBikepeccenxxxxxC
\else
\ifnum#1=57 %
\hatcurLBikepeccenxxxxxD
\else
\ifnum#1=58 %
\hatcurLBikepeccenxxxxxE
\else
??????\fi
\fi
\fi
\fi
\fi
}
\newcommand{\hatcurLBireccen}[1]{\ifnum#1=54 %
\hatcurLBireccenxxxxxA
\else
\ifnum#1=55 %
\hatcurLBireccenxxxxxB
\else
\ifnum#1=56 %
\hatcurLBireccenxxxxxC
\else
\ifnum#1=57 %
\hatcurLBireccenxxxxxD
\else
\ifnum#1=58 %
\hatcurLBireccenxxxxxE
\else
??????\fi
\fi
\fi
\fi
\fi
}
\newcommand{\hatcurLBiReccen}[1]{\ifnum#1=54 %
\hatcurLBiReccenxxxxxA
\else
\ifnum#1=55 %
\hatcurLBiReccenxxxxxB
\else
\ifnum#1=56 %
\hatcurLBiReccenxxxxxC
\else
\ifnum#1=57 %
\hatcurLBiReccenxxxxxD
\else
\ifnum#1=58 %
\hatcurLBiReccenxxxxxE
\else
??????\fi
\fi
\fi
\fi
\fi
}
\newcommand{\hatcurLBizeccen}[1]{\ifnum#1=54 %
\hatcurLBizeccenxxxxxA
\else
\ifnum#1=55 %
\hatcurLBizeccenxxxxxB
\else
\ifnum#1=56 %
\hatcurLBizeccenxxxxxC
\else
\ifnum#1=57 %
\hatcurLBizeccenxxxxxD
\else
\ifnum#1=58 %
\hatcurLBizeccenxxxxxE
\else
??????\fi
\fi
\fi
\fi
\fi
}
\newcommand{\hatcurLCbsqeccen}[1]{\ifnum#1=54 %
\hatcurLCbsqeccenxxxxxA
\else
\ifnum#1=55 %
\hatcurLCbsqeccenxxxxxB
\else
\ifnum#1=56 %
\hatcurLCbsqeccenxxxxxC
\else
\ifnum#1=57 %
\hatcurLCbsqeccenxxxxxD
\else
\ifnum#1=58 %
\hatcurLCbsqeccenxxxxxE
\else
??????\fi
\fi
\fi
\fi
\fi
}
\newcommand{\hatcurLCdipeccen}[1]{\ifnum#1=54 %
\hatcurLCdipeccenxxxxxA
\else
\ifnum#1=55 %
\hatcurLCdipeccenxxxxxB
\else
\ifnum#1=56 %
\hatcurLCdipeccenxxxxxC
\else
\ifnum#1=57 %
\hatcurLCdipeccenxxxxxD
\else
\ifnum#1=58 %
\hatcurLCdipeccenxxxxxE
\else
??????\fi
\fi
\fi
\fi
\fi
}
\newcommand{\hatcurLCdureccen}[1]{\ifnum#1=54 %
\hatcurLCdureccenxxxxxA
\else
\ifnum#1=55 %
\hatcurLCdureccenxxxxxB
\else
\ifnum#1=56 %
\hatcurLCdureccenxxxxxC
\else
\ifnum#1=57 %
\hatcurLCdureccenxxxxxD
\else
\ifnum#1=58 %
\hatcurLCdureccenxxxxxE
\else
??????\fi
\fi
\fi
\fi
\fi
}
\newcommand{\hatcurLCdurhreccen}[1]{\ifnum#1=54 %
\hatcurLCdurhreccenxxxxxA
\else
\ifnum#1=55 %
\hatcurLCdurhreccenxxxxxB
\else
\ifnum#1=56 %
\hatcurLCdurhreccenxxxxxC
\else
\ifnum#1=57 %
\hatcurLCdurhreccenxxxxxD
\else
\ifnum#1=58 %
\hatcurLCdurhreccenxxxxxE
\else
??????\fi
\fi
\fi
\fi
\fi
}
\newcommand{\hatcurLCdurhrshorteccen}[1]{\ifnum#1=54 %
\hatcurLCdurhrshorteccenxxxxxA
\else
\ifnum#1=55 %
\hatcurLCdurhrshorteccenxxxxxB
\else
\ifnum#1=56 %
\hatcurLCdurhrshorteccenxxxxxC
\else
\ifnum#1=57 %
\hatcurLCdurhrshorteccenxxxxxD
\else
\ifnum#1=58 %
\hatcurLCdurhrshorteccenxxxxxE
\else
??????\fi
\fi
\fi
\fi
\fi
}
\newcommand{\hatcurLCdurshorteccen}[1]{\ifnum#1=54 %
\hatcurLCdurshorteccenxxxxxA
\else
\ifnum#1=55 %
\hatcurLCdurshorteccenxxxxxB
\else
\ifnum#1=56 %
\hatcurLCdurshorteccenxxxxxC
\else
\ifnum#1=57 %
\hatcurLCdurshorteccenxxxxxD
\else
\ifnum#1=58 %
\hatcurLCdurshorteccenxxxxxE
\else
??????\fi
\fi
\fi
\fi
\fi
}
\newcommand{\hatcurLChatnetmAeccen}[1]{\ifnum#1=56 %
\hatcurLChatnetmAeccenxxxxxC
\else
??????\fi
}
\newcommand{\hatcurLChatnetmBeccen}[1]{\ifnum#1=56 %
\hatcurLChatnetmBeccenxxxxxC
\else
??????\fi
}
\newcommand{\hatcurLChatnetmeccen}[1]{\ifnum#1=54 %
\hatcurLChatnetmeccenxxxxxA
\else
\ifnum#1=55 %
\hatcurLChatnetmeccenxxxxxB
\else
\ifnum#1=57 %
\hatcurLChatnetmeccenxxxxxD
\else
\ifnum#1=58 %
\hatcurLChatnetmeccenxxxxxE
\else
??????\fi
\fi
\fi
\fi
}
\newcommand{\hatcurLCiblendAeccen}[1]{\ifnum#1=56 %
\hatcurLCiblendAeccenxxxxxC
\else
??????\fi
}
\newcommand{\hatcurLCiblendBeccen}[1]{\ifnum#1=56 %
\hatcurLCiblendBeccenxxxxxC
\else
??????\fi
}
\newcommand{\hatcurLCiblendeccen}[1]{\ifnum#1=54 %
\hatcurLCiblendeccenxxxxxA
\else
\ifnum#1=55 %
\hatcurLCiblendeccenxxxxxB
\else
\ifnum#1=57 %
\hatcurLCiblendeccenxxxxxD
\else
\ifnum#1=58 %
\hatcurLCiblendeccenxxxxxE
\else
??????\fi
\fi
\fi
\fi
}
\newcommand{\hatcurLCimpeccen}[1]{\ifnum#1=54 %
\hatcurLCimpeccenxxxxxA
\else
\ifnum#1=55 %
\hatcurLCimpeccenxxxxxB
\else
\ifnum#1=56 %
\hatcurLCimpeccenxxxxxC
\else
\ifnum#1=57 %
\hatcurLCimpeccenxxxxxD
\else
\ifnum#1=58 %
\hatcurLCimpeccenxxxxxE
\else
??????\fi
\fi
\fi
\fi
\fi
}
\newcommand{\hatcurLCingdureccen}[1]{\ifnum#1=54 %
\hatcurLCingdureccenxxxxxA
\else
\ifnum#1=55 %
\hatcurLCingdureccenxxxxxB
\else
\ifnum#1=56 %
\hatcurLCingdureccenxxxxxC
\else
\ifnum#1=57 %
\hatcurLCingdureccenxxxxxD
\else
\ifnum#1=58 %
\hatcurLCingdureccenxxxxxE
\else
??????\fi
\fi
\fi
\fi
\fi
}
\newcommand{\hatcurLCPeccen}[1]{\ifnum#1=54 %
\hatcurLCPeccenxxxxxA
\else
\ifnum#1=55 %
\hatcurLCPeccenxxxxxB
\else
\ifnum#1=56 %
\hatcurLCPeccenxxxxxC
\else
\ifnum#1=57 %
\hatcurLCPeccenxxxxxD
\else
\ifnum#1=58 %
\hatcurLCPeccenxxxxxE
\else
??????\fi
\fi
\fi
\fi
\fi
}
\newcommand{\hatcurLCPprececcen}[1]{\ifnum#1=54 %
\hatcurLCPprececcenxxxxxA
\else
\ifnum#1=55 %
\hatcurLCPprececcenxxxxxB
\else
\ifnum#1=56 %
\hatcurLCPprececcenxxxxxC
\else
\ifnum#1=57 %
\hatcurLCPprececcenxxxxxD
\else
\ifnum#1=58 %
\hatcurLCPprececcenxxxxxE
\else
??????\fi
\fi
\fi
\fi
\fi
}
\newcommand{\hatcurLCPshorteccen}[1]{\ifnum#1=54 %
\hatcurLCPshorteccenxxxxxA
\else
\ifnum#1=55 %
\hatcurLCPshorteccenxxxxxB
\else
\ifnum#1=56 %
\hatcurLCPshorteccenxxxxxC
\else
\ifnum#1=57 %
\hatcurLCPshorteccenxxxxxD
\else
\ifnum#1=58 %
\hatcurLCPshorteccenxxxxxE
\else
??????\fi
\fi
\fi
\fi
\fi
}
\newcommand{\hatcurLCqeccen}[1]{\ifnum#1=54 %
\hatcurLCqeccenxxxxxA
\else
\ifnum#1=55 %
\hatcurLCqeccenxxxxxB
\else
\ifnum#1=56 %
\hatcurLCqeccenxxxxxC
\else
\ifnum#1=57 %
\hatcurLCqeccenxxxxxD
\else
\ifnum#1=58 %
\hatcurLCqeccenxxxxxE
\else
??????\fi
\fi
\fi
\fi
\fi
}
\newcommand{\hatcurLCqshorteccen}[1]{\ifnum#1=54 %
\hatcurLCqshorteccenxxxxxA
\else
\ifnum#1=55 %
\hatcurLCqshorteccenxxxxxB
\else
\ifnum#1=56 %
\hatcurLCqshorteccenxxxxxC
\else
\ifnum#1=57 %
\hatcurLCqshorteccenxxxxxD
\else
\ifnum#1=58 %
\hatcurLCqshorteccenxxxxxE
\else
??????\fi
\fi
\fi
\fi
\fi
}
\newcommand{\hatcurLCrhoeccen}[1]{\ifnum#1=54 %
\hatcurLCrhoeccenxxxxxA
\else
\ifnum#1=55 %
\hatcurLCrhoeccenxxxxxB
\else
\ifnum#1=56 %
\hatcurLCrhoeccenxxxxxC
\else
\ifnum#1=57 %
\hatcurLCrhoeccenxxxxxD
\else
\ifnum#1=58 %
\hatcurLCrhoeccenxxxxxE
\else
??????\fi
\fi
\fi
\fi
\fi
}
\newcommand{\hatcurLCrprstareccen}[1]{\ifnum#1=54 %
\hatcurLCrprstareccenxxxxxA
\else
\ifnum#1=55 %
\hatcurLCrprstareccenxxxxxB
\else
\ifnum#1=56 %
\hatcurLCrprstareccenxxxxxC
\else
\ifnum#1=57 %
\hatcurLCrprstareccenxxxxxD
\else
\ifnum#1=58 %
\hatcurLCrprstareccenxxxxxE
\else
??????\fi
\fi
\fi
\fi
\fi
}
\newcommand{\hatcurLCTAeccen}[1]{\ifnum#1=54 %
\hatcurLCTAeccenxxxxxA
\else
\ifnum#1=55 %
\hatcurLCTAeccenxxxxxB
\else
\ifnum#1=56 %
\hatcurLCTAeccenxxxxxC
\else
\ifnum#1=57 %
\hatcurLCTAeccenxxxxxD
\else
\ifnum#1=58 %
\hatcurLCTAeccenxxxxxE
\else
??????\fi
\fi
\fi
\fi
\fi
}
\newcommand{\hatcurLCTBeccen}[1]{\ifnum#1=54 %
\hatcurLCTBeccenxxxxxA
\else
\ifnum#1=55 %
\hatcurLCTBeccenxxxxxB
\else
\ifnum#1=56 %
\hatcurLCTBeccenxxxxxC
\else
\ifnum#1=57 %
\hatcurLCTBeccenxxxxxD
\else
\ifnum#1=58 %
\hatcurLCTBeccenxxxxxE
\else
??????\fi
\fi
\fi
\fi
\fi
}
\newcommand{\hatcurLCTeccen}[1]{\ifnum#1=54 %
\hatcurLCTeccenxxxxxA
\else
\ifnum#1=55 %
\hatcurLCTeccenxxxxxB
\else
\ifnum#1=56 %
\hatcurLCTeccenxxxxxC
\else
\ifnum#1=57 %
\hatcurLCTeccenxxxxxD
\else
\ifnum#1=58 %
\hatcurLCTeccenxxxxxE
\else
??????\fi
\fi
\fi
\fi
\fi
}
\newcommand{\hatcurLCzetaeccen}[1]{\ifnum#1=54 %
\hatcurLCzetaeccenxxxxxA
\else
\ifnum#1=55 %
\hatcurLCzetaeccenxxxxxB
\else
\ifnum#1=56 %
\hatcurLCzetaeccenxxxxxC
\else
\ifnum#1=57 %
\hatcurLCzetaeccenxxxxxD
\else
\ifnum#1=58 %
\hatcurLCzetaeccenxxxxxE
\else
??????\fi
\fi
\fi
\fi
\fi
}
\newcommand{\hatcurPPaequiveccen}[1]{\ifnum#1=54 %
\hatcurPPaequiveccenxxxxxA
\else
\ifnum#1=55 %
\hatcurPPaequiveccenxxxxxB
\else
\ifnum#1=56 %
\hatcurPPaequiveccenxxxxxC
\else
\ifnum#1=57 %
\hatcurPPaequiveccenxxxxxD
\else
\ifnum#1=58 %
\hatcurPPaequiveccenxxxxxE
\else
??????\fi
\fi
\fi
\fi
\fi
}
\newcommand{\hatcurPPareccen}[1]{\ifnum#1=54 %
\hatcurPPareccenxxxxxA
\else
\ifnum#1=55 %
\hatcurPPareccenxxxxxB
\else
\ifnum#1=56 %
\hatcurPPareccenxxxxxC
\else
\ifnum#1=57 %
\hatcurPPareccenxxxxxD
\else
\ifnum#1=58 %
\hatcurPPareccenxxxxxE
\else
??????\fi
\fi
\fi
\fi
\fi
}
\newcommand{\hatcurPPareleccen}[1]{\ifnum#1=54 %
\hatcurPPareleccenxxxxxA
\else
\ifnum#1=55 %
\hatcurPPareleccenxxxxxB
\else
\ifnum#1=56 %
\hatcurPPareleccenxxxxxC
\else
\ifnum#1=57 %
\hatcurPPareleccenxxxxxD
\else
\ifnum#1=58 %
\hatcurPPareleccenxxxxxE
\else
??????\fi
\fi
\fi
\fi
\fi
}
\newcommand{\hatcurPPfluxapdimeccen}[1]{\ifnum#1=54 %
\hatcurPPfluxapdimeccenxxxxxA
\else
\ifnum#1=55 %
\hatcurPPfluxapdimeccenxxxxxB
\else
\ifnum#1=56 %
\hatcurPPfluxapdimeccenxxxxxC
\else
\ifnum#1=57 %
\hatcurPPfluxapdimeccenxxxxxD
\else
\ifnum#1=58 %
\hatcurPPfluxapdimeccenxxxxxE
\else
??????\fi
\fi
\fi
\fi
\fi
}
\newcommand{\hatcurPPfluxapeccen}[1]{\ifnum#1=54 %
\hatcurPPfluxapeccenxxxxxA
\else
\ifnum#1=55 %
\hatcurPPfluxapeccenxxxxxB
\else
\ifnum#1=56 %
\hatcurPPfluxapeccenxxxxxC
\else
\ifnum#1=57 %
\hatcurPPfluxapeccenxxxxxD
\else
\ifnum#1=58 %
\hatcurPPfluxapeccenxxxxxE
\else
??????\fi
\fi
\fi
\fi
\fi
}
\newcommand{\hatcurPPfluxavgdimeccen}[1]{\ifnum#1=54 %
\hatcurPPfluxavgdimeccenxxxxxA
\else
\ifnum#1=55 %
\hatcurPPfluxavgdimeccenxxxxxB
\else
\ifnum#1=56 %
\hatcurPPfluxavgdimeccenxxxxxC
\else
\ifnum#1=57 %
\hatcurPPfluxavgdimeccenxxxxxD
\else
\ifnum#1=58 %
\hatcurPPfluxavgdimeccenxxxxxE
\else
??????\fi
\fi
\fi
\fi
\fi
}
\newcommand{\hatcurPPfluxavgeccen}[1]{\ifnum#1=54 %
\hatcurPPfluxavgeccenxxxxxA
\else
\ifnum#1=55 %
\hatcurPPfluxavgeccenxxxxxB
\else
\ifnum#1=56 %
\hatcurPPfluxavgeccenxxxxxC
\else
\ifnum#1=57 %
\hatcurPPfluxavgeccenxxxxxD
\else
\ifnum#1=58 %
\hatcurPPfluxavgeccenxxxxxE
\else
??????\fi
\fi
\fi
\fi
\fi
}
\newcommand{\hatcurPPfluxavglogeccen}[1]{\ifnum#1=54 %
\hatcurPPfluxavglogeccenxxxxxA
\else
\ifnum#1=55 %
\hatcurPPfluxavglogeccenxxxxxB
\else
\ifnum#1=56 %
\hatcurPPfluxavglogeccenxxxxxC
\else
\ifnum#1=57 %
\hatcurPPfluxavglogeccenxxxxxD
\else
\ifnum#1=58 %
\hatcurPPfluxavglogeccenxxxxxE
\else
??????\fi
\fi
\fi
\fi
\fi
}
\newcommand{\hatcurPPfluxperidimeccen}[1]{\ifnum#1=54 %
\hatcurPPfluxperidimeccenxxxxxA
\else
\ifnum#1=55 %
\hatcurPPfluxperidimeccenxxxxxB
\else
\ifnum#1=56 %
\hatcurPPfluxperidimeccenxxxxxC
\else
\ifnum#1=57 %
\hatcurPPfluxperidimeccenxxxxxD
\else
\ifnum#1=58 %
\hatcurPPfluxperidimeccenxxxxxE
\else
??????\fi
\fi
\fi
\fi
\fi
}
\newcommand{\hatcurPPfluxperieccen}[1]{\ifnum#1=54 %
\hatcurPPfluxperieccenxxxxxA
\else
\ifnum#1=55 %
\hatcurPPfluxperieccenxxxxxB
\else
\ifnum#1=56 %
\hatcurPPfluxperieccenxxxxxC
\else
\ifnum#1=57 %
\hatcurPPfluxperieccenxxxxxD
\else
\ifnum#1=58 %
\hatcurPPfluxperieccenxxxxxE
\else
??????\fi
\fi
\fi
\fi
\fi
}
\newcommand{\hatcurPPgeccen}[1]{\ifnum#1=54 %
\hatcurPPgeccenxxxxxA
\else
\ifnum#1=55 %
\hatcurPPgeccenxxxxxB
\else
\ifnum#1=56 %
\hatcurPPgeccenxxxxxC
\else
\ifnum#1=57 %
\hatcurPPgeccenxxxxxD
\else
\ifnum#1=58 %
\hatcurPPgeccenxxxxxE
\else
??????\fi
\fi
\fi
\fi
\fi
}
\newcommand{\hatcurPPieccen}[1]{\ifnum#1=54 %
\hatcurPPieccenxxxxxA
\else
\ifnum#1=55 %
\hatcurPPieccenxxxxxB
\else
\ifnum#1=56 %
\hatcurPPieccenxxxxxC
\else
\ifnum#1=57 %
\hatcurPPieccenxxxxxD
\else
\ifnum#1=58 %
\hatcurPPieccenxxxxxE
\else
??????\fi
\fi
\fi
\fi
\fi
}
\newcommand{\hatcurPPloggeccen}[1]{\ifnum#1=54 %
\hatcurPPloggeccenxxxxxA
\else
\ifnum#1=55 %
\hatcurPPloggeccenxxxxxB
\else
\ifnum#1=56 %
\hatcurPPloggeccenxxxxxC
\else
\ifnum#1=57 %
\hatcurPPloggeccenxxxxxD
\else
\ifnum#1=58 %
\hatcurPPloggeccenxxxxxE
\else
??????\fi
\fi
\fi
\fi
\fi
}
\newcommand{\hatcurPPmeccen}[1]{\ifnum#1=54 %
\hatcurPPmeccenxxxxxA
\else
\ifnum#1=55 %
\hatcurPPmeccenxxxxxB
\else
\ifnum#1=56 %
\hatcurPPmeccenxxxxxC
\else
\ifnum#1=57 %
\hatcurPPmeccenxxxxxD
\else
\ifnum#1=58 %
\hatcurPPmeccenxxxxxE
\else
??????\fi
\fi
\fi
\fi
\fi
}
\newcommand{\hatcurPPmeeccen}[1]{\ifnum#1=54 %
\hatcurPPmeeccenxxxxxA
\else
\ifnum#1=55 %
\hatcurPPmeeccenxxxxxB
\else
\ifnum#1=56 %
\hatcurPPmeeccenxxxxxC
\else
\ifnum#1=57 %
\hatcurPPmeeccenxxxxxD
\else
\ifnum#1=58 %
\hatcurPPmeeccenxxxxxE
\else
??????\fi
\fi
\fi
\fi
\fi
}
\newcommand{\hatcurPPmelongeccen}[1]{\ifnum#1=54 %
\hatcurPPmelongeccenxxxxxA
\else
\ifnum#1=55 %
\hatcurPPmelongeccenxxxxxB
\else
\ifnum#1=56 %
\hatcurPPmelongeccenxxxxxC
\else
\ifnum#1=57 %
\hatcurPPmelongeccenxxxxxD
\else
\ifnum#1=58 %
\hatcurPPmelongeccenxxxxxE
\else
??????\fi
\fi
\fi
\fi
\fi
}
\newcommand{\hatcurPPmeshorteccen}[1]{\ifnum#1=54 %
\hatcurPPmeshorteccenxxxxxA
\else
\ifnum#1=55 %
\hatcurPPmeshorteccenxxxxxB
\else
\ifnum#1=56 %
\hatcurPPmeshorteccenxxxxxC
\else
\ifnum#1=57 %
\hatcurPPmeshorteccenxxxxxD
\else
\ifnum#1=58 %
\hatcurPPmeshorteccenxxxxxE
\else
??????\fi
\fi
\fi
\fi
\fi
}
\newcommand{\hatcurPPmlongeccen}[1]{\ifnum#1=54 %
\hatcurPPmlongeccenxxxxxA
\else
\ifnum#1=55 %
\hatcurPPmlongeccenxxxxxB
\else
\ifnum#1=56 %
\hatcurPPmlongeccenxxxxxC
\else
\ifnum#1=57 %
\hatcurPPmlongeccenxxxxxD
\else
\ifnum#1=58 %
\hatcurPPmlongeccenxxxxxE
\else
??????\fi
\fi
\fi
\fi
\fi
}
\newcommand{\hatcurPPmrcorreccen}[1]{\ifnum#1=54 %
\hatcurPPmrcorreccenxxxxxA
\else
\ifnum#1=55 %
\hatcurPPmrcorreccenxxxxxB
\else
\ifnum#1=56 %
\hatcurPPmrcorreccenxxxxxC
\else
\ifnum#1=57 %
\hatcurPPmrcorreccenxxxxxD
\else
\ifnum#1=58 %
\hatcurPPmrcorreccenxxxxxE
\else
??????\fi
\fi
\fi
\fi
\fi
}
\newcommand{\hatcurPPmshorteccen}[1]{\ifnum#1=54 %
\hatcurPPmshorteccenxxxxxA
\else
\ifnum#1=55 %
\hatcurPPmshorteccenxxxxxB
\else
\ifnum#1=56 %
\hatcurPPmshorteccenxxxxxC
\else
\ifnum#1=57 %
\hatcurPPmshorteccenxxxxxD
\else
\ifnum#1=58 %
\hatcurPPmshorteccenxxxxxE
\else
??????\fi
\fi
\fi
\fi
\fi
}
\newcommand{\hatcurPPperieccen}[1]{\ifnum#1=54 %
\hatcurPPperieccenxxxxxA
\else
\ifnum#1=55 %
\hatcurPPperieccenxxxxxB
\else
\ifnum#1=56 %
\hatcurPPperieccenxxxxxC
\else
\ifnum#1=57 %
\hatcurPPperieccenxxxxxD
\else
\ifnum#1=58 %
\hatcurPPperieccenxxxxxE
\else
??????\fi
\fi
\fi
\fi
\fi
}
\newcommand{\hatcurPPphiconjeccen}[1]{\ifnum#1=54 %
\hatcurPPphiconjeccenxxxxxA
\else
\ifnum#1=55 %
\hatcurPPphiconjeccenxxxxxB
\else
\ifnum#1=56 %
\hatcurPPphiconjeccenxxxxxC
\else
\ifnum#1=57 %
\hatcurPPphiconjeccenxxxxxD
\else
\ifnum#1=58 %
\hatcurPPphiconjeccenxxxxxE
\else
??????\fi
\fi
\fi
\fi
\fi
}
\newcommand{\hatcurPPreccen}[1]{\ifnum#1=54 %
\hatcurPPreccenxxxxxA
\else
\ifnum#1=55 %
\hatcurPPreccenxxxxxB
\else
\ifnum#1=56 %
\hatcurPPreccenxxxxxC
\else
\ifnum#1=57 %
\hatcurPPreccenxxxxxD
\else
\ifnum#1=58 %
\hatcurPPreccenxxxxxE
\else
??????\fi
\fi
\fi
\fi
\fi
}
\newcommand{\hatcurPPreeccen}[1]{\ifnum#1=54 %
\hatcurPPreeccenxxxxxA
\else
\ifnum#1=55 %
\hatcurPPreeccenxxxxxB
\else
\ifnum#1=56 %
\hatcurPPreeccenxxxxxC
\else
\ifnum#1=57 %
\hatcurPPreeccenxxxxxD
\else
\ifnum#1=58 %
\hatcurPPreeccenxxxxxE
\else
??????\fi
\fi
\fi
\fi
\fi
}
\newcommand{\hatcurPPrelongeccen}[1]{\ifnum#1=54 %
\hatcurPPrelongeccenxxxxxA
\else
\ifnum#1=55 %
\hatcurPPrelongeccenxxxxxB
\else
\ifnum#1=56 %
\hatcurPPrelongeccenxxxxxC
\else
\ifnum#1=57 %
\hatcurPPrelongeccenxxxxxD
\else
\ifnum#1=58 %
\hatcurPPrelongeccenxxxxxE
\else
??????\fi
\fi
\fi
\fi
\fi
}
\newcommand{\hatcurPPreshorteccen}[1]{\ifnum#1=54 %
\hatcurPPreshorteccenxxxxxA
\else
\ifnum#1=55 %
\hatcurPPreshorteccenxxxxxB
\else
\ifnum#1=56 %
\hatcurPPreshorteccenxxxxxC
\else
\ifnum#1=57 %
\hatcurPPreshorteccenxxxxxD
\else
\ifnum#1=58 %
\hatcurPPreshorteccenxxxxxE
\else
??????\fi
\fi
\fi
\fi
\fi
}
\newcommand{\hatcurPPrhoeccen}[1]{\ifnum#1=54 %
\hatcurPPrhoeccenxxxxxA
\else
\ifnum#1=55 %
\hatcurPPrhoeccenxxxxxB
\else
\ifnum#1=56 %
\hatcurPPrhoeccenxxxxxC
\else
\ifnum#1=57 %
\hatcurPPrhoeccenxxxxxD
\else
\ifnum#1=58 %
\hatcurPPrhoeccenxxxxxE
\else
??????\fi
\fi
\fi
\fi
\fi
}
\newcommand{\hatcurPPrlongeccen}[1]{\ifnum#1=54 %
\hatcurPPrlongeccenxxxxxA
\else
\ifnum#1=55 %
\hatcurPPrlongeccenxxxxxB
\else
\ifnum#1=56 %
\hatcurPPrlongeccenxxxxxC
\else
\ifnum#1=57 %
\hatcurPPrlongeccenxxxxxD
\else
\ifnum#1=58 %
\hatcurPPrlongeccenxxxxxE
\else
??????\fi
\fi
\fi
\fi
\fi
}
\newcommand{\hatcurPPrshorteccen}[1]{\ifnum#1=54 %
\hatcurPPrshorteccenxxxxxA
\else
\ifnum#1=55 %
\hatcurPPrshorteccenxxxxxB
\else
\ifnum#1=56 %
\hatcurPPrshorteccenxxxxxC
\else
\ifnum#1=57 %
\hatcurPPrshorteccenxxxxxD
\else
\ifnum#1=58 %
\hatcurPPrshorteccenxxxxxE
\else
??????\fi
\fi
\fi
\fi
\fi
}
\newcommand{\hatcurPPtcirceccen}[1]{\ifnum#1=54 %
\hatcurPPtcirceccenxxxxxA
\else
\ifnum#1=55 %
\hatcurPPtcirceccenxxxxxB
\else
\ifnum#1=56 %
\hatcurPPtcirceccenxxxxxC
\else
\ifnum#1=57 %
\hatcurPPtcirceccenxxxxxD
\else
\ifnum#1=58 %
\hatcurPPtcirceccenxxxxxE
\else
??????\fi
\fi
\fi
\fi
\fi
}
\newcommand{\hatcurPPteffeccen}[1]{\ifnum#1=54 %
\hatcurPPteffeccenxxxxxA
\else
\ifnum#1=55 %
\hatcurPPteffeccenxxxxxB
\else
\ifnum#1=56 %
\hatcurPPteffeccenxxxxxC
\else
\ifnum#1=57 %
\hatcurPPteffeccenxxxxxD
\else
\ifnum#1=58 %
\hatcurPPteffeccenxxxxxE
\else
??????\fi
\fi
\fi
\fi
\fi
}
\newcommand{\hatcurPPthetaeccen}[1]{\ifnum#1=54 %
\hatcurPPthetaeccenxxxxxA
\else
\ifnum#1=55 %
\hatcurPPthetaeccenxxxxxB
\else
\ifnum#1=56 %
\hatcurPPthetaeccenxxxxxC
\else
\ifnum#1=57 %
\hatcurPPthetaeccenxxxxxD
\else
\ifnum#1=58 %
\hatcurPPthetaeccenxxxxxE
\else
??????\fi
\fi
\fi
\fi
\fi
}
\newcommand{\hatcurPPtinfalleccen}[1]{\ifnum#1=54 %
\hatcurPPtinfalleccenxxxxxA
\else
\ifnum#1=55 %
\hatcurPPtinfalleccenxxxxxB
\else
\ifnum#1=56 %
\hatcurPPtinfalleccenxxxxxC
\else
\ifnum#1=57 %
\hatcurPPtinfalleccenxxxxxD
\else
\ifnum#1=58 %
\hatcurPPtinfalleccenxxxxxE
\else
??????\fi
\fi
\fi
\fi
\fi
}
\newcommand{\hatcurRVecceneccen}[1]{\ifnum#1=54 %
\hatcurRVecceneccenxxxxxA
\else
\ifnum#1=55 %
\hatcurRVecceneccenxxxxxB
\else
\ifnum#1=56 %
\hatcurRVecceneccenxxxxxC
\else
\ifnum#1=57 %
\hatcurRVecceneccenxxxxxD
\else
\ifnum#1=58 %
\hatcurRVecceneccenxxxxxE
\else
??????\fi
\fi
\fi
\fi
\fi
}
\newcommand{\hatcurRVeccentwosiglimeccen}[1]{\ifnum#1=54 %
\hatcurRVeccentwosiglimeccenxxxxxA
\else
\ifnum#1=55 %
\hatcurRVeccentwosiglimeccenxxxxxB
\else
\ifnum#1=56 %
\hatcurRVeccentwosiglimeccenxxxxxC
\else
\ifnum#1=57 %
\hatcurRVeccentwosiglimeccenxxxxxD
\else
\ifnum#1=58 %
\hatcurRVeccentwosiglimeccenxxxxxE
\else
??????\fi
\fi
\fi
\fi
\fi
}
\newcommand{\hatcurRVfitrmsAeccen}[1]{\ifnum#1=54 %
\hatcurRVfitrmsAeccenxxxxxA
\else
\ifnum#1=56 %
\hatcurRVfitrmsAeccenxxxxxC
\else
\ifnum#1=58 %
\hatcurRVfitrmsAeccenxxxxxE
\else
??????\fi
\fi
\fi
}
\newcommand{\hatcurRVfitrmsBeccen}[1]{\ifnum#1=54 %
\hatcurRVfitrmsBeccenxxxxxA
\else
\ifnum#1=56 %
\hatcurRVfitrmsBeccenxxxxxC
\else
\ifnum#1=58 %
\hatcurRVfitrmsBeccenxxxxxE
\else
??????\fi
\fi
\fi
}
\newcommand{\hatcurRVfitrmseccen}[1]{\ifnum#1=55 %
\hatcurRVfitrmseccenxxxxxB
\else
\ifnum#1=57 %
\hatcurRVfitrmseccenxxxxxD
\else
??????\fi
\fi
}
\newcommand{\hatcurRVgammaAeccen}[1]{\ifnum#1=54 %
\hatcurRVgammaAeccenxxxxxA
\else
\ifnum#1=56 %
\hatcurRVgammaAeccenxxxxxC
\else
\ifnum#1=58 %
\hatcurRVgammaAeccenxxxxxE
\else
??????\fi
\fi
\fi
}
\newcommand{\hatcurRVgammaBeccen}[1]{\ifnum#1=54 %
\hatcurRVgammaBeccenxxxxxA
\else
\ifnum#1=56 %
\hatcurRVgammaBeccenxxxxxC
\else
\ifnum#1=58 %
\hatcurRVgammaBeccenxxxxxE
\else
??????\fi
\fi
\fi
}
\newcommand{\hatcurRVgammaeccen}[1]{\ifnum#1=55 %
\hatcurRVgammaeccenxxxxxB
\else
\ifnum#1=57 %
\hatcurRVgammaeccenxxxxxD
\else
??????\fi
\fi
}
\newcommand{\hatcurRVheccen}[1]{\ifnum#1=54 %
\hatcurRVheccenxxxxxA
\else
\ifnum#1=55 %
\hatcurRVheccenxxxxxB
\else
\ifnum#1=56 %
\hatcurRVheccenxxxxxC
\else
\ifnum#1=57 %
\hatcurRVheccenxxxxxD
\else
\ifnum#1=58 %
\hatcurRVheccenxxxxxE
\else
??????\fi
\fi
\fi
\fi
\fi
}
\newcommand{\hatcurRVjitterAeccen}[1]{\ifnum#1=54 %
\hatcurRVjitterAeccenxxxxxA
\else
\ifnum#1=56 %
\hatcurRVjitterAeccenxxxxxC
\else
\ifnum#1=58 %
\hatcurRVjitterAeccenxxxxxE
\else
??????\fi
\fi
\fi
}
\newcommand{\hatcurRVjitterBeccen}[1]{\ifnum#1=54 %
\hatcurRVjitterBeccenxxxxxA
\else
\ifnum#1=56 %
\hatcurRVjitterBeccenxxxxxC
\else
\ifnum#1=58 %
\hatcurRVjitterBeccenxxxxxE
\else
??????\fi
\fi
\fi
}
\newcommand{\hatcurRVjittereccen}[1]{\ifnum#1=55 %
\hatcurRVjittereccenxxxxxB
\else
\ifnum#1=57 %
\hatcurRVjittereccenxxxxxD
\else
??????\fi
\fi
}
\newcommand{\hatcurRVjittertwosiglimAeccen}[1]{\ifnum#1=54 %
\hatcurRVjittertwosiglimAeccenxxxxxA
\else
\ifnum#1=56 %
\hatcurRVjittertwosiglimAeccenxxxxxC
\else
\ifnum#1=58 %
\hatcurRVjittertwosiglimAeccenxxxxxE
\else
??????\fi
\fi
\fi
}
\newcommand{\hatcurRVjittertwosiglimBeccen}[1]{\ifnum#1=54 %
\hatcurRVjittertwosiglimBeccenxxxxxA
\else
\ifnum#1=56 %
\hatcurRVjittertwosiglimBeccenxxxxxC
\else
\ifnum#1=58 %
\hatcurRVjittertwosiglimBeccenxxxxxE
\else
??????\fi
\fi
\fi
}
\newcommand{\hatcurRVjittertwosiglimeccen}[1]{\ifnum#1=55 %
\hatcurRVjittertwosiglimeccenxxxxxB
\else
\ifnum#1=57 %
\hatcurRVjittertwosiglimeccenxxxxxD
\else
??????\fi
\fi
}
\newcommand{\hatcurRVkeccen}[1]{\ifnum#1=54 %
\hatcurRVkeccenxxxxxA
\else
\ifnum#1=55 %
\hatcurRVkeccenxxxxxB
\else
\ifnum#1=56 %
\hatcurRVkeccenxxxxxC
\else
\ifnum#1=57 %
\hatcurRVkeccenxxxxxD
\else
\ifnum#1=58 %
\hatcurRVkeccenxxxxxE
\else
??????\fi
\fi
\fi
\fi
\fi
}
\newcommand{\hatcurRVKeccen}[1]{\ifnum#1=54 %
\hatcurRVKeccenxxxxxA
\else
\ifnum#1=55 %
\hatcurRVKeccenxxxxxB
\else
\ifnum#1=56 %
\hatcurRVKeccenxxxxxC
\else
\ifnum#1=57 %
\hatcurRVKeccenxxxxxD
\else
\ifnum#1=58 %
\hatcurRVKeccenxxxxxE
\else
??????\fi
\fi
\fi
\fi
\fi
}
\newcommand{\hatcurRVomegaeccen}[1]{\ifnum#1=54 %
\hatcurRVomegaeccenxxxxxA
\else
\ifnum#1=55 %
\hatcurRVomegaeccenxxxxxB
\else
\ifnum#1=56 %
\hatcurRVomegaeccenxxxxxC
\else
\ifnum#1=57 %
\hatcurRVomegaeccenxxxxxD
\else
\ifnum#1=58 %
\hatcurRVomegaeccenxxxxxE
\else
??????\fi
\fi
\fi
\fi
\fi
}
\newcommand{\hatcurRVrheccen}[1]{\ifnum#1=54 %
\hatcurRVrheccenxxxxxA
\else
\ifnum#1=55 %
\hatcurRVrheccenxxxxxB
\else
\ifnum#1=56 %
\hatcurRVrheccenxxxxxC
\else
\ifnum#1=57 %
\hatcurRVrheccenxxxxxD
\else
\ifnum#1=58 %
\hatcurRVrheccenxxxxxE
\else
??????\fi
\fi
\fi
\fi
\fi
}
\newcommand{\hatcurRVrkeccen}[1]{\ifnum#1=54 %
\hatcurRVrkeccenxxxxxA
\else
\ifnum#1=55 %
\hatcurRVrkeccenxxxxxB
\else
\ifnum#1=56 %
\hatcurRVrkeccenxxxxxC
\else
\ifnum#1=57 %
\hatcurRVrkeccenxxxxxD
\else
\ifnum#1=58 %
\hatcurRVrkeccenxxxxxE
\else
??????\fi
\fi
\fi
\fi
\fi
}
\newcommand{\hatcurRVtroneeccen}[1]{\ifnum#1=54 %
\hatcurRVtroneeccenxxxxxA
\else
\ifnum#1=55 %
\hatcurRVtroneeccenxxxxxB
\else
\ifnum#1=56 %
\hatcurRVtroneeccenxxxxxC
\else
\ifnum#1=57 %
\hatcurRVtroneeccenxxxxxD
\else
\ifnum#1=58 %
\hatcurRVtroneeccenxxxxxE
\else
??????\fi
\fi
\fi
\fi
\fi
}
\newcommand{\hatcurRVtrtwoeccen}[1]{\ifnum#1=54 %
\hatcurRVtrtwoeccenxxxxxA
\else
\ifnum#1=55 %
\hatcurRVtrtwoeccenxxxxxB
\else
\ifnum#1=56 %
\hatcurRVtrtwoeccenxxxxxC
\else
\ifnum#1=57 %
\hatcurRVtrtwoeccenxxxxxD
\else
\ifnum#1=58 %
\hatcurRVtrtwoeccenxxxxxE
\else
??????\fi
\fi
\fi
\fi
\fi
}
\newcommand{\hatcurSMEiiloggeccen}[1]{\ifnum#1=54 %
\hatcurSMEiiloggeccenxxxxxA
\else
\ifnum#1=55 %
\hatcurSMEiiloggeccenxxxxxB
\else
\ifnum#1=56 %
\hatcurSMEiiloggeccenxxxxxC
\else
\ifnum#1=57 %
\hatcurSMEiiloggeccenxxxxxD
\else
\ifnum#1=58 %
\hatcurSMEiiloggeccenxxxxxE
\else
??????\fi
\fi
\fi
\fi
\fi
}
\newcommand{\hatcurSMEiiteffeccen}[1]{\ifnum#1=54 %
\hatcurSMEiiteffeccenxxxxxA
\else
\ifnum#1=55 %
\hatcurSMEiiteffeccenxxxxxB
\else
\ifnum#1=56 %
\hatcurSMEiiteffeccenxxxxxC
\else
\ifnum#1=57 %
\hatcurSMEiiteffeccenxxxxxD
\else
\ifnum#1=58 %
\hatcurSMEiiteffeccenxxxxxE
\else
??????\fi
\fi
\fi
\fi
\fi
}
\newcommand{\hatcurSMEiivmaceccen}[1]{\ifnum#1=54 %
\hatcurSMEiivmaceccenxxxxxA
\else
\ifnum#1=55 %
\hatcurSMEiivmaceccenxxxxxB
\else
\ifnum#1=56 %
\hatcurSMEiivmaceccenxxxxxC
\else
\ifnum#1=57 %
\hatcurSMEiivmaceccenxxxxxD
\else
\ifnum#1=58 %
\hatcurSMEiivmaceccenxxxxxE
\else
??????\fi
\fi
\fi
\fi
\fi
}
\newcommand{\hatcurSMEiivmiceccen}[1]{\ifnum#1=54 %
\hatcurSMEiivmiceccenxxxxxA
\else
\ifnum#1=55 %
\hatcurSMEiivmiceccenxxxxxB
\else
\ifnum#1=56 %
\hatcurSMEiivmiceccenxxxxxC
\else
\ifnum#1=57 %
\hatcurSMEiivmiceccenxxxxxD
\else
\ifnum#1=58 %
\hatcurSMEiivmiceccenxxxxxE
\else
??????\fi
\fi
\fi
\fi
\fi
}
\newcommand{\hatcurSMEiivsineccen}[1]{\ifnum#1=54 %
\hatcurSMEiivsineccenxxxxxA
\else
\ifnum#1=55 %
\hatcurSMEiivsineccenxxxxxB
\else
\ifnum#1=56 %
\hatcurSMEiivsineccenxxxxxC
\else
\ifnum#1=57 %
\hatcurSMEiivsineccenxxxxxD
\else
\ifnum#1=58 %
\hatcurSMEiivsineccenxxxxxE
\else
??????\fi
\fi
\fi
\fi
\fi
}
\newcommand{\hatcurSMEiizfeheccen}[1]{\ifnum#1=54 %
\hatcurSMEiizfeheccenxxxxxA
\else
\ifnum#1=55 %
\hatcurSMEiizfeheccenxxxxxB
\else
\ifnum#1=56 %
\hatcurSMEiizfeheccenxxxxxC
\else
\ifnum#1=57 %
\hatcurSMEiizfeheccenxxxxxD
\else
\ifnum#1=58 %
\hatcurSMEiizfeheccenxxxxxE
\else
??????\fi
\fi
\fi
\fi
\fi
}
\newcommand{\hatcurSMEiizfehshorteccen}[1]{\ifnum#1=54 %
\hatcurSMEiizfehshorteccenxxxxxA
\else
\ifnum#1=55 %
\hatcurSMEiizfehshorteccenxxxxxB
\else
\ifnum#1=56 %
\hatcurSMEiizfehshorteccenxxxxxC
\else
\ifnum#1=57 %
\hatcurSMEiizfehshorteccenxxxxxD
\else
\ifnum#1=58 %
\hatcurSMEiizfehshorteccenxxxxxE
\else
??????\fi
\fi
\fi
\fi
\fi
}
\newcommand{\hatcurSMEiloggeccen}[1]{\ifnum#1=54 %
\hatcurSMEiloggeccenxxxxxA
\else
\ifnum#1=55 %
\hatcurSMEiloggeccenxxxxxB
\else
\ifnum#1=56 %
\hatcurSMEiloggeccenxxxxxC
\else
\ifnum#1=57 %
\hatcurSMEiloggeccenxxxxxD
\else
\ifnum#1=58 %
\hatcurSMEiloggeccenxxxxxE
\else
??????\fi
\fi
\fi
\fi
\fi
}
\newcommand{\hatcurSMEiteffeccen}[1]{\ifnum#1=54 %
\hatcurSMEiteffeccenxxxxxA
\else
\ifnum#1=55 %
\hatcurSMEiteffeccenxxxxxB
\else
\ifnum#1=56 %
\hatcurSMEiteffeccenxxxxxC
\else
\ifnum#1=57 %
\hatcurSMEiteffeccenxxxxxD
\else
\ifnum#1=58 %
\hatcurSMEiteffeccenxxxxxE
\else
??????\fi
\fi
\fi
\fi
\fi
}
\newcommand{\hatcurSMEivmaceccen}[1]{\ifnum#1=54 %
\hatcurSMEivmaceccenxxxxxA
\else
\ifnum#1=55 %
\hatcurSMEivmaceccenxxxxxB
\else
\ifnum#1=56 %
\hatcurSMEivmaceccenxxxxxC
\else
\ifnum#1=57 %
\hatcurSMEivmaceccenxxxxxD
\else
\ifnum#1=58 %
\hatcurSMEivmaceccenxxxxxE
\else
??????\fi
\fi
\fi
\fi
\fi
}
\newcommand{\hatcurSMEivmiceccen}[1]{\ifnum#1=54 %
\hatcurSMEivmiceccenxxxxxA
\else
\ifnum#1=55 %
\hatcurSMEivmiceccenxxxxxB
\else
\ifnum#1=56 %
\hatcurSMEivmiceccenxxxxxC
\else
\ifnum#1=57 %
\hatcurSMEivmiceccenxxxxxD
\else
\ifnum#1=58 %
\hatcurSMEivmiceccenxxxxxE
\else
??????\fi
\fi
\fi
\fi
\fi
}
\newcommand{\hatcurSMEivsineccen}[1]{\ifnum#1=54 %
\hatcurSMEivsineccenxxxxxA
\else
\ifnum#1=55 %
\hatcurSMEivsineccenxxxxxB
\else
\ifnum#1=56 %
\hatcurSMEivsineccenxxxxxC
\else
\ifnum#1=57 %
\hatcurSMEivsineccenxxxxxD
\else
\ifnum#1=58 %
\hatcurSMEivsineccenxxxxxE
\else
??????\fi
\fi
\fi
\fi
\fi
}
\newcommand{\hatcurSMEizfeheccen}[1]{\ifnum#1=54 %
\hatcurSMEizfeheccenxxxxxA
\else
\ifnum#1=55 %
\hatcurSMEizfeheccenxxxxxB
\else
\ifnum#1=56 %
\hatcurSMEizfeheccenxxxxxC
\else
\ifnum#1=57 %
\hatcurSMEizfeheccenxxxxxD
\else
\ifnum#1=58 %
\hatcurSMEizfeheccenxxxxxE
\else
??????\fi
\fi
\fi
\fi
\fi
}
\newcommand{\hatcurSMEizfehshorteccen}[1]{\ifnum#1=54 %
\hatcurSMEizfehshorteccenxxxxxA
\else
\ifnum#1=55 %
\hatcurSMEizfehshorteccenxxxxxB
\else
\ifnum#1=56 %
\hatcurSMEizfehshorteccenxxxxxC
\else
\ifnum#1=57 %
\hatcurSMEizfehshorteccenxxxxxD
\else
\ifnum#1=58 %
\hatcurSMEizfehshorteccenxxxxxE
\else
??????\fi
\fi
\fi
\fi
\fi
}
\newcommand{\hatcurXAveccen}[1]{\ifnum#1=54 %
\hatcurXAveccenxxxxxA
\else
\ifnum#1=55 %
\hatcurXAveccenxxxxxB
\else
\ifnum#1=56 %
\hatcurXAveccenxxxxxC
\else
\ifnum#1=57 %
\hatcurXAveccenxxxxxD
\else
\ifnum#1=58 %
\hatcurXAveccenxxxxxE
\else
??????\fi
\fi
\fi
\fi
\fi
}
\newcommand{\hatcurXdisteccen}[1]{\ifnum#1=54 %
\hatcurXdisteccenxxxxxA
\else
\ifnum#1=55 %
\hatcurXdisteccenxxxxxB
\else
\ifnum#1=56 %
\hatcurXdisteccenxxxxxC
\else
\ifnum#1=57 %
\hatcurXdisteccenxxxxxD
\else
\ifnum#1=58 %
\hatcurXdisteccenxxxxxE
\else
??????\fi
\fi
\fi
\fi
\fi
}
\newcommand{\hatcurXdistredeccen}[1]{\ifnum#1=54 %
\hatcurXdistredeccenxxxxxA
\else
\ifnum#1=55 %
\hatcurXdistredeccenxxxxxB
\else
\ifnum#1=56 %
\hatcurXdistredeccenxxxxxC
\else
\ifnum#1=57 %
\hatcurXdistredeccenxxxxxD
\else
\ifnum#1=58 %
\hatcurXdistredeccenxxxxxE
\else
??????\fi
\fi
\fi
\fi
\fi
}
\newcommand{\hatcurXEBVeccen}[1]{\ifnum#1=54 %
\hatcurXEBVeccenxxxxxA
\else
\ifnum#1=55 %
\hatcurXEBVeccenxxxxxB
\else
\ifnum#1=56 %
\hatcurXEBVeccenxxxxxC
\else
\ifnum#1=57 %
\hatcurXEBVeccenxxxxxD
\else
\ifnum#1=58 %
\hatcurXEBVeccenxxxxxE
\else
??????\fi
\fi
\fi
\fi
\fi
}
\newcommand{\hatcurXjhisoredeccen}[1]{\ifnum#1=58 %
\hatcurXjhisoredeccenxxxxxE
\else
??????\fi
}
\newcommand{\hatcurXjkisoredeccen}[1]{\ifnum#1=58 %
\hatcurXjkisoredeccenxxxxxE
\else
??????\fi
}
\newcommand{\hatcurXmhisoredeccen}[1]{\ifnum#1=58 %
\hatcurXmhisoredeccenxxxxxE
\else
??????\fi
}
\newcommand{\hatcurXmiisoredeccen}[1]{\ifnum#1=58 %
\hatcurXmiisoredeccenxxxxxE
\else
??????\fi
}
\newcommand{\hatcurXmjisoredeccen}[1]{\ifnum#1=58 %
\hatcurXmjisoredeccenxxxxxE
\else
??????\fi
}
\newcommand{\hatcurXmkisoredeccen}[1]{\ifnum#1=58 %
\hatcurXmkisoredeccenxxxxxE
\else
??????\fi
}
\newcommand{\hatcurXmvisoredeccen}[1]{\ifnum#1=58 %
\hatcurXmvisoredeccenxxxxxE
\else
??????\fi
}
\newcommand{\hatcurXsecdureccen}[1]{\ifnum#1=54 %
\hatcurXsecdureccenxxxxxA
\else
\ifnum#1=55 %
\hatcurXsecdureccenxxxxxB
\else
\ifnum#1=56 %
\hatcurXsecdureccenxxxxxC
\else
\ifnum#1=57 %
\hatcurXsecdureccenxxxxxD
\else
\ifnum#1=58 %
\hatcurXsecdureccenxxxxxE
\else
??????\fi
\fi
\fi
\fi
\fi
}
\newcommand{\hatcurXsecingdureccen}[1]{\ifnum#1=54 %
\hatcurXsecingdureccenxxxxxA
\else
\ifnum#1=55 %
\hatcurXsecingdureccenxxxxxB
\else
\ifnum#1=56 %
\hatcurXsecingdureccenxxxxxC
\else
\ifnum#1=57 %
\hatcurXsecingdureccenxxxxxD
\else
\ifnum#1=58 %
\hatcurXsecingdureccenxxxxxE
\else
??????\fi
\fi
\fi
\fi
\fi
}
\newcommand{\hatcurXsecondaryeccen}[1]{\ifnum#1=54 %
\hatcurXsecondaryeccenxxxxxA
\else
\ifnum#1=55 %
\hatcurXsecondaryeccenxxxxxB
\else
\ifnum#1=56 %
\hatcurXsecondaryeccenxxxxxC
\else
\ifnum#1=57 %
\hatcurXsecondaryeccenxxxxxD
\else
\ifnum#1=58 %
\hatcurXsecondaryeccenxxxxxE
\else
??????\fi
\fi
\fi
\fi
\fi
}
\newcommand{\hatcurXsecphaseeccen}[1]{\ifnum#1=54 %
\hatcurXsecphaseeccenxxxxxA
\else
\ifnum#1=55 %
\hatcurXsecphaseeccenxxxxxB
\else
\ifnum#1=56 %
\hatcurXsecphaseeccenxxxxxC
\else
\ifnum#1=57 %
\hatcurXsecphaseeccenxxxxxD
\else
\ifnum#1=58 %
\hatcurXsecphaseeccenxxxxxE
\else
??????\fi
\fi
\fi
\fi
\fi
}
\newcommand{\hatcurXviisoredeccen}[1]{\ifnum#1=58 %
\hatcurXviisoredeccenxxxxxE
\else
??????\fi
}
\newcommand{\hatcurXvkisoredeccen}[1]{\ifnum#1=58 %
\hatcurXvkisoredeccenxxxxxE
\else
??????\fi
}

\newcommand{\hatcurxxxxxA}{HATS-54}
\newcommand{\hatcurbxxxxxA}{HATS-54b}
\newcommand{\hatcurcxxxxxA}{HATS-54c}

\newcommand{\hatcurplanetnumxxxxxA}{54}

\newcommand{\hatcurCCtwomassshortxxxxxA}{13223237-4441196}

\newcommand{\hatcurRVgammaabsxxxxxA}{\hatcurRVgammaA{\hatcurplanetnumxxxxxA}}                           % Absolute Gamma velocity

\newcommand{\hatcurRVgammarelxxxxxA}{\hatcurRVgammaA{\hatcurplanetnumxxxxxA}}                           % Relative Gamma velocity. Typically that of the Keck RVs.

\newcommand{\hatcurCCtassvixxxxxA}{\ensuremath{NULL\pm NULL}}                  % TASS V-I

\newcommand{\hatcurSMEversionxxxxxA}{ii}                                       % Final SME version:i or ii?

\newcommand{\hatcurisoshortxxxxxA}{YY}
\newcommand{\hatcurisofullxxxxxA}{Yonsei-Yale (YY)}
\newcommand{\hatcurisocitexxxxxA}{yi:2001}
\newcommand{\hatcurlumindxxxxxA}{\rhostar}
\newcommand{\hatcurjhkfilsetxxxxxA}{ESO}
\newcommand{\hatcurSMEteffxxxxxA}{\ifthenelse{\equal{\hatcurSMEversionxxxxxA}{i}}{\hatcurSMEiteff{\hatcurplanetnumxxxxxA}}{\hatcurSMEiiteff{\hatcurplanetnumxxxxxA}}}
\newcommand{\hatcurSMEzfehxxxxxA}{\ifthenelse{\equal{\hatcurSMEversionxxxxxA}{i}}{\hatcurSMEizfeh{\hatcurplanetnumxxxxxA}}{\hatcurSMEiizfeh{\hatcurplanetnumxxxxxA}}}
\newcommand{\hatcurSMEzfehshortxxxxxA}{\ifthenelse{\equal{\hatcurSMEversionxxxxxA}{i}}{\hatcurSMEizfehshort{\hatcurplanetnumxxxxxA}}{\hatcurSMEiizfehshort{\hatcurplanetnumxxxxxA}}}
\newcommand{\hatcurSMEloggxxxxxA}{\ifthenelse{\equal{\hatcurSMEversionxxxxxA}{i}}{\hatcurSMEilogg{\hatcurplanetnumxxxxxA}}{\hatcurSMEiilogg{\hatcurplanetnumxxxxxA}}}
\newcommand{\hatcurSMEvsinxxxxxA}{\ifthenelse{\equal{\hatcurSMEversionxxxxxA}{i}}{\hatcurSMEivsin{\hatcurplanetnumxxxxxA}}{\hatcurSMEiivsin{\hatcurplanetnumxxxxxA}}}
\newcommand{\hatcurSMEvmacxxxxxA}{\ifthenelse{\equal{\hatcurSMEversionxxxxxA}{i}}{\hatcurSMEivmac{\hatcurplanetnumxxxxxA}}{\hatcurSMEiivmac{\hatcurplanetnumxxxxxA}}}
\newcommand{\hatcurSMEvmicxxxxxA}{\ifthenelse{\equal{\hatcurSMEversionxxxxxA}{i}}{\hatcurSMEivmic{\hatcurplanetnumxxxxxA}}{\hatcurSMEiivmic{\hatcurplanetnumxxxxxA}}}

\newcommand{\hatcurxxxxxB}{HATS-55}
\newcommand{\hatcurbxxxxxB}{HATS-55b}
\newcommand{\hatcurcxxxxxB}{HATS-55c}

\newcommand{\hatcurplanetnumxxxxxB}{55}

\newcommand{\hatcurCCtwomassshortxxxxxB}{07370802-3245195}

\newcommand{\hatcurRVgammaabsxxxxxB}{\hatcurRVgamma{\hatcurplanetnumxxxxxB}}                           % Absolute Gamma velocity

\newcommand{\hatcurRVgammarelxxxxxB}{\hatcurRVgamma{\hatcurplanetnumxxxxxB}}                           % Relative Gamma velocity. Typically that of the Keck RVs.

\newcommand{\hatcurCCtassvixxxxxB}{\ensuremath{NULL\pm NULL}}                  % TASS V-I

\newcommand{\hatcurSMEversionxxxxxB}{ii}                                       % Final SME version:i or ii?

\newcommand{\hatcurisoshortxxxxxB}{YY}
\newcommand{\hatcurisofullxxxxxB}{Yonsei-Yale (YY)}
\newcommand{\hatcurisocitexxxxxB}{yi:2001}
\newcommand{\hatcurlumindxxxxxB}{\rhostar}
\newcommand{\hatcurjhkfilsetxxxxxB}{ESO}
\newcommand{\hatcurSMEteffxxxxxB}{\ifthenelse{\equal{\hatcurSMEversionxxxxxB}{i}}{\hatcurSMEiteff{\hatcurplanetnumxxxxxB}}{\hatcurSMEiiteff{\hatcurplanetnumxxxxxB}}}
\newcommand{\hatcurSMEzfehxxxxxB}{\ifthenelse{\equal{\hatcurSMEversionxxxxxB}{i}}{\hatcurSMEizfeh{\hatcurplanetnumxxxxxB}}{\hatcurSMEiizfeh{\hatcurplanetnumxxxxxB}}}
\newcommand{\hatcurSMEzfehshortxxxxxB}{\ifthenelse{\equal{\hatcurSMEversionxxxxxB}{i}}{\hatcurSMEizfehshort{\hatcurplanetnumxxxxxB}}{\hatcurSMEiizfehshort{\hatcurplanetnumxxxxxB}}}
\newcommand{\hatcurSMEloggxxxxxB}{\ifthenelse{\equal{\hatcurSMEversionxxxxxB}{i}}{\hatcurSMEilogg{\hatcurplanetnumxxxxxB}}{\hatcurSMEiilogg{\hatcurplanetnumxxxxxB}}}
\newcommand{\hatcurSMEvsinxxxxxB}{\ifthenelse{\equal{\hatcurSMEversionxxxxxB}{i}}{\hatcurSMEivsin{\hatcurplanetnumxxxxxB}}{\hatcurSMEiivsin{\hatcurplanetnumxxxxxB}}}
\newcommand{\hatcurSMEvmacxxxxxB}{\ifthenelse{\equal{\hatcurSMEversionxxxxxB}{i}}{\hatcurSMEivmac{\hatcurplanetnumxxxxxB}}{\hatcurSMEiivmac{\hatcurplanetnumxxxxxB}}}
\newcommand{\hatcurSMEvmicxxxxxB}{\ifthenelse{\equal{\hatcurSMEversionxxxxxB}{i}}{\hatcurSMEivmic{\hatcurplanetnumxxxxxB}}{\hatcurSMEiivmic{\hatcurplanetnumxxxxxB}}}

\newcommand{\hatcurxxxxxC}{HATS-56}
\newcommand{\hatcurbxxxxxC}{HATS-56b}
\newcommand{\hatcurcxxxxxC}{HATS-56c}

\newcommand{\hatcurplanetnumxxxxxC}{56}

\newcommand{\hatcurCCtwomassshortxxxxxC}{12003962-4547579}

\newcommand{\hatcurRVgammaabsxxxxxC}{\hatcurRVgammaA{\hatcurplanetnumxxxxxC}}                           % Absolute Gamma velocity

\newcommand{\hatcurRVgammarelxxxxxC}{\hatcurRVgammaA{\hatcurplanetnumxxxxxC}}                           % Relative Gamma velocity. Typically that of the Keck RVs.

\newcommand{\hatcurCCtassvixxxxxC}{\ensuremath{NULL\pm NULL}}                  % TASS V-I

\newcommand{\hatcurSMEversionxxxxxC}{ii}                                       % Final SME version:i or ii?

\newcommand{\hatcurisoshortxxxxxC}{YY}
\newcommand{\hatcurisofullxxxxxC}{Yonsei-Yale (YY)}
\newcommand{\hatcurisocitexxxxxC}{yi:2001}
\newcommand{\hatcurlumindxxxxxC}{\rhostar}
\newcommand{\hatcurjhkfilsetxxxxxC}{ESO}
\newcommand{\hatcurSMEteffxxxxxC}{\ifthenelse{\equal{\hatcurSMEversionxxxxxC}{i}}{\hatcurSMEiteff{\hatcurplanetnumxxxxxC}}{\hatcurSMEiiteff{\hatcurplanetnumxxxxxC}}}
\newcommand{\hatcurSMEzfehxxxxxC}{\ifthenelse{\equal{\hatcurSMEversionxxxxxC}{i}}{\hatcurSMEizfeh{\hatcurplanetnumxxxxxC}}{\hatcurSMEiizfeh{\hatcurplanetnumxxxxxC}}}
\newcommand{\hatcurSMEzfehshortxxxxxC}{\ifthenelse{\equal{\hatcurSMEversionxxxxxC}{i}}{\hatcurSMEizfehshort{\hatcurplanetnumxxxxxC}}{\hatcurSMEiizfehshort{\hatcurplanetnumxxxxxC}}}
\newcommand{\hatcurSMEloggxxxxxC}{\ifthenelse{\equal{\hatcurSMEversionxxxxxC}{i}}{\hatcurSMEilogg{\hatcurplanetnumxxxxxC}}{\hatcurSMEiilogg{\hatcurplanetnumxxxxxC}}}
\newcommand{\hatcurSMEvsinxxxxxC}{\ifthenelse{\equal{\hatcurSMEversionxxxxxC}{i}}{\hatcurSMEivsin{\hatcurplanetnumxxxxxC}}{\hatcurSMEiivsin{\hatcurplanetnumxxxxxC}}}
\newcommand{\hatcurSMEvmacxxxxxC}{\ifthenelse{\equal{\hatcurSMEversionxxxxxC}{i}}{\hatcurSMEivmac{\hatcurplanetnumxxxxxC}}{\hatcurSMEiivmac{\hatcurplanetnumxxxxxC}}}
\newcommand{\hatcurSMEvmicxxxxxC}{\ifthenelse{\equal{\hatcurSMEversionxxxxxC}{i}}{\hatcurSMEivmic{\hatcurplanetnumxxxxxC}}{\hatcurSMEiivmic{\hatcurplanetnumxxxxxC}}}

\newcommand{\hatcurxxxxxD}{HATS-57}
\newcommand{\hatcurbxxxxxD}{HATS-57b}
\newcommand{\hatcurcxxxxxD}{HATS-57c}

\newcommand{\hatcurplanetnumxxxxxD}{57}

\newcommand{\hatcurCCtwomassshortxxxxxD}{04034760-1903242}

\newcommand{\hatcurRVgammaabsxxxxxD}{\hatcurRVgamma{\hatcurplanetnumxxxxxD}}                           % Absolute Gamma velocity

\newcommand{\hatcurRVgammarelxxxxxD}{\hatcurRVgamma{\hatcurplanetnumxxxxxD}}                           % Relative Gamma velocity. Typically that of the Keck RVs.

\newcommand{\hatcurCCtassvixxxxxD}{\ensuremath{NULL\pm NULL}}                  % TASS V-I

\newcommand{\hatcurSMEversionxxxxxD}{ii}                                       % Final SME version:i or ii?

\newcommand{\hatcurisoshortxxxxxD}{YY}
\newcommand{\hatcurisofullxxxxxD}{Yonsei-Yale (YY)}
\newcommand{\hatcurisocitexxxxxD}{yi:2001}
\newcommand{\hatcurlumindxxxxxD}{\rhostar}
\newcommand{\hatcurjhkfilsetxxxxxD}{ESO}
\newcommand{\hatcurSMEteffxxxxxD}{\ifthenelse{\equal{\hatcurSMEversionxxxxxD}{i}}{\hatcurSMEiteff{\hatcurplanetnumxxxxxD}}{\hatcurSMEiiteff{\hatcurplanetnumxxxxxD}}}
\newcommand{\hatcurSMEzfehxxxxxD}{\ifthenelse{\equal{\hatcurSMEversionxxxxxD}{i}}{\hatcurSMEizfeh{\hatcurplanetnumxxxxxD}}{\hatcurSMEiizfeh{\hatcurplanetnumxxxxxD}}}
\newcommand{\hatcurSMEzfehshortxxxxxD}{\ifthenelse{\equal{\hatcurSMEversionxxxxxD}{i}}{\hatcurSMEizfehshort{\hatcurplanetnumxxxxxD}}{\hatcurSMEiizfehshort{\hatcurplanetnumxxxxxD}}}
\newcommand{\hatcurSMEloggxxxxxD}{\ifthenelse{\equal{\hatcurSMEversionxxxxxD}{i}}{\hatcurSMEilogg{\hatcurplanetnumxxxxxD}}{\hatcurSMEiilogg{\hatcurplanetnumxxxxxD}}}
\newcommand{\hatcurSMEvsinxxxxxD}{\ifthenelse{\equal{\hatcurSMEversionxxxxxD}{i}}{\hatcurSMEivsin{\hatcurplanetnumxxxxxD}}{\hatcurSMEiivsin{\hatcurplanetnumxxxxxD}}}
\newcommand{\hatcurSMEvmacxxxxxD}{\ifthenelse{\equal{\hatcurSMEversionxxxxxD}{i}}{\hatcurSMEivmac{\hatcurplanetnumxxxxxD}}{\hatcurSMEiivmac{\hatcurplanetnumxxxxxD}}}
\newcommand{\hatcurSMEvmicxxxxxD}{\ifthenelse{\equal{\hatcurSMEversionxxxxxD}{i}}{\hatcurSMEivmic{\hatcurplanetnumxxxxxD}}{\hatcurSMEiivmic{\hatcurplanetnumxxxxxD}}}

\newcommand{\hatcurxxxxxE}{HATS-58A}
\newcommand{\hatcurbxxxxxE}{HATS-58Ab}
\newcommand{\hatcurcxxxxxE}{HATS-58c}

\newcommand{\hatcurplanetnumxxxxxE}{58}

\newcommand{\hatcurCCtwomassshortxxxxxE}{12270898-4858423}

\newcommand{\hatcurRVgammaabsxxxxxE}{\hatcurRVgammaA{\hatcurplanetnumxxxxxE}}                           % Absolute Gamma velocity

\newcommand{\hatcurRVgammarelxxxxxE}{\hatcurRVgammaA{\hatcurplanetnumxxxxxE}}                           % Relative Gamma velocity. Typically that of the Keck RVs.

\newcommand{\hatcurCCtassvixxxxxE}{\ensuremath{NULL\pm NULL}}                  % TASS V-I

\newcommand{\hatcurSMEversionxxxxxE}{ii}                                       % Final SME version:i or ii?

\newcommand{\hatcurisoshortxxxxxE}{YY}
\newcommand{\hatcurisofullxxxxxE}{Yonsei-Yale (YY)}
\newcommand{\hatcurisocitexxxxxE}{yi:2001}
\newcommand{\hatcurlumindxxxxxE}{\rhostar}
\newcommand{\hatcurjhkfilsetxxxxxE}{ESO}
\newcommand{\hatcurSMEteffxxxxxE}{\ifthenelse{\equal{\hatcurSMEversionxxxxxE}{i}}{\hatcurSMEiteff{\hatcurplanetnumxxxxxE}}{\hatcurSMEiiteff{\hatcurplanetnumxxxxxE}}}
\newcommand{\hatcurSMEzfehxxxxxE}{\ifthenelse{\equal{\hatcurSMEversionxxxxxE}{i}}{\hatcurSMEizfeh{\hatcurplanetnumxxxxxE}}{\hatcurSMEiizfeh{\hatcurplanetnumxxxxxE}}}
\newcommand{\hatcurSMEzfehshortxxxxxE}{\ifthenelse{\equal{\hatcurSMEversionxxxxxE}{i}}{\hatcurSMEizfehshort{\hatcurplanetnumxxxxxE}}{\hatcurSMEiizfehshort{\hatcurplanetnumxxxxxE}}}
\newcommand{\hatcurSMEloggxxxxxE}{\ifthenelse{\equal{\hatcurSMEversionxxxxxE}{i}}{\hatcurSMEilogg{\hatcurplanetnumxxxxxE}}{\hatcurSMEiilogg{\hatcurplanetnumxxxxxE}}}
\newcommand{\hatcurSMEvsinxxxxxE}{\ifthenelse{\equal{\hatcurSMEversionxxxxxE}{i}}{\hatcurSMEivsin{\hatcurplanetnumxxxxxE}}{\hatcurSMEiivsin{\hatcurplanetnumxxxxxE}}}
\newcommand{\hatcurSMEvmacxxxxxE}{\ifthenelse{\equal{\hatcurSMEversionxxxxxE}{i}}{\hatcurSMEivmac{\hatcurplanetnumxxxxxE}}{\hatcurSMEiivmac{\hatcurplanetnumxxxxxE}}}
\newcommand{\hatcurSMEvmicxxxxxE}{\ifthenelse{\equal{\hatcurSMEversionxxxxxE}{i}}{\hatcurSMEivmic{\hatcurplanetnumxxxxxE}}{\hatcurSMEiivmic{\hatcurplanetnumxxxxxE}}}

\newcommand{\hatcur}[1]{\ifnum#1=54 %
\hatcurxxxxxA
\else
\ifnum#1=55 %
\hatcurxxxxxB
\else
\ifnum#1=56 %
\hatcurxxxxxC
\else
\ifnum#1=57 %
\hatcurxxxxxD
\else
\ifnum#1=58 %
\hatcurxxxxxE
\else
??????\fi
\fi
\fi
\fi
\fi
}
\newcommand{\hatcurb}[1]{\ifnum#1=54 %
\hatcurbxxxxxA
\else
\ifnum#1=55 %
\hatcurbxxxxxB
\else
\ifnum#1=56 %
\hatcurbxxxxxC
\else
\ifnum#1=57 %
\hatcurbxxxxxD
\else
\ifnum#1=58 %
\hatcurbxxxxxE
\else
??????\fi
\fi
\fi
\fi
\fi
}
\newcommand{\hatcurc}[1]{\ifnum#1=54 %
\hatcurcxxxxxA
\else
\ifnum#1=55 %
\hatcurcxxxxxB
\else
\ifnum#1=56 %
\hatcurcxxxxxC
\else
\ifnum#1=57 %
\hatcurcxxxxxD
\else
\ifnum#1=58 %
\hatcurcxxxxxE
\else
??????\fi
\fi
\fi
\fi
\fi
}
\newcommand{\hatcurCCtassvi}[1]{\ifnum#1=54 %
\hatcurCCtassvixxxxxA
\else
\ifnum#1=55 %
\hatcurCCtassvixxxxxB
\else
\ifnum#1=56 %
\hatcurCCtassvixxxxxC
\else
\ifnum#1=57 %
\hatcurCCtassvixxxxxD
\else
\ifnum#1=58 %
\hatcurCCtassvixxxxxE
\else
??????\fi
\fi
\fi
\fi
\fi
}
\newcommand{\hatcurCCtwomassshort}[1]{\ifnum#1=54 %
\hatcurCCtwomassshortxxxxxA
\else
\ifnum#1=55 %
\hatcurCCtwomassshortxxxxxB
\else
\ifnum#1=56 %
\hatcurCCtwomassshortxxxxxC
\else
\ifnum#1=57 %
\hatcurCCtwomassshortxxxxxD
\else
\ifnum#1=58 %
\hatcurCCtwomassshortxxxxxE
\else
??????\fi
\fi
\fi
\fi
\fi
}
\newcommand{\hatcurisocite}[1]{\ifnum#1=54 %
\hatcurisocitexxxxxA
\else
\ifnum#1=55 %
\hatcurisocitexxxxxB
\else
\ifnum#1=56 %
\hatcurisocitexxxxxC
\else
\ifnum#1=57 %
\hatcurisocitexxxxxD
\else
\ifnum#1=58 %
\hatcurisocitexxxxxE
\else
??????\fi
\fi
\fi
\fi
\fi
}
\newcommand{\hatcurisofull}[1]{\ifnum#1=54 %
\hatcurisofullxxxxxA
\else
\ifnum#1=55 %
\hatcurisofullxxxxxB
\else
\ifnum#1=56 %
\hatcurisofullxxxxxC
\else
\ifnum#1=57 %
\hatcurisofullxxxxxD
\else
\ifnum#1=58 %
\hatcurisofullxxxxxE
\else
??????\fi
\fi
\fi
\fi
\fi
}
\newcommand{\hatcurisoshort}[1]{\ifnum#1=54 %
\hatcurisoshortxxxxxA
\else
\ifnum#1=55 %
\hatcurisoshortxxxxxB
\else
\ifnum#1=56 %
\hatcurisoshortxxxxxC
\else
\ifnum#1=57 %
\hatcurisoshortxxxxxD
\else
\ifnum#1=58 %
\hatcurisoshortxxxxxE
\else
??????\fi
\fi
\fi
\fi
\fi
}
\newcommand{\hatcurjhkfilset}[1]{\ifnum#1=54 %
\hatcurjhkfilsetxxxxxA
\else
\ifnum#1=55 %
\hatcurjhkfilsetxxxxxB
\else
\ifnum#1=56 %
\hatcurjhkfilsetxxxxxC
\else
\ifnum#1=57 %
\hatcurjhkfilsetxxxxxD
\else
\ifnum#1=58 %
\hatcurjhkfilsetxxxxxE
\else
??????\fi
\fi
\fi
\fi
\fi
}
\newcommand{\hatcurlumind}[1]{\ifnum#1=54 %
\hatcurlumindxxxxxA
\else
\ifnum#1=55 %
\hatcurlumindxxxxxB
\else
\ifnum#1=56 %
\hatcurlumindxxxxxC
\else
\ifnum#1=57 %
\hatcurlumindxxxxxD
\else
\ifnum#1=58 %
\hatcurlumindxxxxxE
\else
??????\fi
\fi
\fi
\fi
\fi
}
\newcommand{\hatcurplanetnum}[1]{\ifnum#1=54 %
\hatcurplanetnumxxxxxA
\else
\ifnum#1=55 %
\hatcurplanetnumxxxxxB
\else
\ifnum#1=56 %
\hatcurplanetnumxxxxxC
\else
\ifnum#1=57 %
\hatcurplanetnumxxxxxD
\else
\ifnum#1=58 %
\hatcurplanetnumxxxxxE
\else
??????\fi
\fi
\fi
\fi
\fi
}
\newcommand{\hatcurRVgammaabs}[1]{\ifnum#1=54 %
\hatcurRVgammaabsxxxxxA
\else
\ifnum#1=55 %
\hatcurRVgammaabsxxxxxB
\else
\ifnum#1=56 %
\hatcurRVgammaabsxxxxxC
\else
\ifnum#1=57 %
\hatcurRVgammaabsxxxxxD
\else
\ifnum#1=58 %
\hatcurRVgammaabsxxxxxE
\else
??????\fi
\fi
\fi
\fi
\fi
}
\newcommand{\hatcurRVgammarel}[1]{\ifnum#1=54 %
\hatcurRVgammarelxxxxxA
\else
\ifnum#1=55 %
\hatcurRVgammarelxxxxxB
\else
\ifnum#1=56 %
\hatcurRVgammarelxxxxxC
\else
\ifnum#1=57 %
\hatcurRVgammarelxxxxxD
\else
\ifnum#1=58 %
\hatcurRVgammarelxxxxxE
\else
??????\fi
\fi
\fi
\fi
\fi
}
\newcommand{\hatcurSMElogg}[1]{\ifnum#1=54 %
\hatcurSMEloggxxxxxA
\else
\ifnum#1=55 %
\hatcurSMEloggxxxxxB
\else
\ifnum#1=56 %
\hatcurSMEloggxxxxxC
\else
\ifnum#1=57 %
\hatcurSMEloggxxxxxD
\else
\ifnum#1=58 %
\hatcurSMEloggxxxxxE
\else
??????\fi
\fi
\fi
\fi
\fi
}
\newcommand{\hatcurSMEteff}[1]{\ifnum#1=54 %
\hatcurSMEteffxxxxxA
\else
\ifnum#1=55 %
\hatcurSMEteffxxxxxB
\else
\ifnum#1=56 %
\hatcurSMEteffxxxxxC
\else
\ifnum#1=57 %
\hatcurSMEteffxxxxxD
\else
\ifnum#1=58 %
\hatcurSMEteffxxxxxE
\else
??????\fi
\fi
\fi
\fi
\fi
}
\newcommand{\hatcurSMEversion}[1]{\ifnum#1=54 %
\hatcurSMEversionxxxxxA
\else
\ifnum#1=55 %
\hatcurSMEversionxxxxxB
\else
\ifnum#1=56 %
\hatcurSMEversionxxxxxC
\else
\ifnum#1=57 %
\hatcurSMEversionxxxxxD
\else
\ifnum#1=58 %
\hatcurSMEversionxxxxxE
\else
??????\fi
\fi
\fi
\fi
\fi
}
\newcommand{\hatcurSMEvmac}[1]{\ifnum#1=54 %
\hatcurSMEvmacxxxxxA
\else
\ifnum#1=55 %
\hatcurSMEvmacxxxxxB
\else
\ifnum#1=56 %
\hatcurSMEvmacxxxxxC
\else
\ifnum#1=57 %
\hatcurSMEvmacxxxxxD
\else
\ifnum#1=58 %
\hatcurSMEvmacxxxxxE
\else
??????\fi
\fi
\fi
\fi
\fi
}
\newcommand{\hatcurSMEvmic}[1]{\ifnum#1=54 %
\hatcurSMEvmicxxxxxA
\else
\ifnum#1=55 %
\hatcurSMEvmicxxxxxB
\else
\ifnum#1=56 %
\hatcurSMEvmicxxxxxC
\else
\ifnum#1=57 %
\hatcurSMEvmicxxxxxD
\else
\ifnum#1=58 %
\hatcurSMEvmicxxxxxE
\else
??????\fi
\fi
\fi
\fi
\fi
}
\newcommand{\hatcurSMEvsin}[1]{\ifnum#1=54 %
\hatcurSMEvsinxxxxxA
\else
\ifnum#1=55 %
\hatcurSMEvsinxxxxxB
\else
\ifnum#1=56 %
\hatcurSMEvsinxxxxxC
\else
\ifnum#1=57 %
\hatcurSMEvsinxxxxxD
\else
\ifnum#1=58 %
\hatcurSMEvsinxxxxxE
\else
??????\fi
\fi
\fi
\fi
\fi
}
\newcommand{\hatcurSMEzfeh}[1]{\ifnum#1=54 %
\hatcurSMEzfehxxxxxA
\else
\ifnum#1=55 %
\hatcurSMEzfehxxxxxB
\else
\ifnum#1=56 %
\hatcurSMEzfehxxxxxC
\else
\ifnum#1=57 %
\hatcurSMEzfehxxxxxD
\else
\ifnum#1=58 %
\hatcurSMEzfehxxxxxE
\else
??????\fi
\fi
\fi
\fi
\fi
}
\newcommand{\hatcurSMEzfehshort}[1]{\ifnum#1=54 %
\hatcurSMEzfehshortxxxxxA
\else
\ifnum#1=55 %
\hatcurSMEzfehshortxxxxxB
\else
\ifnum#1=56 %
\hatcurSMEzfehshortxxxxxC
\else
\ifnum#1=57 %
\hatcurSMEzfehshortxxxxxD
\else
\ifnum#1=58 %
\hatcurSMEzfehshortxxxxxE
\else
??????\fi
\fi
\fi
\fi
\fi
}

\newcounter{planetcounter}

\newboolean{emulateapj}
\setboolean{emulateapj}{true}

\newboolean{rvtablelong}
\setboolean{rvtablelong}{true}

\newboolean{astroph}
\setboolean{astroph}{true}
\shortauthors{Espinoza et al.}
\shorttitle{\hatcur{54}\lowercase{b}--\hatcur{58}\lowercase{b}}
\ifthenelse{\boolean{emulateapj}}{
    
}{
    
}

\begin{document}

%% Titlepage
\title{%%
\hatcur{54}\lowercase{b}--\hatcur{58}\lowercase{b}: five new transiting hot Jupiters including one with a possible temperate companion \footnote{
 The HATSouth network is operated by a collaboration consisting of
Princeton University (PU), the Max Planck Institute f\"ur Astronomie
(MPIA), the Australian National University (ANU), and the Pontificia
Universidad Cat\'olica de Chile (PUC).  The station at Las Campanas
Observatory (LCO) of the Carnegie Institute is operated by PU in
conjunction with PUC, the station at the High Energy Spectroscopic
Survey (H.E.S.S.) site is operated in conjunction with MPIA, and the
station at Siding Spring Observatory (SSO) is operated jointly with
ANU.
 Based in
 part on observations made with the MPG~2.2\,m Telescope at the ESO
 Observatory in La Silla.
}
}

%% Authors
\author[0000-0001-9513-1449]{N. Espinoza}
\altaffiliation{Bernoulli Fellow}
\altaffiliation{IAU-Gruber Fellow}
\affiliation{Max-Planck-Institut f\"ur Astronomie, K\"onigstuhl 17, 69117 Heidelberg, Germany.}

\author[0000-0001-8732-6166]{J. D. Hartman}
\affiliation{Department of Astrophysical Sciences, Princeton University, NJ 08544, USA.}

\author[0000-0001-7204-6727]{G. \'A. Bakos}
\altaffiliation{Packard Fellow}
\altaffiliation{MTA Distinguished Guest Fellow, Konkoly Observatory, Hungary}
\affiliation{Department of Astrophysical Sciences, Princeton University, NJ 08544, USA.}

\author{T. Henning}
\affiliation{Max-Planck-Institut f\"ur Astronomie, K\"onigstuhl 17, 69117 Heidelberg, Germany.}

\author[0000-0001-6023-1335]{D. Bayliss}
\affiliation{Dept. of Physics, University of Warwick, Gibbet Hill Road, Coventry CV4 7AL, UK.}

\author[0000-0002-9832-9271]{J.~Bento}
\affiliation{Research School of Astronomy and Astrophysics, Australian National University, Canberra, ACT 2611, Australia.}

\author[0000-0002-0628-0088]{W.~Bhatti}
\affiliation{Department of Astrophysical Sciences, Princeton University, NJ 08544, USA.}

\author[0000-0002-9158-7315]{R.~Brahm}
\affiliation{Millennium Institute of Astrophysics, Santiago, Chile}
\affiliation{Instituto de Astrof\'isica, Facultad de F\'isica, Pontificia Universidad Cat\'olica de Chile, Av. Vicu\~na Mackenna 4860, 7820436 Macul, Santiago, Chile.}

\author{Z.~Csubry}
\affiliation{Department of Astrophysical Sciences, Princeton University, NJ 08544, USA.}

\author[0000-0001-7070-3842]{V.~Suc}
\affiliation{Instituto de Astrof\'isica, Facultad de F\'isica, Pontificia Universidad Cat\'olica de Chile, Av. Vicu\~na Mackenna 4860, 7820436 Macul, Santiago, Chile.}

\author[0000-0002-5389-3944]{A.~Jord\'an}
\affiliation{Millennium Institute of Astrophysics, Santiago, Chile}
\affiliation{Instituto de Astrof\'isica, Facultad de F\'isica, Pontificia Universidad Cat\'olica de Chile, Av. Vicu\~na Mackenna 4860, 7820436 Macul, Santiago, Chile.}

\author[0000-0002-9428-8732]{L.~Mancini}
\affiliation{Department of Physics, University of Rome Tor Vergata, Via della Ricerca Scientifica 1, I-00133 -- Roma, Italy}
\affiliation{Max-Planck-Institut f\"ur Astronomie, K\"onigstuhl 17, 69117 Heidelberg, Germany.}
\affiliation{INAF -- Astrophysical Observatory of Turin, Via Osservatorio 20, I-10025 -- Pino Torinese, Italy }

\author[0000-0001-5603-6895]{T. G. Tan}
\affiliation{Perth Exoplanet Survey Telescope, Perth, Australia} 

\author[0000-0003-4464-1371]{K.~Penev}
\affiliation{Physics Department, University of Texas at Dallas, 800 W Campbell Rd. MS WT15, Richardson, TX 75080, USA}

\author[0000-0003-2935-7196]{M.~Rabus}
\affiliation{Instituto de Astrof\'isica, Facultad de F\'isica, Pontificia Universidad Cat\'olica de Chile, Av. Vicu\~na Mackenna 4860, 7820436 Macul, Santiago, Chile.}
\affiliation{Max-Planck-Institut f\"ur Astronomie, K\"onigstuhl 17, 69117 Heidelberg, Germany.}

\author[0000-0001-8128-3126]{P.~Sarkis}
\affiliation{Max-Planck-Institut f\"ur Astronomie, K\"onigstuhl 17, 69117 Heidelberg, Germany.}

\author[0000-0002-0455-9384]{M.~de Val-Borro}
\affiliation{Astrochemistry Laboratory, Goddard Space Flight Center, NASA, 8800 Greenbelt Rd, Greenbelt, MD 20771, USA}

\author[0000-0002-3663-3251]{S.~Durkan}
\affiliation{Astrophysics Research Centre, Queens University, Belfast, Belfast, Northern Ireland, UK}

\author{J. L\'az\'ar}
\affiliation{Hungarian Astronomical Association, 1451 Budapest, Hungary}

\author{I. Papp}
\affiliation{Hungarian Astronomical Association, 1451 Budapest, Hungary}

\author{P. S\'ari}
\affiliation{Hungarian Astronomical Association, 1451 Budapest, Hungary}

%% EOF authors

% #####################################################################
%% abstract
\begin{abstract}

\setcounter{footnote}{10}
We report the discovery by the HATSouth project of 5 new transiting hot Jupiters (HATS-54b through HATS-58Ab). \hatcurb{54}, \hatcurb{55} and HATS-58Ab are 
prototypical short period ($P = 2.5-4.2$ days, $R_p\sim1.1-1.2$ \rjup) hot-Jupiters that span effective temperatures from 1350 K to 1750 K, putting them 
in the proposed region of maximum radius inflation efficiency. The HATS-58 system is composed of two stars, HATS-58A and HATS-58B, 
which are detected thanks to Gaia DR2 data and which we account for in the joint modelling of the available data --- with this, we are led to conclude that the 
hot jupiter orbits the brighter HATS-58A star. HATS-57b is a short-period (2.35-day) massive (3.15 \mjup) 1.14 \rjup, dense 
(\hatcurPPrho{57}\,\gcmc) hot-Jupiter, orbiting a very active star (2\% peak-to-peak flux variability). Finally, HATS-56b is a short 
period (4.32-day) highly inflated hot-Jupiter (1.7 \rjup, 0.6 \mjup), which is an excellent target for future atmospheric follow-up, 
especially considering 
the relatively bright nature ($V=11.6$) of its \hatcurISOspec{56} dwarf host star. This latter exoplanet has another very interesting feature: 
the radial velocities show a significant quadratic trend. If we interpret this quadratic trend as arising from the pull of an additional planet in the system, 
we obtain a period of $P_c = 815^{+253}_{-143}$  days for the possible planet HATS-56c, and a minimum mass of $M_c\sin i_c = 5.11 \pm 0.94$ \mjup. 
The candidate planet HATS-56c would have a zero-albedo equilibrium temperature of $T_\textnormal{eq}=332\pm 50$ K, and thus would be orbiting close to the 
habitable zone of HATS-56. Further radial-velocity follow-up, especially over the next two years, is needed to confirm the nature of HATS-56c.

\setcounter{footnote}{0}
\end{abstract}

% #####################################################################
%% keywords
\keywords{
    planetary systems ---
    stars: individual (
\setcounter{planetcounter}{1}
\hatcur{54},
\hatcurCCgsc{54}\loopcommanoperiod
\setcounter{planetcounter}{2}
\hatcur{55},
\hatcurCCgsc{55}\loopcommanoperiod
\setcounter{planetcounter}{3}
\hatcur{56},
\hatcurCCgsc{56}\loopcommanoperiod
\setcounter{planetcounter}{3}
\hatcur{57},
\hatcurCCgsc{57}\loopcommanoperiod
\setcounter{planetcounter}{4}
\hatcur{58},
\hatcurCCgsc{58}\loopcommanoperiod
\setcounter{planetcounter}{5}
) 
    techniques: spectroscopic, photometric
}

%% EOF keywords
%% EOF titlepage

% #####################################################################
%% Introduction
\section{Introduction}
\label{sec:introduction}
With almost 3,000 confirmed exoplanets\footnote{\url{http://www.exoplanets.org/}}, the field of exoplanet discovery and characterization has seen an exponential increase 
in the number of discovered far-away worlds. While space-based dedicated surveys such as \textit{Kepler} \citep{Kepler} have excelled at the detection of small ($R_p<4R_\Earth$) exoplanets, ground-based dedicated surveys such as HATNet \citep{HATN}, HATSouth \citep{HATS}, WASP \citep{WASP}, KELT \citep{KELT} and the recently started 
MASCARA \citep{MASCARA} and NGTS \citep{NGTS} surveys have been pioneering the search of giant exoplanets. This has produced a sample of exoplanets amenable for characterization 
both in terms of radial-velocity follow-up --- which allows us to constrain their densities --- or in terms of atmospheric follow-up --- which allows 
us to have a glimpse at what their atmospheres look like. It has also generated a large sample of well-characterized exoplanets from which we have been able to extract useful information to put our planet formation and evolution theories to test. 

Despite the relatively large number of known exoplanets, less than $10\%$ ($\sim 300$) are well-characterized (i.e., have a mass and radius constrained to better than 20\% precision). 
Discovered mostly from ground-based transit surveys, these --- mostly short-period ($P \lesssim 10$ days), hot --- transiting giant exoplanets have provided unique information that has aided in the understanding of 
the formation, evolution and composition of those far-away worlds. For example, structure modelling coupled with the mass, radius and ages of the warmer (< 1000 K) of these systems has allowed us to understand that they are heavily enriched in 
metals \citep{Thorngren:2016}, which in turn has explicit predictions for their compositions \citep{Espinoza:2017}. This understanding, in turn, has allowed us to calibrate how mass and heavy elements are related, which in turn has been used to elucidate the nature of the observed radius 
inflation of highly irradiated giant exoplanets, bringing us closer to an understanding of the mechanism(s) producing this radius anomaly over a wide range of stellar irradiation, masses 
and sizes \citep{Thorngren:2018, Sestovic:2018}. In terms of formation, short-period giant exoplanets are fundamental probes of the mechanisms that shape their orbits to their 
present-day forms. Although in-situ formation has still not been ruled out \citep{Batygin:2016}, the orbital migration scenario --- either by direct disk migration and/or by interaction with other bodies 
in the system \citep[see, e.g., ][]{Lin:1996, Li:2014, Petrovich:2015} --- is by far the most popular theory to explain the observed short-period orbits of these hot giant exoplanets. All of them have 
discerning features that can be studied with transiting exoplanets, for which one is able to unveil their 3-dimensional orbital shapes if sufficient follow-up is performed. In addition, some 
transiting systems actually reside in systems with other planetary or sub-stellar companions \citep[see, e.g.,][]{Becker:2015, Rey:2018, Sarkis:2018, Yee:2018}, which 
provides new laboratories to study how multiplanetary systems form and evolve.

In this work we present the discovery of five new transiting hot giant exoplanets, one of which is in a possible multiplanetary system with a sub-stellar companion on a possible temperate, eccentric orbit. 
The paper is divided as follows. Section \ref{sec:obs} details our observations, including the HATSouth photometric detection and both photometric and radial-velocity follow-up. Section \ref{sec:analysis} details the 
analysis of the data presented, while in Section 4 we discuss our results. Finally, in Section \ref{sec:conclusions} we present our conclusions.

%% EOF introduction
% #####################################################################
\section{Observations}
\label{sec:obs}
%++++++++++++++++++++++++++++++++++++++++++++++++++++++++++++++++++++++

% =====================================================================
%% Photometric detection
\subsection{Photometric detection}
\label{sec:detection}
%++++++++++++++++++++++++++++++++++++++++++++++++++++++++++++++++++++++
%++++++++++++++++++++++++++++++++++++++++++++++++++++++++++++++++++++++

The photometric detection of the exoplanets presented in this work was made with the HATSouth units based in 
Las Campanas Observatory (LCO; HS-1 and HS-2), at the HESS site in Namibia (HS-3 and HS-4) and at the site 
in Siding Spring Observatory (SSO; HS-5 and HS-6), whose 
operations are described in detail in \cite{bakos:2013:hatsouth}. The details of these observations for each 
of the presented exoplanets can be found in \reftabl{photobs}. 

As with previous results from our group, the data was reduced and analyzed with the procedures detailed in 
\cite{bakos:2013:hatsouth} and \citet{penev:2013:hats1}; briefly, the lightcurves were detrended using the 
trend-filtering algorithm \citep{kovacs:2005:TFA} as described in \cite{bakos:2013:hatsouth}, and then a search 
for periodic, transit-like signals using the Box-fitting Least-Squares algorithm \citep[BLS; see][]{kovacs:2002:BLS} was 
performed. Peaks in the BLS periodogram were found for \hatcur{54}, \hatcur{55}, \hatcur{56}, \hatcur{57} and 
HATS-58 with periods of 2.54, 4.20, 4.32, 2.35 and 4.21 days, respectively, which prompted us to obtain further 
photometric and spectroscopic follow-up in order to confirm the planetary nature of the signals, which we detail 
in the following sections. The phase-folded lightcurves for each planet are presented in Figures~\ref{fig:hatsouth} and~\ref{fig:hatsouthcontd}. 
The data are presented in Table \ref{tab:photobs}.

The lightcurves were also further analyzed in the search for additional periodic signals, either transit-like (with BLS, in the search 
for additional transiting companions in the system) or sinusoidal \citep[with the Generalized Lomb-Scargle --- GLS --- 
periodogram described by][in the search for signals of non-transiting companions and/or intrinsic variability of the 
star]{zechmeister:2009}. For this, the portions of the detected transits were masked out, and GLS and BLS periodograms were produced 
and inspected. No additional signals were found using GLS and BLS in our lightcurves for HATS-54, HATS-55, HATS-56 and HATS-58. However, 
the lightcurve of HATS-57 shows two clear peaks in the GLS periodogram at 6 and 12.8-days. A visual inspection to the lightcurve shows that the 
star is clearly undergoing quasi-periodic modulations with signatures typical to that of starspots going in and out of view, with a peak-to-peak 
variation of $\sim 2$\%. We analyze this signature in detail in Section \ref{sec:stelparam}.

%
%
%% ----------------
\ifthenelse{\boolean{emulateapj}}{
    \begin{figure*}[!ht]
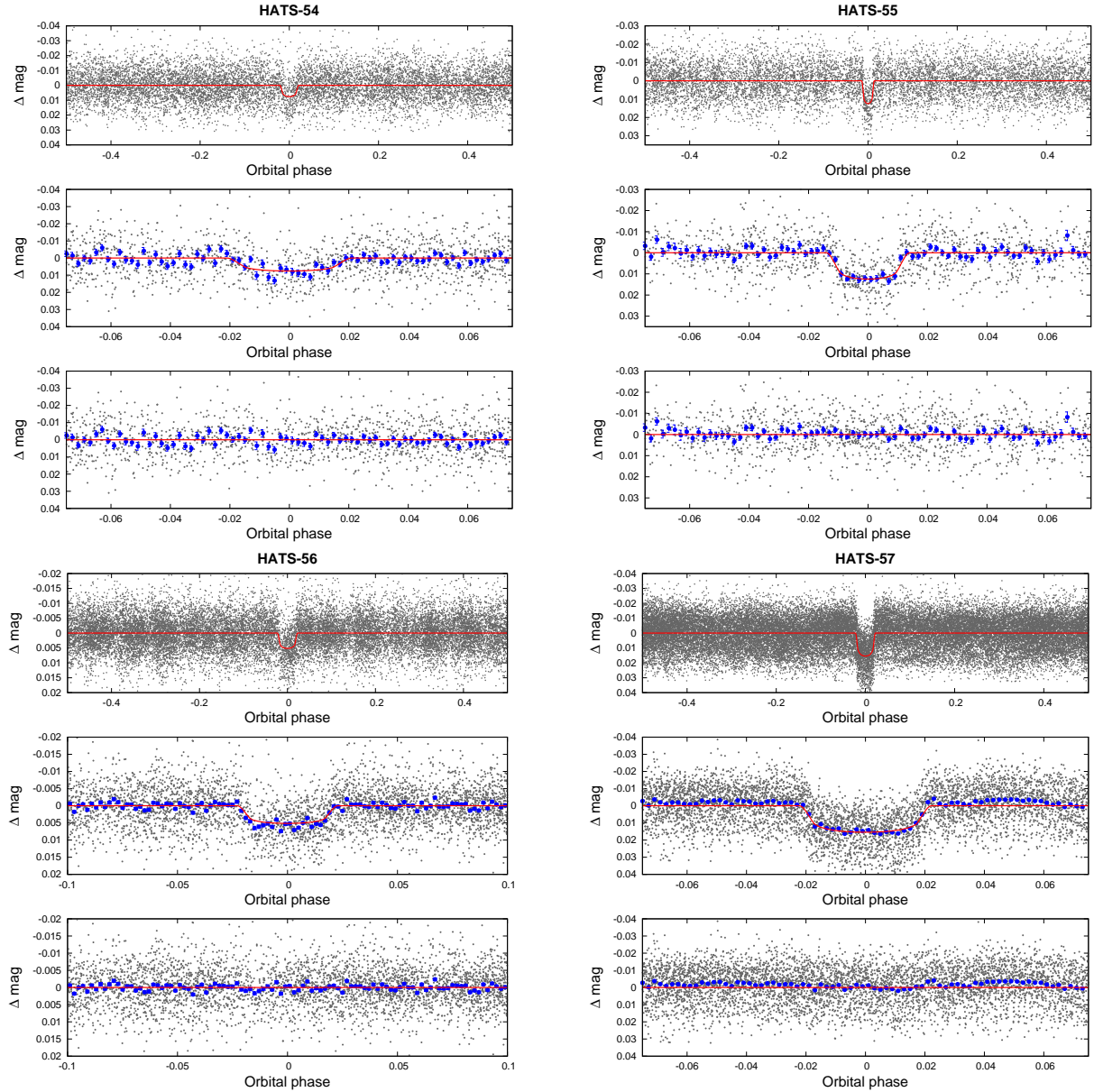

}{
    \begin{figure}[!ht]
}
\plottwo{\hatcurhtr{54}-hs.eps}{\hatcurhtr{55}-hs.eps}
\plottwo{\hatcurhtr{56}-hs.eps}{\hatcurhtr{57}-hs.eps}
\caption{
    Phase-folded unbinned HATSouth light curves for \hatcur{54} (upper left), \hatcur{55} (upper right), \hatcur{56} (bottom left), \hatcur{57} (bottom right). In each case we show three panels. The
    top panel shows the full light curve, the middle panel shows
    the light curve zoomed-in on the transit, and the bottom panel shows the residuals from the best-fit model zoomed-in on the transit. The solid lines show the
    model fits to the light curves. The dark filled circles in the
    middle and bottom panels show the light curves binned in phase with a bin
    size of 0.002.
\label{fig:hatsouth}}
\ifthenelse{\boolean{emulateapj}}{
    \end{figure*}
}{
    \end{figure}
}
%% ----------------

%
%
%% ----------------
\begin{figure}[!ht]
\plotone{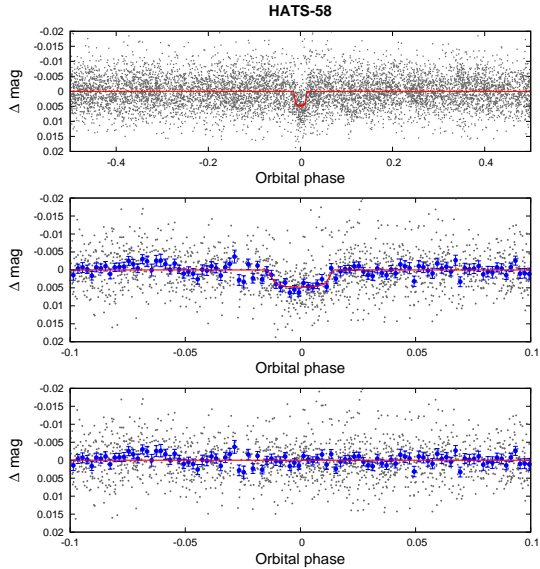}
\caption{
    Same as Figure~\ref{fig:hatsouth}, here we show the phase-folded unbinned HATSouth light curves for HATS-58.
\label{fig:hatsouthcontd}}
\end{figure}
%% ----------------

\startlongtable
%% --------------------------------------------------------------------
%% Table summarizing photometric observations
%%
\ifthenelse{\boolean{emulateapj}}{
    \begin{deluxetable*}{llrrrr}
}{
    \begin{deluxetable}{llrrrr}
}
\tablewidth{0pc}
\tabletypesize{\scriptsize}
\tablecaption{
    Summary of photometric observations
    \label{tab:photobs}
}
\tablehead{
    \multicolumn{1}{c}{Instrument/Field\tablenotemark{a}} &
    \multicolumn{1}{c}{Date(s)} &
    \multicolumn{1}{c}{\# Images} &
    \multicolumn{1}{c}{Cadence\tablenotemark{b}} &
    \multicolumn{1}{c}{Filter} &
    \multicolumn{1}{c}{Precision\tablenotemark{c}} \\
    \multicolumn{1}{c}{} &
    \multicolumn{1}{c}{} &
    \multicolumn{1}{c}{} &
    \multicolumn{1}{c}{(sec)} &
    \multicolumn{1}{c}{} &
    \multicolumn{1}{c}{(mmag)}
}
\startdata
\sidehead{\textbf{\hatcur{54}}}
~~~~HS-2/G700 & 2011 Apr--2012 Jul & 4521 & 292 & $r$ & 9.8 \\
~~~~HS-4/G700 & 2011 Jul--2012 Jul & 3799 & 301 & $r$ & 10.4 \\
~~~~HS-6/G700 & 2012 Jan--2012 Jul & 1425 & 300 & $r$ & 10.7 \\
~~~~Swope~1\,m & 2016 Feb 09 & 89 & 79 & $i$ & 2.2 \\
~~~~PEST~0.3\,m & 2016 Feb 25 & 169 & 132 & $R_{C}$ & 6.3 \\
~~~~CHAT~0.7\,m & 2017 Feb 12 & 50 & 222 & $i$ & 2.1 \\
~~~~LCO~1\,m/SAAO/DomeB & 2017 May 10 & 73 & 221 & $i$ & 1.7 \\
~~~~Swope~1\,m & 2017 May 30 & 137 & 160 & $g$ & 1.9 \\
~~~~LCO~1\,m/SAAO/DomeC & 2017 Jul 05 & 78 & 221 & $i$ & 2.2 \\
~~~~LCO~1\,m/SSO/DomeB & 2017 Jul 13 & 68 & 224 & $i$ & 3.1 \\
\sidehead{\textbf{\hatcur{55}}}
~~~~HS-2/G602 & 2011 Aug--2012 Feb & 4192 & 295 & $r$ & 8.8 \\
~~~~HS-4/G602 & 2011 Aug--2012 Feb & 3047 & 296 & $r$ & 9.3 \\
~~~~HS-6/G602 & 2011 Oct--2012 Feb & 1248 & 303 & $r$ & 8.8 \\
~~~~PEST~0.3\,m & 2015 Feb 14 & 171 & 132 & $R_{C}$ & 5.1 \\
~~~~PETS~0.3\,m & 2015 Mar 03 & 144 & 132 & $R_{C}$ & 4.8 \\
~~~~Swope~1\,m & 2015 Apr 01 & 250 & 59 & $i$ & 3.1 \\
~~~~LCO~1\,m/CTIO/DomeA & 2017 Apr 10 & 69 & 220 & $i$ & 1.8 \\
~~~~LCO~1\,m/CTIO/DomeC & 2017 Apr 10 & 69 & 220 & $i$ & 2.5 \\
\sidehead{\textbf{\hatcur{56}}}
~~~~HS-4/G698 & 2015 May--2015 Jul & 5 & 499 & $r$ & 4.7 \\
~~~~HS-6/G698 & 2015 Dec--2016 Jun & 4846 & 343 & $r$ & 6.6 \\
~~~~HS-2/G698 & 2015 Mar--2016 May & 2487 & 352 & $r$ & 4.6 \\
~~~~HS-4/G698 & 2015 Mar--2016 Jun & 6851 & 324 & $r$ & 5.6 \\
~~~~HS-6/G698 & 2015 Mar--2016 Jun & 5638 & 343 & $r$ & 6.1 \\
~~~~PEST~0.3\,m & 2017 Mar 05 & 182 & 134 & $R_{C}$ & 2.0 \\
~~~~LCO~1\,m/CTIO & 2017 Mar 22 & 139 & 130 & $i$ & 1.1 \\
~~~~LCO~1\,m/SSO & 2017 Mar 27 & 47 & 130 & $i$ & 0.8 \\
\sidehead{\textbf{\hatcur{57}}}
~~~~HS-1/G548 & 2014 Sep--2015 Feb & 5719 & 287 & $r$ & 11.5 \\
~~~~HS-2/G548 & 2014 Jun--2015 Apr & 7689 & 348 & $r$ & 10.4 \\
~~~~HS-3/G548 & 2014 Sep--2015 Mar & 5214 & 353 & $r$ & 10.5 \\
~~~~HS-4/G548 & 2014 Jun--2015 Mar & 5430 & 352 & $r$ & 10.6 \\
~~~~HS-5/G548 & 2014 Sep--2015 Mar & 5041 & 359 & $r$ & 10.6 \\
~~~~HS-6/G548 & 2014 Jul--2015 Mar & 5989 & 351 & $r$ & 10.7 \\
~~~~CHAT~0.7\,m & 2017 Aug 28 & 83 & 143 & $i$ & 1.3 \\
~~~~CHAT~0.7\,m & 2017 Oct 21 & 90 & 146 & $i$ & 1.6 \\
\sidehead{\textbf{HATS-58}}
~~~~HS-1/G699 & 2011 Apr--2012 Aug & 3645 & 290 & $r$ & 4.9 \\
~~~~HS-3/G699 & 2011 Jul--2012 Aug & 3150 & 291 & $r$ & 5.7 \\
~~~~HS-5/G699 & 2011 May--2012 Aug & 750 & 290 & $r$ & 4.7 \\
~~~~PEST~0.3\,m & 2017 Mar 09 & 220 & 132 & $R_{C}$ & 2.2 \\
~~~~PEST~0.3\,m & 2017 Apr 20 & 223 & 132 & $R_{C}$ & 2.2 \\
~~~~LCO~1\,m+SAAO/DomeB & 2017 May 15 & 40 & 130 & $i$ & 0.7 \\
~~~~LCO~1\,m+SSO/DomeB & 2017 Jul 05 & 106 & 134 & $i$ & 2.6 \\
\enddata
\tablenotetext{a}{
    For HATSouth data we list the HATSouth unit, CCD and field name
    from which the observations are taken. HS-1 and -2 are located at
    Las Campanas Observatory in Chile, HS-3 and -4 are located at the
    H.E.S.S. site in Namibia, and HS-5 and -6 are located at Siding
    Spring Observatory in Australia. Each unit has 4 ccds. Each field
    corresponds to one of 838 fixed pointings used to cover the full
    4$\pi$ celestial sphere. All data from a given HATSouth field and
    CCD number are reduced together, while detrending through External
    Parameter Decorrelation (EPD) is done independently for each
    unique unit+CCD+field combination.
}
\tablenotetext{b}{
    The median time between consecutive images rounded to the nearest
    second. Due to factors such as weather, the day--night cycle,
    guiding and focus corrections the cadence is only approximately
    uniform over short timescales.
}
\tablenotetext{c}{
    The RMS of the residuals from the best-fit model.
} \ifthenelse{\boolean{emulateapj}}{
    \end{deluxetable*}
}{
    \end{deluxetable}
}

\subsection{Spectroscopic Observations}
\label{sec:obsspec}

Spectroscopic follow-up was performed on our planet candidates in order to confirm their planetary nature. This spectroscopic follow-up, as 
in previous works, was divided in two types: (1) reconnaissance spectroscopy, usually performed with lower-resolution instruments and which 
serves in order to both get coarse stellar atmospheric parameters (to identify, e.g., if the target is a giant star by the derived value of its 
log-gravity) and identify if there is any large radial-velocity variation (indicative of an eclipsing binary and/or blend), and (2) high-precision 
spectroscopy, used to both obtain better stellar atmospheric parameters and to measure the radial-velocity signature that our candidate planets 
should imprint on the star. 

Reconnaissance spectroscopy was performed with the Wide Field Spectrograph \citep[WiFeS][]{dopita:2007}, located on the Australian National University (ANU) 
2.3m telescope and the CORALIE \citep{queloz:2001} spectrograph, mounted on the 1.2m Euler Telescope at La Silla Observatory (LSO). 
The observing strategy, reduction and data processing of the WiFeS spectra can be found 
in \cite{bayliss:2013:hats3}, whereas the CORALIE data were reduced using the CERES pipeline \citep{ceres:2017}. WiFeS spectra were obtained for 
HATS-54 (4 spectra), HATS-55 (4 spectra), HATS-57 (3 spectra) and HATS-58 (3 spectra), all of which passed our initial screenings in terms of having high 
surface gravities ($\log g \geq 4$) and no large radial-velocity variations ($\leq 1$ km s$^{-1}$). HATS-55 (4 spectra), HATS-56 (1 spectra) and HATS-58 (1 spectra) had 
CORALIE spectra taken, which also helped to rule out false positives with similar standards as for the WiFeS data. 

High-precision spectroscopy, on the other hand, was performed with both the FEROS \citep{kaufer:1998} and HARPS \citep{mayor:2003} spectrographs, which are located 
at the MPG 2.2m telescope and 3.6m ESO telescope, respectively, at LSO. Data obtained from both of those instruments was also reduced with the CERES pipeline. Details 
of all the spectroscopic observations are provided in Table \ref{tab:specobs}. The observed high-precision radial velocities are presented in Table 
\ref{tab:rvs}.

All of our targets showed radial-velocity variations at the periods of the observed transits consistent with being of planetary nature, with no indication of 
being correlated with other stellar parameters (e.g., bisector spans). HATS-56, however, showed an additional long-term trend radial-velocity signal, which shows 
no correlation with other parameters (e.g., bisector span). The phase-folded radial-velocities are presented in Figures \ref{fig:rvbis} and \ref{fig:rvbiscontd}. 
We analyze these in detail in Section \ref{sec:globmod}.

%% --------------------------------------------------------------------
%% Table summarizing spectroscopy observations
%%
\ifthenelse{\boolean{emulateapj}}{
    \begin{deluxetable*}{llrrrrr}
}{
    \begin{deluxetable}{llrrrrrrrr}
}
\tablewidth{0pc}
\tabletypesize{\scriptsize}
\tablecaption{
    Summary of spectroscopy observations.
    \label{tab:specobs}
}
\tablehead{
    \multicolumn{1}{c}{Instrument}          &
    \multicolumn{1}{c}{UT Date(s)}             &
    \multicolumn{1}{c}{\# Spec.}   &
    \multicolumn{1}{c}{Res.}          &
    \multicolumn{1}{c}{S/N Range\tablenotemark{a}}           &
    \multicolumn{1}{c}{$\gamma_{\rm RV}$\tablenotemark{b}} &
    \multicolumn{1}{c}{RV Precision\tablenotemark{c}} \\
    &
    &
    &
    \multicolumn{1}{c}{$\Delta \lambda$/$\lambda$/1000} &
    &
    \multicolumn{1}{c}{(\kms)}              &
    \multicolumn{1}{c}{(\ms)}
}
\startdata
\sidehead{\textbf{\hatcur{54}}}\\
ANU~2.3\,m/WiFeS & 2014 Jun 3 & 1 & 3 & 26 & $\cdots$ & $\cdots$ \\
ANU~2.3\,m/WiFeS & 2014 Jun 3--5 & 3 & 7 & 23--112 & 42.7 & 4000 \\
ESO~3.6\,m/HARPS & 2015 Apr--2017 May & 3 & 115 & 5--12 & 46.060 & 53 \\
MPG~2.2\,m/FEROS & 2015 Jun--2017 Aug & 31 & 48 & 17--44 & 46.127 & 64 \\
\sidehead{\textbf{\hatcur{55}}}\\
ANU~2.3\,m/WiFeS & 2014 Dec 13 & 1 & 3 & 60 & $\cdots$ & $\cdots$ \\
ANU~2.3\,m/WiFeS & 2014 Dec 29--31 & 3 & 7 & 7--103 & -2.3 & 4000 \\
ESO~3.6\,m/HARPS & 2015 Feb--Nov & 8 & 115 & 12--20 & -2.919 & 18 \\
Euler~1.2\,m/Coralie & 2015 Feb--Mar & 4\tablenotemark{d} & 60 & 11--14 & -2.935 & 240 \\
\sidehead{\textbf{\hatcur{56}}}\\
MPG~2.2\,m/FEROS & 2017 Jan--2018 Mar & 56 & 48 & 24--97 & 35.740 & 25 \\
Euler~1.2\,m/Coralie & 2017 Jan 25 & 1\tablenotemark{d} & 60 & 27 & 37.99 & $\cdots$ \\
ESO~3.6\,m/HARPS & 2017 Feb 20--22 & 3 & 115 & 21--36 & 35.730 & 10 \\
\sidehead{\textbf{\hatcur{57}}}\\
ANU~2.3\,m/WiFeS & 2017 Jul 11 & 1 & 3 & 30 & $\cdots$ & $\cdots$ \\
ANU~2.3\,m/WiFeS & 2017 Jul 11--12 & 2 & 7 & 36--59 & -0.5 & 4000 \\
MPG~2.2\,m/FEROS & 2017 Jul--Oct & 15 & 48 & 21--65 & 0.5455 & 28 \\
\sidehead{\textbf{HATS-58}}\\
MPG~2.2\,m/FEROS & 2016 Dec--2017 Mar & 11 & 48 & 47--91 & 19.298 & 58 \\
ANU~2.3\,m/WiFeS & 2016 Dec 20 & 1 & 3 & 54 & $\cdots$ & $\cdots$ \\
ANU~2.3\,m/WiFeS & 2016 Dec 20--22 & 2 & 7 & 52 & 18.7 & 4000 \\
Euler~1.2\,m/Coralie & 2017 Jan 26 & 1\tablenotemark{d} & 60 & 20 & 19.223 & $\cdots$ \\
ESO~3.6\,m/HARPS & 2017 Feb--Apr & 9 & 115 & 23--45 & 19.415 & 12 \\
\enddata 
\tablenotetext{a}{
    S/N per resolution element near 5180\,\AA.
}
\tablenotetext{b}{
    For high-precision RV observations included in the orbit determination this is the zero-point RV from the best-fit orbit. For other instruments it is the mean value. We do not provide this quantity for the lower resolution WiFeS observations which were only used to measure stellar atmospheric parameters.
}
\tablenotetext{c}{
    For high-precision RV observations included in the orbit
    determination this is the scatter in the RV residuals from the
    best-fit orbit (which may include astrophysical jitter), for other
    instruments this is either an estimate of the precision (not
    including jitter), or the measured standard deviation. We do not
    provide this quantity for low-resolution observations from the
    ANU~2.3\,m/WiFeS.
}
\tablenotetext{d}{
    We list here the total number of spectra collected for each instrument, including observations that were excluded from the analysis due to very low S/N or substantial sky contamination. For \hatcur{55} we did not include any of the Coralie observations in the analysis as they had too low RV precision to detect the orbital variation. For \hatcur{56} and HATS-58 we did not include the single Coralie observations in the analysis.
}
\ifthenelse{\boolean{emulateapj}}{
    \end{deluxetable*}
}{
    \end{deluxetable}
}
%% --------------------------------------------------------------------

%% --------------------------------------------------------------------
%% Note: there are two possible versions for this table: one with
%% BJD, RB, RVerr, BS, BSerr, and another one with the former columns
%% plus S and Serr. For the first form it is OK to use the deluxetable
%% environment. 
%% With the second form we need deluxetable*.
%%
%
%
\startlongtable
\tabletypesize{\scriptsize}
\ifthenelse{\boolean{emulateapj}}{
    \begin{deluxetable*}{lrrrrrl}
}{
    \begin{deluxetable}{lrrrrrl}
}
\tablewidth{0pc}
\tablecaption{
    Relative radial velocities and bisector spans for \hatcur{54}--HATS-58.
    \label{tab:rvs}
}
\tablehead{
    \colhead{BJD} &
    \colhead{RV\tablenotemark{a}} &
    \colhead{\ensuremath{\sigma_{\rm RV}}\tablenotemark{b}} &
    \colhead{BS} &
    \colhead{\ensuremath{\sigma_{\rm BS}}} &
    \colhead{Phase} &
    \colhead{Instrument}\\
    \colhead{\hbox{(2,450,000$+$)}} &
    \colhead{(\ms)} &
    \colhead{(\ms)} &
    \colhead{(\ms)} &
    \colhead{(\ms)} &
    \colhead{} &
    \colhead{}
}
\startdata
%%
%% If the table is too long, then we give only sample lines in the
%% submitted version, and the full table is presented in the electronic
%% version.  As regards the astroph version: in such cases we can
%% decide whether to include all data.
%%
\multicolumn{7}{c}{\bf HATS-54} \\
\hline\\
    \input{\hatcurhtr{54}_rvtable.tex}
\cutinhead{\bf HATS-55}
    \input{\hatcurhtr{55}_rvtable.tex}
\cutinhead{\bf HATS-56}
    \input{\hatcurhtr{56}_rvtable.tex}
\cutinhead{\bf HATS-57}
    \input{\hatcurhtr{57}_rvtable.tex}
\cutinhead{\bf HATS-58}
    \input{\hatcurhtr{58}_rvtable.tex}
\enddata
\tablenotetext{a}{
    The zero-point of these velocities is arbitrary. An overall offset
    $\gamma_{\rm rel}$ fitted independently to the velocities from
    each instrument has been subtracted.
}
\tablenotetext{b}{
    Internal errors excluding the component of astrophysical jitter
    considered in \refsecl{globmod}.
}
\ifthenelse{\boolean{rvtablelong}}{
    \tablecomments{
    }
}{
    \tablecomments{
    }
} 
\ifthenelse{\boolean{emulateapj}}{
    \end{deluxetable*}
}{
    \end{deluxetable}
}

%
%% --------------------------------------------------------------------
\ifthenelse{\boolean{emulateapj}}{
    \begin{figure*} [ht]
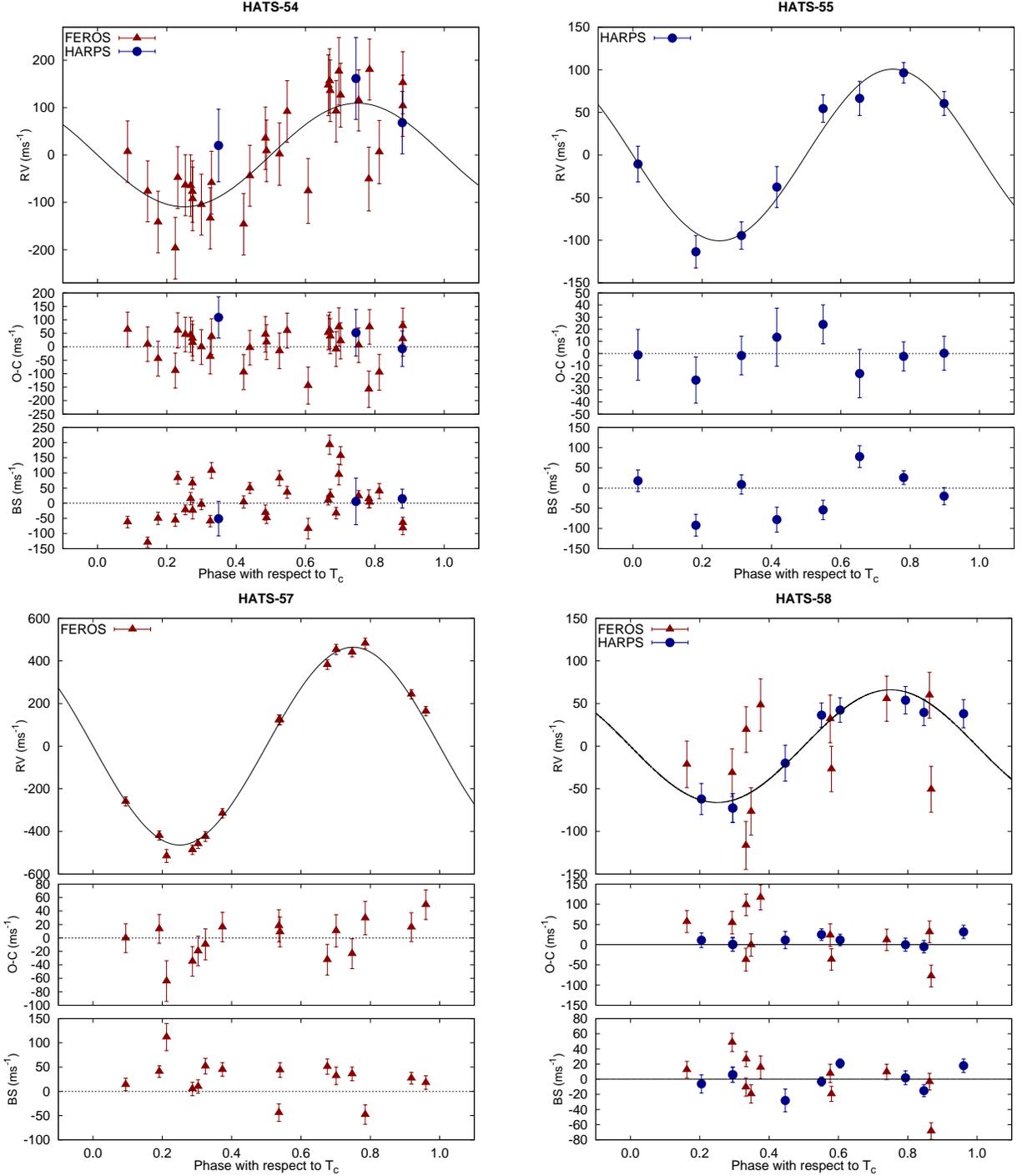

}{
    \begin{figure}[ht]
}
\plottwo{\hatcurhtr{54}-rv.eps}{\hatcurhtr{55}-rv.eps}
\plottwo{\hatcurhtr{57}-rv.eps}{\hatcurhtr{58}-rv.eps}
\caption{
    Phased high-precision RV measurements for \hbox{\hatcur{54}{}} (upper left), \hbox{\hatcur{55}{}} (upper right), \hbox{\hatcur{57}{}} (bottom left), and \hbox{HATS-58} (bottom right). The RVs for \hbox{\hatcur{56}{}} are shown in Figure~\ref{fig:rvbiscontd}. The instruments used are labelled in the plots. In each case we show three panels. The top panel shows the phased measurements together with our best-fit model (see \reftabl{planetparam}) for each system. Zero-phase corresponds to the time of mid-transit. The center-of-mass velocity has been subtracted. The second panel shows the velocity $O\!-\!C$ residuals from the best fit. The error bars include the jitter terms listed in \reftabl{planetparam} added in quadrature to the formal errors for each instrument. The third panel shows the bisector spans (BS). Note the different vertical scales of the panels.
}
\label{fig:rvbis}
\ifthenelse{\boolean{emulateapj}}{
    \end{figure*}
}{
    \end{figure}
}
%% --------------------------------------------------------------------

%
%% --------------------------------------------------------------------
\begin{figure}[ht]
\plotone{\hatcurhtr{56}-rv.eps}
\caption{
    High-precision RV measurements for \hbox{\hatcur{56}}. In the top panel of this figure we show the RVs plotted vs.\ time, together with our best-fit model including 
the orbital wobble of the star due to the planet \hbox{\hatcurb{56}} together with a significant quadratic trend. The bottom three panels are similar to those plotted for 
the other systems in Figure~\ref{fig:rvbis}, except here we have subtracted the quadratic trend from the RVs in the panel showing the phase-folded measurements.
}
\label{fig:rvbiscontd}
\end{figure}
%% --------------------------------------------------------------------

%% --------------------------------------------------------------------

% =====================================================================
\subsection{Photometric follow-up observations}
\label{sec:phot}

Photometric follow-up was obtained for our five systems in order to both refine the transit parameters (including the transit ephemerides) and to rule 
out possible false-positive scenarios (e.g., blended eclipsing binaries, hierarchical triples). The photometric follow-up included data from the 
1m telescopes at the Las Cumbres Observatory Global Telescope (LCOGT) Network \citep{brown:2013:lcogt}, the 0.3m Perth Exoplanet Survey Telescope (PEST), 
the 1m Swope Telescope at Las Campanas Observatory (LCO) and the recently commissioned 0.7m Chilean-Hungarian Automated Telescope (CHAT), also located at 
LCO. The data reduction for the LCOGT telescopes follows the procedures outlined in \cite{Bayliss:2015}, which have been updated for automatization and 
will be detailed in a future publication (Espinoza et al., 2018, in prep.); this latter set of procedures are similar to the ones used to reduce the Swope 
telescope data. The data reduction for the PEST telescope is detailed in \cite{zhou:2014:mebs}. The data reduction for the CHAT telescope follow similar 
procedures to those described for the LCOGT and Swope data; a full description of CHAT, its reduction and scheduling will be detailed in a future publication 
(Jord\'an et al., 2018, in prep.).

\begin{figure*}[!ht]
\plotone{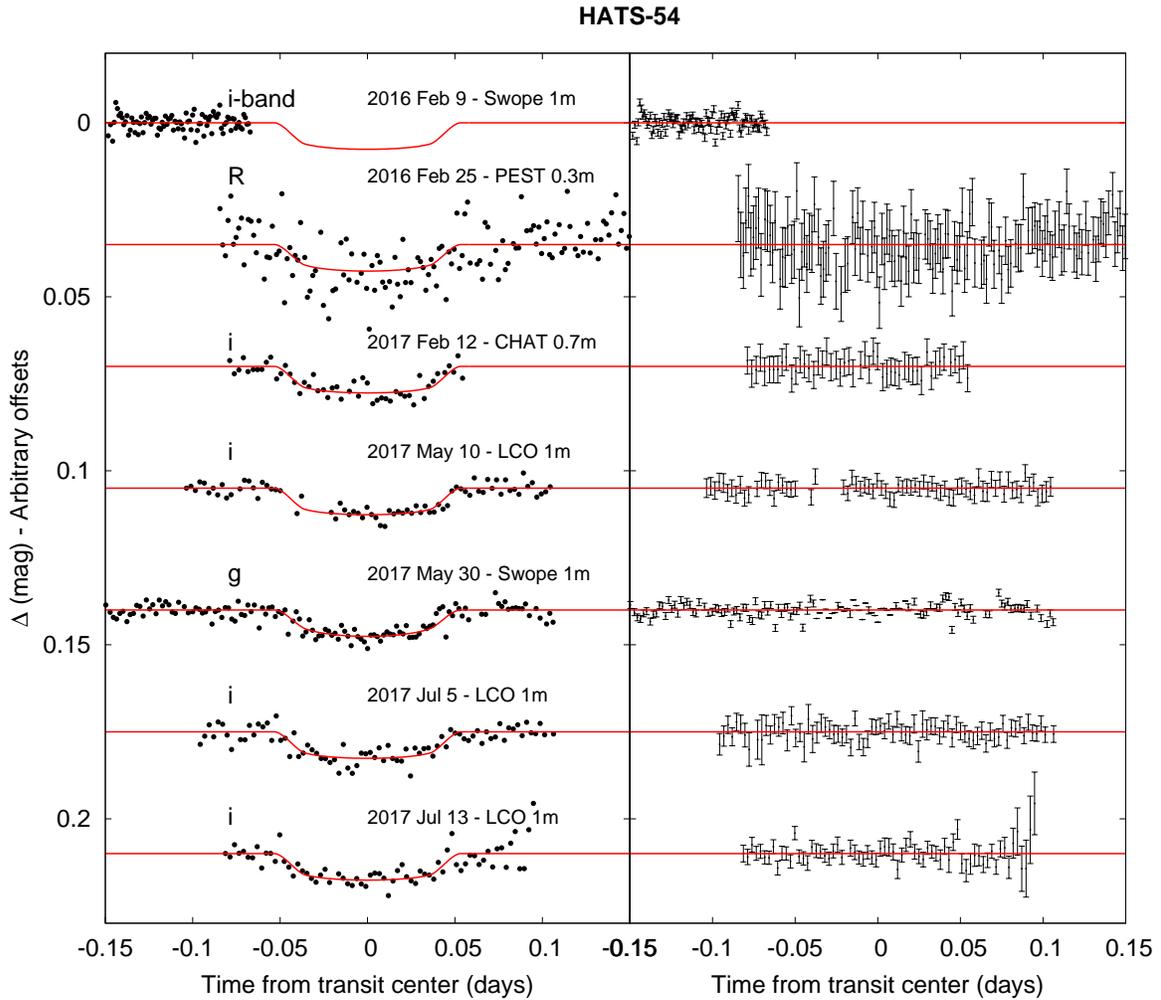}
\caption{
    Unbinned transit \lcs{} for \hatcur{54}.  The light curves have been
    corrected for quadratic trends in time, and linear trends with up
    to three parameters characterizing the shape of the PSF, fitted
    simultaneously with the transit model.
    The dates of the events, filters and instruments used are
    indicated.  Light curves following the first are displaced
    vertically for clarity.  Our best fit from the global modeling
    described in \refsecl{globmod} is shown by the solid lines. The
    residuals from the best-fit model are shown on the right-hand-side in the same
    order as the original light curves.  The error bars represent the
    photon and background shot noise, plus the readout noise.
}
\label{fig:lc:50}
\end{figure*}

\begin{figure*}[!ht]
\plotone{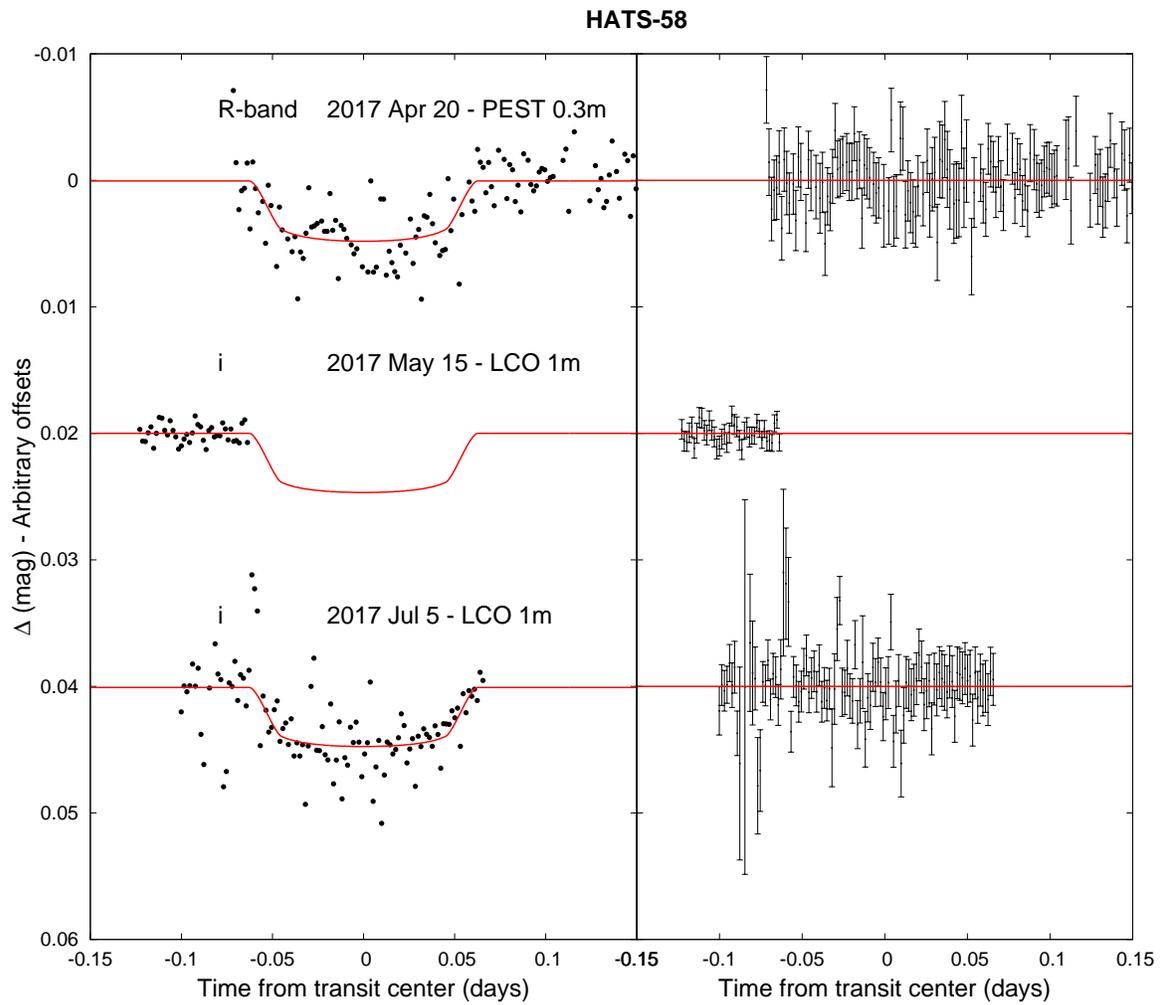}
\caption{
    Same as Fig.~\ref{fig:lc:50}, here we show \lcs{} for HATS-58.
}
\label{fig:lc:54}
\end{figure*}

Photometric follow-up observations were obtained for HATS-54 with all the mentioned instruments between 2016 and 2017, with a total of six transits 
observed in that period (Figure \ref{fig:lc:50}). For HATS-55, transits were observed with PEST, and the Swope and LCO 1m telescopes (Figure \ref{fig:lc:51}). 
This latter dataset is interesting as we observed the same transit of this target from Cerro Tololo Inter-American Observatory (CTIO) using two different 
LCOGT 1m telescopes (on Domes A and C), observing an excellent agreement between both datasets. One transit, a partial transit and an in-transit portion of 
the lightcurve were observed for HATS-56 as well in 2017 from the PEST and LCOGT 1m telescopes (Figure \ref{fig:lc:52}). For HATS-57, photometric follow-up 
was obtained with the CHAT telescope including a partial transit in August 2017 and a full transit in October 2017 (Figure \ref{fig:lc:53}). Finally, 
photometric follow-up was also obtained for HATS-58 in 2017 including two full transits (Figure \ref{fig:lc:54}). The photometric follow-up observations are 
summarized in \reftabl{photobs}.

\setcounter{planetcounter}{1}
%
%% --------------------------------------------------------------------

\begin{figure*}[!ht]
\plotone{\hatcurhtr{55}-lc.eps}
\caption{
    Same as Fig.~\ref{fig:lc:50}, here we show \lcs{} for \hatcur{55}.
}
\label{fig:lc:51}
\end{figure*}

\begin{figure*}[!ht]
\plotone{\hatcurhtr{56}-lc.eps}
\caption{
    Same as Fig.~\ref{fig:lc:50}, here we show \lcs{} for \hatcur{56}.
}
\label{fig:lc:52}
\end{figure*}

\begin{figure*}[!ht]
\plotone{\hatcurhtr{57}-lc.eps}
\caption{
    Same as Fig.~\ref{fig:lc:50}, here we show \lcs{} for \hatcur{57}.
}
\label{fig:lc:53}
\end{figure*}

% This section is for guidance only! It can (and should) be
% significantly rewritten.

%% Observations
%%
\clearpage

%
%
%% --------------------------------------------------------------------
\ifthenelse{\boolean{emulateapj}}{
    \begin{deluxetable*}{llrrrrl}
}{
    \begin{deluxetable}{llrrrrl}
}
\tablewidth{0pc}
\tablecaption{
    Light curve data for \hatcur{54}, \hatcur{55}, \hatcur{56}, \hatcur{57} and HATS-58\label{tab:phfu}.
}
\tablehead{
    \colhead{Object\tablenotemark{a}} &
    \colhead{BJD\tablenotemark{b}} & 
    \colhead{Mag\tablenotemark{c}} & 
    \colhead{\ensuremath{\sigma_{\rm Mag}}} &
    \colhead{Mag(orig)\tablenotemark{d}} & 
    \colhead{Filter} &
    \colhead{Instrument} \\
    \colhead{} &
    \colhead{\hbox{~~~~(2,400,000$+$)~~~~}} & 
    \colhead{} & 
    \colhead{} &
    \colhead{} & 
    \colhead{} &
    \colhead{}
}
\startdata
   HATS-54 & $ 56117.38698 $ & $   0.00004 $ & $   0.00677 $ & $ \cdots $ & $ r$ &         HS\\
   HATS-54 & $ 56018.16380 $ & $   0.01046 $ & $   0.00648 $ & $ \cdots $ & $ r$ &         HS\\
   HATS-54 & $ 55725.58263 $ & $  -0.00472 $ & $   0.01126 $ & $ \cdots $ & $ r$ &         HS\\
   HATS-54 & $ 56025.79655 $ & $  -0.01178 $ & $   0.01018 $ & $ \cdots $ & $ r$ &         HS\\
   HATS-54 & $ 56091.94646 $ & $  -0.00523 $ & $   0.00612 $ & $ \cdots $ & $ r$ &         HS\\
   HATS-54 & $ 56089.40256 $ & $  -0.00491 $ & $   0.00689 $ & $ \cdots $ & $ r$ &         HS\\
   HATS-54 & $ 55941.83994 $ & $   0.01812 $ & $   0.00695 $ & $ \cdots $ & $ r$ &         HS\\
   HATS-54 & $ 56066.50503 $ & $  -0.00225 $ & $   0.00702 $ & $ \cdots $ & $ r$ &         HS\\
   HATS-54 & $ 56061.41691 $ & $  -0.00072 $ & $   0.00768 $ & $ \cdots $ & $ r$ &         HS\\
   HATS-54 & $ 55969.82681 $ & $  -0.01631 $ & $   0.00868 $ & $ \cdots $ & $ r$ &         HS\\

\enddata
\tablenotetext{a}{
    Either \hatcur{54}, \hatcur{55}, \hatcur{56}, \hatcur{57} or HATS-58.%\hatcur{58}.
}
\tablenotetext{b}{
    Barycentric Julian Date is computed directly from the UTC time
    without correction for leap seconds.
}
\tablenotetext{c}{
    The out-of-transit level has been subtracted. For observations
    made with the HATSouth instruments (identified by ``HS'' in the
    ``Instrument'' column) these magnitudes have been corrected for
    trends using the EPD and TFA procedures applied {\em prior} to
    fitting the transit model. This procedure may lead to an
    artificial dilution in the transit depths. The blend factors for
    the HATSouth light curves are listed in
    Table~\ref{tab:planetparam}. For
    observations made with follow-up instruments (anything other than
    ``HS'' in the ``Instrument'' column), the magnitudes have been
    corrected for a quadratic trend in time, and for variations
    correlated with up to three PSF shape parameters, fit simultaneously
    with the transit.
}
\tablenotetext{d}{
    Raw magnitude values without correction for the quadratic trend in
    time, or for trends correlated with the seeing. These are only
    reported for the follow-up observations.
}
\tablecomments{
    This table is available in a machine-readable form in the online
    journal.  A portion is shown here for guidance regarding its form
    and content.
}
\ifthenelse{\boolean{emulateapj}}{
    \end{deluxetable*}
}{
    \end{deluxetable}
}
%% --------------------------------------------------------------------

% =====================================================================
%% Photometric detection
\subsection{Lucky Imaging}
\label{sec:luckyimaging}
%++++++++++++++++++++++++++++++++++++++++++++++++++++++++++++++++++++++
%++++++++++++++++++++++++++++++++++++++++++++++++++++++++++++++++++++++

%% ----------------
\ifthenelse{\boolean{emulateapj}}{
    \begin{figure*}[!ht]
}{
    \begin{figure}[!ht]
}
\plottwo{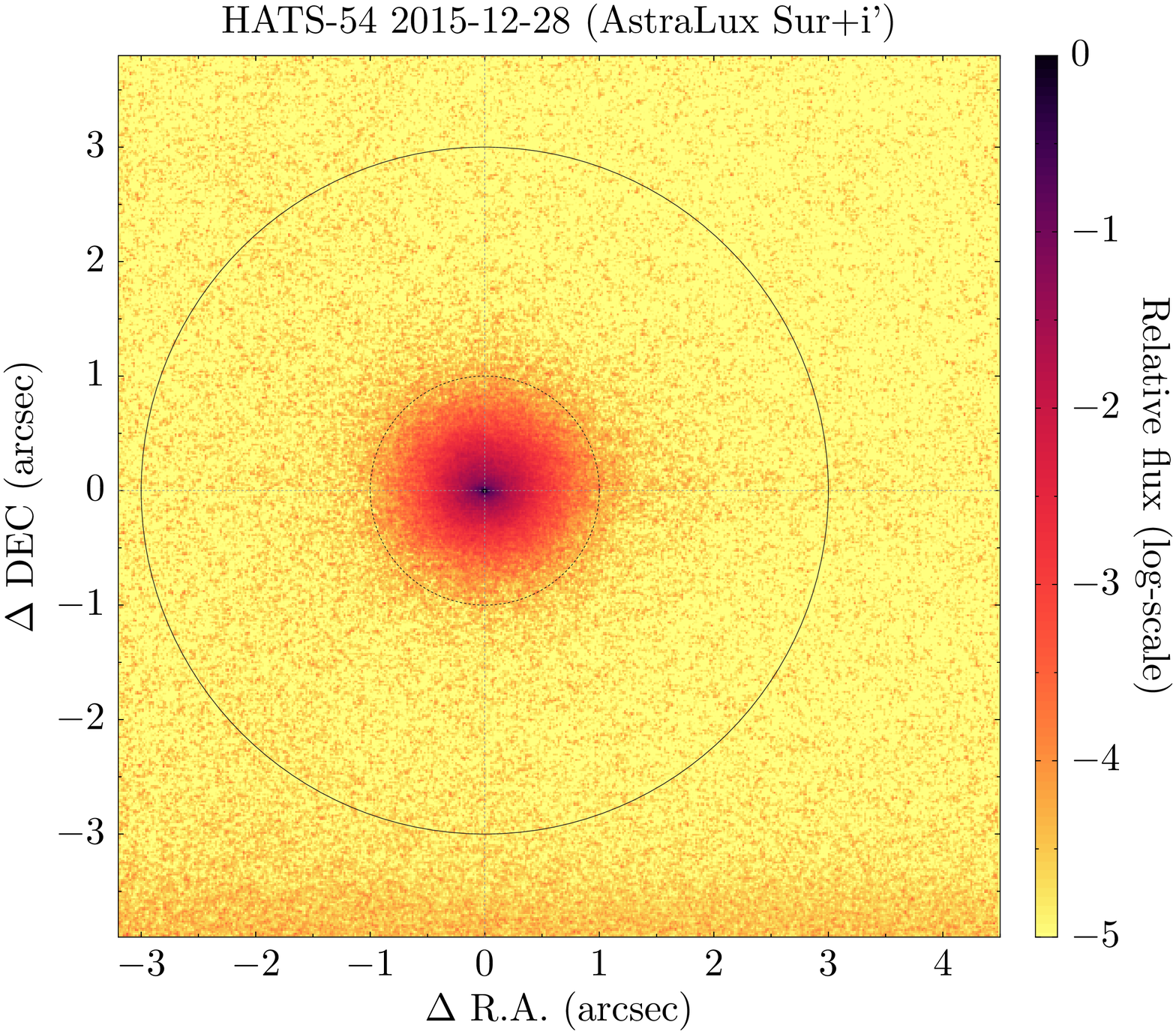}{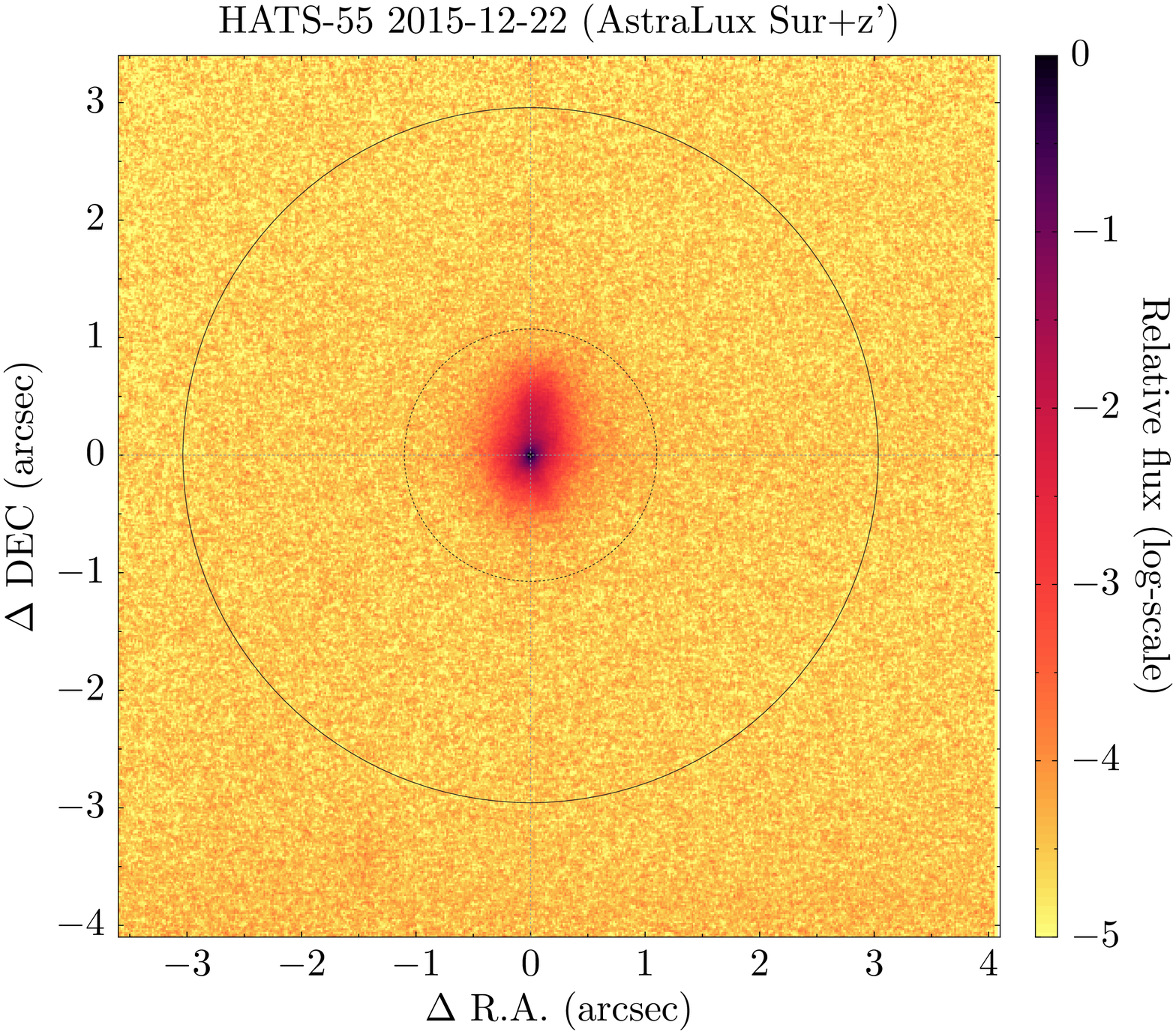}
\caption{
    Astarlux lucky images of \hatcur{54} ({\em left}) and \hatcur{55} ({\em right}). No neighboring sources are detected for \hatcur{54} and \hatcur{55}.
\label{fig:luckyimages}}
\ifthenelse{\boolean{emulateapj}}{
    \end{figure*}
}{
    \end{figure}
}
%% ----------------

%% ----------------
\ifthenelse{\boolean{emulateapj}}{
    \begin{figure*}[!ht]
}{
    \begin{figure}[!ht]
}
\plottwo{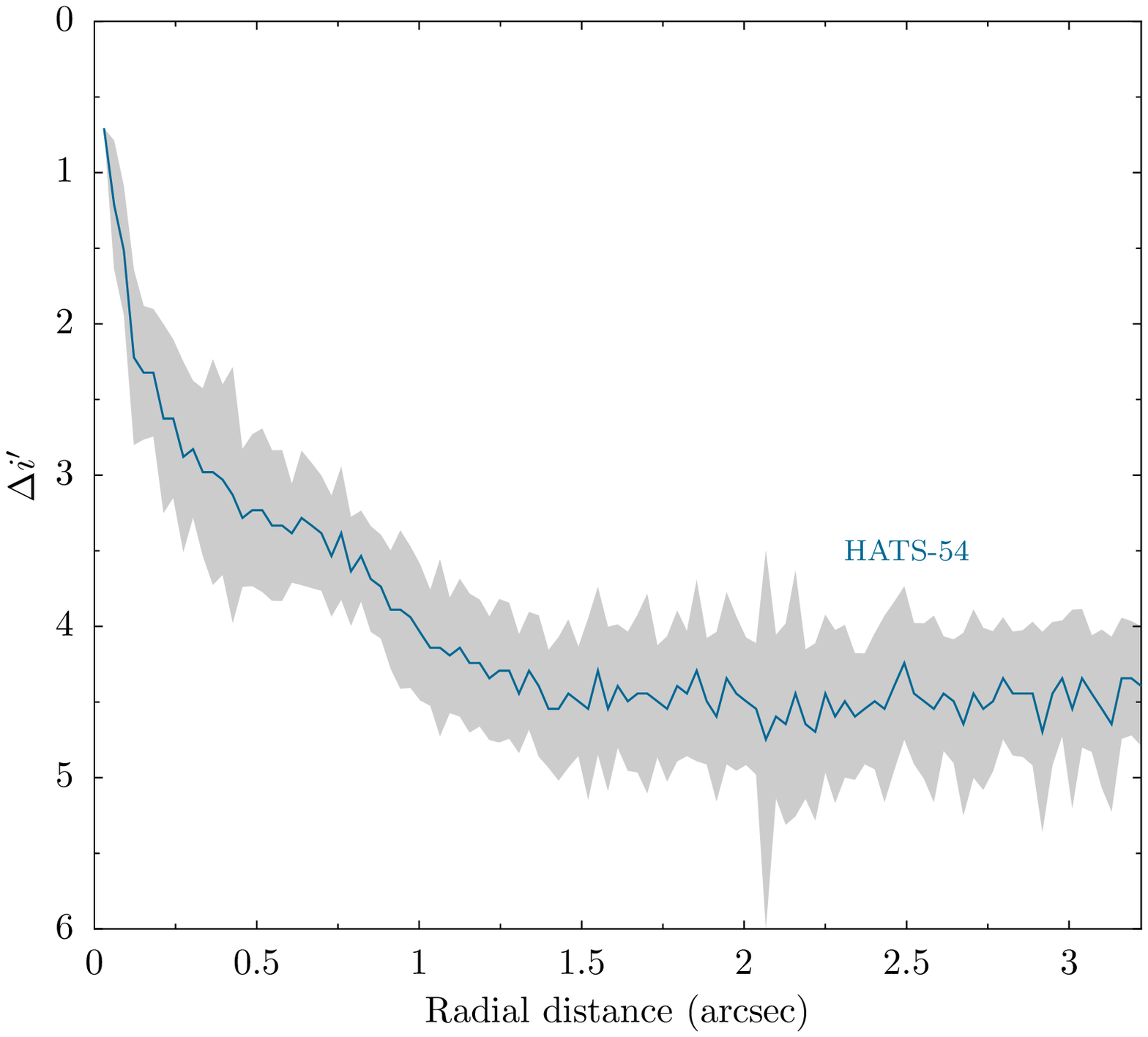}{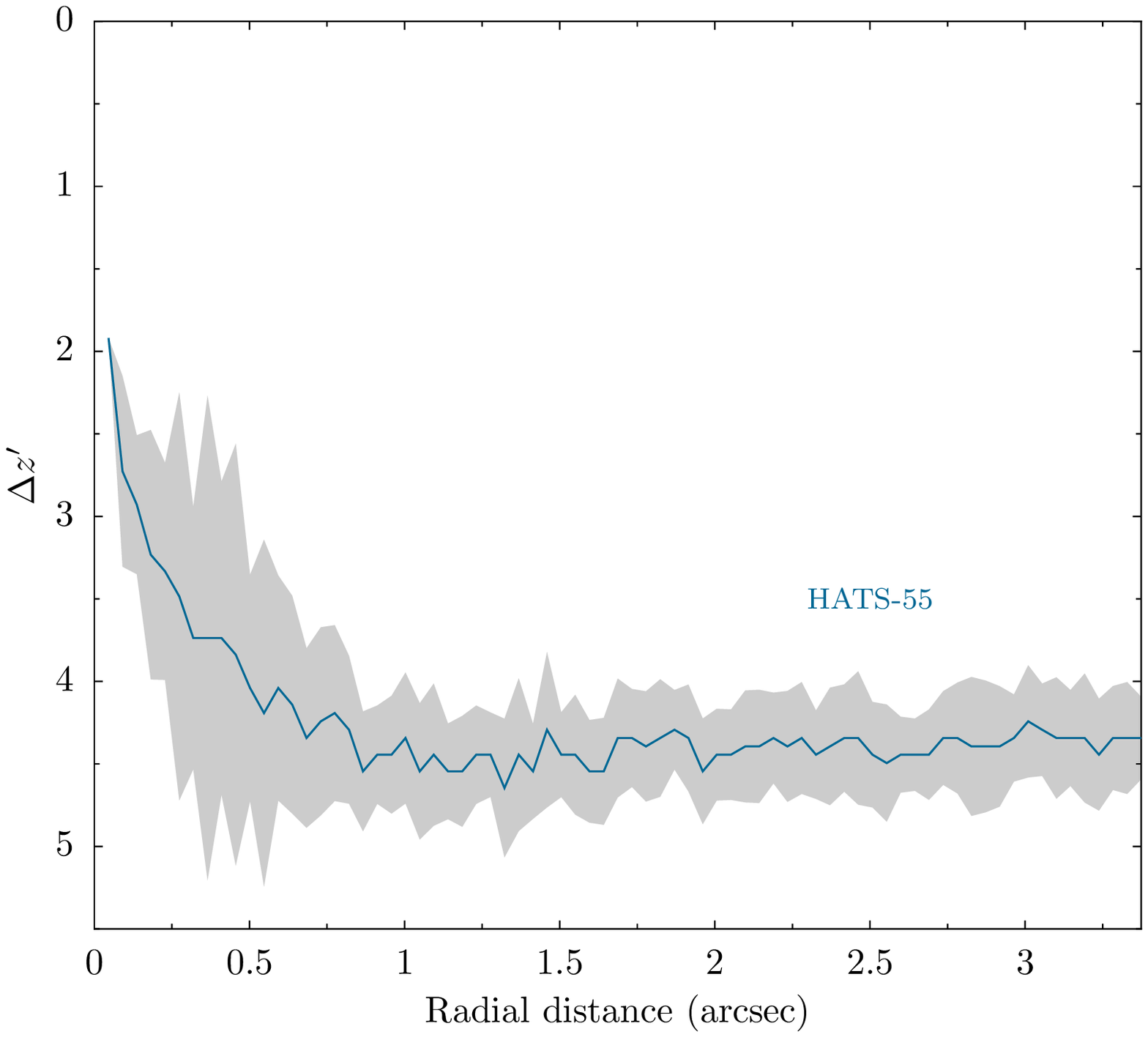}
\caption{
    $5\sigma$ contrast curves for \hatcur{54} ({\em left}), and \hatcur{55} ({\em right}) based on our AstraLux Sur $z^{\prime}-band$ observations. Gray bands show the uncertainty given by the scatter in the contrast in the azimuthal direction at a given radius.
\label{fig:luckyimagecontrastcurves}}
\ifthenelse{\boolean{emulateapj}}{
    \end{figure*}
}{
    \end{figure}
}
%% ----------------

High spatial resolution imaging via ``Lucky imaging" was obtained for \hatcur{54} and \hatcur{55} using Astralux Sur \citep{hippler:2009} at the New Technology Telescope (NTT) 
located in LSO. The data for \hatcur{54} was obtained on December 28, 2015 with the $i'$ band and for \hatcur{55} on December 22, 2015 with the $z'$ band. The stacked images, 
obtained by selecting the best 10\% of all the obtained images, are shown in Figure \ref{fig:luckyimages}, where the plate scale derived in \cite{Janson:2017} of 15.2 
milli-arcseconds (mas)/pixel has been used. We analyzed the images using the algorithms described in \cite{espinoza:2016:hats25hats30}, obtaining an effective full-width 
at half maximum (FWHM) for the stacked \hatcur{54} observations of $42.36\pm5.43$ mas, and for the stacked \hatcur{55} observations of $52.54\pm5.50$ mas. These are 
excellent considering the diffraction limit of the instrument is $\sim 50$ mas according to \cite{hippler:2009}. 5-$\sigma$ contrasts curves were generated with the same 
algorithm, and are presented in Figure \ref{fig:luckyimagecontrastcurves}. No neighboring stars were detected for our targets.

\subsection{Gaia DR2}
\label{sec:gaiadr2}
We queried the coordinates of our target stars into Gaia DR2 \citep{gaiadr2} in order to search for possible companion 
stars detected by the Gaia mission within 5'' from our targets. No companions were found in Gaia for HATS-54 
and HATS-57. We did find companions to our other target stars, which we detail below:
\begin{itemize}
\item \textit{HATS-55.} A very faint source 
($\Delta G = 5.84$) was found at $\Delta \textnormal{RA} = -1.52613'' \pm 0.00032$ and 
$\Delta \textnormal{Dec} = -3.48374'' \pm 0.00039$ from the target. We note that these coordinates 
are observable on the field observed by our AstraLux observations and, actually, once these coordinates 
are known, it is possible to see a faint signal (still within the noise level) in the AstraLux image of 
HATS-55 (Figure \ref{fig:luckyimages}). Performing photometry on the AstraLux image at those coordinates 
we obtain a magnitude difference of $\Delta z'=5.30\pm 0.10$, which is below or 5-sigma contrast level (i.e., 
below the noise level on our image). From Gaia, the proper motion of the target and the companion are 
inconsistent with each other, which implies they are not physically bound.

\item \textit{HATS-56}. A faint ($\Delta G = 3.94$) source was found at 
$\Delta \textnormal{RA} = -1.48296'' \pm 0.00026$ and $\Delta \textnormal{Dec} = 0.59747'' \pm 0.00044$ from 
this target, whose proper motion ($-9.19 \pm 0.57$ mas/yr in RA, $-3.00 \pm 0.74$ mas/yr in Dec) is consistent 
with that of the target ($-8.604 \pm 0.046$ mas/yr in RA, $-2.950 \pm 0.035$ mas/yr in Dec), which could imply 
it is physically bound. However, it is unclear if the Gaia parallax is reliable enough to claim this latter hypothesis 
as true, as it is very uncertain for the faint companion to HATS-56. In any case, the neighbor is faint enough relative 
to the target star that it can be neglected in the analysis. 

\item \textit{HATS-58}. A bright source ($\Delta G = 0.92$ fainter than 
the target star) was found at $\Delta \textnormal{RA} = 0.29733'' \pm 0.00051$ and 
$\Delta \textnormal{Dec} = -0.68025'' \pm 0.00028$ from our target. The proper motion of this object 
measured by Gaia DR2 ($-12.96\pm0.92$ mas/yr in RA, $-2.30 \pm 0.44$ mas/yr in Dec) is consistent to 
the proper motion of our target ($-12.70 \pm 0.30$ mas/yr in RA, $-3.23 \pm 0.16$ mas/yr in Dec) and, 
therefore, we assume they are physically bound. Because of this, from now on in this work we refer to the 
brighter star as HATS-58A and to the fainter companion as HATS-58B. The Gaia photometry gives a very uncertain effective 
temperature for HATS-58B of $5095^{+1842}_{-811}$ K. This neighbor is sufficiently bright relative to the target star that it must be
 taken into account.

\end{itemize}

% #####################################################################
%% Analysis
\section{Analysis}
\label{sec:analysis}

\subsection{Properties of the parent star}
\label{sec:stelparam}

In order to determine the parameters of the parent stars of our planetary candidates, we obtained precise stellar atmospheric parameters using 
ZASPE \citep{brahm:2017:zaspe}, by using the stacked HARPS spectra for \hatcur{55} and the stacked FEROS spectra for the rest of our targets. 
With these atmospheric parameters, we performed a joint analysis with all the available data following the method explained in detail in 
\citet{hartman:2018} (see Section \ref{sec:globmod} for a brief overview) in order to obtain the physical parameters of the stars. With these 
physical parameters at hand, a second ZASPE iteration was performed for all the targets, where the revised value of the log-gravity was used as input 
in order to derive the final atmospheric parameters of the stars; these were then used again in a second iteration of the joint modelling to be detailed 
in Section \ref{sec:globmod} to obtain the final parameters of the stars, which are presented in Table \ref{tab:stellar}. We present the locations of 
our target stars on the absolute $G$ magnitude versus Gaia DR2 BP-RP colors in \reffigl{iso} and \reffigl{isocontd} for all our targets 
except for HATS-58A, for which we present it in the absolute $G$ magnitude versus effective temperature plane as this target did not have a well measured 
BP-RP color. In addition, as will be detailed in Section \ref{sec:globmod}, the analysis for this latter star was special as it is blended with HATS-58B 
in all of our measurements with the exception of Gaia, where the two components of the blend are resolved, as mentioned in the previous section. We account for this 
in our modelling and we were able to obtain a mass for HATS-58B of \hatcurISOmlongB{58} solar-masses.

%As in previous works, with the physical parameters at hand, we compute the distances to the stars by comparing the observed photometry for each 
%target to the predicted magnitudes in each filter using our best-fit isochrones. An $R_{V} = 3.1$ extinction law from \citet{cardelli:1989} 
%was assumed to determine the extinction. For \hatcur{56} and \hatcur{57} we additionally use the parallax measurements available from Gaia DR1 
%\citep{gaiadr1}. The final stellar parameters are presented in Table \ref{tab:stellar}.

As mentioned in Section \ref{sec:detection}, we observe that HATS-57 shows variability at the 2\% level. This variability could be used to estimate the 
rotation period of the star which, combined with the value of $v\sin i_*$ given in Table \ref{tab:stellar}, could in turn give us an estimate of the 
inclination of the star with respect to the line-of-sight, $i_*$. To find the period of this modulation, we model the lightcurve using a 
Gaussian Process (GP) regression. We use the quasi-periodic kernel presented in \cite{celerite} of the form:
\begin{eqnarray*}
k (\tau) = \frac{B}{2+C}e^{-\tau/L}\left[\cos\left(\frac{2\pi\tau}{P_\textnormal{GP}}\right) + (1+C)\right],
\end{eqnarray*}
where $\tau = t_i - t_k$, with $i,k \in [1,2,...N]$, where $N$ is the number of datapoints, and $B$, $C$, $L$ and $P_\textnormal{GP}$
are the hyperparameters of the model, with the latter corresponding to the period of the quasi-periodic oscillations defined by
this kernel. We assume the lightcurve has a zero-point flux and an extra jitter, which we also model. In order to efficiently explore the full parameter space,
we use MultiNest \citep{MultiNest} with the PyMultinest Python wrapper \citep{PyMultiNest} to find the posterior density of the parameters of the
GP. This code, which we call \texttt{GPRotatioNest}, is available at GitHub\footnote{\url{http://www.github.com/nespinoza/GPRotatioNest}}.

\begin{figure*}[!ht]
\plotone{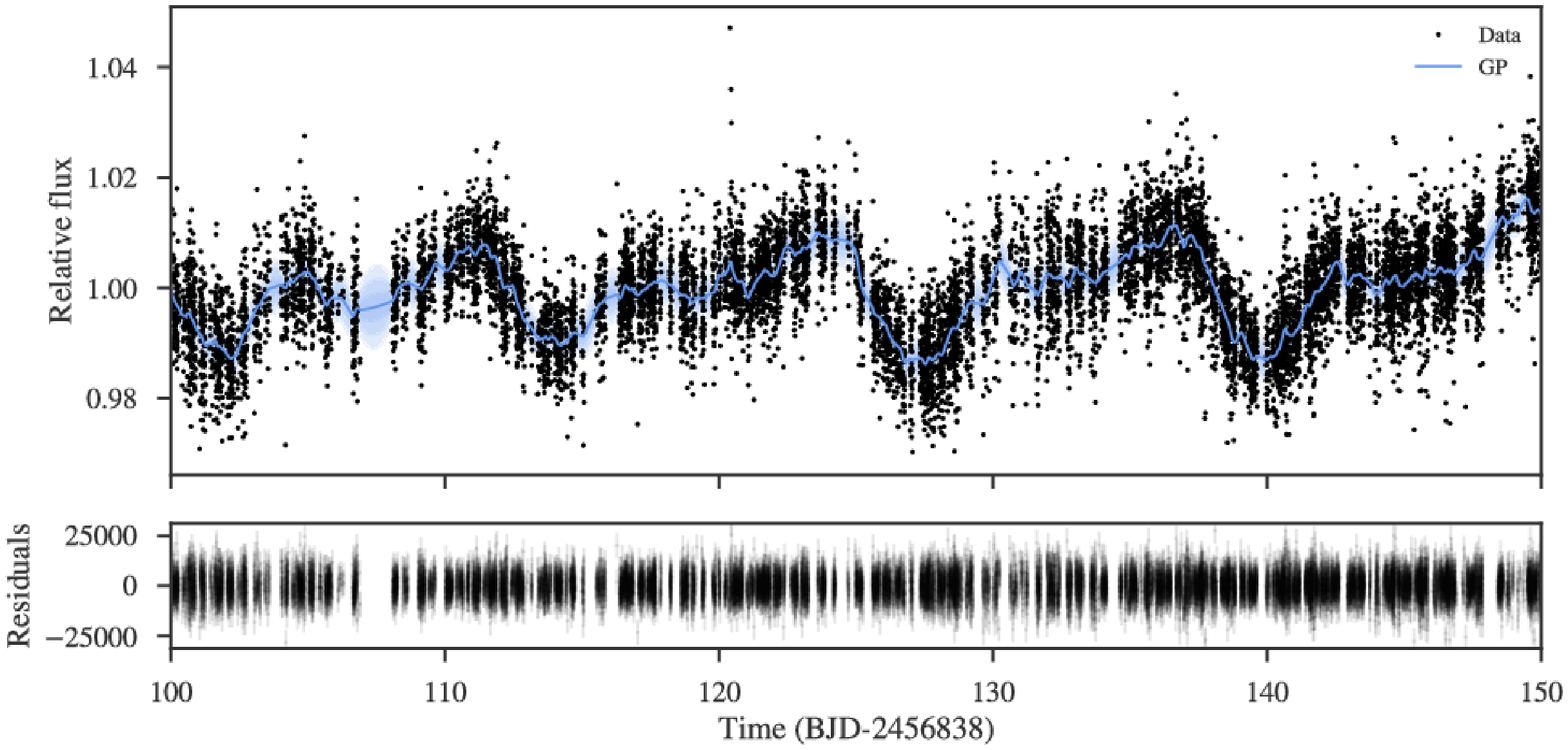}
\caption{
(Top) Portion of the lightcurve of HATS-57 (black points) along with the posterior GP model (blue line --- darker blue bands denote the 3-sigma
credibility bands around it). (Bottom) Residuals between the GP and the data.
\label{fig:gp-hats57}}
\end{figure*}

Using \texttt{GPRotatioNest} on the lightcurve of HATS-57 we find two modes for the period, one at $6.355 \pm 0.018$ days, which is the dominant
peak in the posterior distribution, and another one at $11.27 \pm 0.57$ days. When phasing the data with both periods, it is evident the former does
a significantly better job at coherently adding the periodicity; however, from the same phasing of the data it is obvious that this is \textit{half} the
real periodicity as well. Based on this, we interpret $2P_\textnormal{GP}$, i.e., $12.71\pm0.037$ days, as the rotation period of the star. Figure \ref{fig:gp-hats57}
shows a portion of the data for the lightcurve of HATS-57, along with the prediction from the GP. With this period, the $v\sin i_*$ and radius of the star presented 
in Table \ref{tab:stellar}, we derive an inclination of the star with respect to the line-of sight of $i_*=67.1^{+10.5}_{-10.6}$ degrees. 

%% --------------------------------------------------------------------
\ifthenelse{\boolean{emulateapj}}{
    \begin{figure*}[!ht]
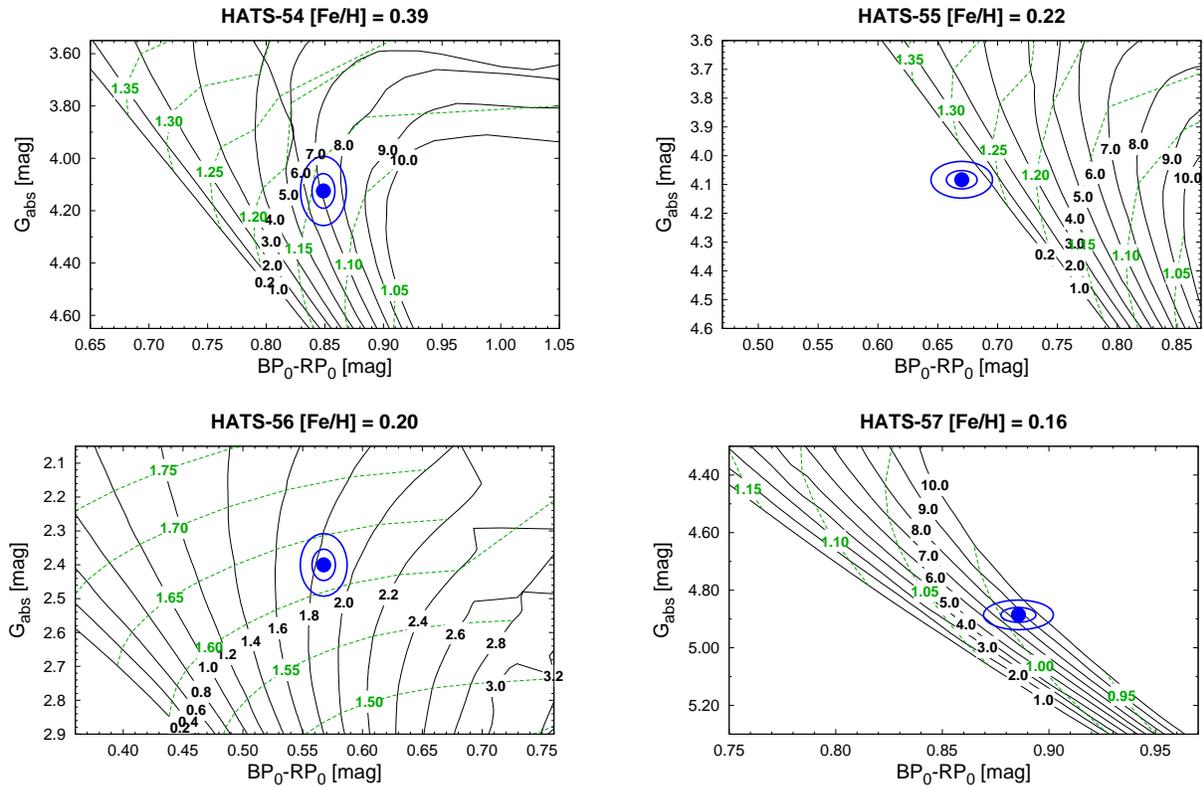

}{
    \begin{figure}[!ht]
}
\plottwo{\hatcurhtr{54}-iso-bprp-gabs.eps}{\hatcurhtr{55}-iso-bprp-gabs.eps}
\plottwo{\hatcurhtr{56}-iso-bprp-gabs.eps}{\hatcurhtr{57}-iso-bprp-gabs.eps}
\caption{
    Model isochrones (black solid lines) from \cite{\hatcurisocite{54}} for the measured
    metallicities of \hatcur{54} (upper left), \hatcur{55} (upper right), \hatcur{56} (bottom left), and \hatcur{57} (bottom right). \hatcur{58} is shown in Figure~\ref{fig:isocontd}. The age of each isochrone in Gyr is labelled in black font. We also show evolutionary tracks for stars of fixed mass (dashed green lines) with the mass of each tracked labelled in solar mass units in green font. The de-reddened BP$_{0}-$RP$_{0}$ colors and absolute G magnitudes from Gaia DR2 are shown 
for each host star are shown using filled blue circles together with
    their 1$\sigma$ and 2$\sigma$ confidence ellipsoids (blue lines). 
}
\label{fig:iso}
\ifthenelse{\boolean{emulateapj}}{
    \end{figure*}
}{
    \end{figure}
}

\begin{figure}[!ht]
\plotone{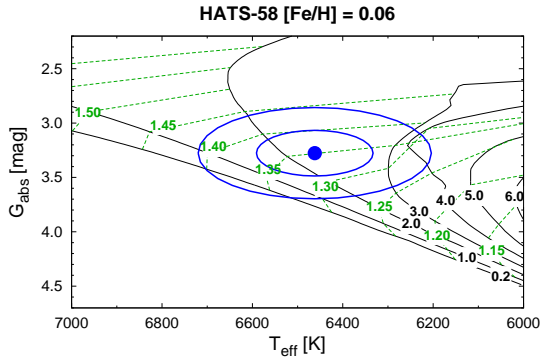}
\caption{
    Same as Figure~\ref{fig:iso}, here we show \hatcur{58}. In this case, however, we use the spectroscopically determined stellar effective temperature value 
    instead of $BP_{0}-RP_{0}$, as this target didn't have a well measured BP-RP color (see text).
}
\label{fig:isocontd}
\end{figure}

%
%
%% --------------------------------------------------------------------
%% Table of stellar parameters. 
%%
\ifthenelse{\boolean{emulateapj}}{
    \begin{deluxetable*}{lcccccl}
}{
    \begin{deluxetable}{lcccccl}
}
\tablewidth{0pc}
\tabletypesize{\tiny        }
\tablecaption{
    Stellar parameters for \hatcur{54}--\hatcur{58}
    \label{tab:stellar}
}
\tablehead{
    \multicolumn{1}{c}{} &
    \multicolumn{1}{c}{\bf HATS-54} &
    \multicolumn{1}{c}{\bf HATS-55} &
    \multicolumn{1}{c}{\bf HATS-56} &
    \multicolumn{1}{c}{\bf HATS-57} &
    \multicolumn{1}{c}{\bf HATS-58} &
    \multicolumn{1}{c}{} \\
    \multicolumn{1}{c}{~~~~~~~~Parameter~~~~~~~~} &
    \multicolumn{1}{c}{Value}                     &
    \multicolumn{1}{c}{Value}                     &
    \multicolumn{1}{c}{Value}                     &
    \multicolumn{1}{c}{Value}                     &
    \multicolumn{1}{c}{Value}                     &
    \multicolumn{1}{c}{Source}
}
\startdata
\noalign{\vskip -3pt}
\sidehead{Astrometric properties and cross-identifications}
~~~~Gaia DR2-ID\dotfill               & \hatcurCCgaiadrtwo{54}  & \hatcurCCgaiadrtwo{55} & \hatcurCCgaiadrtwo{56} & \hatcurCCgaiadrtwo{57} & \hatcurCCgaiadrtwo{58} & \\
~~~~2MASS-ID\dotfill               & \hatcurCCtwomassshort{54}  & \hatcurCCtwomassshort{55} & \hatcurCCtwomassshort{56} & \hatcurCCtwomassshort{57} & \hatcurCCtwomassshort{58} & \\
~~~~GSC-ID\dotfill                 & \hatcurCCgsc{54}      & \hatcurCCgsc{55}     & \hatcurCCgsc{56}     & \hatcurCCgsc{57}     & \hatcurCCgsc{58}     & \\
~~~~R.A. (J2000)\dotfill            & \hatcurCCra{54}       & \hatcurCCra{55}    & \hatcurCCra{56}    & \hatcurCCra{57}    & \hatcurCCra{58}    & Gaia DR2\\
~~~~Dec. (J2000)\dotfill            & \hatcurCCdec{54}      & \hatcurCCdec{55}   & \hatcurCCdec{56}   & \hatcurCCdec{57}   & \hatcurCCdec{58}   & Gaia DR2\\
~~~~$\mu_{\rm R.A.}$ (\masy)              & \hatcurCCpmra{54}     & \hatcurCCpmra{55} & \hatcurCCpmra{56} & \hatcurCCpmra{57} & \hatcurCCpmra{58} & Gaia DR2\\
~~~~$\mu_{\rm Dec.}$ (\masy)              & \hatcurCCpmdec{54}    & \hatcurCCpmdec{55} & \hatcurCCpmdec{56} & \hatcurCCpmdec{57} & \hatcurCCpmdec{58} & Gaia DR2\\
~~~~Parallax (mas)                        & \hatcurCCparallax{54}    & \hatcurCCparallax{55} & \hatcurCCparallax{56} & \hatcurCCparallax{57} & \hatcurCCparallax{58} & Gaia DR2\\
\sidehead{Spectroscopic properties}
% TODO: comment those that are not relevant to the paper.
~~~~$\teffstar$ (K)\dotfill         &  \hatcurSMEteff{54}   & \hatcurSMEteff{55} & \hatcurSMEteff{56} & \hatcurSMEteff{57} & \hatcurSMEteff{58} & ZASPE\tablenotemark{a}\\
~~~~$\feh$\dotfill                  &  \hatcurSMEzfeh{54}   & \hatcurSMEzfeh{55} & \hatcurSMEzfeh{56} & \hatcurSMEzfeh{57} & \hatcurSMEzfeh{58} & ZASPE               \\
~~~~$\vsini$ (\kms)\dotfill         &  \hatcurSMEvsin{54}   & \hatcurSMEvsin{55} & \hatcurSMEvsin{56} & \hatcurSMEvsin{57} & \hatcurSMEvsin{58} & ZASPE                \\
~~~~$\vmac$ (\kms)\dotfill          &  $\hatcurSMEvmac{54}$   & $\hatcurSMEvmac{55}$ & $\hatcurSMEvmac{56}$ & $\hatcurSMEvmac{57}$ & $\hatcurSMEvmac{58}$ & Assumed              \\
~~~~$\vmic$ (\kms)\dotfill          &  $\hatcurSMEvmic{54}$   & $\hatcurSMEvmic{55}$ & $\hatcurSMEvmic{56}$ & $\hatcurSMEvmic{57}$ & $\hatcurSMEvmic{58}$ & Assumed              \\
~~~~$\gamma_{\rm RV}$ (\ms)\dotfill&  \hatcurRVgammaabs{54}  & \hatcurRVgammaabs{55} & \hatcurRVgammaabs{56} & \hatcurRVgammaabs{57} & \hatcurRVgammaabs{58} & FEROS/HARPS\tablenotemark{b}  \\
~~~~$\dot{\gamma}_{\rm RV}$ (\ms\,d$^{-1}$)\dotfill&  $\cdots$  & $\cdots$ & $\hatcurRVtrone{56}$ & $\cdots$ & $\cdots$ & FEROS/HARPS\tablenotemark{c}  \\
~~~~$\ddot{\gamma}_{\rm RV}$ (\ms\,d$^{-2}$)\dotfill& $\cdots$  & $\cdots$ & $\hatcurRVtrtwo{56}$ & $\cdots$ & $\cdots$ & FEROS/HARPS\tablenotemark{c}  \\
\sidehead{Photometric properties}
% TODO: comment those that are not relevant to the paper.
%       Add photometry from other sources, e.g. Tycho-2.
%~~~~$B_T$ (mag)\dotfill             &  10.494 $\pm$ 0.031\phn  & Tycho-2     \\
%~~~~$V_T$ (mag)\dotfill             &  10.038 $\pm$ 0.029\phn  & Tycho-2     \\
~~~~$B$ (mag)\dotfill               &  \hatcurCCtassmB{54}  & \hatcurCCtassmB{55} & \hatcurCCtassmB{56} & \hatcurCCtassmB{57} & \hatcurCCtassmB{58} & APASS\tablenotemark{d} \\
~~~~$V$ (mag)\dotfill               &  \hatcurCCtassmv{54}  & \hatcurCCtassmv{55} & \hatcurCCtassmv{56} & \hatcurCCtassmv{57} & \hatcurCCtassmv{58} & APASS\tablenotemark{d} \\
~~~~$g$ (mag)\dotfill               &  \hatcurCCtassmg{54}  & \hatcurCCtassmg{55} & \hatcurCCtassmg{56} & \hatcurCCtassmg{57} & \hatcurCCtassmg{58} & APASS\tablenotemark{d} \\
~~~~$r$ (mag)\dotfill               &  \hatcurCCtassmr{54}  & \hatcurCCtassmr{55} & \hatcurCCtassmr{56} & \hatcurCCtassmr{57} & \hatcurCCtassmr{58} & APASS\tablenotemark{d} \\
~~~~$i$ (mag)\dotfill               &  \hatcurCCtassmi{54}  & \hatcurCCtassmi{55} & \hatcurCCtassmi{56} & \hatcurCCtassmi{57} & \hatcurCCtassmi{58} & APASS\tablenotemark{d} \\
~~~~$G$ (mag)\dotfill               &  \hatcurCCgaiamG{54} & \hatcurCCgaiamG{55} & \hatcurCCgaiamG{56} & \hatcurCCgaiamG{57} & \hatcurCCgaiamG{58} & Gaia DR2           \\
~~~~$BP$ (mag)\dotfill               &  \hatcurCCgaiamBP{54} & \hatcurCCgaiamBP{55} & \hatcurCCgaiamBP{56} & \hatcurCCgaiamBP{57} & $\cdots$ & Gaia DR2           \\
~~~~$RP$ (mag)\dotfill               &  \hatcurCCgaiamRP{54} & \hatcurCCgaiamRP{55} & \hatcurCCgaiamRP{56} & \hatcurCCgaiamRP{57} & $\cdots$ & Gaia DR2           \\
~~~~$J$ (mag)\dotfill               &  \hatcurCCtwomassJmag{54} & \hatcurCCtwomassJmag{55} & \hatcurCCtwomassJmag{56} & \hatcurCCtwomassJmag{57} & \hatcurCCtwomassJmag{58} & 2MASS           \\
~~~~$H$ (mag)\dotfill               &  \hatcurCCtwomassHmag{54} & \hatcurCCtwomassHmag{55} & \hatcurCCtwomassHmag{56} & \hatcurCCtwomassHmag{57} & \hatcurCCtwomassHmag{58} & 2MASS           \\
~~~~$K_s$ (mag)\dotfill             &  \hatcurCCtwomassKmag{54} & \hatcurCCtwomassKmag{55} & \hatcurCCtwomassKmag{56} & \hatcurCCtwomassKmag{57} & \hatcurCCtwomassKmag{58} & 2MASS           \\
\sidehead{Derived properties}
~~~~$\mstar$ ($\msun$)\dotfill      &  \hatcurISOmlong{54}   & \hatcurISOmlong{55} & \hatcurISOmlong{56} & \hatcurISOmlong{57} & \hatcurISOmlong{58} & Joint fit \tablenotemark{e}\\
~~~~$\rstar$ ($\rsun$)\dotfill      &  \hatcurISOrlong{54}   & \hatcurISOrlong{55} & \hatcurISOrlong{56} & \hatcurISOrlong{57} & \hatcurISOrlong{58} & Joint fit         \\
~~~~$T_\textnormal{eff}$ (K) \dotfill       &  \hatcurISOteff{54}    & \hatcurISOteff{55} & \hatcurISOteff{56} & \hatcurISOteff{57} & \hatcurISOteff{58} & Joint fit         \\
~~~~$\loggstar$ (cgs)\dotfill       &  \hatcurISOlogg{54}    & \hatcurISOlogg{55} & \hatcurISOlogg{56} & \hatcurISOlogg{57} & \hatcurISOlogg{58} & Joint fit         \\ 
~~~~Fe/H (dex) \dotfill       &  \hatcurISOzfeh{54}    & \hatcurISOzfeh{55} & \hatcurISOzfeh{56} & \hatcurISOzfeh{57} & $\cdots$ & Joint fit         \\ 
~~~~$\rhostar$ (\gcmc)\dotfill      &  \hatcurLCrho{54}    & \hatcurLCrho{55} & \hatcurLCrho{56} & \hatcurLCrho{57} & \hatcurLCrho{58} & Joint fit   \\
%~~~~$\rhostar$ (\gcmc) \tablenotemark{f}\dotfill       &  \hatcurISOrho{54}    & \hatcurISOrho{55} & \hatcurISOrho{56} & \hatcurISOrho{57} & \hatcurISOrho{58} & YY+Light curves+ZASPE         \\
~~~~$\lstar$ ($\lsun$)\dotfill      &  \hatcurISOlum{54}     & \hatcurISOlum{55} & \hatcurISOlum{56} & \hatcurISOlum{57} & \hatcurISOlum{58} & Joint fit         \\
%~~~~$M_V$ (mag)\dotfill             &  \hatcurISOmv{54}      & \hatcurISOmv{55} & \hatcurISOmv{56} & \hatcurISOmv{57} & \hatcurISOmv{58} & YY+$\rhostar$+ZASPE         \\
%~~~~$M_K$ (mag,\hatcurjhkfilset{54})\dotfill &  \hatcurISOMK{54} & \hatcurISOMK{55} & \hatcurISOMK{56} & \hatcurISOMK{57} & \hatcurISOMK{58} & YY+$\rhostar$+ZASPE         \\
~~~~Age (Gyr)\dotfill               &  \hatcurISOage{54}     & \hatcurISOage{55} & \hatcurISOage{56} & \hatcurISOage{57} & \hatcurISOage{58} & Joint fit         \\
~~~~$A_{V}$ (mag)\dotfill               &  \hatcurXAv{54}     & \hatcurXAv{55} & \hatcurXAv{56} & \hatcurXAv{57} & \hatcurXAv{58} & Joint fit         \\
~~~~Distance (pc)\dotfill           &  \hatcurXdistred{54}\phn  & \hatcurXdistred{55} & \hatcurXdistred{56} & \hatcurXdistred{57} & \hatcurXdistred{58} & Joint fit\\
\enddata
\tablecomments{
The adopted parameters for all five systems are from a model in which the orbit is assumed to be circular. For HATS-58, all the values refer to the brightest of the components of the two-component stellar system (HATS-58A) --- note 
all the photometry but that of Gaia is blended for this star. See the discussion in Section~\ref{sec:globmod}.
}
\tablenotetext{a}{
    ZASPE = Zonal Atmospherical Stellar Parameter Estimator routine
    for the analysis of high-resolution spectra
    \citep{brahm:2017:zaspe}, applied to the FEROS spectra of each system. These parameters rely primarily on ZASPE, but have a small
    dependence also on the iterative analysis incorporating the
    isochrone search and global modeling of the data.
}
\tablenotetext{b}{
    The listed $\gamma_{\rm RV}$ is from FEROS for \hatcur{54},
    \hatcur{56}, \hatcur{57} and HATS-58. For \hatcur{55} it is
    from HARPS. The error on $\gamma_{\rm RV}$ is determined from the
    orbital fit to the RV measurements, and does not include the
    systematic uncertainty in transforming the velocities to the IAU
    standard system. The velocities have not been corrected for
    gravitational redshifts.
} \tablenotetext{c}{
  For \hatcur{56} the RVs show a significant quadratic trend in addition to the Keplerian orbital variation due to the transiting planet \hatcurb{56} (Fig.~\ref{fig:rvbiscontd}). This trend is modelled as ${\rm RV}(t) = \gamma_{\rm RV} + \dot{\gamma}_{\rm RV}(t - T_{0}) + \ddot{\gamma}_{\rm RV}(t - T_{0})^2$ where $T_{0} = \hatcurLCTA{56}$ is the center time of the first transit observed in the HATSouth light curve.
} \tablenotetext{d}{
    From APASS DR6 for as
    listed in the UCAC 4 catalog \citep{zacharias:2012:ucac4}.  
}
\tablenotetext{e}{
  Obtained through the joint fit detailed in \cite{hartman:2018} and briefly summarized in Section \ref{sec:globmod}.
}
%\tablenotetext{f}{
%%
%    In the case of $\rhostar$ we list two values. The first value is
%    determined from the global fit to the light curves and RV data,
%    without imposing a constraint that the parameters match the
%    stellar evolution models. The second value results from
%    restricting the posterior distribution to combinations of
%    $\rhostar$+$\teffstar$+$\feh$ that match to a \hatcurisoshort{54}
%    stellar model.
%%
%}
\ifthenelse{\boolean{emulateapj}}{
    \end{deluxetable*}
}{
    \end{deluxetable}
}
%% --------------------------------------------------------------------

% =====================================================================
\subsection{Excluding blend scenarios}
\label{sec:blend}

In order to exclude blend scenarios, we carried out an analysis
following \citet{hartman:2012:hat39hat41} and the updates to the procedure 
outlined in \citet{hartman:2018} which allows us to account for 
the information in Gaia DR2 together with all the available photometric and spectroscopic 
data presented in previous sections. We attempt to model the
available photometric data (including light curves and catalog
broad-band photometric measurements) for each object as (1) a hierarchical triple star 
system where the two fainter stars form an eclipsing binary, (2) a blend between
a bright foreground star and a fainter background
eclipsing binary star system, and (3) a bright star with a transiting planet and a 
fainter unresolved stellar companion. The possibilities are then rejected based on that data, 
or based on the radial-velocities and bisector span variations they would imply. 
We constrain the physical properties of the stars in these systems 
using the PARSEC stellar evolutionary models \citep{Marigo:2017} along 
with the MWDUST 3D Galactic extinction model \citep{Bovy:2016}, which is used in order to 
place priors on the extinction coefficient $A_V$. The results for each system are as follows:

\begin{itemize}
\item {\em \hatcur{54}} -- the best-fit blend model, which corresponds to the blend between a bright foreground star and a fainter background 
eclipsing binary system, has a slightly higher $\chi^2$ than the best-fit model of a single star with a planet based solely on the photometry 
($\Delta \chi^2 = 4.7$). However, simulated bisector span and radial-velocity observations for blend models that come close to matching the 
photometry cannot reproduce the observed bisector span and radial-velocity measurements.
\item {\em \hatcur{55}} -- all blend models can be rejected in favor of a model of a single star with a planet based solely on the photometry.
\item {\em \hatcur{56}} -- the best-fit blend model, which corresponds to the blend between a bright foreground star and a fainter background 
eclipsing binary system, has a slightly higher $\chi^2$ than the best-fit model of a single star with a planet based solely on the photometry 
($\Delta \chi^2 = 13.6$). However, as with HATS-54, simulated bisector span and radial-velocity observations for blend models that come close to 
matching the  photometry cannot reproduce the observed bisector span and radial-velocity measurements. In particular, the simulated bisector 
spans show scatters in excess of 100 m/s which we don't observe in our data.
\item {\em \hatcur{57}} -- all blend models can be rejected in favor of a model of a single star with a planet based solely on the photometry.
\item {\em \hatcur{58}} -- The blend analysis in this case was special as all of our data but the Gaia DR2 photometry is blended with the 
companion star HATS-58B. The blend analysis is performed assuming the two sources are a binary and trying each as a potential object that 
either hosts a planet, or is blended with an eclipsing binary. The blend models in which HATS-58A is the blending source are ruled out using 
the photometry alone. The blending model in which HATS-58B is a hierarchical triple star system, however, cannot be ruled out using only 
the photometry. However, this can be rejected based on simulated radial velocities implied by such a system. To perform these simulations, 
we selected a random subset of the links from an MCMC modelling of this scenario and calculated simulated radial-velocities and simulated 
bisector span variations for each scenario. We found the simulated radial-velocities have amplitudes larger than about 2 km/s and the simulated 
bisector span variations have a scater larger than 400 m/s, both of which are inconsistent with our observations. The blending model 
in which HATS-58B is a blend between a bright foreground star and a fainter background eclipsing binary system has actually a lower 
chi-square than the model in which HATS-58A hosts a transiting exoplanet ($\Delta \chi^2 =  -38.6$). However, this scenario can also be 
rejected when the implied radial-velocities and bisector spans for this scenario are compared to our data: they imply radial-velocity 
amplitudes in excess of 1 km/s and bisector span variations with scatters larger than about 700 m/s, both of which are inconsistent with 
our observations. Based solely on the photometry, we cannot differentiate between the scenarios in which either HATS-58A or HATS-58B hosts 
the transiting exoplanet. However, the clean orbital variation measured with HARPS suggests HATS-58A is the star hosting the exoplanet, and 
is the model we select for this system.
\end{itemize}

As is generally the case, we cannot rule out in all of the above detailed cases wether there are additional unresolved faint foreground 
and/or physically associated stars contaminating our measurements. We can, however, put limits to the masses of possible companions stars: 
based on our analysis we place $95\%$ confidence upper limits on the masses of any unresolved stellar companions of $0.28\,\msun$ for HATS-54, 
$0.15\,\msun$ for HATS-55 and $0.41\,\msun$ for HATS-57. For HATS-56, if the faint detected Gaia source is indeed physically bound to it, 
it would have a mass of $0.8058 \pm 0.0076\,\msun$.

% =====================================================================
\subsection{Global modeling of the data}
\label{sec:globmod}

The global modelling of the photometric and RV data was made following the method 
recently introduced in detail in \citet{hartman:2018}, which simultaneously models 
the lightcurves, radial velocities, atmospheric parameters (effective temperature 
and metallicity), the Gaia DR2 parallax and Gaia broad band photometry. 
Lightcurves are modeled using the \citet{mandel:2002}. Radial velocity modelling 
assumes Keplerian orbits, and stellar parameters and parallax are modeled using the 
PARSEC stellar evolution models \citep{Marigo:2017}. A Differential Evolution Markov Chain Monte Carlo (MCMC) procedure 
was used to explore the parameter space and obtain the posterior distributions for our systems. 
This same procedure was applied to all of our targets except for the HATS-58 system, for which a blended object 
(in all of our observations and in non-Gaia broadband photometric measurements) is detected in Gaia 
DR2 at $0.74239\pm 0.00032$ arcseconds from the target. This latter pair of 
blended stars, in turn, have common proper motions and consistent parallaxes which indicate that 
they form a bound system. We model both stars simultaneously in our fits, and do not consider their Gaia BP and RP 
measurements as they are unreliable.

\begin{figure}[!ht]
\plotone{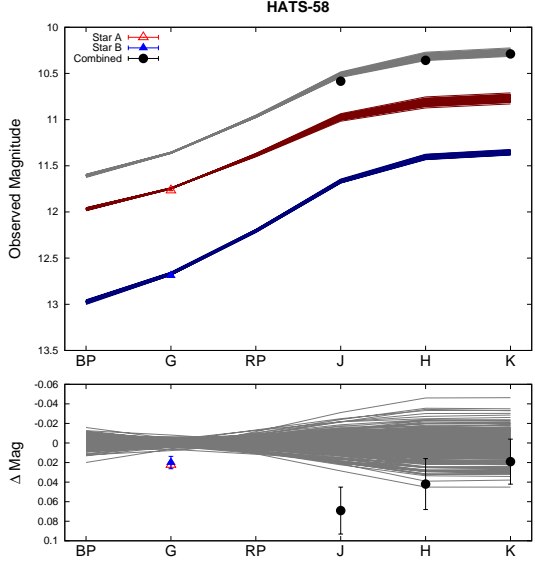}
\caption{
    Same as Figure~\ref{fig:iso}, here for the HATS-58 system. In this case, however, we show the fits for both stellar components (HATS-58A;red and HATS-58B; blue lines), 
    which are blended in the J, H and K 2MASS photometry (black dots), but resolved in Gaia's G band (red and blue triangles).
}
\label{fig:sedcontd}
\end{figure}

Fits using both circular and eccentric models were tried for all of our systems, and the method of \citet{weinberg:2013} 
was used to estimate the Bayesian evidence for each scenario. In all cases the eccentricity is consistent 
with zero. The resulting parameters for each system are listed in \reftabl{planetparam}; the photometric fits 
are shown in Figure \ref{fig:hatsouth} for the HATSouth discovery photometry, Figures \ref{fig:lc:50} through 
\ref{fig:lc:54} for the follow-up lightcurves, Figures \ref{fig:rvbis} and \ref{fig:rvbiscontd} for the RVs, 
\reffigl{iso} and \reffigl{isocontd} show the stellar evolutionary tracks in the Gaia BP-RP vs absolute G magnitude H-R diagram for 
all stars except HATS-58(A), where the same tracks are shown in the effective temperature 
absolute $G$ magnitude plane and, finally, \reffigl{sed} and \reffigl{sedcontd} show the broad-band spectral 
energy distribution (SED) fits to the observed bands, with the latter figure showing the one corresponding to both 
stellar components of the HATS-58 system, HATS-58A and HATS-58B.

For the HATS-58 system we adopt the parameters determined through the blend analysis described in Section 3.2. This analysis 
makes use of the JKTEBOP detached eclipsing binary light curve model \citep{southworth:2004a,southworth:2004b,popper:1981,etzel:1981,nelson:1972} 
in place of the \citet{mandel:2002} transit models. We also treat the stellar masses (for both the planet host and its binary star companion) 
and the system age as jump parameters in this analysis, rather than the inverse half duration of the transit and the stellar effective temperature.  

\ifthenelse{\boolean{emulateapj}}{
    \begin{figure*}[!ht]
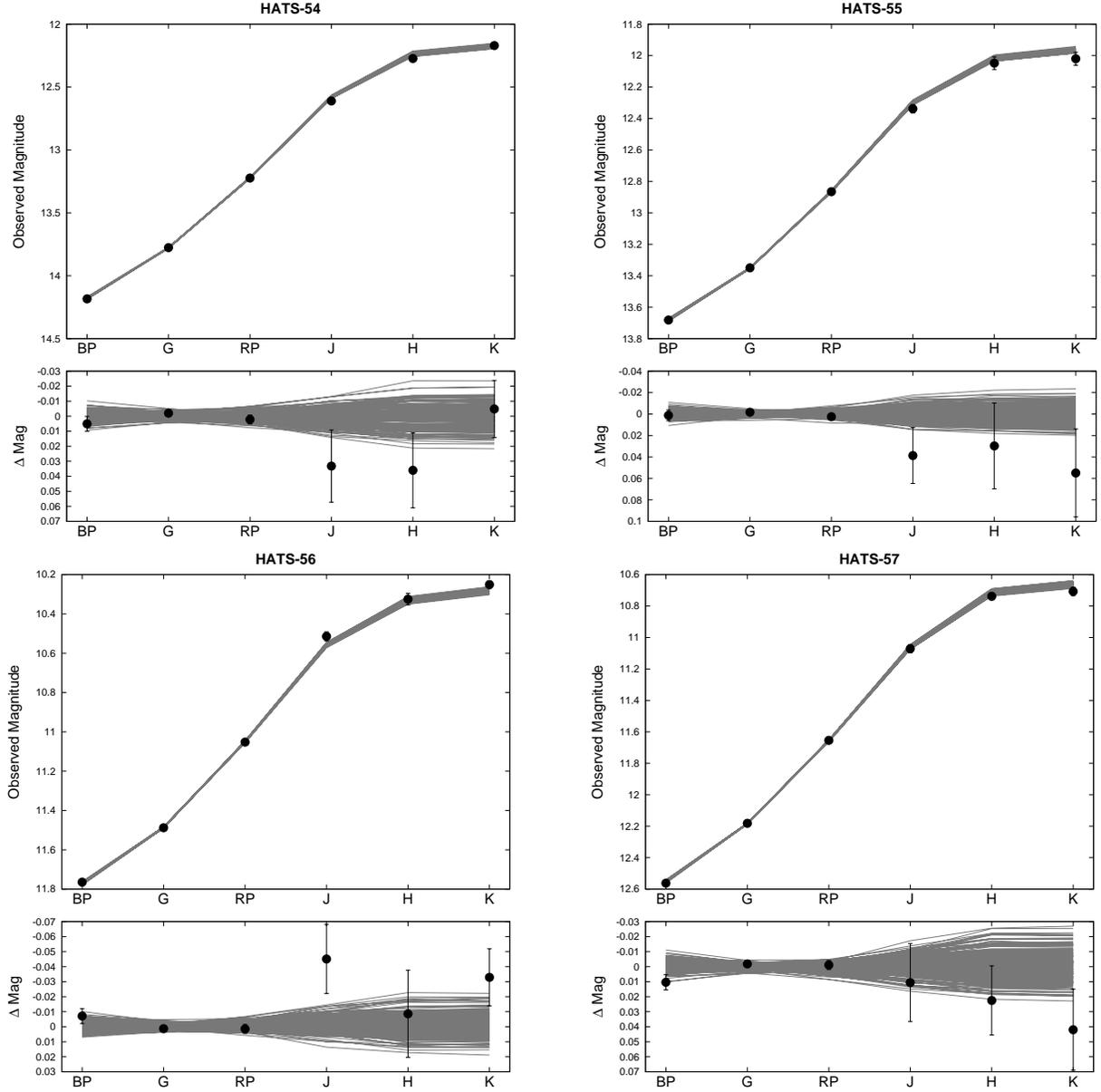

}{
    \begin{figure}[!ht]
}
\plottwo{\hatcurhtr{54}-SED.eps}{\hatcurhtr{55}-SED.eps}
\plottwo{\hatcurhtr{56}-SED.eps}{\hatcurhtr{57}-SED.eps}
\caption{
    Best-fit SED posterior samples from our joint modelling (grey lines) for Gaia's BP, G and RP bands and 2MASS J, H and K bands (black points) for 
    \hatcur{54} (upper left), \hatcur{55} (upper right), \hatcur{56} (bottom left), and \hatcur{57} (bottom right). The one for the HATS-58 system is shown 
    in Figure~\ref{fig:sedcontd}. 
}
\label{fig:sed}
\ifthenelse{\boolean{emulateapj}}{
    \end{figure*}
}{
    \end{figure}
}

As can be seen, HATS-54b, HATS-55b and HATS-58Ab are very similar in terms of densities, being consistent 
with being typical hot-Jupiters. On the other hand, HATS-56b is highly inflated and has a very low density 
of only $\hatcurPPrho{56}$ g cm$^{-3}$, while HATS-57b is massive. We discuss the retrieved parameters of the systems in the next section.

\startlongtable
%
% ---------------------------------------------------------------------
\ifthenelse{\boolean{emulateapj}}{
    \begin{deluxetable*}{lccccc}
}{
    \begin{deluxetable}{lccccc}
}
%\tablewidth{0pc}
\tabletypesize{\tiny}
\tablecaption{Orbital and planetary parameters for \hatcurb{54}--\hatcurb{58}\label{tab:planetparam}}
\tablehead{
    \multicolumn{1}{c}{} &
    \multicolumn{1}{c}{\bf HATS-54b} &
    \multicolumn{1}{c}{\bf HATS-55b} &
    \multicolumn{1}{c}{\bf HATS-56b} &
    \multicolumn{1}{c}{\bf HATS-57b} &
    \multicolumn{1}{c}{\bf HATS-58Ab} \\ 
    \multicolumn{1}{c}{~~~~~~~~~~~~~~~Parameter~~~~~~~~~~~~~~~} &
    \multicolumn{1}{c}{Value} &
    \multicolumn{1}{c}{Value} &
    \multicolumn{1}{c}{Value} &
    \multicolumn{1}{c}{Value} &
    \multicolumn{1}{c}{Value}
}
\startdata
\noalign{\vskip -3pt}
\sidehead{\Lc{} parameters}
~~~$P$ (days)             \dotfill    & $\hatcurLCP{54}$ & $\hatcurLCP{55}$ & $\hatcurLCP{56}$ & $\hatcurLCP{57}$ & $\hatcurLCP{58}$ \\
~~~$T_c$ (${\rm BJD}$)    
      \tablenotemark{a}   \dotfill    & $\hatcurLCT{54}$ & $\hatcurLCT{55}$ & $\hatcurLCT{56}$ & $\hatcurLCT{57}$ & $\hatcurLCT{58}$ \\
~~~$T_{14}$ (days)
      \tablenotemark{a}   \dotfill    & $\hatcurLCdur{54}$ & $\hatcurLCdur{55}$ & $\hatcurLCdur{56}$ & $\hatcurLCdur{57}$ & $\hatcurLCdur{58}$ \\
~~~$T_{12} = T_{34}$ (days)
      \tablenotemark{a}   \dotfill    & $\hatcurLCingdur{54}$ & $\hatcurLCingdur{55}$ & $\hatcurLCingdur{56}$ & $\hatcurLCingdur{57}$ & $\hatcurLCingdur{58}$ \\
~~~$\arstar$              \dotfill    & $\hatcurPPar{54}$ & $\hatcurPPar{55}$ & $\hatcurPPar{56}$ & $\hatcurPPar{57}$ & $\hatcurPPar{58}$ \\
~~~$\zrstar$ \tablenotemark{b}             \dotfill    & $\hatcurLCzeta{54}$\phn & $\hatcurLCzeta{55}$\phn & $\hatcurLCzeta{56}$\phn & $\hatcurLCzeta{57}$\phn & $\hatcurLCzeta{58}$\phn \\
~~~$\rpl/\rstar$          \dotfill    & $\hatcurLCrprstar{54}$ & $\hatcurLCrprstar{55}$ & $\hatcurLCrprstar{56}$ & $\hatcurLCrprstar{57}$ & $\hatcurLCrprstar{58}$ \\
~~~$b^2$                  \dotfill    & $\hatcurLCbsq{54}$ & $\hatcurLCbsq{55}$ & $\hatcurLCbsq{56}$ & $\hatcurLCbsq{57}$ & $\hatcurLCbsq{58}$ \\
~~~$b \equiv a \cos i/\rstar$
                          \dotfill    & $\hatcurLCimp{54}$ & $\hatcurLCimp{55}$ & $\hatcurLCimp{56}$ & $\hatcurLCimp{57}$ & $\hatcurLCimp{58}$ \\
~~~$i$ (deg)              \dotfill    & $\hatcurPPi{54}$\phn & $\hatcurPPi{55}$\phn & $\hatcurPPi{56}$\phn & $\hatcurPPi{57}$\phn & $\hatcurPPi{58}$\phn \\
% If there is a periastron passage
%~~~$T_{peri}$ (days)      \dotfill   & $\hatcurPPperi{54}$           \\
%
\sidehead{HATSouth dilution factors \tablenotemark{d}}
~~~Dilution factor 1 \dotfill & \hatcurLCiblend{54} & \hatcurLCiblend{55} & $\hatcurLCiblendA{56}$ & \hatcurLCiblend{57} & $\cdots$ \\
~~~Dilution factor 2 \dotfill & $\cdots$ & $\cdots$ & $\hatcurLCiblendB{56}$ & $\cdots$ & $\cdots$ \\
\sidehead{Limb-darkening coefficients \tablenotemark{e}}
%% TODO: comment/uncomment LD terms that were used
~~~$c_1,g$                  \dotfill    & $\hatcurLBig{54}$ & $\cdots$ & $\cdots$ & $\cdots$ & $\cdots$ \\
~~~$c_2,g$                  \dotfill    & $\hatcurLBiig{54}$ & $\cdots$ & $\cdots$ & $\cdots$ & $\cdots$ \\
~~~$c_1,r$                  \dotfill    & $\hatcurLBir{54}$ & $\hatcurLBir{55}$ & $\hatcurLBir{56}$ & $\hatcurLBir{57}$ & $\hatcurLBir{58}$ \\
~~~$c_2,r$                  \dotfill    & $\hatcurLBiir{54}$ & $\hatcurLBiir{55}$ & $\hatcurLBiir{56}$ & $\hatcurLBiir{57}$ & $\hatcurLBiir{58}$ \\
~~~$c_1,R$                  \dotfill    & $\hatcurLBiR{54}$ & $\hatcurLBiR{55}$ & $\hatcurLBiR{56}$ & $\cdots$ & $\hatcurLBiR{58}$ \\
~~~$c_2,R$                  \dotfill    & $\hatcurLBiiR{54}$ & $\hatcurLBiiR{55}$ & $\hatcurLBiiR{56}$ & $\cdots$ & $\hatcurLBiiR{58}$ \\
~~~$c_1,i$                  \dotfill    & $\hatcurLBii{54}$ & $\hatcurLBii{55}$ & $\hatcurLBii{56}$ & $\hatcurLBii{57}$ & $\hatcurLBii{58}$ \\
~~~$c_2,i$                  \dotfill    & $\hatcurLBiii{54}$ & $\hatcurLBiii{55}$ & $\hatcurLBiii{56}$ & $\hatcurLBiii{57}$ & $\hatcurLBiii{58}$ \\
\sidehead{RV parameters}
~~~$K$ (\ms)              \dotfill    & $\hatcurRVK{54}$\phn\phn & $\hatcurRVK{55}$\phn\phn & $\hatcurRVK{56}$\phn\phn & $\hatcurRVK{57}$\phn\phn & $\hatcurRVK{58}$\phn\phn \\
% Include if linear drift was also fitted.
%~~~$G_1$ (\ms/day)       \dotfill    & $\hatcurRVlindrift{54}$       \\
%~~~$k_{\rm RV}$\tablenotemark{c} 
%                          \dotfill    & $\hatcurRVk{54}$\phs & $\hatcurRVk{55}$\phs \\
%~~~$h_{\rm RV}$\tablenotemark{c}
%                          \dotfill    & $\hatcurRVkeccen{56}$\phs \\
%~~~$h_{\rm RV}$\tablenotemark{c}
%                          \dotfill    & $\hatcurRVh{54}$ & $\hatcurRVh{55}$ \\
%
% Only if corrected k and h values were given based on isochrones
% and independent luminosity indicator, such as parallax.
%~~~$k_C$\tablenotemark{d} \dotfill   & $\hatcurRVheccen{56}$ \\
%
% Only if corrected k and h values were given based on isochrones
% and independent luminosity indicator, such as parallax.
%~~~$k_C$\tablenotemark{d} \dotfill   & $\hatcurRVkcorr{54}$          \\
%~~~$h_C$\tablenotemark{d} \dotfill   & $\hatcurRVhcorr{54}$          \\
~~~$e$ \tablenotemark{f}               \dotfill    & $\hatcurRVeccentwosiglimeccen{54}$ & $\hatcurRVeccentwosiglimeccen{55}$ & $\hatcurRVeccentwosiglimeccen{56}$ & $\hatcurRVeccentwosiglimeccen{57}$ & $\hatcurRVeccentwosiglimeccen{58}$ \\
%~~~$\omega$ (deg) \dotfill    & $\cdots$ & $\cdots$ & $\hatcurRVomegaeccen{56}$ & $\cdots$ \\
%~~~$\sqrt{e}\cos\omega$               \dotfill    & $\cdots$ & $\cdots$ & $\hatcurRVrkeccen{56}$ & $\cdots$ \\
%~~~$\sqrt{e}\sin\omega$               \dotfill    & $\cdots$ & $\cdots$ & $\hatcurRVrheccen{56}$ & $\cdots$ \\
%~~~$e\cos\omega$               \dotfill    & $\cdots$ & $\cdots$ & $\cdots$ & $\cdots$ \\
%~~~$e\sin\omega$               \dotfill    & $\cdots$ & $\cdots$ & $\cdots$ & $\cdots$ \\
%
~~~RV jitter FEROS (\ms) \tablenotemark{g}       \dotfill    & $\hatcurRVjitterA{54}$ & $\cdots$ & $\hatcurRVjitterA{56}$ & \hatcurRVjitter{57} & \hatcurRVjitterA{58} \\
~~~RV jitter HARPS (\ms)        \dotfill    & $\hatcurRVjittertwosiglimB{54}$ & $\hatcurRVjittertwosiglim{55}$ & \hatcurRVjittertwosiglimB{56} & $\cdots$ & \hatcurRVjittertwosiglimB{58} \\
\sidehead{Planetary parameters}
~~~$\mpl$ ($\mjup$)       \dotfill    & $\hatcurPPmlong{54}$ & $\hatcurPPmlong{55}$ & $\hatcurPPmlong{56}$ & $\hatcurPPmlong{57}$ & $\hatcurPPmlong{58}$ \\
~~~$\rpl$ ($\rjup$)       \dotfill    & $\hatcurPPrlong{54}$ & $\hatcurPPrlong{55}$ & $\hatcurPPrlong{56}$ & $\hatcurPPrlong{57}$ & $\hatcurPPrlong{58}$ \\
~~~$C(\mpl,\rpl)$
    \tablenotemark{h}     \dotfill    & $\hatcurPPmrcorr{54}$ & $\hatcurPPmrcorr{55}$ & $\hatcurPPmrcorr{56}$ & $\hatcurPPmrcorr{57}$ & $\cdots$ \\
~~~$\rhopl$ (\gcmc)       \dotfill    & $\hatcurPPrho{54}$ & $\hatcurPPrho{55}$ & $\hatcurPPrho{56}$ & $\hatcurPPrho{57}$ & $\hatcurPPrho{58}$ \\
~~~$\log g_p$ (cgs)       \dotfill    & $\hatcurPPlogg{54}$ & $\hatcurPPlogg{55}$ & $\hatcurPPlogg{56}$ & $\hatcurPPlogg{57}$ & $\hatcurPPlogg{58}$ \\
~~~$a$ (AU)               \dotfill    & $\hatcurPParel{54}$ & $\hatcurPParel{55}$ & $\hatcurPParel{56}$ & $\hatcurPParel{57}$ & $\hatcurPParel{58}$ \\
~~~$T_{\rm eq}$ (K)        \dotfill   & $\hatcurPPteff{54}$ & $\hatcurPPteff{55}$ & $\hatcurPPteff{56}$ & $\hatcurPPteff{57}$ & $\hatcurPPteff{58}$ \\
~~~$\Theta$ \tablenotemark{i} \dotfill & $\hatcurPPtheta{54}$ & $\hatcurPPtheta{55}$ & $\hatcurPPtheta{56}$ & $\hatcurPPtheta{57}$ & $\hatcurPPtheta{58}$ \\
%
% These are given if the orbit is eccentric
%~~~$F_{per}$ (\ergscmsq) \tablenotemark{e}
%                          \dotfill    & $\hatcurPPfluxperi{54}$      \\
%~~~$F_{ap}$  (\ergscmsq) \tablenotemark{e} 
%                          \dotfill    & $\hatcurPPfluxap{54}$        \\
~~~$\log_{10}\langle F \rangle$ (cgs) \tablenotemark{j}
                          \dotfill    & $\hatcurPPfluxavglog{54}$ & $\hatcurPPfluxavglog{55}$ & $\hatcurPPfluxavglog{56}$ & $\hatcurPPfluxavglog{57}$ & $\hatcurPPfluxavglog{58}$ \\
\enddata
\tablecomments{
For all five systems we adopt a model in which the orbit is assumed to be circular. See the discussion in Section~\ref{sec:globmod}.
}
\tablenotetext{a}{
    Times are in Barycentric Julian Date calculated directly from UTC {\em without} correction for leap seconds.
    \ensuremath{T_c}: Reference epoch of
    mid transit that minimizes the correlation with the orbital
    period.
    \ensuremath{T_{12}}: total transit duration, time
    between first to last contact;
    \ensuremath{T_{12}=T_{34}}: ingress/egress time, time between first
    and second, or third and fourth contact.
}
\tablenotetext{b}{
   Reciprocal of the half duration of the transit used as a jump parameter in our MCMC analysis in place of $\arstar$. It is related to $\arstar$ by the expression $\zrstar = \arstar(2\pi(1+e\sin\omega))/(P\sqrt{1-b^2}\sqrt{1-e^2})$ \citep{bakos:2010:hat11}.
}
\tablenotetext{d}{
    Scaling factor applied to the model transit that is fit to the HATSouth light curves. This factor accounts for dilution of the transit due to blending from neighboring stars and over-filtering of the light curve.  These factors are varied in the fit, with independent values adopted for each HATSouth light curve. The factors listed \hatcur{54}, \hatcur{55}, \hatcur{57} and HATS-58 are for the G700.3, G602.4, G548.3, and G699.1 light curves, respectively. For \hatcur{56} we list the factors for the G698.1 and G698.4 light curves in order.
}
\tablenotetext{e}{
    Values for a quadratic law, adopted from the tabulations by
    \cite{claret:2004} according to the spectroscopic (ZASPE) parameters
    listed in \reftabl{stellar}.
}
\tablenotetext{f}{
    The 95\% confidence upper limit on the eccentricity determined
    when $\sqrt{e}\cos\omega$ and $\sqrt{e}\sin\omega$ are allowed to
    vary in the fit.
}
\tablenotetext{g}{
    Term added in quadrature to the formal RV uncertainties for each
    instrument. This is treated as a free parameter in the fitting
    routine. In cases where the jitter is consistent with zero, we
    list its 95\% confidence upper limit.
}
\tablenotetext{h}{
    Correlation coefficient between the planetary mass \mpl\ and radius
    \rpl\ estimated from the posterior parameter distribution.
}
\tablenotetext{i}{
    The Safronov number is given by $\Theta = \frac{1}{2}(V_{\rm
    esc}/V_{\rm orb})^2 = (a/\rpl)(\mpl / \mstar )$
    \citep[see][]{hansen:2007}.
}
\tablenotetext{j}{
    Incoming flux per unit surface area, averaged over the orbit.
}
\ifthenelse{\boolean{emulateapj}}{
    \end{deluxetable*}
}{
    \end{deluxetable}
}

% ---------------------------------------------------------------------

%% EOF Analysis

% #####################################################################
%% Discussion
\section{Discussion}
\label{sec:dis}

\begin{figure*}
\epsscale{1.15}
\plottwo{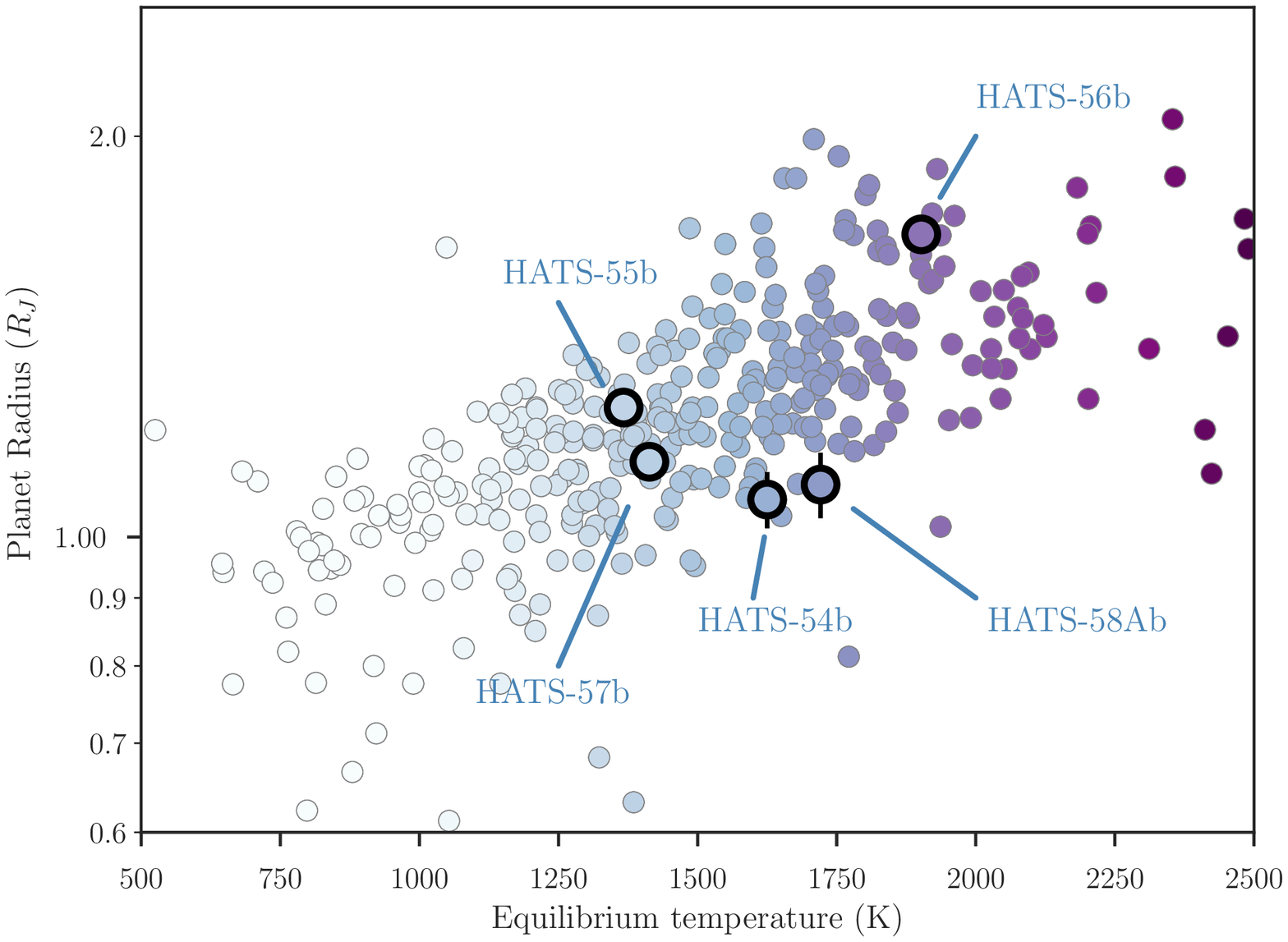}{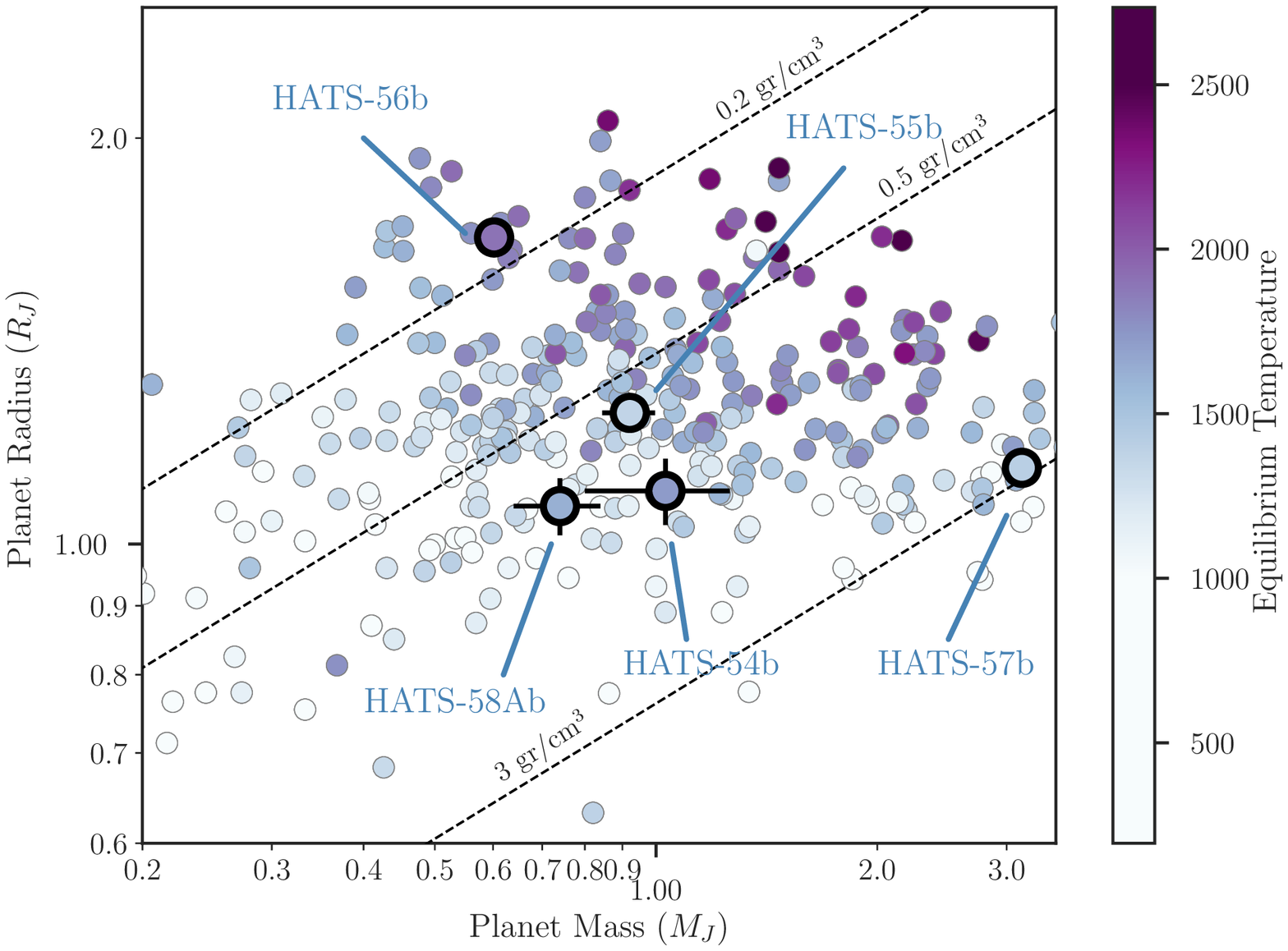}
\caption{
    Equilibrium temperature-radius and mass-radius diagrams of known exoplanets obtained from TEPcat \citep{tepcat}. Colored points with 
    errorbars indicate HATS-54b to HATS-58Ab, with colors indicating the planetary effective temperature of our newly 
    discovered transiting exoplanets. The color is consistent between both diagrams.
}
\label{fig:mrd}
\end{figure*}

Figure \ref{fig:mrd} puts our newly discovered exoplanets in the context of known and well-studied exoplanets (with radii and masses 
estimated to better than 20\%) in both the equilibrium temperature/radius and the mass/radius diagrams. As can be observed, the parameters of HATS-54b, HATS-55b 
and HATS-58Ab make them consistent with being part of the well-represented population of inflated hot-Jupiters, with HATS-54b and 
HATS-58Ab falling in terms of equilibrium temperature on the interesting regime of maximum heating efficiency for inflation 
proposed by \cite{TF:2018}. In addition, as discussed in Section \ref{sec:gaiadr2}, both HATS-56 and HATS-58 are most likely 
systems composed of at least two stars. On one hand, given the separation observed by Gaia DR2 between HATS-55 and the 
companion of $3.80336''\pm 0.00038$, and the calculated distance to the system of $623.6\pm6.2$ pc, the projected separation of the stars assuming they are bound is $2361\pm 23$ AU. On the other hand, given the separation observed by Gaia DR2 between 
HATS-58A and HATS-58B of $0.74238'' \pm 0.00033$ and the calculated distance to the system of $492 \pm 21$ pc, the projected 
separation of the stars assuming they are bound is $365\pm 15$ AU. 

HATS-57b, on the other hand, is a dense ($\hatcurPPrho{57}$ gr cm$^{-3}$) and quite massive hot-Jupiter which seems to 
fall within the expected size given its equilibrium temperature, specially 
if one considers that inflation is slightly less pronounced for more massive planets \citep{Sestovic:2018}. The planet's radius 
and mass are consistent with the models of \cite{TF:2018} for HATS-57b's equilibrium temperature of $\hatcurPPteff{57}$ K, suggesting 
that the inflation mechanism is indeed operating in HATS-57b just like in every other hot-Jupiter with a similar equilibrium temperature. 
Interestingly, the expected amplitude of the Rossiter-Mclaughlin (RM) effect on this system is of order $v\sin i (R_p/R_s)^2\sim 60$ m s$^{-1}$; 
this is about one half of the total observed uncertainties on the RVs observed in our high-precision RV follow-up and thus this could 
be a good system to characterize with this effect. The system is particularly interesting because according to the derived stellar period 
in Section \ref{sec:stelparam}, the star show hints of being slightly misaligned with 
respect to the plane of the sky ($22.9^{+10.5}_{-10.6}$ degrees). Given the nearly 
edge-on inclination of the planetary system with respect to the plane of the sky ($i=87.88\pm0.40$ degrees), this hints that this may be 
a misaligned system, a hypothesis which can be tested with RM measurements.

Finally, HATS-56b is highly inflated and possesses a very low density of $\hatcurPPrho{56}$ gr cm$^{-3}$. Its inflated nature 
is, however, not rare given its relatively large equilibrium temperature of $\hatcurPPteff{56}$ K, which in turn makes it a very good 
candidate for future atmospheric follow-up, especially given the brightness of the host star ($V=11.6$). The expected atmospheric scale-height 
for HATS-56b is around 1100 km, which in turn implies an expected signal in transmission between 120-360 ppm, around 70\% the expected transmission 
signal for HD 209458b. An additional very interesting feature of this hot-Jupiter is that it shows a 
significant quadratic trend in its radial velocities (see Figure \ref{fig:rvbiscontd}), that could imply an additional companion. 
In order to see what this latter interpretation would mean if it actually were another planet around HATS-56, we used \texttt{juliet} 
(Espinoza, Kossakowski \& Brahm, in prep.; code available via GitHub\footnote{\url{https://github.com/nespinoza/juliet}}), a tool that allows not only 
to fit multi-planetary systems but to estimate the bayesian evidence, $Z$, of different models, in order to fit a 
two-planet solution to the radial-velocities. To do this, \texttt{juliet} couples \texttt{radvel} \citep{radvel} with MultiNest in order to perform 
the posterior sampling and to calculate said bayesian evidences. We used the already derived properties of HATS-56b (defined mainly by its transits) as inputs. 
We fix in our two-planet fit the eccentricity (to zero), and give as priors the posteriors on the period and time-of-transit center of HATS-56b 
presented in \reftabl{planetparam} and perform a two-planet fit to the radial-velocity data in order to explore the parameter space using wide 
priors on the parameters for the unknown second planetary properties (a Jeffreys prior for the period from 5 to 10,000 days, a time of transit center 
uniform between 2457700 and 2467700, a uniform prior for the semi-amplitude between 0 and 1000 m/s), and wide priors for the semi-amplitude of the 
known transiting planet (uniform between 0 and 100 m/s), allowing eccentric orbits for the outer planet. 

\begin{figure*}
\epsscale{1.15}
\plotone{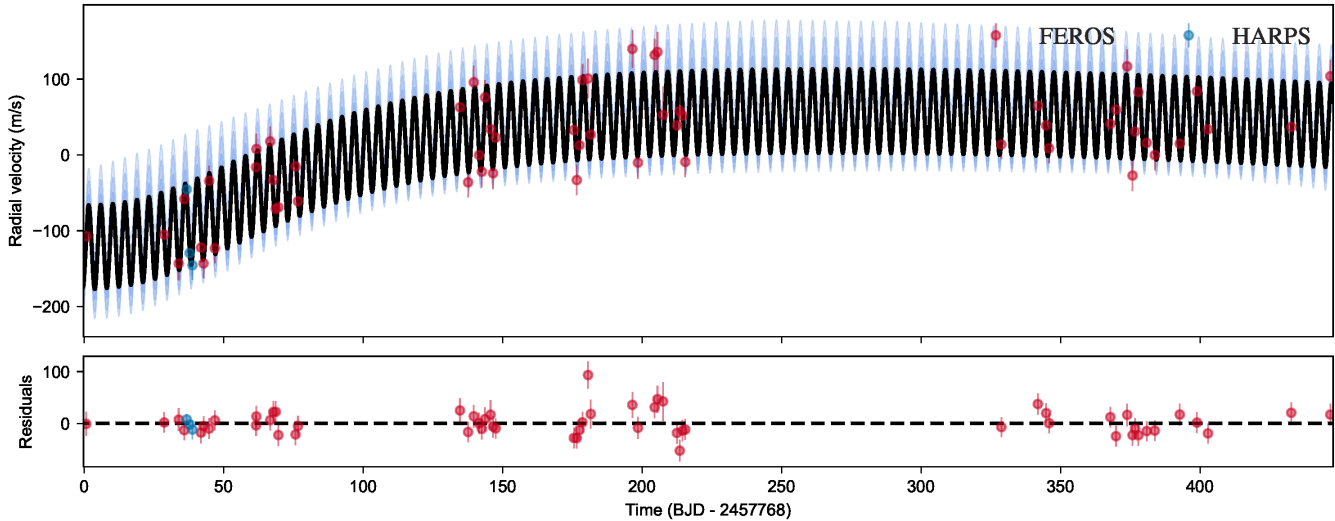}
\caption{
    Two-planet modelling of the radial velocities (red points FEROS, blue points HARPS) for HATS-56. The best-fit radial-velocity model is shown in 
solid black line with blue bands denoting the 68, 95 and 99\% credibility interval. Bottom panel shows the residuals of the fit.
}
\label{fig:hats-56c}
\end{figure*}

Figure \ref{fig:hats-56c} shows our modelling of the radial-velocity assuming a two-planet model for them. As 
expected, we recover the same semi-amplitude for HATS-56b derived in previous sections, while for the possible planet HATS-56c we obtain a highly uncertain 
period of $P_c = 815^{+253}_{-143}$ days, and a time of transit-center of $2462738^{+1624}_{-882}$ days (BJD UTC), coupled with a possible 
eccentric orbit with $e_c = 0.46\pm 0.07$ and $\omega_c = 177^{+2.3}_{-3.1}$ radians, and a semi-amplitude of $K_c = 94^{+13}_{-10}$ m s$^{-1}$. 
It is interesting to note that this model is favored over a fit with a simple quadratic trend ($\ln Z > 5$ in favor of the two Keplerians). 
These values imply a minimum mass for the possible planet c of $M_c\sin i_c = 5.11 \pm 0.94 M_J$. Perhaps the most interesting feature of the possible 
planet HATS-56c is its derived distance from the star and, hence, its equilibrium temperature. We use the very tight constraint on the stellar density for 
the star and the derived period for this possible planet to derive a value $a/R_* = 194^{+38}_{-43}$ from Kepler's third law 
($1.99\pm 0.43$ AU). Combining this with the stellar effective temperature, we obtain a zero-albedo equilibrium temperature for the possible 
planet HATS-56c of $T_\textnormal{eq}=332\pm 50$ K, which would imply a temperate companion that would fall very close to the habitable zone of the 
star. If confirmed, HATS-56c would be a very interesting system to study due to the possibility that satellites 
orbiting it could present habitable conditions in terms of the stellar irradiation.

It is interesting to note that on top of the exciting feature of the HATS-56 system being a multi-component system, Gaia DR2 data reveals an additional 
(possibly bound, stellar) companion to HATS-56. Given the observed separation of this companion from HATS-56 of $1.59879''\pm 0.00029$ and the derived 
distance to the system of $577.1 \pm 9.6$ pc, if physically bound the companion would be at least at a distance of $922\pm 15$ AU (which is inconsistent 
with the derived distance of HATS-56c). This is inconsistent with the derived range of distances that could give rise to the observed RV long-term trend 
and as such this cannot explain it given our data.

Confirmation of the candidate exoplanet HATS-56c could be performed if further radial-velocity follow-up is performed within the next $\sim 2$ years. 
The expected time of periastron passage, taking our best-fit model for the candidate, is expected to occur around mid-2020, but monitoring the decrease 
of the radial-velocity curve as it approaches this point will be very important to both predict the exact time of periastron passage (in order to increase 
the sampling of the radial-velocity follow-up) and to constrain the exact shape of the radial-velocity curve, which has useful information for constraining 
the orbit of the possible planet. Regarding possible transits, given the current uncertainty on the period and time of transit-center, catching a possible 
transit event of the candidate planet is rather difficult. We inspected the HAT-South photometry but we found it is not precise enough to provide any 
constraints on possible events; the errors on the transit parameters are so large, that it is very difficult to analyze the lightcurve, especially 
considering that we are most likely searching for only one transit. With further radial-velocity follow-up, however, this search could be made even 
in the current HAT-South lighcurves, which could be joined with other photometric surveys in order to search for the possible transit signature of 
HATS-56c. In fact, the \textit{TESS} mission \citep{tess} will observe HATS-56 during its passage through Sector 2, and this could provide a brief but interesting 
search for this extra possible signal

\section{Conclusions}
\label{sec:conclusions}

In this work we have presented the discovery of HATS-54b through HATS-58Ab. HATS-54b, HATS-55b and HATS-58Ab are typical hot Jupiters in many aspects, 
but sample the interesting effective temperature range were the maximum heating efficiency for inflation is proposed to occur \citep{TF:2018}. HATS-56b 
and HATS-57b, however, are special: the latter is dense hot jupiter which could be a good target for RM observations 
and orbits an apparently active star showing peak-to-peak variability on the order of 2\%, whereas HATS-56b is not only an excellent target for future 
atmospheric follow-up but also for future radial-velocity monitoring in order to confirm the planetary nature of an evident long-term radial velocity 
signal observed during our high resolution spectroscopic follow-up. If we assume this latter signal is actually from an additional planet in the system, 
this would be a super-Jupiter with a minimum mass of $M_c\sin i_c = 5.11 \pm 0.94 M_J$, and could orbit close to the habitable zone of HATS-56, which 
would be interesting in terms of habitability if there are satellites orbiting the possible planet HATS-56c. Radial-velocity monitoring of this system 
during the next two years will be very useful in order to both constrain the time of periastron passage (expected to occur on 2020) and to 
constrain the possible times of transit center of this external possible companion.
% #####################################################################
%% Acknowledgements
\acknowledgements 

Development of the HATSouth
project was funded by NSF MRI grant NSF/AST-0723074, operations have
been supported by NASA grants NNX09AB29G, NNX12AH91H, and NNX17AB61G, and follow-up
observations have received partial support from grant NSF/AST-1108686.
%%
%J.H.\ acknowledges support from NASA grant NNX14AE87G.
N.E.\ acknowledges support from the Gruber Foundation. 
A.J.\ acknowledges support from FONDECYT project 1171208, CONICYT project BASAL AFB-170002, and by the Ministry for the Economy, Development, and Tourism's Programa Iniciativa Cient\'{i}fica Milenio through grant IC\,120009, awarded to the Millennium Institute of Astrophysics (MAS).
M.R. acknowledges support from CONICYT project Basal AFB-170002.
%A.J.\ acknowledges support from FONDECYT project 1171208, BASAL CATA
%PFB-06, and project IC120009 ``Millennium Institute of Astrophysics
%(MAS)'' of the Millenium Science Initiative, Chilean Ministry of
%Economy. 
R.B.\ acknowledges support from project
IC120009 ``Millenium Institute of Astrophysics (MAS)'' of the
Millennium Science Initiative, Chilean Ministry of Economy.
%R.B. acknowledges support by the Ministry of Economy, Development, and Tourism’s Millennium Science Initiative through grant IC120009, awarded to The Millennium Institute of Astrophysics (MAS)
V.S.\ acknowledges support form BASAL CATA PFB-06.  
A.V. is supported by the NSF Graduate Research Fellowship, Grant No. DGE 1144152.
This work is based on observations made with ESO Telescopes at the La
Silla Observatory.
This paper also uses observations obtained with facilities of the Las
Cumbres Observatory Global Telescope.
%%
%Work at the Australian National University is supported by ARC Laureate
%Fellowship Grant FL0992131.
%%
We acknowledge the use of the AAVSO Photometric All-Sky Survey (APASS),
funded by the Robert Martin Ayers Sciences Fund, and the SIMBAD
database, operated at CDS, Strasbourg, France.
Operations at the MPG~2.2\,m Telescope are jointly performed by the
Max Planck Gesellschaft and the European Southern Observatory.  The
imaging system GROND has been built by the high-energy group of MPE in
collaboration with the LSW Tautenburg and ESO\@.  We thank the MPG 2.2m telescope support team for their technical assistance during observations."
This work has made use of data from the European Space Agency (ESA)
mission {\it Gaia} (\url{https://www.cosmos.esa.int/gaia}), processed by
the {\it Gaia} Data Processing and Analysis Consortium (DPAC,
\url{https://www.cosmos.esa.int/web/gaia/dpac/consortium}). Funding
for the DPAC has been provided by national institutions, in particular
the institutions participating in the {\it Gaia} Multilateral Agreement.
%%
%% EOF Acknowledgements

% #####################################################################
%% Bibliography
\clearpage
\bibliographystyle{aasjournal}
\bibliography{hatsbib}

\clearpage

\end{document}